\newif\ifarxiv \arxivtrue
\ifarxiv
\documentclass{article}
\else
\documentclass[runningheads, orivec]{llncs}
\fi

\usepackage{mysetting}
\usepackage{float}
\usepackage[hypertexnames=false]{hyperref}
\usepackage{amssymb}
\usepackage{amsfonts}
\ifarxiv\usepackage{thm-restate}
\usepackage[numbers]{natbib}
\renewcommand{\citet}{\cite}
\else
\newcommand{\citet}{\cite}
\fi

\begin{document}
  \title{Unifying Function- and Argument-First Bidirectional Type Systems\ifarxiv\thanks{%
      This is a full version of the paper that has been accepted for publication in the Proceedings of Asian Symposium on Programming Languages and Systems (APLAS 2026), Lecture Notes in Computer Science (LNCS), Springer. This version discusses the algorithmic typing in detail with its full definition,
  a heuristics of guide insertion, and possible extensions of \lang.  The definitions omitted from the short version can also be found.}\fi}

\mycomment{
  \author{Takuma Yoshioka}
  \orcid{0009-0001-5949-6288}
  \affiliation{%
    \institution{Kyoto University}
    \city{Kyoto}
    \country{Japan}
  }
  \email{yoshioka@fos.kuis.kyoto-u.ac.jp}

  \author{Taro Sekiyama}
  \orcid{0000-0001-9286-230X}
  \affiliation{%
    \institution{National Institute of Informatics \& SOKENDAI}
    \city{Tokyo}
    \country{Japan}
  }
  \email{tsekiyama@acm.org}

  \author{Atsushi Igarashi}
  \orcid{0000-0002-5143-9764}
  \affiliation{%
    \institution{Kyoto University}
    \city{Kyoto}
    \country{Japan}
  }
  \email{igarashi@kuis.kyoto-u.ac.jp}
}

\ifarxiv
\providecommand{\doi}[1]{%
    \href{https://doi.org/\detokenize{#1}}{%
      \url{https://doi.org/#1}}%
}
\author{Takuma Yoshioka \\ Kyoto University, Japan \and
  Taro Sekiyama \\   National Institute of Informatics \& SOKENDAI, Tokyo, Japan \and
  Atsushi Igarashi  \\ Kyoto University, Japan}
  \else
\author{Takuma Yoshioka\inst{1}\orcidID{0009-0001-5949-6288} \and
  Taro Sekiyama\inst{2}\orcidID{0000-0001-9286-230X} \and
  Atsushi Igarashi\inst{1}\orcidID{0000-0002-5143-9764}}

\authorrunning{T. Yoshioka et al.}

\institute{Kyoto University, Kyoto, Japan \and
  National Institute of Informatics \& SOKENDAI, Tokyo, Japan}
\fi

\maketitle

\begin{abstract}
  Bidirectional typing mixes type synthesis and type checking into a single process.
Existing bidirectional type systems can be classified into two styles based on
whether, given a function application, a bidirectional typing algorithm
synthesizes the function's type first and typechecks the argument against the synthesized argument type, or
it synthesizes the arguments' types first and typechecks the function against the synthesized arguments' types.
We call the former \emph{function-first} and the latter \emph{argument-first}.
Not only do the two styles significantly differ in how the type
systems and typing algorithms are formalized, but also they lead to
incompatible typeabilities, forcing a language designer to select one
style and to give up the other's typeabilities.

In this paper, we unify the two styles and develop \lang with a new
bidirectional type system for higher-rank polymorphism.
Key ideas of the unification are twofold.
Each function application is annotated with a bit of
information to represent whether function- or argument-first typing is
used, to allow a language designer (or even a programmer) to switch between
the two styles at their discretion.
We reformulate the function- and argument-first type systems by using
ideas from \emph{colored types} (Odersky et al., 2001) and \emph{boxy
  types} (Vytiniotis et al., 2006), which can specify which part of a
type should be synthesized or used for checking in a flexible manner.
We also develop a typing algorithm based on the worklist approach by Zhao et al.

The (declarative) type system of \lang is shown to be sound and to
subsume two representative function- and argument-first bidirectional
type systems.
Our typing algorithm is shown to be sound with respect to the type system of \lang and
complete with respect to representative function- and argument-first systems.
We mechanically prove the metatheorems using the Abella theorem prover.

\mycomment{
  \emph{Bidirectional typing} mixes the two typing procedures,
  \emph{type synthesis} and \emph{type checking}, into a single procedure.
  Given a program $\ottnt{e}$,
  type synthesis returns its type $\ottnt{A}$, and
  type checking also takes $\ottnt{A}$ as input, verifying that $\ottnt{e}$ has $\ottnt{A}$.
  Bidirectional typing has been used to implement advanced typing features, including
  dependent types, higher-rank polymorphism, and type error localization.

  \mycomment{
    \emph{Bidirectional typing} mixes the two typing processes:
    \emph{type synthesis}, which takes a program as input and returns its type, and
    \emph{type checking}, which takes a program and a type as input and verifies
    that the program has the given type, into a single process.
    It has been used to implement complex typing features, such as
    dependent types, higher-rank polymorphism, and total type error localization and recovery.
  }

  \mycomment{
    \emph{Bidirectional typing} is a technique to implement type systems,
    especially the systems with complex typing features, such as
    dependent types, higher-rank polymorphism, and total type error localization and recovery.
    Its core idea is to combine the two typing processes:
    \emph{type synthesis}, which takes a program as input and returns its type, and
    \emph{type checking}, which takes a program and a type as input and verifies
    that the program has the given type, into a single process.
    Making use of already known type information in type checking
    help reduce cumbersome type annotations that typically used to
    lead the type synthesis process to determine the type of a given program.
  }

  Existing bidirectional type systems have different formalizations,
  resulting in incompatible typeabilities.
  One of the significant differences in formalization is whether, given a function application,
  a bidirectional typing procedure synthesizes the function's type first and typechecks the argument, or
  it synthesizes the argument's type first and typechecks the function.
  We refer to the former as \emph{function-first} and the latter as \emph{argument-first}.
  This difference in formalization leads to incompatible typeabilities, forcing a language designer
  to select one of the bidirectional type systems and
  to give up the others' typeabilities.
  \mycomment{
    Existing bidirectional type systems depend on distinct formalizations;
    even worse, their typeabilities are incompatible.
    This situation forces a language designer to select one of the bidirectional type systems and
    to give up the others' typeabilities.
    One of the significant incompatibilities is whether, given a function application,
    a bidirectional typing process synthesizes the function's type first and typechecks the argument, or
    it synthesizes the argument's type first and typechecks the function.
    We refer to the former as \emph{function-first} and the latter as \emph{argument-first}.
  }

  We propose a unified language \lang, its type system, and a typing algorithm for it.
  The type system of \lang subsumes
  the function- and argument-first styles of bidirectional typing using \emph{boxy types},
  which specify which part of a type should be synthesized or used for checking.
  This feature of \lang allows a language designer or a programmer
  to switch between two different typing styles at their discretion.
  Furthermore, the type system of \lang supports higher-rank polymorphism, demonstrating that
  boxy types also help to unify the function- and argument-first styles
  in a setting with a complex typing feature.

  \mycomment{
    The important metaproperties of \lang consist of two parts.
    First, the type system of \lang subsumes the existing bidirectional type systems,
    each of which is either the function- or argument-first style.
    Second, our typing algorithm for \lang does not exceed the type system of \lang
    regarding typeability and is sufficient to subsume the existing bidirectional systems.
    We mechanically prove these theorems in the Abella theorem prover.
  }

  The important metaproperties of \lang consist of
  declarative and algorithmic parts.
  On the declarative side,
  the type system of \lang subsumes the existing bidirectional type systems,
  each of which is either the function- or argument-first style.
  On the algorithmic side,
  our typing algorithm for \lang is weaker than the type system of \lang in terms of typeability.
  Still, it is sufficient to subsume the existing bidirectional type systems.
  We mechanically prove these theorems in the Abella theorem prover.

  \mycomment{
    We mechanically prove that the type system of \lang
    subsumes the existing bidirectional type systems,
    each of which is either the function- or argument-first style.
    We also prove that
    our typing algorithm for \lang does not exceed the type system of \lang
    regarding typeability and is sufficient to subsume the existing bidirectional systems.
  }

  \mycomment{
    Higher-rank polymorphism augments the typeability of a type system
    by allowing polymorphic types to appear anywhere in types.
    Bidirectional typing is one of the well-studied ways to support higher-rank polymorphism.
    However, existing bidirectional type systems depend on distinct formalizations;
    even worse, their typeabilities are incompatible.
    This situation forces a language designer to select one of the bidirectional type systems and
    to give up the others' typeabilities.
    One of the significant incompatibilities is whether, given a function application,
    a bidirectional type system synthesizes the function's type first and typechecks the argument, or
    it synthesizes the argument's type first and typechecks the function.
    We propose a unified type system that
    subsumes function- and argument-first styles of bidirectional typing, adopting boxy types.
    Our type system allows a language designer to switch between the two different typing styles
    at their discretion.
    We are working on proofs that our type system is sound (resp. complete)
    with respect to System F extended with type coercions (resp. each of the existing systems).
  }
}

\end{abstract}

\mycomment{

  \ccsdesc[500]{Do Not Use This Code~Generate the Correct Terms for Your Paper}
  \ccsdesc[300]{Do Not Use This Code~Generate the Correct Terms for Your Paper}
  \ccsdesc{Do Not Use This Code~Generate the Correct Terms for Your Paper}
  \ccsdesc[100]{Do Not Use This Code~Generate the Correct Terms for Your Paper}

  \keywords{Do, Not, Use, This, Code, Put, the, Correct, Terms, for, Your, Paper}
}

\section{Introduction}

\mycomment{
  Typing a program attempts to assign a type to every part of it, typically at compile time,
  according to a \emph{type system}'s specification.
  A \emph{well-typed} program, which means typing it succeeds,
  raises no error to some extent at runtime.
  For example, consider the function $\mathsf{add42}$ that takes the integer and
  returns the integer added $42$ to the argument.
  The function application $\mathsf{add42}\ 1$ is well-typed
  because $1$ has the type $\mathsf{int}$ as required by $\mathsf{add42}$.
  In contrast, typing the function application $\mathsf{add42}\ \texttt{"1"}$ fails
  because the string $\texttt{"1"}$ cannot have the type $\mathsf{int}$ in many typical type systems.

  The process of typing given programs can be classified into two kinds
  from the perspective of whether the type is an input or an output.
  \emph{Type checking} takes a type as an input and judges
  whether a given program has that type.
  \emph{Type synthesis}, on the other hand, returns the type of the given program.
  In broad terms, type synthesis is relatively more difficult than type checking.
  While many programming languages use type annotations to make type synthesis easier,
  especially when they have complex types,
  writing type annotations can be cumbersome for programmers.
}

\TY{
  Should we avoid mentioning
  the burden of type annotations and
  the undecidability of type inference?
  It could lead to a misunderstanding that
  this paper focuses on implementing type inference.
}

\emph{Bidirectional typing}~\cite{pierce_local_1998} mixes
\emph{type checking}, which takes a type as input and determines whether a given program has that type,
and \emph{type synthesis}, which returns the type of a given program, into a single process.
For example, given a function application $\mathit{f} \, \ottsym{(}   \lambda  \mathit{x}  .  \mathit{x}   \ottsym{)}$,
a bidirectional type system may
synthesize the type, say, $ \ottsym{(}    \mathsf{int}    \rightarrow    \mathsf{int}    \ottsym{)}   \rightarrow    \mathsf{unit}  $ for the function $\mathit{f}$, then
check that the argument $ \lambda  \mathit{x}  .  \mathit{x} $ has the type $  \mathsf{int}    \rightarrow    \mathsf{int}  $.
Bidirectional typing has been used for
implementing complex typing features, such as
dependent types~\cite{sjoberg_programming_2015,altenkirch_pisigma_2010},
higher-rank polymorphism~\cite{jones_practical_2007,dunfield_complete_2013}, and
total type error localization and recovery~\cite{zhao_total_2024}.

However, each existing bidirectional type system fixes
the built-in typing order, making their typeabilities incompatible.
Typing function applications is one of the main problems.
Many bidirectional type systems~\cite{pierce_local_1998,jones_practical_2007,dunfield_complete_2013}
prefer to synthesize the type of the function first, then
extract the argument type from the synthesized type, and
check that the argument has the extracted type.
We call this typing style \emph{function-first}.
By contrast, another system~\cite{xie_let_2018} prefers
to synthesize the type of the argument first, which we call \emph{argument-first}.
The argument-first style checks that the given function has
the synthesized type as its argument type.
The gap between the function- and argument-first styles makes them incompatible,
thereby forcing the language designer to select one and give up the other's typeability.
Furthermore, supporting (higher-rank) polymorphism also distinguishes between the two formalizations,
in terms of when polymorphic types are instantiated.

We propose a language \lang
that subsumes function- and argument-first bidirectional typing,
and a typing algorithm for that system.\footnote{
  BH stands for ``both-handed'' because
  our language flexibly switches between the function- and argument-first styles.
}
The type system of \lang chooses either of the two styles
based on ``guides'', statically given \emph{at each function application}\footnote{
  Contextual typing~\cite{xue_contextual_2024}
  can also choose between the two styles,
  but it cannot switch between them for every function application.
  See \zcref{subsec:gen_bidi} for more details.
}.
The language designer can choose the function-first (resp. argument-first)
bidirectional typing at their discretion
by inserting the guide for the function-first (resp. argument-first) style.
For example, our framework allows the argument-first style
only if a given argument is a variable and the function-first style otherwise.
The existing two styles can be obtained as instances of the type
system, by fixing all the guides in a program to the function-first or argument-first.
Thanks to the guides, we can separate the problem of determining the typing order
from a bidirectional type system and leave that decision to the language designer.
To show that this separation works well not only in a simple setting,
\lang supports higher-rank polymorphism.

A challenge with incorporating function- and argument-first bidirectional typing
is to address \emph{partially} known type information.
For example, suppose that, given the expression $\mathit{f} \, \mathit{x} \, \mathit{y}$, a typing process
first synthesizes the type of $\mathit{y}$, as in the argument-first style,
secondly synthesizes the type of $\mathit{f}$ using $\mathit{y}$'s type, and then
checks that $\mathit{x}$ has the first argument type of $\mathit{f}$.
At the point of synthesizing $\mathit{f}$'s type,
the known information on $\mathit{f}$ is that
$\mathit{f}$ takes two arguments, and the second argument type is the same as the type of $\mathit{y}$.
Therefore, we need a mechanism to use \emph{partially} known type information,
but neither the function- nor argument-first style represents such type information.

Actually, the idea of such partially known type information
during type synthesis/checking has been studied in the literature~\cite{odersky_colored_2001,vytiniotis_boxy_2006}.
In Odersky et al.~\citet{odersky_colored_2001}, each type constructor in a type expression is assigned one of the two \emph{colors}, which stand for which part should be synthesized,
while, in Vytiniotis et al.~\citet{vytiniotis_boxy_2006}, boxes in a type expression
specify which part of a type should be synthesized.
For instance, a colored type
$\textcolor{blue}{ \mathsf{unit} } \textcolor{red}{ \rightarrow   \mathsf{unit} }$ and
a boxy type $ \boxed{  \mathsf{unit}  }   \rightarrow   \mathsf{int} $ mean that
the argument type should be synthesized, resulting in $ \mathsf{unit} $, and
the other parts are given for checking.
The notation used in this paper is based on boxy types,
because we adopt the restriction on boxy types that
no synthesized part of a type occurs within a checking part.\footnote{%
  In colored types, synthesized and checked parts can be arbitrarily nested.
}
The distinction between synthesized and checked parts in a type expression
is helpful to unify function- and argument-first bidirectional typing.
For example, recall the above typing process for the expression $\mathit{f} \, \mathit{x} \, \mathit{y}$.
The boxy type $ \boxed{ \ottnt{A_{{\mathrm{1}}}} }   \rightarrow  \ottnt{A_{{\mathrm{2}}}}  \rightarrow   \boxed{ \ottnt{B} } $ indicates that the part outside the boxes consists of
the known type information for synthesizing $\mathit{f}$'s type, that is,
the second argument $\mathit{y}$'s type is $\ottnt{A_{{\mathrm{2}}}}$.
\mycomment{
  For example, suppose that, given the expression $\mathit{f} \, \mathit{x} \, \mathit{y}$, a typing process
  first synthesizes the type of $\mathit{y}$, as the argument-first style,
  secondly synthesizes the type of $\mathit{f}$ using the second argument type, and then
  checks that $\mathit{x}$ has the first argument type of $\mathit{f}$.
  At the point of synthesizing $\mathit{f}$'s type,
  the known information of $\mathit{f}$ is that
  $\mathit{f}$ takes two arguments, and the second argument type is the same as the type of $\mathit{y}$.
  The boxy type $ \boxed{ \ottnt{A_{{\mathrm{1}}}} }   \rightarrow  \ottnt{A_{{\mathrm{2}}}}  \rightarrow   \boxed{ \ottnt{B} } $ represents this known type information
  because the parts outside the boxes are given type information
  (suppose that the synthesized type of $\mathit{y}$ is $\ottnt{A_{{\mathrm{2}}}}$).
  The synthesized type $\ottnt{A_{{\mathrm{1}}}}$ is used in type checking for $\mathit{x}$.
}

\mycomment{
  While boxy types help make the function- and argument-first styles coexist in one system,
  naturally arises the question of whether they are adaptable to advanced typing features.
  \lang supports \emph{higher-rank polymorphism}~\cite{odersky_putting_1996}
  to answer this question affirmatively.

  Higher-rank polymorphism allows a polymorphic type to occur anywhere in types.
  For example, the $\mathsf{runST}$ function in Haskell
  required by the encapsulated state monad~\cite{launchbury_lazy_1994} should have the type
  \[
    \forall a . (\forall s . \mathsf{ST}\ s\ a) \rightarrow a~.
  \]
  Peyton Jones et al.~\citet{jones_practical_2007} show
  other examples that require higher-rank polymorphism.
  While higher-rank polymorphism brings a powerful typeability,
  the type synthesis for higher-rank polymorphism is generally undecidable~\cite{wells_typability_1999}.
  \lang allows for writing type annotations
  on argument variables of lambda abstractions to support higher-rank polymorphism,
  as some existing bidirectional type systems do.
}

The type system of \lang is \emph{declarative} in that it guesses types for universal quantifiers,
and \emph{predicative} in that the guessed types are only monotypes.
Although the bidirectional type system in the simply typed setting~\cite{dunfield_bidirectional_2022}
immediately leads to its typing algorithm,
bidirectional type systems~\cite{jones_practical_2007,dunfield_complete_2013,xie_let_2018}
for higher-rank polymorphism usually rely on guessing oracles.
Furthermore, those systems restrict the guessed types to monomorphic types,
making their typing algorithms simple and decidable.
The type system of \lang follows these design decisions\footnote{
  \zcref{subsec:impre} discusses the choice to follow
  ML-style type systems supporting first-class polymorphism.
}.

We provide a typing algorithm for the declarative type system. %
Our typing algorithm is based on the \emph{worklist approach}
introduced by Zhao et al.~\citet{zhao_mechanical_2019}.
Their typing algorithm is for
the function-first style bidirectional type system proposed by
Dunfield and Krishnaswami~\citet{dunfield_complete_2013}.
Still, we show that its core idea is also applicable to the argument-first style,
and even to the unified style.
\TY{How deeply shall we mention technical aspects?}

The metatheory of \lang is classified into declarative and algorithmic parts.
From the declarative point of view, the type system of \lang is
sound with respect to System F extended with type coercions~\cite{cretin_power_2012}, and
complete with respect to the function-first bidirectional type system by
Dunfield and Krishnaswami~\citet{dunfield_complete_2013} and
the argument-first one by Xie and Oliveira~\citet{xie_let_2018}.
From the algorithmic point of view,
our typing algorithm is sound with respect to the type system of \lang.
Although it is not quite complete with respect to the type system of \lang,
it is complete with respect to weaker systems---the two bidirectional type systems of
Dunfield and Krishnaswami~\citet{dunfield_complete_2013} and
Xie and Oliveira~\citet{xie_let_2018}.

The contributions of this work are summarized as follows.
\begin{itemize}
  \item We provide a unified type system that
        subsumes both the function- and argument-first bidirectional type systems
        with higher-rank polymorphism using boxy types.
        Our language \lang serves as an intermediate language in that
        a language designer can use their own approach to determine
        whether to choose function- or argument-first bidirectional typing and
        reflect the choice in compiling their language to \lang.

  \item We provide a sound typing algorithm
        with respect to the type system of \lang.
        It is also complete with respect to existing
        function- and argument-first bidirectional type systems.
        This algorithm is independent of
        which of the two styles a language designer chooses.

  \item We mechanically formalize both declarative and algorithmic typing and
    prove the aforementioned soundness and completeness theorems,
    as well as termination of the algorithm, using the Abella theorem prover~\cite{gacek_abella_2008}.
\end{itemize}
In addition, we have a prototype implementation of our typing algorithm with
a heuristic elaboration of an ML-like language to \lang.
It is written in OCaml and available on Zenodo at \url{https://doi.org/10.5281/zenodo.22218838}, together with the Abella proofs.

The rest of the paper is organized as follows.
\zcref{sec:overview} reviews the technical background of our work and
overviews our approach.
\zcref{sec:decl} presents \lang, its declarative type system, and its metatheory.
\zcref{sec:algo_short} summarizes our typing algorithm for \lang and
its metatheory.
\zcref{sec:related} discusses related work and
\zcref{sec:future} concludes. %
\ifarxiv\else%
  Interested readers are referred to a full version of this
  paper~\cite{YoshiokaSI26}, which describes the details of the typing algorithm
  with its design principles, a guide-inserting elaboration, possible extensions, and omitted definitions.
\fi

\section{Overview}\label{sec:overview}

\subsection{Function- and Argument-First Bidirectional Typing}\label{subsec:ffaf}

A bidirectional type system has two modes of typing:
\emph{synthesis} and \emph{checking modes}.
A judgment of the synthesis mode takes the form of $\Psi  \vdash  \ottnt{e}  \Rightarrow  \ottnt{A}$,
where $\Psi$, $\ottnt{e}$, and $\ottnt{A}$ are
a typing context, an expression, and a type, respectively.
This judgment means that, given $\Psi$ and $\ottnt{e}$, the type $\ottnt{A}$ is synthesized.
By contrast, a judgment of the checking mode takes the form of $\Psi  \vdash  \ottnt{e}  \Leftarrow  \ottnt{A}$,
meaning that, given $\Psi$, $\ottnt{e}$, and $\ottnt{A}$, it can be checked that
$\ottnt{e}$ has the type $\ottnt{A}$ under $\Psi$.

Bidirectional typing switches between the two modes, usually based on term constructors.
For example,
a typing rule~\cite{dunfield_bidirectional_2022} for function applications is:
\[\small
  \inferrule{
  \Psi  \vdash  \ottnt{e_{{\mathrm{1}}}}  \Rightarrow   \ottnt{A}   \rightarrow   \ottnt{B}  \\ \Psi  \vdash  \ottnt{e_{{\mathrm{2}}}}  \Leftarrow  \ottnt{A}
  }{
    \Psi  \vdash  \ottnt{e_{{\mathrm{1}}}} \, \ottnt{e_{{\mathrm{2}}}}  \Rightarrow  \ottnt{B}
  }\ .
\]
One can read off a recursive algorithm for type synthesis/checking, if
the two modes are specified appropriately: this rule means that, given
a typing context $\Psi$ and a function application $\ottnt{e_{{\mathrm{1}}}} \, \ottnt{e_{{\mathrm{2}}}}$, the
algorithm first synthesizes the type $ \ottnt{A}   \rightarrow   \ottnt{B} $ of $\ottnt{e_{{\mathrm{1}}}}$, then
checks that the argument $\ottnt{e_{{\mathrm{2}}}}$ has the type $\ottnt{A}$ required by the
function $\ottnt{e_{{\mathrm{1}}}}$, and finally outputs the type $\ottnt{B}$ as a result of
the type synthesis for $\ottnt{e_{{\mathrm{1}}}} \, \ottnt{e_{{\mathrm{2}}}}$.

Bidirectional typing rules for function applications are classified into two styles.
One style is, given a function application, to first synthesize the type of the function.
We call this style \emph{function-first}.
In this sense, the typing rule shown above is in the function-first style.
Many bidirectional type systems~\cite{pierce_local_1998,jones_practical_2007,dunfield_complete_2013}
can be classified as the function-first style.
We select the system by Dunfield and Krishnaswami~\citet{dunfield_complete_2013},
referred to as the DK system for short, as a subject of our research
because of its simplicity. %
Another style is to first synthesize the type of the argument. %
We call this style \emph{argument-first}.
The argument-first style of bidirectional typing originates from
the work by Xie and Oliveira~\citet{xie_let_2018}.
We refer to their type system (without let polymorphism) as the XO system for short\footnote{
  We leave supporting let polymorphism as future work.
}.

The formalization of the argument-first style is very different from that of the function-first style.
The XO system has no typechecking rule.
Instead, it has an \emph{application context} $\Xi$
in its typing judgment, as $\Psi  \mid  \Xi  \vdash_{ \mathrm{XO} }  \ottnt{e}  \Rightarrow  \ottnt{A}$,
to reuse the known type information.
An application context stacks the synthesized types of the arguments.
To see the role of an application context,
we show the XO typing rules for function applications and lambda abstractions as follows.
\[\small
  \inferrule{
  \Psi  \mid   \emptyset   \vdash_{ \mathrm{XO} }  \ottnt{e_{{\mathrm{2}}}}  \Rightarrow  \ottnt{A} \\ \Psi  \mid  \Xi  \ottsym{,}  \ottnt{A}  \vdash_{ \mathrm{XO} }  \ottnt{e_{{\mathrm{1}}}}  \Rightarrow   \ottnt{A}   \rightarrow   \ottnt{B} 
  }{
    \Psi  \mid  \Xi  \vdash_{ \mathrm{XO} }  \ottnt{e_{{\mathrm{1}}}} \, \ottnt{e_{{\mathrm{2}}}}  \Rightarrow  \ottnt{B}
  }
  \qquad
  \inferrule{
    \Psi  \ottsym{,}  \mathit{x}  \ottsym{:}  \ottnt{A}  \mid  \Xi  \vdash_{ \mathrm{XO} }  \ottnt{e}  \Rightarrow  \ottnt{B}
  }{
    \Psi  \mid  \Xi  \ottsym{,}  \ottnt{A}  \vdash_{ \mathrm{XO} }   \lambda  \mathit{x}  .  \ottnt{e}   \Rightarrow   \ottnt{A}   \rightarrow   \ottnt{B} 
  }
\]
The typing rule for function applications in the XO system stacks
the type synthesized from the argument\footnote{
  We omit a type generalization from the original typing rule
  because let polymorphism is out of scope. %
}.
It first synthesizes the type $\ottnt{A}$ of the argument $\ottnt{e_{{\mathrm{2}}}}$;
pushes the synthesized type into the application context $\Xi$; and
synthesizes the type $ \ottnt{A}   \rightarrow   \ottnt{B} $ of the function $\ottnt{e_{{\mathrm{1}}}}$
using the information about the argument types accumulated
in $\Xi  \ottsym{,}  \ottnt{A}$.
On the other hand, the typing rule for lambda abstractions
uses stacked type information to determine the argument type. %

\mycomment{
The typing rule for function applications in the XO system stacks
a synthesized type from the argument, as shown below\footnote{
  We omit a type generalization from the original typing rule
  because let polymorphism is out of scope of our work.
}.
\[
  \inferrule{
  \Psi  \mid   \emptyset   \vdash_{ \mathrm{XO} }  \ottnt{e_{{\mathrm{2}}}}  \Rightarrow  \ottnt{A} \\ \Psi  \mid  \Xi  \ottsym{,}  \ottnt{A}  \vdash_{ \mathrm{XO} }  \ottnt{e_{{\mathrm{1}}}}  \Rightarrow   \ottnt{A}   \rightarrow   \ottnt{B} 
  }{
    \Psi  \mid  \Xi  \vdash_{ \mathrm{XO} }  \ottnt{e_{{\mathrm{1}}}} \, \ottnt{e_{{\mathrm{2}}}}  \Rightarrow  \ottnt{B}
  }
\]
This rule first synthesizes the type $\ottnt{A}$ of the argument $\ottnt{e_{{\mathrm{2}}}}$;
pushes the synthesized type into the application context $\Xi$; and
synthesizes the type $ \ottnt{A}   \rightarrow   \ottnt{B} $ of the function $\ottnt{e_{{\mathrm{1}}}}$
using the information about argument types accumulated in
the application context $\Xi  \ottsym{,}  \ottnt{A}$.
On the other hand, the following typing rule
uses stacked type information to determine the argument type of a lambda abstraction.
\[
  \inferrule{
    \Psi  \ottsym{,}  \mathit{x}  \ottsym{:}  \ottnt{A}  \mid  \Xi  \vdash_{ \mathrm{XO} }  \ottnt{e}  \Rightarrow  \ottnt{B}
  }{
    \Psi  \mid  \Xi  \ottsym{,}  \ottnt{A}  \vdash_{ \mathrm{XO} }   \lambda  \mathit{x}  .  \ottnt{e}   \Rightarrow   \ottnt{A}   \rightarrow   \ottnt{B} 
  }
\]
}

While the function- and argument-first styles are formalized in different ways
even in a simple setting,
their difference becomes larger when considering (higher-rank) polymorphism.
For example, consider the function application $ \mathsf{id}  \, 42$
where $ \mathsf{id} $ has the type $  \forall  \ottmv{a}  .  \ottmv{a}    \rightarrow   \ottmv{a} $~.
The DK system first synthesizes the type of $ \mathsf{id} $ and then
guesses a monotype to instantiate it for typechecking $42$.
The XO system, in contrast, first synthesizes the type of $42$ and then
guesses a monotype to instantiate the type of $ \mathsf{id} $, ensuring the argument type is $ \mathsf{int} $.
In short, if a function has a polymorphic type,
the function-first (resp. argument-first) style guesses a type to instantiate it
in the derivation for the argument (resp. function).

\mycomment{
The typing derivation for $ \mathsf{id}  \, 42$ in the DK system is as follows.
\[
  \inferrule* {
    \Psi  \vdash_{ \mathrm{DK} }   \mathsf{id}   \Rightarrow    \forall  \ottmv{a}  .  \ottmv{a}    \rightarrow   \ottmv{a}  \\
      \forall  \ottmv{a}  .  \ottmv{a}    \rightarrow   \ottmv{a}  \text{ is instantiated to }   \mathsf{int}    \rightarrow    \mathsf{int}   \\
    \Psi  \vdash_{ \mathrm{DK} }  42  \Leftarrow   \mathsf{int} 
  }{
    \Psi  \vdash_{ \mathrm{DK} }   \mathsf{id}  \, 42  \Rightarrow   \mathsf{int} 
  }
\]
The DK system first synthesizes the type of $ \mathsf{id} $ and
\emph{guesses} the type to instantiate it.
In contrast, the typing derivation for $ \mathsf{id}  \, 42$ in the XO system is as follows.
\[
  \inferrule* {
  \Psi  \mid   \emptyset   \vdash_{ \mathrm{XO} }  42  \Rightarrow   \mathsf{int}  \\ \Psi  \mid   \mathsf{int}   \vdash_{ \mathrm{XO} }   \mathsf{id}   \Rightarrow    \mathsf{int}    \rightarrow    \mathsf{int}  
  }{
    \Psi  \mid   \emptyset   \vdash_{ \mathrm{XO} }   \mathsf{id}  \, 42  \Rightarrow   \mathsf{int} 
  }
\]
The XO system first synthesizes the type of $42$ and
\emph{reuses} the synthesized type to instantiate the type of $ \mathsf{id} $.
}

More crucially, the typeabilities of the DK and XO systems are incompatible
with higher-rank polymorphism.
Consider the following two expressions:
\begin{align}
  \small
   & \ottsym{(}   \lambda  \mathit{f}  \ottsym{:}   \ottsym{(}    \forall  \ottmv{a}  .  \ottmv{a}    \rightarrow   \ottmv{a}   \ottsym{)}   \rightarrow   \ottsym{(}    \mathsf{int}   \times   \mathsf{bool}    \ottsym{)}   .  \mathit{f}   \ottsym{)}  \appright  \ottsym{(}   \lambda  \mathit{g}  .  \ottsym{(}  \mathit{g} \, 42  \ottsym{,}  \mathit{g} \,  \mathsf{true}   \ottsym{)}   \ottsym{)}\label{app_non_guide1} \\
   & \ottsym{(}   \lambda  \mathit{f}  .  \mathit{f}   \ottsym{)}  \appleft  \ottsym{(}   \lambda  \mathit{g}  \ottsym{:}    \forall  \ottmv{a}  .  \ottmv{a}    \rightarrow   \ottmv{a}   .  \ottsym{(}  \mathit{g} \, 42  \ottsym{,}  \mathit{g} \,  \mathsf{true}   \ottsym{)}   \ottsym{)} \label{app_non_guide2}
\end{align}
Note that both systems only infer monotypes for parameters of lambda abstractions,
following the Hindley--Milner type inference.
The former expression is well-typed under the DK system but ill-typed under the XO system.
The XO system fails to synthesize the desired type of $\mathit{g}$,
which should be polymorphic for the expression to be well-typed,
but it can only assign a monomorphic type to it.
The latter expression is well-typed under the XO system but ill-typed under the DK system,
because the DK system fails to synthesize the desired type of $\mathit{f}$, which should be polymorphic.
This example shows that their typeabilities are incompatible,
as Xie and Oliveira~\citet{xie_let_2018} pointed out.

\mycomment{
\subsection{Bidirectional Typing for Higher-Rank Polymorphism}\label{subsec:hrp}

Bidirectional typing is extended with higher-rank polymorphism
in different ways~\cite{jones_practical_2007,dunfield_complete_2013,xie_let_2018}.
The central concern to support higher-rank polymorphism is
a typing rule for function applications.
In particular, the function-first style rule for function applications
cannot be straightforwardly extended to accommodate higher-rank polymorphism.
In the presence of higher-rank polymorphism,
the synthesized type for a function can be universally quantified,
e.g., $  \forall  \ottmv{a}  .  \ottmv{a}    \rightarrow   \ottmv{a} $.
The function-first style requires extracting the argument type from the function's type
to check that the argument has the required argument type by the function.
However, we cannot easily extract the argument type from a universally quantified type.
In contrast, because the argument-first style does not require extracting
the argument type from the function's type,
it fits higher-rank polymorphism well.

The DK system has an \emph{application judgment} to support polymorphic types.
The typing rule for function applications in the DK system is as follows.
\[
  \inferrule{
  \Psi  \vdash_{ \mathrm{DK} }  \ottnt{e_{{\mathrm{1}}}}  \Rightarrow  \ottnt{A} \\  \Psi   \vdash_{ \mathrm{DK} }   \ottnt{A}  \bullet  \ottnt{e_{{\mathrm{2}}}}  \Rightarrow\!\!\!\Rightarrow  \ottnt{B} 
  }{
    \Psi  \vdash_{ \mathrm{DK} }  \ottnt{e_{{\mathrm{1}}}} \, \ottnt{e_{{\mathrm{2}}}}  \Rightarrow  \ottnt{B}
  }~.
\]
The application judgment $ \Psi   \vdash_{ \mathrm{DK} }   \ottnt{A}  \bullet  \ottnt{e_{{\mathrm{2}}}}  \Rightarrow\!\!\!\Rightarrow  \ottnt{B} $ in this rule means that,
if we get a function type $ \ottnt{A'}   \rightarrow   \ottnt{B} $ by instantiating
all the universal quantifiers at the top of the type $\ottnt{A}$,
we check that $\ottnt{e_{{\mathrm{2}}}}$ has the type $\ottnt{A'}$ and output $\ottnt{B}$.

More formally, the application judgment is defined by
\begin{mathpar}
  \inferrule{
  \Psi  \vdash  \tau \\  \Psi   \vdash_{ \mathrm{DK} }    [  \tau  /  \ottmv{a}  ]  \ottnt{A}   \bullet  \ottnt{e}  \Rightarrow\!\!\!\Rightarrow  \ottnt{B} 
  }{
     \Psi   \vdash_{ \mathrm{DK} }    \forall  \ottmv{a}  .  \ottnt{A}   \bullet  \ottnt{e}  \Rightarrow\!\!\!\Rightarrow  \ottnt{B} 
  }

  \inferrule{
    \Psi  \vdash_{ \mathrm{DK} }  \ottnt{e}  \Leftarrow  \ottnt{A}
  }{
     \Psi   \vdash_{ \mathrm{DK} }    \ottnt{A}   \rightarrow   \ottnt{B}   \bullet  \ottnt{e}  \Rightarrow\!\!\!\Rightarrow  \ottnt{B} 
  }
\end{mathpar}
The former rule guesses a type $\tau$ to instantiate the universally quantified type variable $\ottmv{a}$.
The latter rule checks that the expression $\ottnt{e}$ has the argument type $\ottnt{A}$ of the inferred type, and
outputs the return type $\ottnt{B}$. %

The incompatibility between the typeabilities of DK and XO systems
arises as the most crucial difference, considering higher-rank polymorphism.
Consider the following two expressions:
\begin{itemize}
  \item $\ottsym{(}   \lambda  \mathit{f}  \ottsym{:}   \ottsym{(}    \forall  \ottmv{a}  .  \ottmv{a}    \rightarrow   \ottmv{a}   \ottsym{)}   \rightarrow   \ottsym{(}    \mathsf{int}   \times   \mathsf{bool}    \ottsym{)}   .  \mathit{f}   \ottsym{)} \, \ottsym{(}   \lambda  \mathit{g}  .  \ottsym{(}  \mathit{g} \, 42  \ottsym{,}  \mathit{g} \,  \mathsf{true}   \ottsym{)}   \ottsym{)}$ and
  \item $\ottsym{(}   \lambda  \mathit{f}  .  \mathit{f}   \ottsym{)} \, \ottsym{(}   \lambda  \mathit{g}  \ottsym{:}    \forall  \ottmv{a}  .  \ottmv{a}    \rightarrow   \ottmv{a}   .  \ottsym{(}  \mathit{g} \, 42  \ottsym{,}  \mathit{g} \,  \mathsf{true}   \ottsym{)}   \ottsym{)}$~.
\end{itemize}
Note that both the DK and XO systems only infer monotypes for parameters of lambda abstractions,
following the Hindley--Milner type inference.
The former expression is well-typed under the DK system but ill-typed under the XO system,
because the XO system fails to synthesize the type of $\mathit{g}$, which should be polymorphic
for the expression to be well-typed.
The latter expression is well-typed under the XO system but ill-typed under the DK system,
because the DK system fails to synthesize the type of $\mathit{f}$, which should be polymorphic.
This example suggests that their typeabilities are incompatible,
as Xie and Oliveira~\citet{xie_let_2018} pointed out.
}

\subsection{Our Work: Unifying Function- and Argument-First Styles}\label{subsec:our_work}

We aim to unify the function- and argument-first bidirectional type systems,
specifically the DK and XO systems, into a single system
that incorporates their different typeabilities.
This feature helps augment the typeabilities of existing bidirectional type systems.

We clarify a challenge in mixing function- and argument-first bidirectional typing
before showing our approach.
As shown in \zcref{subsec:ffaf},
if a lambda abstraction whose argument has a polymorphic type is supplied as an argument,
the function- and argument-first styles differ in typeability,
i.e., where we must write type annotations.
Therefore, the situation in which two or more arguments are lambda abstractions
of this kind is troublesome.
For example, suppose that,
given the following declarations and the function application $\ottsym{(}  \mathit{f_{{\mathrm{1}}}} \, \mathit{f_{{\mathrm{2}}}}  \ottsym{)} \, \mathit{f_{{\mathrm{3}}}}$,
we use the function-first style for the left application $\mathit{f_{{\mathrm{1}}}} \, \mathit{f_{{\mathrm{2}}}}$ and
the argument-first style for the right application $\ottsym{(}  \mathit{f_{{\mathrm{1}}}} \, \mathit{f_{{\mathrm{2}}}}  \ottsym{)} \, \mathit{f_{{\mathrm{3}}}}$~.
\begin{align*}
  \mathit{f_{{\mathrm{1}}}} & =  \lambda  \mathit{g_{{\mathrm{1}}}}  \ottsym{:}   \ottsym{(}    \forall  \ottmv{a}  .  \ottmv{a}    \rightarrow   \ottmv{a}   \ottsym{)}   \rightarrow   \ottsym{(}    \mathsf{int}   \times   \mathsf{bool}    \ottsym{)}   .   \lambda  \mathit{g_{{\mathrm{2}}}}  .  \ottsym{(}  \mathit{g_{{\mathrm{1}}}} \, \ottsym{(}   \lambda  \mathit{x}  .  \mathit{x}   \ottsym{)}  \ottsym{,}  \mathit{g_{{\mathrm{2}}}} \, \ottsym{(}   \lambda  \mathit{x}  .  \mathit{x}   \ottsym{)}  \ottsym{)}                                                           \\
  \mathit{f_{{\mathrm{2}}}} & =  \lambda  \mathit{h_{{\mathrm{1}}}}  .  \ottsym{(}  \mathit{h_{{\mathrm{1}}}} \, 42  \ottsym{,}  \mathit{h_{{\mathrm{1}}}} \,  \mathsf{true}   \ottsym{)}                                                 , \qquad
                                                                                          \mathit{f_{{\mathrm{3}}}} =  \lambda  \mathit{h_{{\mathrm{2}}}}  \ottsym{:}    \forall  \ottmv{a}  .  \ottmv{a}    \rightarrow   \ottmv{a}   .  \ottsym{(}  \mathit{h_{{\mathrm{2}}}} \,  \mathsf{false}   \ottsym{,}  \mathit{h_{{\mathrm{2}}}} \, 24  \ottsym{)} 
\end{align*}
Because both the DK and XO systems guess only monotypes for parameters of lambda abstractions,
we must reuse the synthesized type information about $\mathit{f_{{\mathrm{3}}}}$ to
synthesize the type of $\mathit{f_{{\mathrm{1}}}}$.
However, neither the DK nor the XO system can address the situation where, given a function,
we must synthesize its first argument type and typecheck its second argument type.
Therefore, we need a more flexible representation for partially known type information.

Our key idea to address this challenge is to adapt \emph{boxy types}~\cite{vytiniotis_boxy_2006},
which were originally proposed for the function-first style,
to the argument-first style.
A boxy type specifies that
partial type information is synthesized and that
the other part is typechecked
by enclosing type information to be synthesized within a box.
For example, the boxy type $ \boxed{  \mathsf{unit}  }   \rightarrow   \mathsf{int} $ intuitively indicates that
$ \mathsf{unit} $ is synthesized, and that
the type information outside the box, namely the function arrow and the return type $ \mathsf{int} $, is given.
\mycomment{
  A boxy type specifies which part of the type should be
  synthesized or used for checking.
  For example, the boxy type $ \boxed{  \mathsf{unit}  }   \rightarrow   \mathsf{int} $ intuitively means that
  the argument type $ \mathsf{unit} $ is synthesized, given the function arrow and the return type $ \mathsf{int} $.
  In this syntax, we enclose type information to be synthesized within a box,
  whereas type information outside the boxes is used for checking.
}
We can mix the function- and argument-first bidirectional typing using boxy types.
Recall the expression $\ottsym{(}  \mathit{f_{{\mathrm{1}}}} \, \mathit{f_{{\mathrm{2}}}}  \ottsym{)} \, \mathit{f_{{\mathrm{3}}}}$. %
We can use the following boxy type
for $\mathit{f_{{\mathrm{1}}}}$ %
to accept the program, where $ \mathsf{ID} $ is $  \forall  \ottmv{a}  .  \ottmv{a}    \rightarrow   \ottmv{a} $~.
\[\small
   \boxed{   \mathsf{ID}    \rightarrow   \ottsym{(}    \mathsf{int}   \times   \mathsf{bool}    \ottsym{)}  }   \rightarrow  \ottsym{(}   \mathsf{ID}   \rightarrow  \ottsym{(}    \mathsf{bool}   \times   \mathsf{int}    \ottsym{)}  \ottsym{)}  \rightarrow   \boxed{  \ottsym{(}    \mathsf{int}   \times   \mathsf{bool}    \ottsym{)}  \times  \ottsym{(}    \mathsf{bool}   \times   \mathsf{int}    \ottsym{)}  } 
\]

Introducing boxy types enables us to define
function- and argument-first typing rules for function applications
without an application context.
We use the metavariables $\ottnt{A}$ and ${A^\Box}$ to denote
an ordinary type, i.e., a type with no box, and a boxy type, respectively.
Then, our typing rules are as follows.

\begin{center}
  \begin{tabular}{cc}
    \begin{minipage}{0.45\linewidth}
      \begin{equation}
        \small
        \inferrule{
          \Psi  \vdash  \ottnt{e_{{\mathrm{1}}}}  \ottsym{:}   \boxed{ \ottnt{A} }   \rightarrow  {A^\Box} \\
          \Psi  \vdash  \ottnt{e_{{\mathrm{2}}}}  \ottsym{:}  \ottnt{A}
        }{
          \Psi  \vdash  \ottnt{e_{{\mathrm{1}}}} \, \ottnt{e_{{\mathrm{2}}}}  \ottsym{:}  {A^\Box}
        } \label{app_r}
      \end{equation}
    \end{minipage} &
    \begin{minipage}{0.45\linewidth}
      \begin{equation}
        \small
        \inferrule{
          \Psi  \vdash  \ottnt{e_{{\mathrm{1}}}}  \ottsym{:}  \ottnt{A}  \rightarrow  {A^\Box} \\
          \Psi  \vdash  \ottnt{e_{{\mathrm{2}}}}  \ottsym{:}   \boxed{ \ottnt{A} } 
        }{
          \Psi  \vdash  \ottnt{e_{{\mathrm{1}}}} \, \ottnt{e_{{\mathrm{2}}}}  \ottsym{:}  {A^\Box}
        } \label{app_l}
      \end{equation}
    \end{minipage}
  \end{tabular}
\end{center}
The former rule \eqref{app_r} corresponds to the function-first typing.
It synthesizes the argument type of the function $\ottnt{e_{{\mathrm{1}}}}$ and
checks that $\ottnt{e_{{\mathrm{2}}}}$ has the synthesized type $\ottnt{A}$.
Algorithmically, $ \boxed{ \ottnt{A} } $ denotes a placeholder that
should be replaced with a synthesized type, and $\ottnt{A}$ is \emph{the} synthesized one.
For example, supposing $ \mathit{f}  \ottsym{:}    \forall  \ottmv{a}  .  \ottmv{a}    \rightarrow   \ottmv{a}   \in  \Psi $, we have the following typing derivation.
\[\small
  \inferrule* {
  \Psi  \vdash  \mathit{f}  \ottsym{:}   \boxed{  \mathsf{unit}  }   \rightarrow   \mathsf{unit}  \\ \Psi  \vdash  \ottsym{()}  \ottsym{:}   \mathsf{unit} 
  }{
    \Psi  \vdash  \mathit{f} \, \ottsym{()}  \ottsym{:}   \mathsf{unit} 
  }
\]
First, the non-boxed type $ \mathsf{unit} $ in the conclusion means that
$\mathit{f} \, \ottsym{()}$ is typechecked against $ \mathsf{unit} $.
Then, the polymorphic type $  \forall  \ottmv{a}  .  \ottmv{a}    \rightarrow   \ottmv{a} $ of $\mathit{f}$ is instantiated with $ \mathsf{unit} $
because $\mathit{f}$'s return type is required to be $ \mathsf{unit} $.
Finally, the argument $\ottsym{()}$ is typechecked against $ \mathsf{unit} $.

The latter typing rule \eqref{app_l}, which is for the argument-first style,
first synthesizes a type $\ottnt{A}$ of $\ottnt{e_{{\mathrm{2}}}}$.
It checks that $\ottnt{e_{{\mathrm{1}}}}$ has a function type with the synthesized argument type $\ottnt{A}$,
as expressed by the boxy type $\ottnt{A}  \rightarrow  {A^\Box}$.
Recall that a type with no box indicates that it is fully given information.
Unlike the XO system, a boxy type, not an application context,
stacks the synthesized argument types.

The remaining issue is: given a function application,
should we choose the rule \eqref{app_r} or \eqref{app_l}?
Here, our central interest is a declarative type system,
rather than an algorithmic one that determines how to choose them.
Therefore, we assume that every function application has an annotation
so that the language designer can choose the rule as they wish.

We introduce \emph{application guides} to decide
whether we use the function- or argument-first typing rule.
An application guide is either
a \emph{function-first guide} $ \appright $ or
an \emph{argument-first guide} $ \appleft $.
If a function application takes the form of $\ottnt{e_{{\mathrm{1}}}}  \appright  \ottnt{e_{{\mathrm{2}}}}$ (resp. $\ottnt{e_{{\mathrm{1}}}}  \appleft  \ottnt{e_{{\mathrm{2}}}}$),
we choose the function-first (resp. argument-first) rule.
These constructs allow a language designer or even a programmer to
switch their typing style between the two rules.
For example, both programs shown in \zcref{subsec:ffaf} are made well-typed
by properly inserting application guides like:
{
  \small
  \begin{align}
     & \ottsym{(}   \lambda  \mathit{f}  \ottsym{:}   \ottsym{(}    \forall  \ottmv{a}  .  \ottmv{a}    \rightarrow   \ottmv{a}   \ottsym{)}   \rightarrow   \ottsym{(}    \mathsf{int}   \times   \mathsf{bool}    \ottsym{)}   .  \mathit{f}   \ottsym{)}  \appright  \ottsym{(}   \lambda  \mathit{g}  .  \ottsym{(}  \mathit{g} \, 42  \ottsym{,}  \mathit{g} \,  \mathsf{true}   \ottsym{)}   \ottsym{)}\label{app_guide1} \\
     & \ottsym{(}   \lambda  \mathit{f}  .  \mathit{f}   \ottsym{)}  \appleft  \ottsym{(}   \lambda  \mathit{g}  \ottsym{:}    \forall  \ottmv{a}  .  \ottmv{a}    \rightarrow   \ottmv{a}   .  \ottsym{(}  \mathit{g} \, 42  \ottsym{,}  \mathit{g} \,  \mathsf{true}   \ottsym{)}   \ottsym{)} \label{app_guide2}
  \end{align}
}

Adopting application guides enables \lang to serve as a fundamental intermediate language.
Although writing a guide for every function application is cumbersome,
a language designer can define their own scheme to
choose either the function- or argument-first style along with their language.
For example, if a language designer wants to use argument-first typing
only when an argument is a variable,
they may introduce the following translation.
\[
  \small
  \begin{array}{r@{\ }c@{\ }l@{\qquad}r@{\ }c@{\ }ll}
               \ottnt{e} \, \mathit{x} &  \mathbin{=}  & \ottnt{e}  \appleft  \mathit{x}   &
               \ottnt{e_{{\mathrm{1}}}} \, \ottnt{e_{{\mathrm{2}}}}                              &  \mathbin{=}  & \ottnt{e_{{\mathrm{1}}}}  \appright  \ottnt{e_{{\mathrm{2}}}} & (\totherwise)
  \end{array}
\]

Our prototype implementation includes
a more practical insertion of application guides.
It tries to make the best use of known type information in type checking.
For example, our insertion algorithm heuristically chooses the function-first style
if a variable $\mathit{f}$ takes an argument $\ottnt{e}$,
because the type of $\mathit{f}$ is known when typing $\mathit{f} \, \ottnt{e}$.
Similarly, it uses the argument-first style if an argument type is fully known.
Using our heuristics transforms
the two expressions \eqref{app_non_guide1} and \eqref{app_non_guide2} into
\eqref{app_guide1} and \eqref{app_guide2}, respectively.
In other words, it enables programmers to enjoy the benefits of
the function- and argument-first styles without writing application guides directly.
However, evaluating how well our elaboration algorithm works in practice
is left for future work.
Further details of our heuristics
can be found in \arxiv{\zcref{sec:marking}}{the full version of this paper}.

\section{\texorpdfstring{\lang}{lambdaBH} and Its Type System}\label{sec:decl}

This section shows our language \lang and its declarative type system.
We discuss only subtyping and typing---interested readers are referred
to \arxiv{\zcref{sec:full_defs}}{the full version of this paper} for
the omitted definitions, including well-formedness of types.

The type system of \lang is declarative in the sense that it guesses monotypes
for instantiating polymorphic types or for parameters of lambda abstractions.
We design this type system so that its synthesis result is uniquely determined
if the guessing oracle outputs the same monotype.
\mycomment{
  For example, consider the situation where a typing algorithm tries to
  infer the type of the variable $\mathit{x}$, given the typing context $\mathit{x}  \ottsym{:}   \forall  \ottmv{a}  .  \ottmv{a}   \rightarrow  \ottmv{a}$.
  An inferred result could be
  $ \tau   \rightarrow   \tau $, where $\tau$ is a guessed type, or $  \forall  \ottmv{a}  .  \ottmv{a}    \rightarrow   \ottmv{a} $.
  However, we prevent $ \tau   \rightarrow   \tau $ from being an inferred type
  to simplify a typing algorithm.
  Note that
  the inferred result for $\mathit{x}$ would be $ \tau   \rightarrow   \tau $
  if $\mathit{x}$ were required to be a function type.
  The typing algorithm is simple even in this situation
  because it represents the guessed type $\tau$ as a fresh unification variable,
  i.e., the output is unique.
}

\subsection{Syntax}\label{subsec:syntax}

\begin{figure}[t]
  \small
  \[
    \begin{array}{rrcl}
      \textbf{Expressions}     & \ottnt{e}        &  \Coloneqq  &
      \ottsym{()}  \mid  \mathit{x}  \mid   \lambda  \mathit{x}  .  \ottnt{e}   \mid   \lambda  \mathit{x}  \ottsym{:}  \ottnt{A}  .  \ottnt{e}   \mid 
      \ottnt{e_{{\mathrm{1}}}}  \appright  \ottnt{e_{{\mathrm{2}}}}  \mid  \ottnt{e_{{\mathrm{1}}}}  \appleft  \ottnt{e_{{\mathrm{2}}}} \\
      \textbf{Types}           & \ottnt{A}, \ottnt{B} &  \Coloneqq  &
       \mathsf{unit}   \mid  \ottmv{a}  \mid   \ottnt{A}   \rightarrow   \ottnt{B}   \mid   \forall  \ottmv{a}  .  \ottnt{A}      \\
      \textbf{Monotypes}       & \tau, \sigma &  \Coloneqq  &
       \mathsf{unit}   \mid  \ottmv{a}  \mid   \tau   \rightarrow   \sigma                             \\
      \textbf{Boxy types}      & {A^\Box}, {B^\Box} &  \Coloneqq  &
       \boxed{ \ottnt{A} }   \mid   \mathsf{unit}   \mid  \ottmv{a}  \mid 
      {A^\Box}  \rightarrow  {B^\Box}  \mid   \forall  \ottmv{a}  .  \ottnt{A}                         \\
      \textbf{Typing contexts} & \Psi       &  \Coloneqq  &
       \emptyset   \mid  \Psi  \ottsym{,}  \ottmv{a}  \mid  \Psi  \ottsym{,}  \mathit{x}  \ottsym{:}  \ottnt{A}                  \\
    \end{array}
  \]
  \caption{Syntax}\label{fig:syntax}
\end{figure}

We show the syntax of \lang in \zcref{fig:syntax}.
\emph{Expressions}, ranged over by $\ottnt{e}$, consist of
the unit term $\ottsym{()}$, variables $\mathit{x}$,
lambda abstractions $ \lambda  \mathit{x}  .  \ottnt{e} $,
lambda abstractions with type annotations $ \lambda  \mathit{x}  \ottsym{:}  \ottnt{A}  .  \ottnt{e} $, and
two kinds of function applications $\ottnt{e_{{\mathrm{1}}}}  \appright  \ottnt{e_{{\mathrm{2}}}}$ and $\ottnt{e_{{\mathrm{1}}}}  \appleft  \ottnt{e_{{\mathrm{2}}}}$.
We use type annotations in lambda abstractions
to introduce variables of higher-rank polymorphic types.
For example, we can write a function that takes a polymorphic identity function,
like $ \lambda  \mathit{f}  \ottsym{:}    \forall  \ottmv{a}  .  \ottmv{a}    \rightarrow   \ottmv{a}   .  \ottnt{e} $~.
Function applications involve application guides, as explained in \zcref{sec:overview}.
\emph{Types}, ranged over by $\ottnt{A}$ and $\ottnt{B}$, consist of
the unit type $ \mathsf{unit} $, type variables $\ottmv{a}$,
function types $ \ottnt{A}   \rightarrow   \ottnt{B} $, and universal types $ \forall  \ottmv{a}  .  \ottnt{A} $~.
\emph{Monotypes}, ranged over by $\tau$ and $\sigma$, are the types
with no universal type.

The syntax for boxy types, ranged over by ${A^\Box}$ and ${B^\Box}$,
reflects the intuition explained in \zcref{subsec:our_work}.
\mycomment{
  Firstly, we prevent a synthesized type from including type information to be used for checking,
  contrary to \emph{colored types}~\cite{odersky_colored_2001},
  which is similar to boxy types.
  For example, we prevent synthesizing the ``boxy'' type
  $\boxed{ \mathsf{unit}   \rightarrow  \langle  \mathsf{unit}  \rangle}$,
  supposing that $\langle \ottnt{A} \rangle$ means that
  the type $\ottnt{A}$ should be used for checking.
  It intuitively means that, given the return type $ \mathsf{unit} $,
  the typing process synthesizes the function arrow and argument type $ \mathsf{unit} $.
  This typing process seems odd
  because the given information about the return type immediately leads to the function arrow, i.e.,
  the boxy type should be $ \boxed{  \mathsf{unit}  }   \rightarrow   \mathsf{unit} $.
  Therefore, we exclude the type $\boxed{ \mathsf{unit}   \rightarrow  \langle  \mathsf{unit}  \rangle}$
  from boxy types.
  We recognize no drawback of this restriction,
  but the formal comparison between boxy types and colored types is left for future work.
}
Note that we only allow a type $\ottnt{A}$, with no box, inside a universal type.
Consider the typing judgment
\[
  \mathit{x}  \ottsym{:}    \forall  \ottmv{a}  .   \forall  \ottmv{b}  .  \ottmv{b}     \rightarrow   \ottmv{b}   \vdash  \mathit{x}  \ottsym{:}   \forall  \ottmv{a}  .   \boxed{ \text{?} }  
\]
to see why this restriction is justified.
This judgment would mean that
we synthesize the type denoted as ``$ \text{?} $'' under the universal type $\forall$.
However, the typing process cannot \emph{uniquely} determine the synthesis result.
For example, both filling the hole with $  \forall  \ottmv{b}  .  \ottmv{b}    \rightarrow   \ottmv{b} $ and with $  \forall  \ottmv{b}  .  \ottmv{a}    \rightarrow   \ottmv{a} $
are valid,
because our subtyping system can derive the following two subtyping judgments.
\noindent\begin{center}
  \begin{tabular}{cc}
    \begin{minipage}{0.45\linewidth}
      \begin{equation}
        \small
        \inferrule*{
          \inferrule*{
            \inferrule*{
              \inferrule*{
                \cdots
              }{
                 \ottmv{a}  \ottsym{,}  \ottmv{b}   \vdash  \subL{   \forall  \ottmv{b}  .  \ottmv{b}    \rightarrow   \ottmv{b}  }{  \ottmv{b}   \rightarrow   \ottmv{b}  } 
              }
            }{
               \ottmv{a}   \vdash  \subL{   \forall  \ottmv{b}  .  \ottmv{b}    \rightarrow   \ottmv{b}  }{  \forall  \ottmv{b}  .   \ottmv{b}   \rightarrow   \ottmv{b}   } 
            }
          }{
             \ottmv{a}   \vdash  \subL{   \forall  \ottmv{a}  .   \forall  \ottmv{b}  .  \ottmv{b}     \rightarrow   \ottmv{b}  }{  \forall  \ottmv{b}  .   \ottmv{b}   \rightarrow   \ottmv{b}   } 
          }
        }{
            \emptyset    \vdash  \subL{   \forall  \ottmv{a}  .   \forall  \ottmv{b}  .  \ottmv{b}     \rightarrow   \ottmv{b}  }{  \forall  \ottmv{a}  .    \forall  \ottmv{b}  .  \ottmv{b}    \rightarrow   \ottmv{b}   } 
        }
      \end{equation}
    \end{minipage} &
    \begin{minipage}{0.45\linewidth}
      \begin{equation}
        \small
        \inferrule*{
          \inferrule*{
            \inferrule*{
              \inferrule*{
                \cdots
              }{
                 \ottmv{a}  \ottsym{,}  \ottmv{b}   \vdash  \subL{   \forall  \ottmv{b}  .  \ottmv{b}    \rightarrow   \ottmv{b}  }{ \ottmv{a}  \rightarrow  \ottmv{a} } 
              }
            }{
               \ottmv{a}   \vdash  \subL{   \forall  \ottmv{b}  .  \ottmv{b}    \rightarrow   \ottmv{b}  }{  \forall  \ottmv{b}  .  \ottmv{a}   \rightarrow  \ottmv{a} } 
            }
          }{
             \ottmv{a}   \vdash  \subL{   \forall  \ottmv{a}  .   \forall  \ottmv{b}  .  \ottmv{b}     \rightarrow   \ottmv{b}  }{  \forall  \ottmv{b}  .  \ottmv{a}   \rightarrow  \ottmv{a} } 
          }
        }{
            \emptyset    \vdash  \subL{   \forall  \ottmv{a}  .   \forall  \ottmv{b}  .  \ottmv{b}     \rightarrow   \ottmv{b}  }{  \forall  \ottmv{a}  .   \forall  \ottmv{b}  .  \ottmv{a}    \rightarrow  \ottmv{a} } 
        }
      \end{equation}
    \end{minipage}
  \end{tabular}
\end{center}
Therefore,
we require all type information within a universal type to be given.

\mycomment{
  We introduce a key idea of mixed-direction types before presenting their syntax.
  Mixed-direction types originate from colored types~\cite{odersky_colored_2001}.
  They specify which part of a type should be inferred or checked.
  In a mixed-direction type, a part with a checking marker (${}^\Leftarrow$)
  means that the part is information about given types, i.e., should be checked.
  On the other hand, a part without a checking marker means that
  the part is an inferred type.
  For example, suppose that the mixed-direction type ${A^\Box}$ is $\texttt{\textcolor{red}{<<no parses (char 3): <= ***(unit -> <= unit) >>}}$.
  In this case, the given type information of ${A^\Box}$ is that
  ${A^\Box}$ is a function type and its return type is $ \mathsf{unit} $;
  and the inferred type information is that the return type of ${A^\Box}$ is $ \mathsf{unit} $.
  We prevent an inferred type from including information about given types.
  This restriction is similar to the work of Vytiniotis et al.~\citet{vytiniotis_boxy_2006}.

  The syntax for mixed-direction types is straightforward except for
  a universal quantification checking $\texttt{\textcolor{red}{<<no parses (char 3): <= ***forall a . A >>}}$,
  following the intuition explained above.
  We use a type $\ottnt{A}$, not a mixed-direction type, inside a universal quantification.
  The following informal discussion justifies this restriction.
  Consider the judgment
  \[
    \texttt{\textcolor{red}{<<no parses (char 41): x : forall a . forall b . b -> b \mbox{$\mid$}- x : <***= forall a . \mbox{?} >>}}.
  \]
  This judgment means that
  we infer the type under the universal quantifier ${}^\Leftarrow \forall a$
  to fill the hole denoted as $ \text{?} $.
  We cannot determine a type as a result of this inference.
  For example, both filling the hole with $ \ottmv{a}   \rightarrow   \ottmv{a} $ and filling it with $  \forall  \ottmv{b}  .  \ottmv{b}    \rightarrow   \ottmv{b} $
  are valid, because the subtyping judgments
  \[
     \emptyset   \vdash_{ \mathrm{DK} }    \forall  \ottmv{a}  .   \forall  \ottmv{b}  .  \ottmv{b}     \rightarrow   \ottmv{b}   \mathbin{<:}    \forall  \ottmv{a}  .  \ottmv{a}    \rightarrow   \ottmv{a} 
  \]
  and
  \[
     \emptyset   \vdash_{ \mathrm{DK} }    \forall  \ottmv{a}  .   \forall  \ottmv{b}  .  \ottmv{b}     \rightarrow   \ottmv{b}   \mathbin{<:}    \forall  \ottmv{a}  .   \forall  \ottmv{b}  .  \ottmv{b}     \rightarrow   \ottmv{b} 
  \]
  hold in the DK system.
  Thus, we conservatively require that
  all the type information inside a universal quantification should be given.
  To avoid defining an additional syntax category for
  the mixed-direction types that include no type $\ottnt{A}$,
  we simply write $\ottnt{A}$ inside a universal quantification.
}

Typing contexts are standard\footnote{
  We use $\Psi$ rather than $\Gamma$ to denote a typing context,
  following Zhao et al.~\cite{zhao_mechanical_2019}.
}.
Note that,
following existing systems~\cite{odersky_colored_2001,vytiniotis_boxy_2006},
typing contexts are always assumed to be complete inputs.  Thus,
the types in a typing context cannot include boxes.

\subsection{Subtyping}

\begin{figure}[t]
  \small
  \noindent\fbox{$ \Psi   \vdash  \subL{ \ottnt{A} }{ {B^\Box} } $}\quad\fbox{$ \Psi   \vdash  \subR{ {A^\Box} }{ \ottnt{B} } $}
  \hfill\mbox{}
  \begin{mathpar}
    \inferrule{\Psi  \vdash  \ottnt{A}}{ \Psi   \vdash  \subL{ \ottnt{A} }{  \boxed{ \ottnt{A} }  } }\ (\SLBox)

    \inferrule{\Psi  \vdash  \ottnt{A}}{ \Psi   \vdash  \subR{  \boxed{ \ottnt{A} }  }{ \ottnt{A} } }\ (\SRBox)

    \inferrule{
      \Psi  \vdash   \mathsf{unit} 
    }{
       \Psi   \vdash  \subL{  \mathsf{unit}  }{  \mathsf{unit}  } 
    }\ (\SLUnit)

    \inferrule{
      \Psi  \vdash   \mathsf{unit} 
    }{
       \Psi   \vdash  \subR{  \mathsf{unit}  }{  \mathsf{unit}  } 
    }\ (\SRUnit)

    \inferrule{
      \Psi  \vdash  \ottmv{a}
    }{
       \Psi   \vdash  \subL{ \ottmv{a} }{ \ottmv{a} } 
    }\ (\SLTVar)

    \inferrule{
      \Psi  \vdash  \ottmv{a}
    }{
       \Psi   \vdash  \subR{ \ottmv{a} }{ \ottmv{a} } 
    }\ (\SRTVar)

    \inferrule{
     \Psi   \vdash  \subR{ {A^\Box} }{ \ottnt{A} }  \\  \Psi   \vdash  \subL{ \ottnt{B} }{ {B^\Box} } 
    }{
       \Psi   \vdash  \subL{  \ottnt{A}   \rightarrow   \ottnt{B}  }{ {A^\Box}  \rightarrow  {B^\Box} } 
    }\ (\SLFun)

    \inferrule{
     \Psi   \vdash  \subL{ \ottnt{A} }{ {A^\Box} }  \\  \Psi   \vdash  \subR{ {B^\Box} }{ \ottnt{B} } 
    }{
       \Psi   \vdash  \subR{ {A^\Box}  \rightarrow  {B^\Box} }{  \ottnt{A}   \rightarrow   \ottnt{B}  } 
    }\ (\SRFun)

    \inferrule{
       \Psi  \ottsym{,}  \ottmv{a}   \vdash  \subL{ \ottnt{A} }{ \ottnt{B} } 
    }{
       \Psi   \vdash  \subL{ \ottnt{A} }{  \forall  \ottmv{a}  .  \ottnt{B}  } 
    }\ (\SLAllR)

    \inferrule{
    \mathsf{not}\square \, \ottsym{(}  {A^\Box}  \ottsym{)} \\  \Psi  \ottsym{,}  \ottmv{a}   \vdash  \subR{ {A^\Box} }{ \ottnt{B} } 
    }{
       \Psi   \vdash  \subR{ {A^\Box} }{  \forall  \ottmv{a}  .  \ottnt{B}  } 
    }\ (\SRAllR)

    \inferrule{
    \mathsf{not}\square \, \ottsym{(}  {B^\Box}  \ottsym{)} \\
    \Psi  \vdash  \tau \\  \Psi   \vdash  \subL{  [  \tau  /  \ottmv{a}  ]  \ottnt{A}  }{ {B^\Box} } 
    }{
       \Psi   \vdash  \subL{  \forall  \ottmv{a}  .  \ottnt{A}  }{ {B^\Box} } 
    }\ (\SLAllL)

    \inferrule{
    \Psi  \vdash  \tau \\  \Psi   \vdash  \subR{  [  \tau  /  \ottmv{a}  ]  \ottnt{A}  }{ \ottnt{B} } 
    }{
       \Psi   \vdash  \subR{  \forall  \ottmv{a}  .  \ottnt{A}  }{ \ottnt{B} } 
    }\ (\SRAllL)
  \end{mathpar}
  \caption{Subtyping}\label{fig:subtyping}
\end{figure}

We have two subtyping judgments, $ \Psi   \vdash  \subL{ \ottnt{A} }{ {B^\Box} } $ and $ \Psi   \vdash  \subR{ {A^\Box} }{ \ottnt{B} } $~.
The box as a subscript indicates on which side boxes can appear.
In the subtyping judgment $ \Psi   \vdash  \subL{ \ottnt{A} }{ {B^\Box} } $,
the input consists of the typing context $\Psi$, the type $\ottnt{A}$, and
the part of the boxy type ${B^\Box}$ outside boxes,
and the output is the part of the boxy type ${B^\Box}$ inside boxes.
For example, the subtyping judgment $  \emptyset    \vdash  \subL{   \forall  \ottmv{a}  .  \ottmv{a}    \rightarrow   \ottmv{a}  }{  \boxed{  \mathsf{unit}  }   \rightarrow   \mathsf{unit}  } $
means that, given the empty typing context, the subtype $  \forall  \ottmv{a}  .  \ottmv{a}    \rightarrow   \ottmv{a} $, and
the information that the supertype should be a function type
with the return type $ \mathsf{unit} $,
our subtyping system outputs that the argument type of the supertype is $ \mathsf{unit} $.
The subtyping judgment $ \Psi   \vdash  \subR{ {A^\Box} }{ \ottnt{B} } $ is the reversed version of $ \Psi   \vdash  \subL{ \ottnt{A} }{ {B^\Box} } $
in the sense that the boxy type is a subtype, not a supertype.
We show the subtyping rules in \zcref{fig:subtyping}.
We use the names prefixed with ``SL'' (resp. ``SR'') for
the rules of the subtyping judgment $ \Psi   \vdash  \subL{ \ottnt{A} }{ {B^\Box} } $ (resp. $ \Psi   \vdash  \subR{ {A^\Box} }{ \ottnt{B} } $).

The rules \SLBox and \SRBox are for
cases with no given type information in a boxy type.
Thus, they return the type $\ottnt{A}$ without changing it.
The rules \SLUnit, \SRUnit, \SLTVar, \SRTVar, \SLFun, and \SRFun are straightforward.
The outermost type constructor of the boxy type is the same as that of the type.

The rules \SLAllR and \SRAllR generalize a polymorphic variable,
similarly to the corresponding rules in the DK and XO systems.
Note that these rules never instantiate the type in a box
because the type is guessed and should be reused without changing.
\SRAllR has the premise condition $\mathsf{not}\square \, \ottsym{(}  {A^\Box}  \ottsym{)}$, which
means that the outermost constructor is not a box. %
We need this condition to ensure that
the synthesis result, namely the type in a box, is unique
if the guessing oracle outputs the same monotype.
Without this condition, the subtyping judgments
$  \emptyset    \vdash  \subR{  \boxed{  \mathsf{unit}  }  }{  \forall  \ottmv{a}  .   \mathsf{unit}   } $ and
$  \emptyset    \vdash  \subR{  \boxed{  \forall  \ottmv{a}  .   \mathsf{unit}   }  }{  \forall  \ottmv{a}  .   \mathsf{unit}   } $ would hold,
but they have different synthesis results, $ \mathsf{unit} $ and $ \forall  \ottmv{a}  .   \mathsf{unit}  $.
The condition $\mathsf{not}\square \, \ottsym{(}  {A^\Box}  \ottsym{)}$ excludes the former judgment, and
only the latter is valid via \SRBox.

The rules \SLAllL and \SRAllL instantiate quantified type variables
with guessed monotypes.
These rules are also similar to the corresponding rules in the DK and XO systems.
In both rules, guessed and substituted types must be monomorphic
because our system is predicative.
The condition $\mathsf{not}\square \, \ottsym{(}  {B^\Box}  \ottsym{)}$ in the rule \SLAllL
is required for the same reason as $\mathsf{not}\square \, \ottsym{(}  {A^\Box}  \ottsym{)}$ in \SRAllR.

\subsection{Typing}\label{subsec:typing}

\begin{figure}[t]
  \noindent\fbox{$\Psi  \vdash  \ottnt{e}  \ottsym{:}  {A^\Box}$}\hfill\vspace{1ex}
  \small
  \begin{mathpar}
    \inferrule{
       \Psi   \vdash  \subL{  \mathsf{unit}  }{ {A^\Box} } 
    }{
      \Psi  \vdash  \ottsym{()}  \ottsym{:}  {A^\Box}
    }\ (\TUnit)

    \inferrule{
     \mathit{x}  \ottsym{:}  \ottnt{A}  \in  \Psi  \\  \Psi   \vdash  \subL{ \ottnt{A} }{ {A^\Box} } 
    }{
      \Psi  \vdash  \mathit{x}  \ottsym{:}  {A^\Box}
    }\ (\TVar)

    \inferrule{
      \Psi  \ottsym{,}  \ottmv{a}  \vdash  \ottnt{e}  \ottsym{:}  \ottnt{A}
    }{
      \Psi  \vdash  \ottnt{e}  \ottsym{:}   \forall  \ottmv{a}  .  \ottnt{A} 
    }\ (\TAll)

    \inferrule{
    \Psi  \vdash  \tau \\ \Psi  \ottsym{,}  \mathit{x}  \ottsym{:}  \tau  \vdash  \ottnt{e}  \ottsym{:}   \boxed{ \ottnt{B} } 
    }{
      \Psi  \vdash   \lambda  \mathit{x}  .  \ottnt{e}   \ottsym{:}   \boxed{  \tau   \rightarrow   \ottnt{B}  } 
    }\ (\TAbsBox)

    \inferrule{
      \Psi  \ottsym{,}  \mathit{x}  \ottsym{:}  \ottnt{A}  \vdash  \ottnt{e}  \ottsym{:}  {B^\Box}
    }{
      \Psi  \vdash   \lambda  \mathit{x}  .  \ottnt{e}   \ottsym{:}  \ottnt{A}  \rightarrow  {B^\Box}
    }\ (\TAbsFunA)

    \inferrule{
     \Psi   \vdash  \subR{ {A^\Box} }{ \tau }  \\ \Psi  \ottsym{,}  \mathit{x}  \ottsym{:}  \tau  \vdash  \ottnt{e}  \ottsym{:}  {B^\Box}
    }{
      \Psi  \vdash   \lambda  \mathit{x}  .  \ottnt{e}   \ottsym{:}  {A^\Box}  \rightarrow  {B^\Box}
    }\ (\TAbsFunB)
    \\
    \inferrule{
      \Psi  \ottsym{,}  \mathit{x}  \ottsym{:}  \ottnt{A}  \vdash  \ottnt{e}  \ottsym{:}   \boxed{ \ottnt{B} } 
    }{
      \Psi  \vdash   \lambda  \mathit{x}  \ottsym{:}  \ottnt{A}  .  \ottnt{e}   \ottsym{:}   \boxed{  \ottnt{A}   \rightarrow   \ottnt{B}  } 
    }\ (\TAAbsBox)

    \inferrule{
     \Psi   \vdash  \subR{ {A^\Box} }{ \ottnt{A} }  \\ \Psi  \ottsym{,}  \mathit{x}  \ottsym{:}  \ottnt{A}  \vdash  \ottnt{e}  \ottsym{:}  {B^\Box}
    }{
      \Psi  \vdash   \lambda  \mathit{x}  \ottsym{:}  \ottnt{A}  .  \ottnt{e}   \ottsym{:}  {A^\Box}  \rightarrow  {B^\Box}
    }\ (\TAAbsFun)
    \\
    \inferrule{
    \Psi  \vdash  \ottnt{e_{{\mathrm{1}}}}  \ottsym{:}   \boxed{ \ottnt{A} }   \rightarrow  {A^\Box} \\ \Psi  \vdash  \ottnt{e_{{\mathrm{2}}}}  \ottsym{:}  \ottnt{A}
    }{
      \Psi  \vdash  \ottnt{e_{{\mathrm{1}}}}  \appright  \ottnt{e_{{\mathrm{2}}}}  \ottsym{:}  {A^\Box}
    }\ (\TAppR)

    \inferrule{
    \Psi  \vdash  \ottnt{e_{{\mathrm{1}}}}  \ottsym{:}  \ottnt{A}  \rightarrow  {A^\Box} \\ \Psi  \vdash  \ottnt{e_{{\mathrm{2}}}}  \ottsym{:}   \boxed{ \ottnt{A} } 
    }{
      \Psi  \vdash  \ottnt{e_{{\mathrm{1}}}}  \appleft  \ottnt{e_{{\mathrm{2}}}}  \ottsym{:}  {A^\Box}
    }\ (\TAppL)
  \end{mathpar}
  \caption{Typing}\label{fig:typing}
\end{figure}

Our typing judgment takes the form of $\Psi  \vdash  \ottnt{e}  \ottsym{:}  {A^\Box}$~.
Note that we only have a single judgment form
because using boxy types enables us to unify type synthesis and checking,
with $\Psi  \vdash  \ottnt{e}  \ottsym{:}   \boxed{ \ottnt{A} } $ and $\Psi  \vdash  \ottnt{e}  \ottsym{:}  \ottnt{A}$ corresponding to
type synthesis and checking respectively.
In the typing judgment $\Psi  \vdash  \ottnt{e}  \ottsym{:}  {A^\Box}$,
the input consists of $\Psi$, $\ottnt{e}$, and
the part of the boxy type ${A^\Box}$ \emph{outside} boxes.
The output is a part of the boxy type ${A^\Box}$ \emph{inside} boxes.
For example, the typing judgment $\mathit{x}  \ottsym{:}    \forall  \ottmv{a}  .  \ottmv{a}    \rightarrow   \ottmv{a}   \vdash  \mathit{x}  \ottsym{:}   \boxed{  \mathsf{unit}  }   \rightarrow   \mathsf{unit} $
means that, given the typing context $\mathit{x}  \ottsym{:}   \forall  \ottmv{a}  .  \ottmv{a}   \rightarrow  \ottmv{a}$, the expression $\mathit{x}$, and
the information that $\mathit{x}$ must have a function type with the return type $ \mathsf{unit} $,
our typing system outputs that the argument type is $ \mathsf{unit} $.

The rules \TUnit and \TVar are standard except that
we use $ \Psi   \vdash  \subL{  \mathsf{unit}  }{ {A^\Box} } $ and $ \Psi   \vdash  \subL{ \ottnt{A} }{ {A^\Box} } $
in their respective premises.
These premises enable our type system to represent
type synthesis and checking in one typing rule.
The rule \TAll is also standard and similar to the corresponding rule in the DK system.

There are three rules \TAbsBox, \TAbsFunA, and \TAbsFunB for
lambda abstractions without type annotations
and two rules \TAAbsBox and \TAAbsFun for those with.
\TAbsBox applies in the case where no type information is given, i.e.,
the entire type must be synthesized.
In this case, a monotype $\tau$ is guessed as the argument type,
similarly to the DK, XO, and other bidirectional type
systems~\cite{jones_practical_2007,vytiniotis_boxy_2006,zhao_mechanical_2019}
that support higher-rank polymorphism.
\TAbsFunA applies when the argument type is fully given (by $\ottnt{A}$), in which case
we need no guess.
\TAbsFunB applies when the argument type is partially given (by ${A^\Box}$).
The condition $\mathsf{hasBox} \, \ottsym{(}  {A^\Box}  \ottsym{)}$ means that there is no type $\ottnt{A}$ such that ${A^\Box} = \ottnt{A}$---in other words, at least one box appears in ${A^\Box}$---to avoid the overlap with \TAbsFunA.
In this case, we need to guess an argument type $\tau$
and check that the guess matches ${A^\Box}$ by using the subtyping judgment $ \Psi   \vdash  \subR{ {A^\Box} }{ \tau } $.
\TAAbsBox is for the case where the type has to be fully synthesized.
\TAAbsFun is for the case where a part of the type is given.  The second premise means that
$ \lambda  \mathit{x}  \ottsym{:}  \ottnt{A}  .  \ottnt{e} $ can be given type $\ottnt{A}  \rightarrow  {B^\Box}$, which has to be a subtype of ${A^\Box}  \rightarrow  {B^\Box}$, hence the first premise.

The rule \TAppR is for function applications with the function-first guide.
It means that
the argument type of the function is first synthesized, and
the argument is checked against that type. %

The rule \TAppL is for function applications with the argument-first guide.
This rule means that
the type $\ottnt{A}$ of the argument is first synthesized, and
the function is checked against the type that includes the synthesized argument type.

\subsection{Metatheory}\label{subsec:decl_metatheory}

\paragraph{Soundness.}\label{subsubsec:sound}

The type system of \lang is sound with respect to \sysfi~\cite{cretin_power_2012},
an extension of System F with type coercions. %
We choose \sysfi, not System F,
because we require a base calculus to be equipped with a mechanism
to encode our subtyping derivations.

We briefly explain the auxiliary functions before stating the soundness theorem.
$ |  {A^\Box}  | _ \square $ is the type that results from removing the boxes from the boxy type ${A^\Box}$.
$ |  \ottnt{e}  | $ is the expression in the untyped lambda-calculus
that results from eliminating type annotations and application guides
from $\ottnt{e}$.
This translation preserves the reduction of the source language
because neither type annotations nor application guides influence the reduction.
Given the expression $\ottnt{M}$ of \sysfi,
$ |  \ottnt{M}  |_{ \mathrm{Coer} } $ is the expression in the untyped lambda-calculus that results by eliminating
type annotations, type abstractions, and coercions.
This translation preserves the original semantics,
as proven in Cretin and R{\'e}my~\citet{cretin_power_2012}.

\begin{theorem}[Soundness w.r.t.\ \sysfi]\label{thm:sound_sysfi}
  If $\Psi  \vdash  \ottnt{e}  \ottsym{:}  {A^\Box}$, then there exists some $\ottnt{M}$ such that
  $\Psi  \vdash_{ \mathrm{F}_{\iota} }  \ottnt{M}  \ottsym{:}   |  {A^\Box}  | _ \square $ and $  |  \ottnt{e}  |     \mathbin{=}     |  \ottnt{M}  |_{ \mathrm{Coer} }  $.
\end{theorem}

\mycomment{
  We define $ |  {A^\Box}  | _ \square $, $ |  \ottnt{e}  | $, and $ |  \ottnt{e}  |_{ \mathrm{Coer} } $ as follows.
  \[
    \begin{array}{r@{\ }c@{\ }l}
                  |  \ottnt{A}  | _ \square                                                       & = & \ottnt{A}                  \\
                 \texttt{\textcolor{red}{<<no parses (char 3): \mbox{$\mid$}<=*** unit\mbox{$\mid$} >>}}                                                        & = &  \mathsf{unit}                     \\
                 \texttt{\textcolor{red}{<<no parses (char 3): \mbox{$\mid$}<=*** a\mbox{$\mid$} >>}}                                                           & = & \ottmv{a}                       \\
                 \texttt{\textcolor{red}{<<no parses (char 4): \mbox{$\mid$} <=*** (P -> Q) \mbox{$\mid$} >>}}                                                  & = &   |  {A^\Box}  | _ \square    \rightarrow    |  {B^\Box}  | _ \square                \\
                 \texttt{\textcolor{red}{<<no parses (char 4): \mbox{$\mid$} <=*** forall a. A \mbox{$\mid$} >>}}                                               & = &  \forall  \ottmv{a}  .  \ottnt{A}             \\[2ex]
                  |  \ottsym{()}  |                                                        & = & \ottsym{()}              \\
                  |  \mathit{x}  |                                                               & = & \mathit{x}                       \\
                  |   \lambda  \mathit{x}  .  \ottnt{e}   |                                                            & = &  \lambda  \mathit{x}  .   |  \ottnt{e}  |                   \\
                  |   \lambda  \mathit{x}  \ottsym{:}  \ottnt{A}  .  \ottnt{e}   |                                                          & = &  \lambda  \mathit{x}  .   |  \ottnt{e}  |                   \\
                  |  \ottnt{e_{{\mathrm{1}}}}  \appright  \ottnt{e_{{\mathrm{2}}}}  |                                                        & = &  |  \ottnt{e_{{\mathrm{1}}}}  |  \,  |  \ottnt{e_{{\mathrm{2}}}}  |                \\
                  |  \ottnt{e_{{\mathrm{1}}}}  \appleft  \ottnt{e_{{\mathrm{2}}}}  |                                                        & = &  |  \ottnt{e_{{\mathrm{1}}}}  |  \,  |  \ottnt{e_{{\mathrm{2}}}}  |                \\[2ex]
                  |  \ottsym{()}  |_{ \mathrm{Coer} }                                                        & = & \ottsym{()}        \\
                  |  \mathit{x}  |_{ \mathrm{Coer} }                                                         & = & \mathit{x}                       \\
                  |   \lambda  \mathit{x}  \ottsym{:}  \ottnt{A}  .  \ottnt{e}   |_{ \mathrm{Coer} }                                                 & = &  \lambda  \mathit{x}  .   |  \ottnt{e}  |_{ \mathrm{Coer} }             \\
                  |  \ottnt{e_{{\mathrm{1}}}} \, \ottnt{e_{{\mathrm{2}}}}  |_{ \mathrm{Coer} }                                                     & = &  |  \ottnt{e_{{\mathrm{1}}}}  |_{ \mathrm{Coer} }  \,  |  \ottnt{e_{{\mathrm{2}}}}  |_{ \mathrm{Coer} }       \\
                  |   \Lambda  \ottmv{a}  .  \ottnt{e}   |_{ \mathrm{Coer} }  & = &  |  \ottnt{e}  |_{ \mathrm{Coer} }  \\
                  |   \ottnt{C}  \langle  \ottnt{e}  \rangle   |_{ \mathrm{Coer} }                                                     & = &  |  \ottnt{e}  |_{ \mathrm{Coer} } 
    \end{array}
  \]
}

\mycomment{
  We briefly explain the auxiliary definitions in the statement of \zcref{thm:sound_sysfi}.
  $ |  {A^\Box}  | _ \square $ is the type that results from removing checking marks from the mixed-direction type ${A^\Box}$.
  $ |  \ottnt{e}  | $ is the expression that results from eliminating type annotations and application signposts
  from the expression $\ottnt{e}$.
  $ |  \ottnt{e}  |_{ \mathrm{Coer} } $ is the expression that results from eliminating
  type annotations, type abstractions, and coercions from the expression $\ottnt{e}$.
}

\paragraph{Completeness.}\label{subsubsec:complete}

The type system of \lang is complete with respect to both the DK and XO systems.

We introduce a few notations to state the completeness theorem regarding the DK system.
We use the translation $ |  \ottnt{e}  |_{ \mathrm{DK} } $ from the expression $\ottnt{e}$ in the DK system to
the expression in \lang.
This translation simply bridges the syntactic gap between the DK system and \lang.
Since the DK system adopts function-first typing, we translate a function application in the DK system
to a function application with the function-first guide $ \appright $.
Furthermore, because the DK system allows for annotating any expression with a type,
we translate a type-annotated expression $\ottnt{e}  \ottsym{:}  \ottnt{A}$ to a function application $\ottsym{(}   \lambda  \mathit{x}  \ottsym{:}  \ottnt{A}  .  \mathit{x}   \ottsym{)}  \appright  \ottnt{e}$.
These changes do not affect the reduction of the DK language.

\begin{theorem}[Completeness w.r.t.\ the DK System]\label{thm:complete_dk}
  \mbox{}\begin{itemize}
    \item If $\Psi  \vdash_{ \mathrm{DK} }  \ottnt{e}  \Rightarrow  \ottnt{A}$, then there exists some $\ottnt{B}$ such that
          $\Psi  \vdash_{ \mathrm{DK} }  \ottnt{B}  \mathbin{<:}  \ottnt{A}$ and $\Psi  \vdash   |  \ottnt{e}  |_{ \mathrm{DK} }   \ottsym{:}   \boxed{ \ottnt{B} } $.
    \item If $\Psi  \vdash_{ \mathrm{DK} }  \ottnt{e}  \Leftarrow  \ottnt{A}$, then $\Psi  \vdash   |  \ottnt{e}  |_{ \mathrm{DK} }   \ottsym{:}  \ottnt{A}$.
  \end{itemize}
\end{theorem}

\mycomment{
  In the complete theorem, the translation $ |  \ottnt{e}  |_{ \mathrm{DK} } $ is defined as follows.
  \[
    \begin{array}{r@{\ }c@{\ }l}
                  |  \ottsym{()}  |_{ \mathrm{DK} }  & = & \ottsym{()}        \\
                  |  \mathit{x}  |_{ \mathrm{DK} }        & = & \mathit{x}                 \\
                  |   \lambda  \mathit{x}  .  \ottnt{e}   |_{ \mathrm{DK} }     & = &  \lambda  \mathit{x}  .   |  \ottnt{e}  |_{ \mathrm{DK} }            \\
                  |  \ottnt{e_{{\mathrm{1}}}} \, \ottnt{e_{{\mathrm{2}}}}  |_{ \mathrm{DK} }    & = &  |  \ottnt{e_{{\mathrm{1}}}}  |_{ \mathrm{DK} }   \appright   |  \ottnt{e_{{\mathrm{2}}}}  |_{ \mathrm{DK} }   \\
                  |  \ottnt{e}  \ottsym{:}  \ottnt{A}  |_{ \mathrm{DK} }    & = & \ottsym{(}   \lambda  \mathit{x}  \ottsym{:}  \ottnt{A}  .  \mathit{x}   \ottsym{)}  \appright   |  \ottnt{e}  |_{ \mathrm{DK} } 
    \end{array}
  \]
}

A notable point is that we use the subtyping relation $\Psi  \vdash_{ \mathrm{DK} }  \ottnt{B}  \mathbin{<:}  \ottnt{A}$, not the equality on types.
This choice comes from the limitation of the DK system.
The DK system only synthesizes a monotype for a lambda abstraction.
For example, it can derive only the former of the following two judgments.
\begin{align*}\small
   & \mathit{f}  \ottsym{:}    \forall  \ottmv{a}  .  \ottmv{a}    \rightarrow   \ottmv{a}   \vdash_{ \mathrm{DK} }   \lambda  \mathit{x}  .  \mathit{f}   \Rightarrow    \mathsf{unit}    \rightarrow     \mathsf{unit}    \rightarrow    \mathsf{unit}         \\
   & \mathit{f}  \ottsym{:}    \forall  \ottmv{a}  .  \ottmv{a}    \rightarrow   \ottmv{a}   \vdash_{ \mathrm{DK} }   \lambda  \mathit{x}  .  \mathit{f}   \Rightarrow     \mathsf{unit}    \rightarrow    \forall  \ottmv{a}  .  \ottmv{a}     \rightarrow   \ottmv{a} 
\end{align*}
\mycomment{
  For example, it can derive the following judgment
  \[
    \mathit{f}  \ottsym{:}    \forall  \ottmv{a}  .  \ottmv{a}    \rightarrow   \ottmv{a}   \vdash_{ \mathrm{DK} }   \lambda  \mathit{x}  .  \mathit{f}   \Rightarrow    \mathsf{unit}    \rightarrow     \mathsf{unit}    \rightarrow    \mathsf{unit}   ,
  \]
  but cannot
  \[
    \mathit{f}  \ottsym{:}    \forall  \ottmv{a}  .  \ottmv{a}    \rightarrow   \ottmv{a}   \vdash_{ \mathrm{DK} }   \lambda  \mathit{x}  .  \mathit{f}   \Rightarrow     \mathsf{unit}    \rightarrow    \forall  \ottmv{a}  .  \ottmv{a}     \rightarrow   \ottmv{a} .
  \]
}
However, the XO system and ours can derive only the latter. %
Thus, the completeness theorem must state that
the type system of \lang may synthesize a more general type
than the DK system does.

To state the completeness theorem regarding the XO system,
we use the two auxiliary functions $ |  \ottnt{e}  |_{ \mathrm{XO} } $ and $ \mathsf{FoldCurry} ( \Xi ,  \ottnt{A} ) $.
We translate the expression $\ottnt{e}$ in the XO system to
the expression $ |  \ottnt{e}  |_{ \mathrm{XO} } $ in \lang by marking every function application
with $ \appleft $.
Because the XO system stacks the synthesized type information about the arguments
in an application context $\Xi$ as explained in \zcref{subsec:ffaf},
we embed such information in a boxy type using $ \mathsf{FoldCurry} ( \Xi ,  \ottnt{A} ) $ defined as follows.
\[
  \begin{array}{r@{\ }c@{\ }l@{\qquad}r@{\ }c@{\ }l}
                \mathsf{FoldCurry} (  \emptyset  ,  \ottnt{A} )  & = &  \boxed{ \ottnt{A} }                 &
                \mathsf{FoldCurry} ( \ottsym{(}  \Xi  \ottsym{,}  \ottnt{B}  \ottsym{)} ,   \ottnt{B}   \rightarrow   \ottnt{A}  )            & = & \ottnt{B}  \rightarrow   \mathsf{FoldCurry} ( \Xi ,  \ottnt{A} ) 
  \end{array}
\]
Then, the completeness theorem is stated as follows:
\begin{theorem}[Completeness w.r.t.\ the XO System]\label{thm:complete_xo}
  If $\Psi  \mid  \Xi  \vdash_{ \mathrm{XO} }  \ottnt{e}  \Rightarrow  \ottnt{A}$, then $\Psi  \vdash   |  \ottnt{e}  |_{ \mathrm{XO} }   \ottsym{:}   \mathsf{FoldCurry} ( \Xi ,  \ottnt{A} ) $.
\end{theorem}

\section{Algorithmic Typing}\label{sec:algo_short}

This section briefly explains our typing algorithm.
Its formal definition can be found in
\arxiv{\zcref{sec:algo}}{the full version of this paper}.

\subsection{Typing Algorithm based on the Worklist Approach}

Our typing algorithm is based on the \emph{worklist approach}.
It was originally proposed by Zhao et al.~\cite{zhao_formalization_2018}
to  formalize an algorithm for polymorphic subtyping~\cite{odersky_putting_1996}.
Afterward, Zhao et al.~\cite{zhao_mechanical_2019} extended it to the DK system.

A \emph{worklist} is a typing context extended with
\emph{existential variables} and subtyping/typing judgments, defined as follows.
\[
  \small
  \Gamma  \Coloneqq   \emptyset   \mid  \Gamma  \ottsym{,}  \ottmv{a}  \mid  \Gamma  \ottsym{,}  \monoalpha  \mid  \Gamma  \ottsym{,}  \mathit{x}  \ottsym{:}  \ottnt{A}  \mid 
  \Gamma  \Vdash   \subL{ \ottnt{A} }{ {B^\Box} }   \mid  \Gamma  \Vdash   \subR{ {A^\Box} }{ \ottnt{B} }   \mid  \Gamma  \Vdash  \ottnt{e}  \ottsym{:}  {A^\Box} \ \cdots
\]
An existential variable $\monoalpha$ in $\Gamma_{{\mathrm{1}}}  \ottsym{,}  \monoalpha  \ottsym{,}  \Gamma_{{\mathrm{2}}}$ is solved to
only a monotype that is well-formed under $\Gamma_{{\mathrm{1}}}$.
In other words, a typing algorithm based on the worklist approach
propagates the guessed monotype through an existential variable.

A typing algorithm using a worklist is defined as a reduction $\Gamma_{{\mathrm{1}}}  \rightsquigarrow  \Gamma_{{\mathrm{2}}}$.
It means that a typing algorithm, given a worklist $\Gamma_{{\mathrm{1}}}$,
pops the rightmost judgment in $\Gamma_{{\mathrm{1}}}$ and processes it in one step,
resulting in $\Gamma_{{\mathrm{2}}}$.
For example, let us consider the following reduction.
\[
  \small
   \emptyset   \Vdash   \subL{  \ottnt{A}   \rightarrow   \ottnt{B}  }{ {A^\Box}  \rightarrow  {B^\Box} }  \quad  \rightsquigarrow  \quad
   \emptyset   \Vdash   \subL{ \ottnt{B} }{ {B^\Box} }   \Vdash   \subR{ {A^\Box} }{ \ottnt{A} } 
\]
The left-hand side worklist means that the first judgment of a typing algorithm is to check that
the type $ \ottnt{A}   \rightarrow   \ottnt{B} $ is a subtype of ${A^\Box}  \rightarrow  {B^\Box}$ under the empty typing context.
This reduction means that we need to check that
${A^\Box}$ (resp. $\ottnt{B}$) is a subtype of $\ottnt{A}$ (resp. ${B^\Box}$)
to check that $ \ottnt{A}   \rightarrow   \ottnt{B} $ is a subtype of ${A^\Box}  \rightarrow  {B^\Box}$.
In this sense, a worklist can be seen as a ``flattened'' version of a corresponding derivation tree.
After we learn that the two subtyping relations hold,
the typing process removes the completed work items and
reaches the empty worklist $ \emptyset $, indicating that
the typing process has successfully finished.
Generally speaking,
a typing process successfully ends if $\Gamma  \rightsquigarrow^*   \emptyset $ holds,
where $ \rightsquigarrow^* $ is defined as the reflexive and transitive closure of $ \rightsquigarrow $.

We introduce \emph{polymorphic existential variables} to adapt
the worklist approach to boxy types.
A boxy type may have multiple ``output channels''; e.g.,
$ \boxed{  \mathsf{unit}  }   \rightarrow   \boxed{   \forall  \ottmv{a}  .  \ottmv{a}    \rightarrow   \ottmv{a}  } $ means that
the output types are $ \mathsf{unit} $ and $  \forall  \ottmv{a}  .  \ottmv{a}    \rightarrow   \ottmv{a} $.
Therefore, our typing algorithm must propagate
the (polymorphic) types that fill up boxes
from one work item to the remaining work items.
We assign a polymorphic existential variable to each box in a boxy type
to support this propagation.
This idea, assigning a special variable to each box,
is also used in the type inference algorithm of
the original work on boxy types~\cite{vytiniotis_boxy_2006}.

We use the following reduction example to explain
how polymorphic existential variables work in our typing algorithm.
\[
  \small
  \Gamma  \ottsym{,}  \triangleright  \ottsym{\{}  \vec{\polybeta}  \ottsym{\}}  \Vdash  \ottnt{e_{{\mathrm{1}}}}  \appright  \ottnt{e_{{\mathrm{2}}}}  \ottsym{:}  {A^\Box} \quad  \rightsquigarrow  \quad
  \Gamma  \ottsym{,}  \triangleright  \ottsym{\{}  \vec{\polybeta}  \ottsym{,}  \polyalpha  \ottsym{\}}  \Vdash  \ottnt{e_{{\mathrm{2}}}}  \ottsym{:}  \polyalpha  \Vdash  \ottnt{e_{{\mathrm{1}}}}  \ottsym{:}   \boxed{ \polyalpha }   \rightarrow  {A^\Box}
\]
$\triangleright  \ottsym{\{}  \vec{\polybeta}  \ottsym{\}}$ in the former worklist accumulates
polymorphic existential variables.
After this reduction, our typing algorithm synthesizes
an argument type, which can be polymorphic, for $\ottnt{e_{{\mathrm{1}}}}$, and
uses it to typecheck $\ottnt{e_{{\mathrm{2}}}}$.
To propagate the synthesized type,
this reduction introduces a new polymorphic existential variable $\polyalpha$.
We use an existential variable and a polymorphic one
to guess a monotype and to synthesize a (polymorphic) type in a box,
respectively.

\mycomment{
  A \emph{worklist} $\Gamma$ is a typing context that includes
  unification variables and subtyping/typing judgments.
  A unification variable has its own scope in a worklist.
  For example, a unification variable $\monoalpha$ in $\Gamma_{{\mathrm{1}}}  \ottsym{,}  \monoalpha  \ottsym{,}  \Gamma_{{\mathrm{2}}}$
  is solved to only a type well-formed under $\Gamma_{{\mathrm{1}}}$.

  A typing algorithm using a worklist is defined as reduction $\Gamma_{{\mathrm{1}}}  \rightsquigarrow  \Gamma_{{\mathrm{2}}}$.
  Through reduction steps of a worklist,
  (sub)typing judgments are poped to determine whether it holds.
  For example, $\Gamma  \Vdash   \subL{  \ottnt{A}   \rightarrow   \ottnt{B}  }{ {A^\Box}  \rightarrow  {B^\Box} } $ is reduced to
  $\Gamma  \Vdash   \subL{ \ottnt{B} }{ {B^\Box} }   \Vdash   \subR{ {A^\Box} }{ \ottnt{A} } $ in our setting, where
  $ \Vdash $ is just a worklist separator only used for (sub)typing judgments.
  A typing process successfully ends if $\Gamma  \rightsquigarrow^*   \emptyset $ holds,
  where $ \rightsquigarrow^* $ is defined as the reflexive and transitive closure of $ \rightsquigarrow $.
}

\subsection{Metatheory} \label{sec:algo_metatheory}

Our major metatheorems about the typing algorithm are:
(1) the soundness w.r.t.\ the declarative type system of \lang,
(2) the completeness w.r.t.\ the DK and XO systems, and
(3) its decidability.

To relate our typing algorithm and the type system of \lang,
we use a \emph{declarative worklist} $\Omega$ and
\emph{worklist instantiation} $\Gamma  \xrightarrow{ \mathrm{inst} }  \Omega$,
following Zhao et al.~\cite{zhao_mechanical_2019}.
Declarative worklists and their reduction correspond to
the declarative type system of \lang.
For example, $\Omega  \Vdash   \subL{ \ottnt{A} }{ {B^\Box} } $ is reduced to $\Omega$
only if $  |  \Omega  |    \vdash  \subL{ \ottnt{A} }{ {B^\Box} } $ holds, where $ |  \Omega  | $
is a typing context obtained by eliminating all (sub)typing judgments from $\Omega$.
Worklist instantiation substitutes well-formed types for
all (polymorphic) existential variables.
Namely, if $\Gamma  \xrightarrow{ \mathrm{inst} }  \Omega$ and $\Omega  \rightsquigarrow^*   \emptyset $ hold,
there exists a solution for (polymorphic) existential variables that
makes the corresponding declarative typing hold.

\paragraph{Soundness.}\label{subsubsec:algo_sound}

Our typing algorithm is sound with respect to the type system of \lang.
We denote the well-formedness of $\Gamma$ by $\vdash  \Gamma$.
\ifarxiv
  \begin{restatable}[Soundness w.r.t. the Declarative System of {\lang}]{theorem}{algsoundness}
    \label{thm:alg-soundness}
  If $\vdash  \Gamma$ and $\Gamma  \rightsquigarrow^*   \emptyset $,
  then there exists some $\Omega$ such that $\Gamma  \xrightarrow{ \mathrm{inst} }  \Omega$ and $\Omega  \rightsquigarrow^*   \emptyset $.
\end{restatable}
  \else
\begin{theorem}[Soundness w.r.t. the Declarative System of {\lang}]
  If $\vdash  \Gamma$ and $\Gamma  \rightsquigarrow^*   \emptyset $,
  then there exists some $\Omega$ such that $\Gamma  \xrightarrow{ \mathrm{inst} }  \Omega$ and $\Omega  \rightsquigarrow^*   \emptyset $.
\end{theorem}
\fi
As a corollary, we show the soundness for each specific work item.
\ifarxiv
\begin{restatable}[Soundness for Each Work Item]{corollary}{algsoundnesscor}
  \begin{enumerate}
    \item If $\vdash  \ottsym{(}  \Gamma  \Vdash   \subL{ \ottnt{A} }{ {B^\Box} }   \ottsym{)}$ and $\Gamma  \Vdash   \subL{ \ottnt{A} }{ {B^\Box} }   \rightsquigarrow^*   \emptyset $,
          then there exist some $\ottnt{A'}$, ${B^\Box}'$, and $\Omega$ such that
          $\Gamma  \Vdash   \subL{ \ottnt{A} }{ {B^\Box} }   \xrightarrow{ \mathrm{inst} }  \Omega  \Vdash   \subL{ \ottnt{A'} }{ {B^\Box}' } $ and $  |  \Omega  |    \vdash  \subL{ \ottnt{A'} }{ {B^\Box}' } $.
    \item If $\vdash  \ottsym{(}  \Gamma  \Vdash   \subR{ {A^\Box} }{ \ottnt{B} }   \ottsym{)}$ and $\Gamma  \Vdash   \subR{ {A^\Box} }{ \ottnt{B} }   \rightsquigarrow^*   \emptyset $,
          then there exist some ${A^\Box}'$, $\ottnt{B'}$, and $\Omega$ such that
          $\Gamma  \Vdash   \subR{ {A^\Box} }{ \ottnt{B} }   \xrightarrow{ \mathrm{inst} }  \Omega  \Vdash   \subR{ {A^\Box}' }{ \ottnt{B'} } $ and $  |  \Omega  |    \vdash  \subR{ {A^\Box}' }{ \ottnt{B'} } $.
    \item If $\vdash  \ottsym{(}  \Gamma  \Vdash  \ottnt{e}  \ottsym{:}  {A^\Box}  \ottsym{)}$ and $\Gamma  \Vdash  \ottnt{e}  \ottsym{:}  {A^\Box}  \rightsquigarrow^*   \emptyset $,
          then there exist some $\ottnt{e'}$, ${A^\Box}'$, and $\Omega$ such that
          $\Gamma  \Vdash  \ottnt{e}  \ottsym{:}  {A^\Box}  \xrightarrow{ \mathrm{inst} }  \Omega  \Vdash  \ottnt{e'}  \ottsym{:}  {A^\Box}'$ and $ |  \Omega  |   \vdash  \ottnt{e'}  \ottsym{:}  {A^\Box}'$.
  \end{enumerate}
\end{restatable}
  \else
\begin{corollary}[Soundness for Each Work Item]
  \begin{enumerate}
    \item If $\vdash  \ottsym{(}  \Gamma  \Vdash   \subL{ \ottnt{A} }{ {B^\Box} }   \ottsym{)}$ and $\Gamma  \Vdash   \subL{ \ottnt{A} }{ {B^\Box} }   \rightsquigarrow^*   \emptyset $,
          then there exist some $\ottnt{A'}$, ${B^\Box}'$, and $\Omega$ such that
          $\Gamma  \Vdash   \subL{ \ottnt{A} }{ {B^\Box} }   \xrightarrow{ \mathrm{inst} }  \Omega  \Vdash   \subL{ \ottnt{A'} }{ {B^\Box}' } $ and $  |  \Omega  |    \vdash  \subL{ \ottnt{A'} }{ {B^\Box}' } $.
    \item If $\vdash  \ottsym{(}  \Gamma  \Vdash   \subR{ {A^\Box} }{ \ottnt{B} }   \ottsym{)}$ and $\Gamma  \Vdash   \subR{ {A^\Box} }{ \ottnt{B} }   \rightsquigarrow^*   \emptyset $,
          then there exist some ${A^\Box}'$, $\ottnt{B'}$, and $\Omega$ such that
          $\Gamma  \Vdash   \subR{ {A^\Box} }{ \ottnt{B} }   \xrightarrow{ \mathrm{inst} }  \Omega  \Vdash   \subR{ {A^\Box}' }{ \ottnt{B'} } $ and $  |  \Omega  |    \vdash  \subR{ {A^\Box}' }{ \ottnt{B'} } $.
    \item If $\vdash  \ottsym{(}  \Gamma  \Vdash  \ottnt{e}  \ottsym{:}  {A^\Box}  \ottsym{)}$ and $\Gamma  \Vdash  \ottnt{e}  \ottsym{:}  {A^\Box}  \rightsquigarrow^*   \emptyset $,
          then there exist some $\ottnt{e'}$, ${A^\Box}'$, and $\Omega$ such that
          $\Gamma  \Vdash  \ottnt{e}  \ottsym{:}  {A^\Box}  \xrightarrow{ \mathrm{inst} }  \Omega  \Vdash  \ottnt{e'}  \ottsym{:}  {A^\Box}'$ and $ |  \Omega  |   \vdash  \ottnt{e'}  \ottsym{:}  {A^\Box}'$.
  \end{enumerate}
\end{corollary}
\fi

\paragraph{Incompleteness with respect to the Declarative System.}\label{subsubsec:algo_incomplete}

One may expect that our typing algorithm is complete with respect to
the declarative system.
\ifarxiv
\begin{restatable}{statement}{algocompleteness}
  If $\vdash  \Gamma$ and $\Gamma  \xrightarrow{ \mathrm{inst} }  \Omega$ and $\Omega  \rightsquigarrow^*   \emptyset $, then $\Gamma  \rightsquigarrow^*   \emptyset $.
\end{restatable}
  \else
\begin{statement}
  If $\vdash  \Gamma$ and $\Gamma  \xrightarrow{ \mathrm{inst} }  \Omega$ and $\Omega  \rightsquigarrow^*   \emptyset $, then $\Gamma  \rightsquigarrow^*   \emptyset $.
\end{statement}
\fi
However, the following $\Gamma$ and $\Omega$ show a counterexample.
\[
  \small
  \begin{array}{rcl@{\qquad}rcl}
               \Gamma & = & (\triangleright  \ottsym{\{}  \polyalpha  \ottsym{\}}  \Vdash  \ottsym{()}  \ottsym{:}   \boxed{ \polyalpha }   \Vdash  \ottsym{()}  \ottsym{:}   \boxed{ \polyalpha } )     &
               \Omega             & = & ( \emptyset   \Vdash  \ottsym{()}  \ottsym{:}   \boxed{  \mathsf{unit}  }   \Vdash  \ottsym{()}  \ottsym{:}   \boxed{  \mathsf{unit}  } )
  \end{array}
\]
Our typing algorithm reduces $\Gamma$ to $\triangleright  \ottsym{\{}  \ottsym{\}}  \Vdash  \ottsym{()}  \ottsym{:}   \boxed{  \mathsf{unit}  } $.
However, this reduction does not proceed further
because the type inside a box must be a polymorphic existential variable
in our typing algorithm.

We can still prove a weaker statement by excluding ``invalid'' inputs of the following two kinds:
\begin{enumerate}
  \item A type other than a polymorphic existential variable in a box; and
  \item Two or more occurrences of a polymorphic existential variable in a box.
\end{enumerate}
We introduce two judgments, $\vdash^\downarrow  \Gamma$ and $\ottsym{\{}  \vec{\polyalpha}  \ottsym{\}}  \vdash  \Gamma  \dashv  \ottsym{\{}  \vec{\polybeta}  \ottsym{\}}$,
to exclude them.
(See \arxiv{\zcref{sec:full_defs}}{the full version of this paper}
to see the full definitions of these judgments.)
The former judgment $\vdash^\downarrow  \Gamma$ imposes a stricter well-formedness than $\vdash  \Gamma$.
It excludes the first kind of invalid inputs.
The latter judgment $\ottsym{\{}  \vec{\polyalpha}  \ottsym{\}}  \vdash  \Gamma  \dashv  \ottsym{\{}  \vec{\polybeta}  \ottsym{\}}$ accumulates
polymorphic existential variables in boxes into the variable set $\ottsym{\{}  \vec{\polyalpha}  \ottsym{\}}$,
resulting in $\ottsym{\{}  \vec{\polybeta}  \ottsym{\}}$.
During this accumulation process, the judgment checks that
a previously encountered polymorphic existential variable never occurs in a box.
Therefore, $\ottsym{\{}  \ottsym{\}}  \vdash  \Gamma  \dashv  \ottsym{\{}  \vec{\polyalpha}  \ottsym{\}}$ excludes the second kind of invalid inputs.
\AI{it would be nice to add intuitive reading of the two judgments.}
Under the restrictions, our typing algorithm is ``complete'' with respect to
the declarative typing of \lang.
\arxiv{
  \begin{restatable}{lemma}{PartialComplete}\label{thm:partial_complete}
    If $\vdash^\downarrow  \Gamma$ and $\ottsym{\{}  \ottsym{\}}  \vdash  \Gamma  \dashv  \ottsym{\{}  \vec{\polyalpha}  \ottsym{\}}$ and $\Gamma  \xrightarrow{ \mathrm{inst} }  \Omega$ and $\Omega  \rightsquigarrow^*   \emptyset $,
    then $\Gamma  \rightsquigarrow^*   \emptyset $.
  \end{restatable}
}{
  \begin{lemma}\label{thm:partial_complete}
    If $\vdash^\downarrow  \Gamma$ and $\ottsym{\{}  \ottsym{\}}  \vdash  \Gamma  \dashv  \ottsym{\{}  \vec{\polyalpha}  \ottsym{\}}$ and $\Gamma  \xrightarrow{ \mathrm{inst} }  \Omega$ and $\Omega  \rightsquigarrow^*   \emptyset $,
    then $\Gamma  \rightsquigarrow^*   \emptyset $.
  \end{lemma}
}

\paragraph{Completeness with respect to the DK and XO systems.}\label{subsubsec:algo_complete}

Our typing algorithm is complete with respect to both the DK and XO systems.
We show that
the restriction posed on the input in \zcref{thm:partial_complete}
is not so strict as to reject the well-typed program in the DK system or the XO system.
\ifarxiv
\begin{restatable}[Completeness w.r.t.\ the DK and XO systems]{corollary}{algocompletenessDKXO}\mbox{}
  \begin{itemize}
    \item If $\Psi  \vdash_{ \mathrm{DK} }  \ottnt{e}  \Rightarrow  \ottnt{A}$, then $\Psi  \ottsym{,}  \triangleright  \ottsym{\{}  \polyalpha  \ottsym{\}}  \Vdash   |  \ottnt{e}  |_{ \mathrm{DK} }   \ottsym{:}   \boxed{ \polyalpha }   \rightsquigarrow^*   \emptyset $.
    \item If $\Psi  \vdash_{ \mathrm{DK} }  \ottnt{e}  \Leftarrow  \ottnt{A}$, then $\Psi  \ottsym{,}  \triangleright  \ottsym{\{}  \ottsym{\}}  \Vdash   |  \ottnt{e}  |_{ \mathrm{DK} }   \ottsym{:}  \ottnt{A}  \rightsquigarrow^*   \emptyset $.
    \item If $\Psi  \mid  \Xi  \vdash_{ \mathrm{XO} }  \ottnt{e}  \Rightarrow  \ottnt{A}$, then $\Psi  \ottsym{,}  \triangleright  \ottsym{\{}  \polyalpha  \ottsym{\}}  \Vdash   |  \ottnt{e}  |_{ \mathrm{XO} }   \ottsym{:}   \mathsf{FoldCurry} ( \Xi ,  \polyalpha )   \rightsquigarrow^*   \emptyset $.
  \end{itemize}
\end{restatable}
  \else
\begin{corollary}[Completeness w.r.t.\ the DK and XO systems]
  \begin{itemize}
    \item If $\Psi  \vdash_{ \mathrm{DK} }  \ottnt{e}  \Rightarrow  \ottnt{A}$, then $\Psi  \ottsym{,}  \triangleright  \ottsym{\{}  \polyalpha  \ottsym{\}}  \Vdash   |  \ottnt{e}  |_{ \mathrm{DK} }   \ottsym{:}   \boxed{ \polyalpha }   \rightsquigarrow^*   \emptyset $.
    \item If $\Psi  \vdash_{ \mathrm{DK} }  \ottnt{e}  \Leftarrow  \ottnt{A}$, then $\Psi  \ottsym{,}  \triangleright  \ottsym{\{}  \ottsym{\}}  \Vdash   |  \ottnt{e}  |_{ \mathrm{DK} }   \ottsym{:}  \ottnt{A}  \rightsquigarrow^*   \emptyset $.
    \item If $\Psi  \mid  \Xi  \vdash_{ \mathrm{XO} }  \ottnt{e}  \Rightarrow  \ottnt{A}$, then $\Psi  \ottsym{,}  \triangleright  \ottsym{\{}  \polyalpha  \ottsym{\}}  \Vdash   |  \ottnt{e}  |_{ \mathrm{XO} }   \ottsym{:}   \mathsf{FoldCurry} ( \Xi ,  \polyalpha )   \rightsquigarrow^*   \emptyset $.
  \end{itemize}
\end{corollary}
\fi
This corollary indicates that
our typing algorithm successfully types the well-typed program under the DK or the XO system
by properly inserting the polymorphic existential variable.

\mycomment{
  Our typing algorithm is complete with respect to both the DK and XO systems,
  while the completeness with respect to the type system of \lang does not hold.
  To prove this completeness, we show that, under a restriction that excludes
  a counterexample to the completeness w.r.t. the declarative type system of \lang,
  our typing algorithm is complete with respect to the type system of \lang.
  Two judgments, $\vdash^\downarrow  \Gamma$ and $\ottsym{\{}  \ottsym{\}}  \vdash  \Gamma  \dashv  \ottsym{\{}  \vec{\polyalpha}  \ottsym{\}}$, represent this restriction.
  See the Appendix to see the conterexample and
  the full definitions of these judgments.
  \begin{lemma}\label{thm:partial_complete}
    If $\vdash^\downarrow  \Gamma$ and $\ottsym{\{}  \ottsym{\}}  \vdash  \Gamma  \dashv  \ottsym{\{}  \vec{\polyalpha}  \ottsym{\}}$ and $\Gamma  \xrightarrow{ \mathrm{inst} }  \Omega$ and $\Omega  \rightsquigarrow^*   \emptyset $,
    then $\Gamma  \rightsquigarrow^*   \emptyset $.
  \end{lemma}
  \mycomment{
    The statements $\vdash^\downarrow  \Gamma$ and $\ottsym{\{}  \ottsym{\}}  \vdash  \Gamma  \dashv  \ottsym{\{}  \vec{\polyalpha}  \ottsym{\}}$ restrict
    the input provided for our typing algorithm.
    The rejected input is classified into two groups.
    The former is just invalid for our typing algorithm.
    For example, one of them is $ \emptyset   \Vdash   \subL{  \forall  \ottmv{a}  .  \ottnt{A}  }{ \ottnt{B} } $.
    Because our typing algorithm introduces an existential variable
    just earlier than a scope delimiter,
    we need at least one scope delimiter earlier than any work.
    Another example is $\Gamma  \Vdash  \ottnt{e_{{\mathrm{1}}}}  \ottsym{:}   \boxed{ \polyalpha }   \Vdash  \ottnt{e_{{\mathrm{2}}}}  \ottsym{:}   \boxed{ \polyalpha } $.
    In this worklist, the types of $\ottnt{e_{{\mathrm{1}}}}$ and $\ottnt{e_{{\mathrm{2}}}}$ are synthesized,
    but the synthesized types will be oddly passed to the same polymorphic existential variable $\polyalpha$.
    The latter group includes the worklist that causes the cyclic scoping problem.
  }

  We show that
  the restriction posed on the input of our typing algorithm in \zcref{thm:partial_complete}
  is not too strict to reject the well-typed program in the DK system or the XO system.
  \begin{corollary}[Completeness of Algorithm w.r.t.\ the DK and XO systems]\mbox{}
    \begin{itemize}
      \item If $\Psi  \vdash_{ \mathrm{DK} }  \ottnt{e}  \Rightarrow  \ottnt{A}$, then $\Psi  \ottsym{,}  \triangleright  \ottsym{\{}  \polyalpha  \ottsym{\}}  \Vdash   |  \ottnt{e}  |_{ \mathrm{DK} }   \ottsym{:}   \boxed{ \polyalpha }   \rightsquigarrow^*   \emptyset $.
      \item If $\Psi  \vdash_{ \mathrm{DK} }  \ottnt{e}  \Leftarrow  \ottnt{A}$, then $\Psi  \ottsym{,}  \triangleright  \ottsym{\{}  \ottsym{\}}  \Vdash   |  \ottnt{e}  |_{ \mathrm{DK} }   \ottsym{:}  \ottnt{A}  \rightsquigarrow^*   \emptyset $.
      \item If $\Psi  \mid  \Xi  \vdash_{ \mathrm{XO} }  \ottnt{e}  \Rightarrow  \ottnt{A}$, then $\Psi  \ottsym{,}  \triangleright  \ottsym{\{}  \polyalpha  \ottsym{\}}  \Vdash   |  \ottnt{e}  |_{ \mathrm{XO} }   \ottsym{:}   \mathsf{FoldCurry} ( \Xi ,  \polyalpha )   \rightsquigarrow^*   \emptyset $.
    \end{itemize}
  \end{corollary}
  This corollary indicates that
  our typing algorithm successfully types the well-typed program under the DK or the XO system
  by properly inserting the polymorphic existential variable.
}

\paragraph{Decidability.}\label{subsubsec:algo_decidability}

Our typing algorithm is decidable.
\ifarxiv
\begin{restatable}{theorem}{algodecidability}
  If $\vdash^\downarrow  \Gamma$ and $\ottsym{\{}  \ottsym{\}}  \vdash  \Gamma  \dashv  \ottsym{\{}  \vec{\polyalpha}  \ottsym{\}}$,
  then $\Gamma  \rightsquigarrow^*   \emptyset $ or
  there exists some non-empty $\Gamma'$
  such that $\Gamma  \rightsquigarrow^*  \Gamma'$ and $\Gamma' \not  \rightsquigarrow $.
\end{restatable}
  \else
\begin{theorem}
  If $\vdash^\downarrow  \Gamma$ and $\ottsym{\{}  \ottsym{\}}  \vdash  \Gamma  \dashv  \ottsym{\{}  \vec{\polyalpha}  \ottsym{\}}$,
  then $\Gamma  \rightsquigarrow^*   \emptyset $ or
  there exists some non-empty $\Gamma'$
  such that $\Gamma  \rightsquigarrow^*  \Gamma'$ and $\Gamma' \not  \rightsquigarrow $.
\end{theorem}
\fi
The proof strategy follows that of
the decidability shown by Zhao et al.~\cite{zhao_mechanical_2019}.

\section{Related Work}\label{sec:related}

\subsection{Predicative Systems}

Peyton Jones et al.~\citet{jones_practical_2007} developed
a function-first style bidirectional type system
supporting higher-rank polymorphism.
They force a synthesized type to have no top-level quantifier
by instantiating higher-rank types.
However, not only does their system lack the typeability that the argument-first style has,
but this requirement introduces a further limitation regarding typeability.
For example, their system cannot synthesize the type $   \mathsf{unit}    \rightarrow    \forall  \ottmv{a}  .  \ottmv{a}     \rightarrow   \ottmv{a} $,
given the expression $ \lambda  \mathit{x}  \ottsym{:}   \mathsf{unit}   .  \mathit{f} $ and the typing context $\mathit{f}  \ottsym{:}   \forall  \ottmv{a}  .  \ottmv{a}   \rightarrow  \ottmv{a}$.
Having no restriction on the forms of synthesized types, our system can synthesize this type,
as the XO system can.

Dunfield and Krishnaswami~\citet{dunfield_complete_2013} proposed
an \emph{application judgment},
which extracts the argument type from a (polymorphic) function type.
They added it to a function-first style bidirectional type system
to support higher-rank polymorphism. %
We show that boxy types enable our type system to subsume
their type system in \zcref{subsec:decl_metatheory}.

Xie and Oliveira~\citet{xie_let_2018} introduced argument-first bidirectional typing
using an application context, as discussed in \zcref{sec:overview}.
We show that boxy types can subsume the notion of
an application context in \zcref{subsec:decl_metatheory}.
However, their system also supports let-polymorphism and
is a conservative extension of the Hindley--Milner type system,
in contrast to the type system of \lang.

\subsection{Impredicative Systems}\label{subsec:impre}

Work on first-class polymorphism for ML-like languages includes
$\mathsf{Poly}\-\mathsf{ML}$~\cite{DBLP:journals/iandc/GarrigueR99},
$\mathsf{ML}^{\mathsf{F}}$~\cite{DBLP:conf/icfp/BotlanR03}, and
$\mathsf{FPH}$~\cite{DBLP:conf/icfp/VytiniotisWJ08}.  These systems
are based on the Damas--Milner style, while supporting impredicative
polymorphism.  Nevertheless, their annotation-free inference does not
cover every program accepted by argument-first typing.  For example,
if $\mathit{f}$ has type $  \forall  \ottmv{a}  .  \ottmv{a}    \rightarrow   \ottmv{a} $, the corresponding
unannotated program $\ottsym{(}   \lambda  \mathit{x}  .  \ottsym{(}  \mathit{x} \, \ottsym{()}  \ottsym{,}  \mathit{x} \, 42  \ottsym{)}   \ottsym{)} \, \mathit{f}$ is rejected: these
systems do not use the argument type to check the body of an abstraction under
the assumption $\mathit{x}  \ottsym{:}   \forall  \ottmv{a}  .  \ottmv{a}   \rightarrow  \ottmv{a}$.  Poly-ML and
$\mathsf{ML}^{\mathsf{F}}$ require polymorphism to be made explicit
when a lambda-bound variable is used polymorphically, while
$\mathsf{FPH}$ infers monotypes for unannotated lambda-bound variables.

Serrano et al.~\citet{serrano_quick_2020} proposed an algorithm called \emph{Quick Look}
to support impredicativity,
combining bidirectional typing and constraint-based type inference.
They showed that Quick Look can be easily integrated
with the existing type inference algorithms.
While their system lacks some typeability that the XO system has,
similarly to the system of Peyton Jones et al.~\citet{jones_practical_2007},
Quick Look can be a good guide to extending
our system with impredicativity.

Zhao and Oliveira~\citet{zhao_elementary_2022} extended the DK system
with top and bottom types and impredicative explicit type applications.
Incorporating their change to the DK system would be
a first step toward extending our system with impredicativity.

\TY{Mention the boxy type paper, too?}

\subsection{Generalizing Bidirectional Typing}\label{subsec:gen_bidi}

Vytiniotis et al.~\citet{vytiniotis_boxy_2006} proposed \emph{boxy types},
on which our formalization is based.
While their type system is impredicative and
a conservative extension of the Hindley--Milner type system,
it forces a synthesized type to have no top-level quantifier
by instantiating higher-rank types, as Peyton Jones et al.~\citet{jones_practical_2007} do.
Since the XO system has no such restriction,
we need to develop our language \lang to subsume the XO system.
We expect that the way Vytiniotis et al.\ support impredicative polymorphism
could serve as a guide to extending the type system of \lang to impredicative polymorphism.

Odersky et al.~\citet{odersky_colored_2001} refined
local type inference~\cite{pierce_local_1998} and
introduced \emph{colored types} to reduce type annotations.
Colored types specify how type information flows and
allow type information intended for checking to occur in a synthesized type.
Boxy types are similar to their colored types,
except that they prevent type information that should be checked from arising in a synthesized type.
However, because the declarative type system with colored types is based on $F_{<:}$,
our typing and subtyping rules differ from theirs.
\AI{Write more about the differences?}

\emph{Contextual typing}~\cite{xue_contextual_2024} combined
function- and argument-first styles of bidirectional typing to some degree.
However, contextual typing has no feature to freely switch between
function- and argument-first typing, i.e.,
it cannot synthesize the type for the function application $\ottsym{(}  \mathit{f}  \appright  \mathit{x_{{\mathrm{1}}}}  \ottsym{)}  \appleft  \mathit{x_{{\mathrm{2}}}}$ in our syntax.
Moreover, it does not support any kind of polymorphism,
while it has been shown to scale to
subtyping, intersection types, overloading, and records.

Xue et al.~\citet{xue_local_2026} proposed
\emph{Contextual System F} based on contextual typing.
Contextual System F is a variant of implicit System F and more flexibly switches
between function- and argument-first bidirectional typing than
Xue and Oliveira~\citet{xue_contextual_2024}.
However, it is even more restrictive in how it switches than the type system of \lang.
For example, given the function application $\ottsym{(}  \mathit{f}  \appright  \mathit{x_{{\mathrm{1}}}}  \ottsym{)}  \appleft  \mathit{x_{{\mathrm{2}}}}$ written in our syntax,
the type system of \lang synthesizes the first argument type of $\mathit{f}$
using the synthesized type of $\mathit{x_{{\mathrm{2}}}}$ to proceed with type checking for $\mathit{x_{{\mathrm{1}}}}$.
Still, Contextual System F disallows this typing flow and
applies type synthesis or checking to arguments only in left-to-right order,
i.e., the typing procedure of Contextual System F always flows from $\mathit{x_{{\mathrm{1}}}}$ to $\mathit{x_{{\mathrm{2}}}}$
in the example.
While this restriction in Contextual System F helps realize
implicit and impredicative polymorphism both in declarative and algorithmic systems,
it prevents Contextual System F from accepting some programs
that are well-typed under either the DK or XO system, and
the algorithmic typing of Contextual System F sticks to the restriction.
Incorporating the typeability of Contextual System F,
especially impredicativity, is an interesting future direction of our work.

\mycomment{
  \subsection{Worklist Approach}

  Zhao et al.~\citet{zhao_formalization_2018} first proposed the worklist approach
  to mechanically formalizing the algorithmic typing for
  polymorphic subtyping of Odersky et al.~\citet{odersky_putting_1996}, and
  Zhao et al.~\citet{zhao_mechanical_2019} extended this approach to the DK system.
  Our typing algorithm is based on the latter worklist approach,
  but does not use its \emph{judgment chains}.

  A judgment chain does not fit boxy types, while it works well with the DK system.
  A judgment chain has a work item $\ottnt{e} \Rightarrow_a \omega$ for type synthesis.
  It synthesizes the type of $\ottnt{e}$ and substitutes the synthesized type for $\ottmv{a}$ in $\omega$.
  Since a typing judgment using boxy types has multiple output channels,
  a judgment chain must deal with this feature to support boxy types.
  For example, consider the following declarative typing judgment of \lang.
  \begin{align}
     \emptyset   \vdash   \lambda  \mathit{x}  \ottsym{:}   \mathsf{unit}   .   \lambda  \mathit{y}  \ottsym{:}   \mathsf{unit}   .  \mathit{y}    \ottsym{:}   \boxed{  \mathsf{unit}  }   \rightarrow   \boxed{   \mathsf{unit}    \rightarrow    \mathsf{unit}   }  \label{d_eg}
  \end{align}
  This judgment returns two types, $ \mathsf{unit} $ and $  \mathsf{unit}    \rightarrow    \mathsf{unit}  $~.
  Therefore, the corresponding worklist that adopts judgment chains could be:
  \begin{align}
     \emptyset   \Vdash   \lambda  \mathit{x}  \ottsym{:}   \mathsf{unit}   .   \lambda  \mathit{y}  \ottsym{:}   \mathsf{unit}   .  \mathit{y}   :  \boxed{ \ottmv{a} }   \rightarrow   \boxed{ \ottmv{b} } 
    \Rightarrow_{a, b} \omega \label{w_eg}
  \end{align}
  This worklist would be reduced to the following one, considering the typing rule \TAAbsBox.
  \[
    \mathit{x}  \ottsym{:}   \mathsf{unit}   \Vdash   \subR{  \boxed{ \ottmv{a} }  }{  \mathsf{unit}  }  \Rightarrow_{a} (
     \lambda  \mathit{y}  \ottsym{:}   \mathsf{unit}   .  \mathit{y}   \ottsym{:}   \boxed{ \ottmv{b} }  \Rightarrow_b \omega
    )
  \]
  Note that the subtyping work item also has output.
  This reduction has two problems.
  One is that the subtyping work item $ \subR{  \boxed{ \ottmv{a} }  }{  \mathsf{unit}  } $ is under the worklist $\mathit{x}  \ottsym{:}   \mathsf{unit} $~.
  This situation does not reflect the premise $ \emptyset   \vdash   \subR{  \boxed{  \mathsf{unit}  }  }{  \mathsf{unit}  } $
  of the declarative derivation for \eqref{d_eg}.
  Second is that a typing algorithm must ``divide'' the type variable list $a, b$
  of the judgment chain in \eqref{w_eg} into $a$ and $b$.
  Introducing this division mechanism seems to complicate
  a typing algorithm and its formalization.
  Thus, we avoid using judgment chains.

  Some studies have extended the worklist approach to supporting advanced typing features, such as
  the top and bottom types and impredicative explicit type applications~\cite{zhao_elementary_2022},
  bounded quantification~\cite{cui_greedy_2023}, and
  intersection and union types~\cite{jiang_bidirectional_2025}.
  These studies would guide us to extend
  our typing algorithm using boxy types to advanced typing features.
}

\mycomment{
  \subsection{Level-Based Type Inference}\label{subsec:level}

  A \emph{rank} or \emph{level}~\cite{remy1992extension,fan_practical_2025}, a natural number attached to a type variable,
  helps manage the scope of type/unification variables,
  serving as a scope delimiter in our algorithmic typing.
  We briefly explain how ranked type variables work in the following.

  R{\'e}my~\citet{remy1992extension} assigned a \emph{rank}
  to every type variable to implement more efficient let-generalization of types.
  A rank equipped with a type variable indicates
  how many let-expressions are introduced before the type variable.
  In their type system,
  we increase the rank when we enter a let-expression and
  decrease it when we leave its scope.
  Their type inference process uses ranks to determine
  which type variables in the type of a let-bound expression should be generalized.
  Their typing rule for let-expressions is as follows.
  \[
    \inferrule{
      \Psi \vdash^{n + 1} \ottnt{e_{{\mathrm{1}}}} : \ottnt{A} \\
      \Psi, x : \forall \mathrm{FTV}^{n+1} (\ottnt{A}) . \ottnt{A} \vdash^{n} \ottnt{e_{{\mathrm{2}}}} : \ottnt{B}
    }{
      \Psi \vdash^{n} \mathbf{let}\ x = \ottnt{e_{{\mathrm{1}}}}\ \mathbf{in} \ottnt{e_{{\mathrm{2}}}} : \ottnt{B}
    }
  \]
  This rule only generalizes the free type variables that have the rank $(n+1)$
  in the type $\ottnt{A}$ using $\mathrm{FTV}^{n+1} (\ottnt{A})$.
  This generalization feature accelerates the type inference process.

  Fan et al.~\citet{fan_practical_2025} applied the notion of the rank,
  which they call the \emph{level},
  to function-first bidirectional typing with higher-rank polymorphism.
  Their bidirectional typing rule for checking the expression against a polymorphic type
  is as follows.
  \[
    \inferrule{
    \Psi \vdash^{n+1} \ottnt{e}  \Leftarrow  \ottnt{A} \\ \text{$a$ has the level $n+1$}
    }{
      \Psi \vdash^{n} \ottnt{e}  \Leftarrow   \forall  \ottmv{a}  .  \ottnt{A} 
    }
  \]
  Note that this rule increases the rank (level) $n$ to $n + 1$.
  Their typing algorithm uses ranks not only for let-generalization
  but also for management of the scope of unification/type variables.
  It solves the $n$-ranked unification variable to only those types
  in which every unification or type variable has at most $n$ rank.
  In other words, every unification variable must be solved to
  the well-formed type under the context in which it is first introduced.
  In this sense, their rank increment corresponds to
  our introduction of the scope delimiter $\triangleright  \ottsym{\{}  \ottsym{\}}$.
}

\section{Conclusion and Future Work}\label{sec:future}

In this paper, we propose a unified language \lang and its declarative type system that
subsumes function- and argument-first bidirectional type systems and
supports higher-rank polymorphism using boxy types.
\lang serves as an intermediate language and leaves design space to
allow a language designer to define their own approach to choosing
either function- or argument-first typing.
We formalize them and prove that
the type system of \lang is complete with respect to both the DK and XO systems and
is sound with respect to \sysfi.

Moreover, we present a typing algorithm for \lang using the worklist approach.
This algorithm is independent of the heuristics that a language designer
adopts to determine whether they use function- or argument-first typing.
We mechanically prove that
our typing algorithm is sound with respect to the type system of \lang.
Our typing algorithm is not complete with respect to the type system of \lang
\AI{due to the cyclic scoping problem}, but
we show that it is complete with respect to both the DK and XO systems.

Future work on the declarative side includes extending \lang
to support product types, let-polymorphism, and impredicativity.
Since the type of an argument in a function application
is fully synthesized or checked in the type system of \lang,
the following derivation with a product type is not supported.
\[\small
  \inferrule{
  \Psi  \vdash  \mathit{f}  \ottsym{:}    \boxed{ \ottnt{A_{{\mathrm{1}}}} }   \times  \ottnt{A_{{\mathrm{2}}}}   \rightarrow  \ottnt{B} \\ \Psi  \vdash  \ottsym{(}  \ottnt{e_{{\mathrm{1}}}}  \ottsym{,}  \ottnt{e_{{\mathrm{2}}}}  \ottsym{)}  \ottsym{:}   \ottnt{A_{{\mathrm{1}}}}  \times   \boxed{ \ottnt{A_{{\mathrm{2}}}} }  
  }{
    \Psi  \vdash  \mathit{f} \, \ottsym{(}  \ottnt{e_{{\mathrm{1}}}}  \ottsym{,}  \ottnt{e_{{\mathrm{2}}}}  \ottsym{)}  \ottsym{:}  \ottnt{B}
  }
\]
A challenging future direction is to support this kind of typing.
Supporting let-polymorphism enables \lang to subsume the original XO system completely.
Supporting explicit type applications is another interesting future direction.
The works of Xie and Oliveira~\cite{xie_let_2018} and Zhao et al.~\cite{zhao_elementary_2022}
guide us to extend \lang with explicit type applications to allow for impredicative type instantiations.
The other existing work~\cite{vytiniotis_boxy_2006,xue_local_2026} could also help
support impredicativity.

Future work on the algorithmic side includes
combining our typing algorithm with
a level-based type-inference method~\cite{remy1992extension,fan_practical_2025} and
providing a typeability benchmark.
Because a level-based type inference has been used to implement practical programming languages,
adopting it for \lang helps bring \lang's typeability to practical implementations,
requiring minimal changes.
To evaluate the practical usefulness of our framework,
we need to explore how stable our typing is with
certain program transformations.
For example, studying whether a well-typed program in \lang
remains well-typed after
removing type annotations, $\beta$-reduction, and $\eta$-conversion
is an interesting future direction.

\mycomment{
  The future work includes extending \lang to let-polymorphism.
  This extension enables \lang to subsume the original XO system completely.
  The existing ways~\cite{vytiniotis_boxy_2006,xie_let_2018,fan_practical_2025}
  to realize let-generalization could help support let-polymorphism.
  Another future direction is to combine our typing algorithm
  with a level-based type inference.
  Because a level-based type inference has been used to implement practical programming languages,
  combining it with boxy types would help augment the typeability of an existing language
  with minimal changes to its implementation.
}

\mycomment{
  In this paper, we present a declarative system that
  unifies function- and argument-first bidirectional type systems and
  supports higher-rank polymorphism
  using mixed-direction types.
  We mechanically formalize our system and
  prove that our declarative system is sound (resp. complete)
  with respect to \sysfi (resp. the DK system).

  The future work includes proving that our system is complete with respect to the XO system.
  Developing an algorithmic type system corresponding to our declarative system is
  one of the most important directions.
  Let-polymorphism will be an interesting extension for our system to
  completely subsume the XO system.
}

\ifarxiv
  \bibliographystyle{plainnat}
\else
  \bibliographystyle{splncs04}
\fi

\bibliography{paper}

\arxiv{
  \setcounter{secnumdepth}{3}
  \appendix
  \section{Design Direction Towards Algorithmic Typing}\label{sec:algo_pre}

This section overviews our typing algorithm for \lang.
Our typing algorithm is based on the \emph{worklist approach}.
It was originally proposed by Zhao et al.~\citet{zhao_formalization_2018}
to mechanically formalize an algorithm for polymorphic subtyping~\cite{odersky_putting_1996} and
subsequently extended by Zhao et al.~\citet{zhao_mechanical_2019} to the DK system.
\zcref{subsec:worklist} reviews the notion of the latter worklist approach and
\zcref{subsec:algo_design} discusses key design ideas behind our algorithm
to extend the worklist approach to boxy types.

\subsection{Review: Worklist Approach}\label{subsec:worklist}

A \emph{worklist} is a typing context extended with
\emph{existential variables} and subtyping/typing judgments.
An existential variable should be solved to a monotype similarly to a unification variable,
but its scope is managed in a worklist.
This approach enables us
to avoid accidentally solving unification variables, which
violates the scope condition imposed by the declarative type system.

\mycomment{
  A core idea of the worklist approach is to introduce an \emph{existential variable},
  which should be solved to a monotype, similarly to a unification variable,
  but whose scope is strictly managed, and
  to contain subtyping/typing judgments are contained in a single \emph{worklist}.
  This approach enables us
  to avoid accidentally solving unification variables, which results in undesirable typing, and
  to easily propagate a unification result from one judgment to another.
}

A worklist $\Gamma$ is defined as follows.
\[
  \Gamma  \Coloneqq   \emptyset   \mid  \Gamma  \ottsym{,}  \ottmv{a}  \mid  \Gamma  \ottsym{,}  \monoalpha  \mid  \Gamma  \ottsym{,}  \mathit{x}  \ottsym{:}  \ottnt{A}  \mid 
  \Gamma  \Vdash  \ottnt{A}  \mathbin{<:}  \ottnt{B}  \mid  \cdots
\]
A worklist $\Gamma$ can contain an existential variable $\monoalpha$ and
a subtyping judgment $\ottnt{A}  \mathbin{<:}  \ottnt{B}$\footnote{
  We only consider judgments without boxy types for a while.
}. %
An existential variable $\monoalpha$ in $\Gamma_{{\mathrm{1}}}  \ottsym{,}  \monoalpha  \ottsym{,}  \Gamma_{{\mathrm{2}}}$ is solved to
only a type that is well-formed under $\Gamma_{{\mathrm{1}}}$.
Note that an existential variable can occur in a type.

A typing algorithm using a worklist is defined as a reduction $\Gamma_{{\mathrm{1}}}  \rightsquigarrow  \Gamma_{{\mathrm{2}}}$,
which means that a typing algorithm, given a worklist $\Gamma_{{\mathrm{1}}}$,
pops the rightmost judgment in $\Gamma_{{\mathrm{1}}}$ and processes it in one step,
resulting in $\Gamma_{{\mathrm{2}}}$.
For example, let us consider the following reduction.
\[
   \emptyset   \Vdash   \ottnt{A_{{\mathrm{1}}}}   \rightarrow   \ottnt{B_{{\mathrm{1}}}}   \mathbin{<:}   \ottnt{A_{{\mathrm{2}}}}   \rightarrow   \ottnt{B_{{\mathrm{2}}}}  \quad  \rightsquigarrow  \quad
   \emptyset   \Vdash  \ottnt{B_{{\mathrm{1}}}}  \mathbin{<:}  \ottnt{B_{{\mathrm{2}}}}  \Vdash  \ottnt{A_{{\mathrm{2}}}}  \mathbin{<:}  \ottnt{A_{{\mathrm{1}}}}
\]
The left-hand side worklist means that the first judgment of a typing algorithm is to check that
the type $ \ottnt{A_{{\mathrm{1}}}}   \rightarrow   \ottnt{B_{{\mathrm{1}}}} $ is a subtype of $ \ottnt{A_{{\mathrm{2}}}}   \rightarrow   \ottnt{B_{{\mathrm{2}}}} $ under the empty typing context.
\mycomment{
  \[
     \emptyset   \Vdash   \ottnt{A_{{\mathrm{1}}}}   \rightarrow   \ottnt{B_{{\mathrm{1}}}}   \mathbin{<:}   \ottnt{A_{{\mathrm{2}}}}   \rightarrow   \ottnt{B_{{\mathrm{2}}}} 
  \]
  This worklist means that the first judgment of a typing algorithm is to check that
  the type $ \ottnt{A_{{\mathrm{1}}}}   \rightarrow   \ottnt{B_{{\mathrm{1}}}} $ is a subtype of $ \ottnt{A_{{\mathrm{2}}}}   \rightarrow   \ottnt{B_{{\mathrm{2}}}} $ under the empty typing context.
  The worklist is reduced to
  \[
     \emptyset   \Vdash  \ottnt{B_{{\mathrm{1}}}}  \mathbin{<:}  \ottnt{B_{{\mathrm{2}}}}  \Vdash  \ottnt{A_{{\mathrm{2}}}}  \mathbin{<:}  \ottnt{A_{{\mathrm{1}}}}~.
  \]
}
This reduction means that we need to check that
$\ottnt{A_{{\mathrm{2}}}}$ (resp. $\ottnt{B_{{\mathrm{1}}}}$) is a subtype of $\ottnt{A_{{\mathrm{1}}}}$ (resp. $\ottnt{B_{{\mathrm{2}}}}$)
to check that $ \ottnt{A_{{\mathrm{1}}}}   \rightarrow   \ottnt{B_{{\mathrm{1}}}} $ is a subtype of $ \ottnt{A_{{\mathrm{2}}}}   \rightarrow   \ottnt{B_{{\mathrm{2}}}} $.
In this sense, a worklist can be seen as a ``flattened'' version of a corresponding derivation tree.
After we learn that the two subtyping relations hold,
the typing process removes the completed work items and
reaches the empty worklist $ \emptyset $, indicating that
the typing process has successfully finished.

Consider the following worklist to understand why existential variables are not globally declared.
\[
  \Gamma  \Vdash    \forall  \ottmv{a}  .   \mathsf{unit}     \rightarrow   \ottmv{a}   \mathbin{<:}    \mathsf{unit}    \rightarrow    \forall  \ottmv{b}  .  \ottmv{b}  
\]
A reduction of this worklist must get stuck because
the corresponding declarative typing judgment
\[
  \Psi  \vdash    \forall  \ottmv{a}  .   \mathsf{unit}     \rightarrow   \ottmv{a}   \mathbin{<:}    \mathsf{unit}    \rightarrow    \forall  \ottmv{b}  .  \ottmv{b}  
\]
is not derivable for any $\Psi$. %
\TY{How much shall we explain that this is not derivable?}
The worklist is reduced to
\[
  \Gamma  \ottsym{,}  \monoalpha  \Vdash    \mathsf{unit}    \rightarrow   \monoalpha   \mathbin{<:}    \mathsf{unit}    \rightarrow    \forall  \ottmv{b}  .  \ottmv{b}  
\]
by introducing the existential variable $\monoalpha$ to represent the monotype $\tau$ used to
instantiate the quantified type variable in the declarative typing.
The worklist $\Gamma  \ottsym{,}  \monoalpha$ means that $\monoalpha$
can be solved to only a type that is well-formed under $\Gamma$.
In a few steps, we reach the following reduction
(after removing the successful work item $ \mathsf{unit}   \mathbin{<:}   \mathsf{unit} $)
which introduces the type variable $\ottmv{b}$.
\[
  \Gamma  \ottsym{,}  \monoalpha  \Vdash  \monoalpha  \mathbin{<:}   \forall  \ottmv{b}  .  \ottmv{b} 
  \quad  \rightsquigarrow  \quad \Gamma  \ottsym{,}  \monoalpha  \ottsym{,}  \ottmv{b}  \Vdash  \monoalpha  \mathbin{<:}  \ottmv{b}
\]
\mycomment{
  In a few steps, we reach the following worklist,
  (after removing the successful work item $ \mathsf{unit}   \mathbin{<:}   \mathsf{unit} $)
  \[
    \Gamma  \ottsym{,}  \monoalpha  \Vdash  \monoalpha  \mathbin{<:}   \forall  \ottmv{b}  .  \ottmv{b} \
  \]
  and then introduce the type variable $\ottmv{b}$ in the next step as follows:
  \[
    \Gamma  \ottsym{,}  \monoalpha  \ottsym{,}  \ottmv{b}  \Vdash  \monoalpha  \mathbin{<:}  \ottmv{b}\ .
  \]
}
Because the existential variable $\monoalpha$ can be solved only to a type well-formed under $\Gamma$,
unlike globally declared variables,
it cannot be solved to $\ottmv{b}$.
Thus, the typing process fails as required owing to
the scope management of existential variables.

\mycomment{
  Consider the following worklist to understand
  why a worklist helps propagate a unification result from one judgment to another.
  \[
    \monoalpha  \ottsym{,}  \monobeta  \Vdash    \mathsf{unit}    \rightarrow   \monobeta   \mathbin{<:}   \monoalpha   \rightarrow   \monoalpha 
  \]
  The reduction of it is as follows.
  \[
    (\monoalpha  \ottsym{,}  \monobeta  \Vdash    \mathsf{unit}    \rightarrow   \monobeta   \mathbin{<:}   \monoalpha   \rightarrow   \monoalpha )
     \rightsquigarrow  (\monoalpha  \ottsym{,}  \monobeta  \Vdash  \monobeta  \mathbin{<:}  \monoalpha  \Vdash  \monoalpha  \mathbin{<:}   \mathsf{unit} )
     \rightsquigarrow  (\monobeta  \Vdash  \monobeta  \mathbin{<:}   \mathsf{unit} )
     \rightsquigarrow   \emptyset 
  \]
  In the second step, we solve the existential variable $\monoalpha$ to $ \mathsf{unit} $.
  The typing algorithm substitutes unification results into the remaining work,
  $\monobeta  \mathbin{<:}  \monoalpha$ in this case, instead of threading them.
  Therefore, the reduction result of the second step has the work item $\monobeta  \mathbin{<:}   \mathsf{unit} $,
  by substituting $\monoalpha$ with $ \mathsf{unit} $.
  This treatment of propagating unification results avoids
  duplicating them in a typing process.
}

\subsection{Polymorphic Existential Variables and Scope Delimiters}\label{subsec:algo_design}

Our typing algorithm differs from that of Zhao et al.~\citet{zhao_mechanical_2019}
mainly in two aspects:
using \emph{polymorphic existential variables}, and
introducing \emph{scope delimiters}.
We use the former %
to support boxy types.
To manage their scope, we introduce the latter. %

\mycomment{
  move it to related work or later.
  \subsubsection{The Absence of Judgment Chains}\label{subsubsec:omega}

  Zhao et al.~\citet{zhao_mechanical_2019} introduced a \emph{judgment chain} $\omega$
  to extend the initial worklist approach~\cite{zhao_formalization_2018} with
  type judgment, especially with type synthesis.
  Because type synthesis returns type,
  a typing algorithm must propagate the output type to other work items.
  For example, the declarative typing rule for function applications in the DK system
  reuses the synthesized type of the given function for typechecking the argument.
  However, a worklist has no mechanism to transfer the type from one work item to another.
  A judgment chain implements this mechanism.

  A judgment chain $\omega$ is defined as follows.
  \[
    \omega  \Coloneqq  \ottnt{A}  \mathbin{<:}  \ottnt{B}  \mid  \ottnt{e} \Leftarrow \ottnt{A}  \mid 
    \ottnt{e} \Rightarrow_a \omega  \mid  \ottnt{A} \bullet \ottnt{e} \RRightarrow_a \omega
  \]
  A work item $\ottnt{e} \Rightarrow_a \omega$ corresponds to a type synthesis judgment.
  A typing algorithm takes the expression $\ottnt{e}$ as input,
  synthesizes the type $\ottnt{A}$ of the expression, and
  substitutes the synthesized type $\ottnt{A}$ into the type variable $\ottmv{a}$
  in the subsequent judgment chain $\omega$.
  A work item $\ottnt{A} \bullet \ottnt{e} \RRightarrow_a \omega$ corresponds to an application judgment,
  similarly to a type synthesis work.
  The typing algorithm proposed by Zhao et al.~\citet{zhao_mechanical_2019}
  has the following reduction rule.
  \[
    \Gamma  \Vdash  \ottnt{e_{{\mathrm{1}}}} \, \ottnt{e_{{\mathrm{2}}}} \Rightarrow_a \omega  \rightsquigarrow 
    \Gamma  \Vdash  \ottnt{e_{{\mathrm{1}}}} \Rightarrow_b (b \bullet \ottnt{e_{{\mathrm{2}}}} \RRightarrow_a \omega)
  \]
  This rule corresponds to the declarative typing rule for function applications in the DK system.

  A judgment chain does not fit boxy types, while it works well with the DK system.
  A typing judgment using boxy types has multiple outputs
  in contrast to a type synthesis judgment.
  For example, the following declarative typing judgment of \lang returns two types,
  $ \mathsf{unit} $ and $  \mathsf{unit}    \rightarrow    \mathsf{unit}  $.
  \begin{align}
     \emptyset   \vdash   \lambda  \mathit{x}  \ottsym{:}   \mathsf{unit}   .   \lambda  \mathit{y}  \ottsym{:}   \mathsf{unit}   .  \mathit{y}    \ottsym{:}   \boxed{  \mathsf{unit}  }   \rightarrow   \boxed{   \mathsf{unit}    \rightarrow    \mathsf{unit}   }  \label{d_eg}
  \end{align}
  However, a judgment chain only propagates one type.
  We might extend a judgment chain to support multiple outputs,
  but it seems to be troublesome.
  Consider the above declarative typing judgment of \lang again.
  The corresponding worklist to it could be:
  \begin{align}
     \emptyset   \Vdash   \lambda  \mathit{x}  \ottsym{:}   \mathsf{unit}   .   \lambda  \mathit{y}  \ottsym{:}   \mathsf{unit}   .  \mathit{y}   :  \boxed{ \ottmv{a} }   \rightarrow   \boxed{ \ottmv{b} } 
    \Rightarrow_{a, b} \omega \label{w_eg}
  \end{align}
  This worklist would be reduced to the following one, considering the typing rule \TAAbsBox.
  \[
    \mathit{x}  \ottsym{:}   \mathsf{unit}   \Vdash   \subR{  \boxed{ \ottmv{a} }  }{  \mathsf{unit}  }  \Rightarrow_{a} (
     \lambda  \mathit{y}  \ottsym{:}   \mathsf{unit}   .  \mathit{y}   \ottsym{:}   \boxed{ \ottmv{b} }  \Rightarrow_b \omega
    )
  \]
  Note that the subtyping work item also has output.
  This reduction has two problems.
  One is that the subtyping work item $ \subR{  \boxed{ \ottmv{a} }  }{  \mathsf{unit}  } $ is under the worklist $\mathit{x}  \ottsym{:}   \mathsf{unit} $.
  This situation does not reflect the premise $ \emptyset   \vdash   \subR{  \boxed{  \mathsf{unit}  }  }{  \mathsf{unit}  } $
  of the declarative derivation for \eqref{d_eg}.
  Second is that a typing algorithm must ``divide'' the type variables $a$ and $b$
  of the judgment chain in \eqref{w_eg} into $a$ and $b$.
  Introducing this division mechanism seems to complicate
  a typing algorithm and its formalization.
  Thus, we avoid using judgment chains.
}

\subsubsection{Polymorphic Existential Variables}\label{subsubsec:pEVar}

We introduce \emph{polymorphic existential variables} to adapt
the worklist approach to boxy types.
A boxy type may have multiple ``output channels''; e.g.,
$ \boxed{  \mathsf{unit}  }   \rightarrow   \boxed{   \mathsf{unit}    \rightarrow    \mathsf{unit}   } $ means that
the output types are the argument type $ \mathsf{unit} $ and the return type $  \mathsf{unit}    \rightarrow    \mathsf{unit}  $.
Therefore, our typing algorithm must propagate the types that fill up boxes
from one work item to the remaining work items.
We assign a polymorphic existential variable to each box in a boxy type
to support this propagation.
This idea is also used in the type inference algorithm of
the original work on boxy types~\cite{vytiniotis_boxy_2006}.

We propagate a solved type in the box of a given boxy type
through a polymorphic existential variable $\polyalpha$,
which can be solved to a polymorphic type, in contrast to an existential variable,
which propagates guessed monotypes.
Roughly speaking, our algorithmic typing rule for function applications
with the function-first guide is as follows, where we use a work item $\ottnt{e}  \ottsym{:}  {A^\Box}$ for typing.
\[
  \Gamma  \Vdash  \ottnt{e_{{\mathrm{1}}}}  \appright  \ottnt{e_{{\mathrm{2}}}}  \ottsym{:}  {A^\Box} \quad  \rightsquigarrow  \quad  \Gamma  \ottsym{,}  \polyalpha  \Vdash  \ottnt{e_{{\mathrm{2}}}}  \ottsym{:}  \polyalpha  \Vdash  \ottnt{e_{{\mathrm{1}}}}  \ottsym{:}   \boxed{ \polyalpha }   \rightarrow  {A^\Box}
\]
The aspects that differ from the corresponding declarative typing rule \TAppR are:
we introduce a polymorphic existential variable $\polyalpha$ and
replace the boxed type $\ottnt{A}$ in \TAppR with it. %
The polymorphic existential variable $\polyalpha$ acts as a placeholder for
the result of type synthesis.
The work item $\ottnt{e_{{\mathrm{1}}}}  \ottsym{:}   \boxed{ \polyalpha }   \rightarrow  {A^\Box}$ means that
our typing algorithm synthesizes the argument type of $\ottnt{e_{{\mathrm{1}}}}$ and
substitutes the synthesized type for $\polyalpha$.
The work item $\ottnt{e_{{\mathrm{2}}}}  \ottsym{:}  \polyalpha$ means that the expression $\ottnt{e_{{\mathrm{2}}}}$
is typechecked against the synthesized argument type of $\ottnt{e_{{\mathrm{1}}}}$.
Note that we have no rule that solves a polymorphic existential variable outside boxes.
A non-boxed polymorphic existential variable, e.g., $\polyalpha$ in $\ottnt{e_{{\mathrm{2}}}}  \ottsym{:}  \polyalpha$,
should have been solved when the work item comes to the rightmost position of the worklist.

\mycomment{
  For example, consider the following declarative judgment in \lang.
  \[
    \mathit{f}  \ottsym{:}   \ottsym{(}   \forall  \ottmv{a}  .   \mathsf{unit}    \ottsym{)}   \rightarrow    \mathsf{unit}    \vdash  \mathit{f}  \appright  \ottsym{()}  \ottsym{:}   \mathsf{unit} 
  \]
  A typing algorithm for this judgment works as follows,
  introducing the polymorphic existential variable $\polyalpha$.
  \begin{align*}
     & \mathit{f}  \ottsym{:}   \ottsym{(}   \forall  \ottmv{a}  .   \mathsf{unit}    \ottsym{)}   \rightarrow    \mathsf{unit}    \Vdash  \mathit{f}  \appright  \ottsym{()}  \ottsym{:}   \mathsf{unit}                           \\
     &  \rightsquigarrow  \mathit{f}  \ottsym{:}   \ottsym{(}   \forall  \ottmv{a}  .   \mathsf{unit}    \ottsym{)}   \rightarrow    \mathsf{unit}    \ottsym{,}  \polyalpha  \Vdash  \ottsym{()}  \ottsym{:}  \polyalpha  \Vdash  \mathit{f}  \ottsym{:}   \boxed{ \polyalpha }   \rightarrow   \mathsf{unit} 
  \end{align*}
  The resulting worklist suggests that a typing algorithm subsequently works as follows.
  A typing algorithm synthesizes the argument type of $\mathit{f}$,
  substitutes the synthesized type into the polymorphic existential variable $\polyalpha$, and then
  checks that the unit $\ottsym{()}$ has the argument type of $\mathit{f}$, i.e., $ \forall  \ottmv{a}  .   \mathsf{unit}  $.
}

\subsubsection{Scope Delimiters}\label{subsubsec:scope_delimiter}

We introduce \emph{scope delimiters} to manage scoping of existential variables
in a setting with polymorphic existential variables.
The notion of scope delimiters is closely related to
ranks or levels of type variables~\cite{remy1992extension,fan_practical_2025}.

First, we explain why using polymorphic existential variables complicates
the scope management of existential variables.
Consider the following two declarative typing derivations as examples.

\begin{center}
  \begin{tabular}{c@{\qquad}c}
    \begin{minipage}{0.45\linewidth}
      \begin{equation}
        \begin{aligned}
          \inferrule* {
            \inferrule* {
              \Gamma  \ottsym{,}  \ottmv{b}  \vdash  \tau\\
              \Gamma  \ottsym{,}  \ottmv{b}  \vdash   \subL{  \tau   \rightarrow   \tau  }{  \ottmv{b}   \rightarrow   \ottmv{b}  } 
            }{
              \Gamma  \ottsym{,}  \ottmv{b}  \vdash   \subL{   \forall  \ottmv{a}  .  \ottmv{a}    \rightarrow   \ottmv{a}  }{  \ottmv{b}   \rightarrow   \ottmv{b}  } 
            }
          }{
            \Gamma  \vdash   \subL{   \forall  \ottmv{a}  .  \ottmv{a}    \rightarrow   \ottmv{a}  }{  \forall  \ottmv{b}  .   \ottmv{b}   \rightarrow   \ottmv{b}   } 
          }
        \end{aligned}
        \label{d_sub1}
      \end{equation}
    \end{minipage} &
    \begin{minipage}{0.45\linewidth}
      \begin{equation}
        \begin{aligned}
          \inferrule* {
            \Gamma  \vdash  \tau \\
            \inferrule* {
              \Gamma  \vdash   \subR{  \boxed{ \tau }  }{ \tau }  \\
              \Gamma  \vdash   \subL{ \tau }{  \boxed{ \tau }  } 
            }{
              \Gamma  \vdash   \subL{  \tau   \rightarrow   \tau  }{  \boxed{ \tau }   \rightarrow   \boxed{ \tau }  } 
            }
          }{
            \Gamma  \vdash   \subL{   \forall  \ottmv{a}  .  \ottmv{a}    \rightarrow   \ottmv{a}  }{  \boxed{ \tau }   \rightarrow   \boxed{ \tau }  } 
          }
        \end{aligned}
        \label{d_sub2}
      \end{equation}
    \end{minipage}
  \end{tabular}
\end{center}

\noindent
A typing algorithm for \lang must support them
because they hold in the declarative type system of \lang.
Regarding the derivation \eqref{d_sub1}, the reduction of the corresponding worklist is as follows.
\[
  \begin{array}{rcl}
    \Gamma  \Vdash   \subL{   \forall  \ottmv{a}  .  \ottmv{a}    \rightarrow   \ottmv{a}  }{  \forall  \ottmv{b}  .   \ottmv{b}   \rightarrow   \ottmv{b}   } 
     &  \rightsquigarrow  & \Gamma  \ottsym{,}  \ottmv{b}  \Vdash   \subL{   \forall  \ottmv{a}  .  \ottmv{a}    \rightarrow   \ottmv{a}  }{  \ottmv{b}   \rightarrow   \ottmv{b}  }  \\
     &  \rightsquigarrow  & \Gamma  \ottsym{,}  \ottmv{b}  \ottsym{,}  \monogamma  \Vdash   \subL{  \monogamma   \rightarrow   \monogamma  }{  \ottmv{b}   \rightarrow   \ottmv{b}  } 
  \end{array}
\]
The existential variable $\monogamma$ is introduced in the second step
just after the type variable $\ottmv{b}$,
because the monotype $\tau$ in \eqref{d_sub1} is well-formed under $\Gamma  \ottsym{,}  \ottmv{b}$.
However, introducing existential variables at the rightmost position in a worklist
does not always work well with polymorphic existential variables.
The reduction of the worklist corresponding to the derivation \eqref{d_sub2} fails
if we introduce an existential variable at the rightmost in a worklist.
\[
  \begin{array}{rcl}
    \Gamma  \ottsym{,}  \polyalpha  \ottsym{,}  \polybeta  \Vdash   \subL{   \forall  \ottmv{a}  .  \ottmv{a}    \rightarrow   \ottmv{a}  }{  \boxed{ \polyalpha }   \rightarrow   \boxed{ \polybeta }  } 
     &  \rightsquigarrow  & \Gamma  \ottsym{,}  \polyalpha  \ottsym{,}  \polybeta  \ottsym{,}  \monogamma  \Vdash   \subL{  \monogamma   \rightarrow   \monogamma  }{  \boxed{ \polyalpha }   \rightarrow   \boxed{ \polybeta }  }                      \\
     &  \rightsquigarrow  & \Gamma  \ottsym{,}  \polyalpha  \ottsym{,}  \polybeta  \ottsym{,}  \monogamma  \Vdash   \subL{ \monogamma }{  \boxed{ \polybeta }  }   \Vdash   \subR{  \boxed{ \polyalpha }  }{ \monogamma }  \quad \not  \rightsquigarrow 
  \end{array}
\]
Note that the existential variable $\monogamma$ corresponds to the guessed monotype $\tau$
in the declarative subtyping judgments $\Gamma  \vdash   \subR{  \boxed{ \tau }  }{ \tau } $ and $\Gamma  \vdash   \subL{ \tau }{  \boxed{ \tau }  } $.
Because the existential variable $\monogamma$ is after $\polyalpha$ and $\polybeta$ in this worklist,
we cannot solve them to $\monogamma$.
Therefore, our typing algorithm inserts $\monogamma$
\emph{before} the polymorphic existential variables $\polyalpha$ and $\polybeta$ to solve them.
\[
  \Gamma  \ottsym{,}  \polyalpha  \ottsym{,}  \polybeta  \Vdash   \subL{   \forall  \ottmv{a}  .  \ottmv{a}    \rightarrow   \ottmv{a}  }{  \boxed{ \polyalpha }   \rightarrow   \boxed{ \polybeta }  } 
  \quad  \rightsquigarrow  \quad \Gamma  \ottsym{,}  \monogamma  \ottsym{,}  \polyalpha  \ottsym{,}  \polybeta  \Vdash   \subL{  \monogamma   \rightarrow   \monogamma  }{  \boxed{ \polyalpha }   \rightarrow   \boxed{ \polybeta }  } 
\]

We introduce \emph{scope delimiters} to tell
our typing algorithm where existential variables should be inserted.
A scope delimiter takes the form of $\triangleright  \ottsym{\{}  \vec{\polyalpha}  \ottsym{\}}$,
where $\vec{\polyalpha}$ is a sequence of polymorphic existential variables.
Our typing algorithm always introduces a scope delimiter alongside a type variable and
inserts an existential variable just before the rightmost scope delimiter.
For example, a typing algorithm corresponding to the derivation \eqref{d_sub1} works as follows.
\begin{align*}
  \Gamma  \Vdash   \subL{   \forall  \ottmv{a}  .  \ottmv{a}    \rightarrow   \ottmv{a}  }{  \forall  \ottmv{b}  .   \ottmv{b}   \rightarrow   \ottmv{b}   } 
   & \quad  \rightsquigarrow  \quad \Gamma  \ottsym{,}  \ottmv{b}  \ottsym{,}  \triangleright  \ottsym{\{}  \ottsym{\}}  \Vdash   \subL{   \forall  \ottmv{a}  .  \ottmv{a}    \rightarrow   \ottmv{a}  }{  \ottmv{b}   \rightarrow   \ottmv{b}  }  \\
   & \quad  \rightsquigarrow  \quad \Gamma  \ottsym{,}  \ottmv{b}  \ottsym{,}  \monogamma  \ottsym{,}  \triangleright  \ottsym{\{}  \ottsym{\}}  \Vdash   \subL{  \monogamma   \rightarrow   \monogamma  }{  \ottmv{b}   \rightarrow   \ottmv{b}  } 
\end{align*}
Because a type variable and a scope delimiter are introduced simultaneously,
we can use the type variable to solve existential variables inserted just before the scope delimiter.
To see why we insert existential variables before the rightmost scope delimiter,
consider the following typing process corresponding to the derivation \eqref{d_sub2}, where
we suppose that the worklist $\Gamma$ includes a type variable introduced
along with the scope delimiter $\triangleright  \ottsym{\{}  \polyalpha  \ottsym{,}  \polybeta  \ottsym{\}}$.
\[
  \Gamma  \ottsym{,}  \triangleright  \ottsym{\{}  \polyalpha  \ottsym{,}  \polybeta  \ottsym{\}}  \Vdash   \subL{   \forall  \ottmv{a}  .  \ottmv{a}    \rightarrow   \ottmv{a}  }{  \boxed{ \polyalpha }   \rightarrow   \boxed{ \polybeta }  } 
  \quad  \rightsquigarrow  \quad \Gamma  \ottsym{,}  \monogamma  \ottsym{,}  \triangleright  \ottsym{\{}  \polyalpha  \ottsym{,}  \polybeta  \ottsym{\}}  \Vdash   \subL{  \monogamma   \rightarrow   \monogamma  }{  \boxed{ \polyalpha }   \rightarrow   \boxed{ \polybeta }  } 
\]
The scope delimiter and its introduction mechanism enable us
to insert $\monogamma$ before $\triangleright  \ottsym{\{}  \polyalpha  \ottsym{,}  \polybeta  \ottsym{\}}$.

\mycomment{
  An existential variable is always inserted just earlier than the outermost scope delimiter.
  For example, a typing algorithm for judgment \eqref{d_sub1} works as follows,
  using a scope delimiter.
  \[
    (\triangleright  \ottsym{\{}  \polyalpha  \ottsym{,}  \polybeta  \ottsym{\}}  \Vdash   \subL{   \forall  \ottmv{a}  .  \ottmv{a}    \rightarrow   \ottmv{a}  }{  \boxed{ \polyalpha }   \rightarrow   \boxed{ \polybeta }  } )
     \rightsquigarrow  (\monogamma  \ottsym{,}  \triangleright  \ottsym{\{}  \polyalpha  \ottsym{,}  \polybeta  \ottsym{\}}  \Vdash   \subL{  \monogamma   \rightarrow   \monogamma  }{  \boxed{ \polyalpha }   \rightarrow   \boxed{ \polybeta }  } )
  \]
  The existential variable $\monogamma$ is inserted just earlier
  than the scope delimiter $\triangleright  \ottsym{\{}  \polyalpha  \ottsym{,}  \polybeta  \ottsym{\}}$.
  A typing algorithm for judgment \eqref{d_sub2} also inserts
  the existential variable $\monogamma$ in the same way.
  \[
    ( \emptyset   \Vdash   \subL{   \forall  \ottmv{a}  .  \ottmv{a}    \rightarrow   \ottmv{a}  }{  \forall  \ottmv{b}  .   \ottmv{b}   \rightarrow   \ottmv{b}   } )
     \rightsquigarrow  (\ottmv{b}  \ottsym{,}  \triangleright  \ottsym{\{}  \ottsym{\}}  \Vdash   \subL{   \forall  \ottmv{a}  .  \ottmv{a}    \rightarrow   \ottmv{a}  }{  \ottmv{b}   \rightarrow   \ottmv{b}  } )
     \rightsquigarrow  (\ottmv{b}  \ottsym{,}  \monogamma  \ottsym{,}  \triangleright  \ottsym{\{}  \ottsym{\}}  \Vdash   \subL{  \monogamma   \rightarrow   \monogamma  }{  \ottmv{b}   \rightarrow   \ottmv{b}  } )
  \]
  The scope delimiter $\triangleright  \ottsym{\{}  \ottsym{\}}$ is inserted
  at the same time as the type variable $\ottmv{b}$ is introduced.
  This introduction
  ensures that we can use the type variable to instantiate
  the subsequently introduced existential variables.
}

\mycomment{
  We define scope delimiters as a sequence of polymorphic existential variables,
  rather than just a symbol to delimit a worklist, for practical reasons,
  rather than theoretical ones. %
  Consider the following worklist, supposing that
  we split the scope delimiter $\triangleright  \ottsym{\{}  \vec{\polyalpha}  \ottsym{\}}$ into $\triangleright$ and $\vec{\polyalpha}$.
  \[
    \Gamma_{{\mathrm{1}}}  \ottsym{,}  \polyalpha  \ottsym{,}  \Gamma_{{\mathrm{2}}}  \Vdash   \lambda  \mathit{x}  .  \mathit{x}   \ottsym{:}   \boxed{ \polyalpha } 
  \]
  In this case, a typing algorithm must search the rightmost $\triangleright$
  in the worklist $\Gamma_{{\mathrm{1}}}$ to know where it should insert
  an existential variable for the argument type of $\mathit{x}$.
  We avoid such searching to simplify our typing algorithm.
  In our typing algorithm, the worklist above should be:
  \[
    \Gamma_{{\mathrm{1}}}  \ottsym{,}  \triangleright  \ottsym{\{}   \ldots   \ottsym{,}  \polyalpha  \ottsym{,}   \ldots   \ottsym{\}}  \ottsym{,}  \Gamma_{{\mathrm{2}}}  \Vdash   \lambda  \mathit{x}  .  \mathit{x}   \ottsym{:}   \boxed{ \polyalpha } 
  \]
}

  \section{Algorithmic Typing}\label{sec:algo}

\mycomment{
  This section firstly shows a technical background and key design ideas
  of our typing algorithm for \lang,
  before proposing the typing algorithm and showing the metatheory about it.
  Our typing algorithm is based on the \emph{worklist approach},
  which was originally proposed by Zhao et al.~\citet{zhao_formalization_2018}
  to mechanically formalize an algorithm for polymorphic subtyping~\cite{odersky_putting_1996} and
  subsequently extended by Zhao et al.~\citet{zhao_mechanical_2019} for the DK system.
  We explain the notion of the latter worklist approach in \zcref{subsec:worklist}.
  We step-by-step show key design ideas of our typing algorithm
  to extend the worklist approach for boxy types in \zcref{subsec:algo_design}.
  Proceed to \zcref{subsec:algo_def} for readers who wish to view
  the definition of our typing algorithm.
}

This section proposes the typing algorithm for \lang and shows the metatheorems about it.

\subsection{Algorithmic Type System}\label{subsec:algo_def}

Here, we explain the updated syntax for the worklist approach and our typing algorithm for \lang.

\subsubsection{Syntax}\label{subsubsec:algo_syntax}

\begin{figure}[t]
  \small
  \[
    \begin{array}{rrcl}
      \textbf{Types}          & \ottnt{A}, \ottnt{B} &  \Coloneqq  &
       \mathsf{unit}   \mid  \ottmv{a}  \mid  \monoalpha \mycomment{ \mid  \polyalpha}  \mid   \ottnt{A}   \rightarrow   \ottnt{B}   \mid   \forall  \ottmv{a}  .  \ottnt{A}  \\
      \textbf{Boxy types}     & {A^\Box}, {B^\Box} &  \Coloneqq  &
       \boxed{ \polyalpha }   \mid   \mathsf{unit}   \mid  \ottmv{a}  \mid  \monoalpha  \mid  \polyalpha  \mid 
      {A^\Box}  \rightarrow  {B^\Box}  \mid   \forall  \ottmv{a}  .  \ottnt{A}                                \\
      \textbf{Work items}     & \omega        &  \Coloneqq  &
       \subL{ \ottnt{A} }{ {A^\Box} }   \mid   \subR{ {A^\Box} }{ \ottnt{A} }   \mid  \ottnt{e}  \ottsym{:}  {A^\Box}                                                      \\
      \textbf{Worklists}      & \Gamma        &  \Coloneqq  &
       \emptyset   \mid  \Gamma  \ottsym{,}  \ottmv{a}  \mid  \Gamma  \ottsym{,}  \monoalpha  \mid  \Gamma  \ottsym{,}  \triangleright  \ottsym{\{}  \vec{\polyalpha}  \ottsym{\}}  \mid 
      \Gamma  \ottsym{,}  \mathit{x}  \ottsym{:}  \ottnt{A}  \mid  \Gamma  \Vdash  \omega                               \\
      \textbf{Mini worklists} & \Delta        &  \Coloneqq  &
       \emptyset   \mid  \Delta  \ottsym{,}  \mathit{x}  \ottsym{:}  \ottnt{A}  \mid  \Delta  \Vdash  \omega                                                    \\
    \end{array}
  \]
  \caption{Syntax for the algorithmic system}\label{fig:algo_syntax}
\end{figure}

We show the syntax for our typing algorithm in \zcref{fig:algo_syntax}.
\TY{Shall we grey out the extended bits?}
We omit the syntax of expressions,
because they are the same as those shown in \zcref{fig:syntax}.

We add an existential variable to types and boxy types, and
a polymorphic existential variable to boxy types.
Note that we allow only a polymorphic type variable to be in a box.

\mycomment{
  Types are extended with
  an existential variable $\monoalpha$. %
  Their scope is managed in a worklist.
  An existential variable $\monoalpha$ should be solved to only a monomorphic type.

  Boxy types are extended with
  an existential variable and a polymorphic existential variable,
  similarly to types.
  A polymorphic existential variable $\polyalpha$
  can be solved to a polymorphic type
  to propagate the synthesis results for boxes in a boxy type,
  as explained in \zcref{subsubsec:pEVar}.
  Note that we allow only a polymorphic type variable to be in a box.
}
\mycomment{
  Note that we allow a boxy type to take the form of $ \boxed{  \mathsf{unit}  } $, for example.
  From the algorithmic point of view, such a boxy type is meaningless
  because the type inside a box is output, and a typing algorithm should find that type.
  Indeed, our typing algorithm has no rule that contains the boxy type
  rather than the form of $ \boxed{ \polyalpha } $.
  However, we allow for such a meaningless boxy type
  to define the syntax using HOAS easily.
}

Work items, ranged over by $\omega$, consist of
two kinds of subtyping work items $ \subL{ \ottnt{A} }{ {B^\Box} } $ and $ \subR{ {A^\Box} }{ \ottnt{B} } $, and
a typing work item $\ottnt{e}  \ottsym{:}  {A^\Box}$.
These two subtyping work items $ \subL{ \ottnt{A} }{ {B^\Box} } $ and $ \subR{ {A^\Box} }{ \ottnt{B} } $ correspond to
$ \Psi   \vdash  \subL{ \ottnt{A} }{ {B^\Box} } $ and $ \Psi   \vdash  \subR{ {A^\Box} }{ \ottnt{B} } $, respectively.
A typing work item $\ottnt{e}  \ottsym{:}  {A^\Box}$ corresponds to
$\Psi  \vdash  \ottnt{e}  \ottsym{:}  {A^\Box}$.
Our typing algorithm removes these work items from a given worklist and processes them.

Worklists, ranged over by $\Gamma$, consist of
the empty worklist $ \emptyset $, the type variable binding $\Gamma  \ottsym{,}  \ottmv{a}$,
the existential variable binding $\Gamma  \ottsym{,}  \monoalpha$, the scope delimiter binding $\Gamma  \ottsym{,}  \triangleright  \ottsym{\{}  \vec{\polyalpha}  \ottsym{\}}$,
and the work item binding $\Gamma  \Vdash  \omega$.
\mycomment{
  We introduce the scope delimiter $\triangleright  \ottsym{\{}  \vec{\polyalpha}  \ottsym{\}}$ to
  manage the scope of (polymorphic) existential variables properly
  as explained in \zcref{subsubsec:scope_delimiter}.
}
The work item binding $\Gamma  \Vdash  \omega$ intuitively means that
a typing algorithm tackles the work item $\omega$ under the worklist $\Gamma$,
as explained in \zcref{sec:algo_pre}.
For example, our typing algorithm solves
both $\polyalpha$ and $\polybeta$ to $ \mathsf{unit} $
for the worklist $\triangleright  \ottsym{\{}  \polyalpha  \ottsym{,}  \polybeta  \ottsym{\}}  \Vdash   \subL{   \mathsf{unit}    \rightarrow    \mathsf{unit}   }{  \boxed{ \polyalpha }   \rightarrow   \boxed{ \polybeta }  } $.

Mini worklists, ranged over by $\Delta$, are worklists without type variable declarations.
We only use a mini worklist to specify the rightmost scope delimiter in a worklist
by writing $\Gamma  \ottsym{,}  \triangleright  \ottsym{\{}  \vec{\polyalpha}  \ottsym{\}}  \ottsym{,}  \Delta$.
\mycomment{
  They do not affect the well-formedness of (boxy) types.
  For example, if the type $\ottnt{A}$ is well-formed under the worklist $\Gamma  \ottsym{,}  \Delta$, then
  $\ottnt{A}$ is also well-formed under the worklist $\Gamma$.
  A couple of our algorithmic typing rules use a mini worklist to
  keep the well-formedness of types in the declarative rules.
}

\subsubsection{Algorithmic Typing}\label{subsubsec:algo_typing}

Our algorithmic typing is defined as the reduction of worklists.
Algorithmic typing rules have the form of $\Gamma  \rightsquigarrow  \Gamma'$.
We write $\Gamma  \rightsquigarrow^*  \Gamma'$ to denote multiple reduction steps, i.e.,
$ \rightsquigarrow^* $ is defined as the reflexive and transitive closure of $ \rightsquigarrow $.
Using the multiple-step reduction,
$\Gamma  \rightsquigarrow^*   \emptyset $ represents a successful algorithmic typing.
Note that every newly introduced variable is supposed to be fresh.

We break down our algorithmic typing rules into several parts and explain each.

\paragraph{Binding removal}
We show our binding removal rules in \zcref{fig:binding_removal}.
\begin{figure}[t]
  \small
  \[
    \begin{array}{rcll@{\qquad}rcll}
                 \Gamma  \ottsym{,}  \ottmv{a} &  \rightsquigarrow  & \Gamma & (\ATyVar) & \Gamma  \ottsym{,}  \monoalpha    &  \rightsquigarrow  & \Gamma & (\AEVar) \\
                 \Gamma  \ottsym{,}  \triangleright  \ottsym{\{}  \vec{\polyalpha}  \ottsym{\}}      &  \rightsquigarrow  & \Gamma & (\ADelim) & \Gamma  \ottsym{,}  \mathit{x}  \ottsym{:}  \ottnt{A} &  \rightsquigarrow  & \Gamma & (\AVar)
    \end{array}
  \]
  \caption{Binding removal rules}\label{fig:binding_removal}
\end{figure}
These rules drop variable declarations that are no longer used,
as they are out of scope for the remaining work items in $\Gamma$.

\paragraph{Algorithmic subtyping}
We show our algorithmic subtyping rules in \zcref{fig:algo_subtyping}.
\begin{figure}[t]
  \renewcommand{\arraystretch}{1.2}
  \small
  \[
    \begin{array}{rcll}
                 \Gamma_{{\mathrm{1}}}  \ottsym{,}  \triangleright  \ottsym{\{}  \vec{\polybeta}_{{\mathrm{1}}}  \ottsym{,}  \polyalpha  \ottsym{,}  \vec{\polybeta}_{{\mathrm{2}}}  \ottsym{\}}  \ottsym{,}  \Gamma_{{\mathrm{2}}}  \Vdash   \subL{ \ottnt{A} }{  \boxed{ \polyalpha }  }  &  \rightsquigarrow     &
                 \Gamma_{{\mathrm{1}}}  \ottsym{,}  \triangleright  \ottsym{\{}  \vec{\polybeta}_{{\mathrm{1}}}  \ottsym{,}  \vec{\polybeta}_{{\mathrm{2}}}  \ottsym{\}}  \ottsym{,}   [  \ottnt{A}  /  \polyalpha  ]  \Gamma_{{\mathrm{2}}}              & (\ASLBox)                \\
                                                                 &           &
                 \tif   \mathrm{FTV}( \ottnt{A} )   \subseteq \mathrm{dom}( \Gamma_{{\mathrm{1}}} )                 &                          \\
                 \Gamma_{{\mathrm{1}}}  \ottsym{,}  \triangleright  \ottsym{\{}  \vec{\polybeta}_{{\mathrm{1}}}  \ottsym{,}  \polyalpha  \ottsym{,}  \vec{\polybeta}_{{\mathrm{2}}}  \ottsym{\}}  \ottsym{,}  \Gamma_{{\mathrm{2}}}  \Vdash   \subR{  \boxed{ \polyalpha }  }{ \ottnt{A} }  &  \rightsquigarrow     &
                 \Gamma_{{\mathrm{1}}}  \ottsym{,}  \triangleright  \ottsym{\{}  \vec{\polybeta}_{{\mathrm{1}}}  \ottsym{,}  \vec{\polybeta}_{{\mathrm{2}}}  \ottsym{\}}  \ottsym{,}   [  \ottnt{A}  /  \polyalpha  ]  \Gamma_{{\mathrm{2}}}              & (\ASRBox)                \\
                                                                 &           &
                 \tif   \mathrm{FTV}( \ottnt{A} )   \subseteq \mathrm{dom}( \Gamma_{{\mathrm{1}}} )                  &                          \\
    \end{array}
  \]
  \[
    \begin{array}{rcll@{\qquad}rcll}
                 \Gamma  \Vdash   \subL{  \mathsf{unit}  }{  \mathsf{unit}  }  &  \rightsquigarrow      &
                 \Gamma                   & (\ASLUnit) &
                 \Gamma  \Vdash   \subR{  \mathsf{unit}  }{  \mathsf{unit}  }  &  \rightsquigarrow      &
                 \Gamma                   & (\ASRUnit)                \\
                 \Gamma  \Vdash   \subL{ \ottmv{a} }{ \ottmv{a} }       &  \rightsquigarrow      &
                 \Gamma                   & (\ASLTVar) &
                 \Gamma  \Vdash   \subR{ \ottmv{a} }{ \ottmv{a} }       &  \rightsquigarrow      &
                 \Gamma                   & (\ASRTVar)                \\
                 \Gamma  \Vdash   \subL{ \monoalpha }{ \monoalpha }      &  \rightsquigarrow      &
                 \Gamma                   & (\ASLEVar) &
                 \Gamma  \Vdash   \subR{ \monoalpha }{ \monoalpha }      &  \rightsquigarrow      &
                 \Gamma                   & (\ASREVar)
    \end{array}
  \]
  \[
    \begin{array}{rcll}
                 \Gamma  \Vdash   \subL{  \ottnt{A}   \rightarrow   \ottnt{B}  }{ {A^\Box}  \rightarrow  {B^\Box} }                             &  \rightsquigarrow      &
                 \Gamma  \Vdash   \subL{ \ottnt{B} }{ {B^\Box} }   \Vdash   \subR{ {A^\Box} }{ \ottnt{A} }                          & (\ASLFun)                  \\
                 \Gamma  \Vdash   \subR{ {A^\Box}  \rightarrow  {B^\Box} }{  \ottnt{A}   \rightarrow   \ottnt{B}  }                             &  \rightsquigarrow      &
                 \Gamma  \Vdash   \subR{ {B^\Box} }{ \ottnt{B} }   \Vdash   \subL{ \ottnt{A} }{ {A^\Box} }                          & (\ASRFun)                  \\
                 \Gamma  \Vdash   \subL{ \ottnt{A} }{  \forall  \ottmv{b}  .  \ottnt{B}  }                            &  \rightsquigarrow      &
                 \Gamma  \ottsym{,}  \ottmv{b}  \ottsym{,}  \triangleright  \ottsym{\{}  \ottsym{\}}  \Vdash   \subL{ \ottnt{A} }{ \ottnt{B} }                             & (\ASLAllR)                 \\
                 \Gamma  \Vdash   \subR{ {A^\Box} }{  \forall  \ottmv{b}  .  \ottnt{B}  }                           &  \rightsquigarrow      &
                 \Gamma  \ottsym{,}  \ottmv{b}  \ottsym{,}  \triangleright  \ottsym{\{}  \ottsym{\}}  \Vdash   \subR{ {A^\Box} }{ \ottnt{B} }  \quad \tif \mathsf{not}\square \, \ottsym{(}  {A^\Box}  \ottsym{)} & (\ASRAllR)                 \\
                 \Gamma  \ottsym{,}  \triangleright  \ottsym{\{}  \vec{\polyalpha}  \ottsym{\}}  \ottsym{,}  \Delta  \Vdash   \subL{  \forall  \ottmv{a}  .  \ottnt{A}  }{ {B^\Box} }               &  \rightsquigarrow      &
                 \Gamma  \ottsym{,}  \monoalpha  \ottsym{,}  \triangleright  \ottsym{\{}  \vec{\polyalpha}  \ottsym{\}}  \ottsym{,}  \Delta  \Vdash   \subL{  [  \monoalpha  /  \ottmv{a}  ]  \ottnt{A}  }{ {B^\Box} }                & (\ASLAllL)                 \\
                                                                        &            &
                 \tif \mathsf{not}\forall \, \ottsym{(}  {B^\Box}  \ottsym{)} \tand \mathsf{not}\square \, \ottsym{(}  {B^\Box}  \ottsym{)}            &                            \\
                 \Gamma  \ottsym{,}  \triangleright  \ottsym{\{}  \vec{\polyalpha}  \ottsym{\}}  \ottsym{,}  \Delta  \Vdash   \subR{  \forall  \ottmv{a}  .  \ottnt{A}  }{ \ottnt{B} }              &  \rightsquigarrow      &
                 \Gamma  \ottsym{,}  \monoalpha  \ottsym{,}  \triangleright  \ottsym{\{}  \vec{\polyalpha}  \ottsym{\}}  \ottsym{,}  \Delta  \Vdash   \subR{  [  \monoalpha  /  \ottmv{a}  ]  \ottnt{A}  }{ \ottnt{B} }                & (\ASRAllL)                 \\
                                                                        &            &
                 \tif \mathsf{not}\forall \, \ottsym{(}  \ottnt{B}  \ottsym{)}                                 &
    \end{array}
  \]
  \caption{Algorithmic subtyping rules}\label{fig:algo_subtyping}
\end{figure}
Almost all of these subtyping rules are similar to the declarative counterparts.
For example, the rule \ASLFun,
which decomposes the two given function (boxy) types and
adds two subtyping work items to the given worklist,
corresponds to \SLFun.
Note that
our typing algorithm introduces the scope delimiter $\triangleright  \ottsym{\{}  \ottsym{\}}$ in \ASLAllR and \ASRAllR
as explained in \zcref{subsubsec:scope_delimiter}.

\mycomment{
  The four rules, \ASLBox, \ASRBox, \ASLAllL, and \ASRAllL, differ from
  the declarative rules \SLBox, \SRBox, \SLAllL, and \SRAllL, respectively.
  We explain these differences in the following.
}

Our typing algorithm solves a polymorphic existential variable $\polyalpha$ in the box
to a type $\ottnt{A}$ using the rule \ASLBox, when it encounters the work item $ \subL{ \ottnt{A} }{  \boxed{ \polyalpha }  } $.
This rule corresponds to the rule \SLBox,
but has the side condition $  \mathrm{FTV}( \ottnt{A} )   \subseteq \mathrm{dom}( \Gamma_{{\mathrm{1}}} ) $, meaning that
the free type and existential variables of the type $\ottnt{A}$ are bound in $\Gamma_{{\mathrm{1}}}$.
Using this condition ensures that
the polymorphic existential variable $\polyalpha$ is solved to
the well-formed type $\ottnt{A}$ under its scope $\Gamma_{{\mathrm{1}}}$.
We use the condition $  \mathrm{FTV}( \ottnt{A} )   \subseteq \mathrm{dom}( \Gamma ) $
instead of the well-formedness of the type $\ottnt{A}$ under $\Gamma_{{\mathrm{1}}}$
to simplify an implementation of our typing algorithm.
Because checking the well-formedness of $\ottnt{A}$ under $\Gamma_{{\mathrm{1}}}$
demands the well-formedness of $\Gamma_{{\mathrm{1}}}$,
checking the condition $  \mathrm{FTV}( \ottnt{A} )   \subseteq \mathrm{dom}( \Gamma_{{\mathrm{1}}} ) $ is easier than
checking the well-formedness of $\ottnt{A}$ under $\Gamma_{{\mathrm{1}}}$.
The rule \ASRBox is similar.

The rule \ASLAllL differs from the declarative rule \SLAllL in two respects.
The first point is that it has the side condition $\mathsf{not}\forall \, \ottsym{(}  {B^\Box}  \ottsym{)}$, meaning that
the boxy type ${B^\Box}$ does not take the form of $ \forall  \ottmv{a}  .  \ottnt{A} $~.
This condition enables our typing algorithm to determine the rule to use uniquely.
For example, given the following worklist, our typing algorithm always uses the rule \ASLAllR.
\[
  \Gamma  \Vdash   \subL{   \forall  \ottmv{a}  .  \ottmv{a}    \rightarrow   \ottmv{a}  }{  \forall  \ottmv{b}  .   \ottmv{b}   \rightarrow   \ottmv{b}   }  %
\]
This approach is similar to
the work by Zhao et al.~\citet{zhao_mechanical_2019}.
We may impose similar conditions on other rules for the same reason.
The second point is that the rule \ASLAllL introduces the existential variable $\monoalpha$ and
uses it to instantiate a quantified type variable.
The introduced existential variable $\monoalpha$ will be solved to
the monotype that is well-formed under the worklist $\Gamma$.
Because the mini worklist $\Delta$ has no type nor existential variable binding, and
no monotype has polymorphic existential variables,
the monotype to which $\monoalpha$ is solved is well-formed under the worklist $\Gamma$
if it is well-formed under the worklist $\Gamma  \ottsym{,}  \triangleright  \ottsym{\{}  \vec{\polyalpha}  \ottsym{\}}  \ottsym{,}  \Delta$.
This implication keeps the correspondence between our typing algorithm and
the declarative type system of \lang.
\mycomment{
  One may be concerned that
  a polymorphic existential variable in ${B^\Box}$ can occur in $\Gamma$.
  In this situation, our typing algorithm might get stuck later.
  However, this kind of situation never happens in reducing the worklist $\Gamma$,
  as long as $\Gamma$ does not fall into this case, which justifies our typing algorithm.
  The rule \ASRAllL has the same matter.
}

We show an example corresponding to the declarative judgment
$  \emptyset    \vdash  \subL{   \forall  \ottmv{a}  .  \ottmv{a}    \rightarrow   \ottmv{a}  }{  \boxed{  \mathsf{unit}  }   \rightarrow   \boxed{  \mathsf{unit}  }  } $~.
\[
  \begin{array}{rll}
    \triangleright  \ottsym{\{}  \polyalpha  \ottsym{,}  \polybeta  \ottsym{\}}  \Vdash   \subL{   \forall  \ottmv{a}  .  \ottmv{a}    \rightarrow   \ottmv{a}  }{  \boxed{ \polyalpha }   \rightarrow   \boxed{ \polybeta }  } 
     &  \rightsquigarrow   & \monoalpha  \ottsym{,}  \triangleright  \ottsym{\{}  \polyalpha  \ottsym{,}  \polybeta  \ottsym{\}}  \Vdash   \subL{  \monoalpha   \rightarrow   \monoalpha  }{  \boxed{ \polyalpha }   \rightarrow   \boxed{ \polybeta }  }    \\
     &  \rightsquigarrow   & \monoalpha  \ottsym{,}  \triangleright  \ottsym{\{}  \polyalpha  \ottsym{,}  \polybeta  \ottsym{\}}  \Vdash   \subL{ \monoalpha }{  \boxed{ \polybeta }  }   \Vdash   \subR{  \boxed{ \polyalpha }  }{ \monoalpha }  \\
     &  \rightsquigarrow^*  &  \emptyset 
  \end{array}
\]
The first step introduces the existential variable $\monoalpha$ using \ASLAllL.
The second step decomposes the function types using \ASLFun.
After these steps,
both polymorphic existential variables $\polyalpha$ and $\polybeta$ will be solved to $\monoalpha$.
As the final result, we reach the empty worklist,
indicating that this typing process has successfully ended.

\paragraph{Solving existential variables}

We show the algorithmic subtyping rules that solve existential variables
in \zcref{fig:algo_subtyping_ex}.
We omit the rules for $ \subtypingright $
because they are similar to the rules for $ \subtypingleft $. %
\begin{figure}[t]
  \renewcommand{\arraystretch}{1.2}
  \small
  \[
    \begin{array}{rcl@{\quad}l}
                 \Gamma_{{\mathrm{1}}}  \ottsym{,}  \monoalpha  \ottsym{,}  \Gamma_{{\mathrm{2}}}  \ottsym{,}  \monobeta  \ottsym{,}  \Gamma_{{\mathrm{3}}}  \Vdash   \subL{ \monoalpha }{ \monobeta }  &  \rightsquigarrow         &
                 \Gamma_{{\mathrm{1}}}  \ottsym{,}  \monoalpha  \ottsym{,}  \Gamma_{{\mathrm{2}}}  \ottsym{,}   [  \monoalpha  /  \monobeta  ]  \Gamma_{{\mathrm{3}}}            & (\ASLExExA)                  \\
                 \Gamma_{{\mathrm{1}}}  \ottsym{,}  \monoalpha  \ottsym{,}  \Gamma_{{\mathrm{2}}}  \ottsym{,}  \monobeta  \ottsym{,}  \Gamma_{{\mathrm{3}}}  \Vdash   \subL{ \monobeta }{ \monoalpha }  &  \rightsquigarrow         &
                 \Gamma_{{\mathrm{1}}}  \ottsym{,}  \monoalpha  \ottsym{,}  \Gamma_{{\mathrm{2}}}  \ottsym{,}   [  \monoalpha  /  \monobeta  ]  \Gamma_{{\mathrm{3}}}            & (\ASLExExB)                  \\
                 \Gamma_{{\mathrm{1}}}  \ottsym{,}  \ottmv{a}  \ottsym{,}  \Gamma_{{\mathrm{2}}}  \ottsym{,}  \monoalpha  \ottsym{,}  \Gamma_{{\mathrm{3}}}  \Vdash   \subL{ \ottmv{a} }{ \monoalpha }    &  \rightsquigarrow         &
                 \Gamma_{{\mathrm{1}}}  \ottsym{,}  \ottmv{a}  \ottsym{,}  \Gamma_{{\mathrm{2}}}  \ottsym{,}   [  \ottmv{a}  /  \monoalpha  ]  \Gamma_{{\mathrm{3}}}              & (\ASLExTVarA)                \\
                 \Gamma_{{\mathrm{1}}}  \ottsym{,}  \ottmv{a}  \ottsym{,}  \Gamma_{{\mathrm{2}}}  \ottsym{,}  \monoalpha  \ottsym{,}  \Gamma_{{\mathrm{3}}}  \Vdash   \subL{ \monoalpha }{ \ottmv{a} }    &  \rightsquigarrow         &
                 \Gamma_{{\mathrm{1}}}  \ottsym{,}  \ottmv{a}  \ottsym{,}  \Gamma_{{\mathrm{2}}}  \ottsym{,}   [  \ottmv{a}  /  \monoalpha  ]  \Gamma_{{\mathrm{3}}}              & (\ASLExTVarB)                \\
                 \Gamma_{{\mathrm{1}}}  \ottsym{,}  \monoalpha  \ottsym{,}  \Gamma_{{\mathrm{2}}}  \Vdash   \subL{  \mathsf{unit}  }{ \monoalpha }        &  \rightsquigarrow         &
                 \Gamma_{{\mathrm{1}}}  \ottsym{,}   [   \mathsf{unit}   /  \monoalpha  ]  \Gamma_{{\mathrm{2}}}                  & (\ASLExUnitA)                \\
                 \Gamma_{{\mathrm{1}}}  \ottsym{,}  \monoalpha  \ottsym{,}  \Gamma_{{\mathrm{2}}}  \Vdash   \subL{ \monoalpha }{  \mathsf{unit}  }        &  \rightsquigarrow         &
                 \Gamma_{{\mathrm{1}}}  \ottsym{,}   [   \mathsf{unit}   /  \monoalpha  ]  \Gamma_{{\mathrm{2}}}                  & (\ASLExUnitB)
    \end{array}
  \]
  \[
    \begin{array}{l@{\quad}l}
                 \Gamma_{{\mathrm{1}}}  \ottsym{,}  \monoalpha  \ottsym{,}  \Gamma_{{\mathrm{2}}}  \Vdash   \subL{  \ottnt{A}   \rightarrow   \ottnt{B}  }{ \monoalpha }                                       &                            \\
                 \quad  \rightsquigarrow  \quad
                              \Gamma_{{\mathrm{1}}}  \ottsym{,}  \monoalpha_{{\mathrm{1}}}  \ottsym{,}  \monoalpha_{{\mathrm{2}}}  \ottsym{,}   [   \monoalpha_{{\mathrm{1}}}   \rightarrow   \monoalpha_{{\mathrm{2}}}   /  \monoalpha  ]  \ottsym{(}  \Gamma_{{\mathrm{2}}}  \Vdash   \subL{ \ottnt{B} }{ \monoalpha_{{\mathrm{2}}} }   \Vdash   \subR{ \monoalpha_{{\mathrm{1}}} }{ \ottnt{A} }   \ottsym{)}  & (\ASLExFunA) \\
                 \quad \phantom{{} \rightsquigarrow {}} \quad \tif  \monoalpha  \notin   \mathrm{FTV}( \ottnt{A}  \rightarrow  \ottnt{B} )                                                     \\
                 \Gamma_{{\mathrm{1}}}  \ottsym{,}  \monoalpha  \ottsym{,}  \Gamma_{{\mathrm{2}}}  \Vdash   \subL{ \monoalpha }{ {A^\Box}  \rightarrow  {B^\Box} }                                       &                           \\
                 \quad  \rightsquigarrow  \quad
                              \Gamma_{{\mathrm{1}}}  \ottsym{,}  \monoalpha_{{\mathrm{1}}}  \ottsym{,}  \monoalpha_{{\mathrm{2}}}  \ottsym{,}   [   \monoalpha_{{\mathrm{1}}}   \rightarrow   \monoalpha_{{\mathrm{2}}}   /  \monoalpha  ]  \ottsym{(}  \Gamma_{{\mathrm{2}}}  \Vdash   \subL{ \monoalpha_{{\mathrm{2}}} }{ {B^\Box} }   \Vdash   \subR{ {A^\Box} }{ \monoalpha_{{\mathrm{1}}} }   \ottsym{)}   & (\ASLExFunB) \\
                 \quad \phantom{{} \rightsquigarrow {}} \quad \tif  \monoalpha  \notin   \mathrm{FTV}( {A^\Box}  \rightarrow  {B^\Box} )  
    \end{array}
  \]
  \caption{Algorithmic subtyping rules solving existential variables}\label{fig:algo_subtyping_ex}
\end{figure}

\mycomment{
  The rules, \ASLExExA, \ASLExExB, \ASLExTVarA, and \ASLExTVarB,
  solve the existential variable to the existential or type variable in its opposite site.
  The occurrence order of the existential variable in the worklist
  is crucial in these rules.

  The rules \ASLExUnitA and \ASLExUnitB just substitute $ \mathsf{unit} $ for the given existential variable.
}

The first six rules instantiate the existential variable with the type on the opposite side.
The occurrence order of the existential variable in the worklist
is crucial in these rules.

The rules \ASLExFunA and \ASLExFunB ``decompose'' the given existential variable $\monoalpha$
into $\monoalpha_{{\mathrm{1}}}$ and $\monoalpha_{{\mathrm{2}}}$ for the argument and return type, respectively.
We use the side conditions $ \monoalpha  \notin   \mathrm{FTV}( \ottnt{A}  \rightarrow  \ottnt{B} )  $ and $ \monoalpha  \notin   \mathrm{FTV}( {A^\Box}  \rightarrow  {B^\Box} )  $
for the occurs check as usual.

We show an example corresponding to the declarative judgment
$  \emptyset    \vdash  \subL{   \forall  \ottmv{a}  .  \ottmv{a}    \rightarrow   \ottmv{a}  }{  \mathsf{unit}   \rightarrow   \boxed{  \mathsf{unit}  }  } $~.
\[
  \begin{array}{rll}
    \triangleright  \ottsym{\{}  \polyalpha  \ottsym{\}}  \Vdash   \subL{   \forall  \ottmv{a}  .  \ottmv{a}    \rightarrow   \ottmv{a}  }{  \mathsf{unit}   \rightarrow   \boxed{ \polyalpha }  } 
     &  \rightsquigarrow  & \monoalpha  \ottsym{,}  \triangleright  \ottsym{\{}  \polyalpha  \ottsym{\}}  \Vdash   \subL{  \monoalpha   \rightarrow   \monoalpha  }{  \mathsf{unit}   \rightarrow   \boxed{ \polyalpha }  }    \\
     &  \rightsquigarrow  & \monoalpha  \ottsym{,}  \triangleright  \ottsym{\{}  \polyalpha  \ottsym{\}}  \Vdash   \subL{ \monoalpha }{  \boxed{ \polyalpha }  }   \Vdash   \subR{  \mathsf{unit}  }{ \monoalpha }  \\
     &  \rightsquigarrow  & \triangleright  \ottsym{\{}  \polyalpha  \ottsym{\}}  \Vdash   \subL{  \mathsf{unit}  }{  \boxed{ \polyalpha }  }                    \\
  \end{array}
\]
The first two steps are similar to the aforementioned example.
The third step solves the existential variable $\monoalpha$ to $ \mathsf{unit} $.
After that, the polymorphic existential variable is solved to $ \mathsf{unit} $ as a result.

\paragraph{Algorithmic typing}

We show our algorithmic typing rules in \zcref{fig:algo_typing}.
\begin{figure}[t]
  \renewcommand{\arraystretch}{1.2}
  \small
  \[
    \begin{array}{rcl@{\quad}l}
                 \Gamma  \Vdash  \ottsym{()}  \ottsym{:}  {A^\Box}                                                    &  \rightsquigarrow     &
                 \Gamma  \Vdash   \subL{  \mathsf{unit}  }{ {A^\Box} }  \quad \tif \mathsf{not}\forall \, \ottsym{(}  {A^\Box}  \ottsym{)}                   & (\ATUnit)                \\
                 \Gamma  \Vdash  \mathit{x}  \ottsym{:}  {A^\Box}                                                     &  \rightsquigarrow     &
                 \Gamma  \Vdash   \subL{ \ottnt{A} }{ {A^\Box} }  \quad \tif  \mathit{x}  \ottsym{:}  \ottnt{A}  \in  \Gamma  \tand \mathsf{not}\forall \, \ottsym{(}  {A^\Box}  \ottsym{)} & (\ATVar)                 \\
                 \Gamma  \Vdash  \ottnt{e}  \ottsym{:}   \forall  \ottmv{a}  .  \ottnt{A}                                           &  \rightsquigarrow     &
                 \Gamma  \ottsym{,}  \ottmv{a}  \ottsym{,}  \triangleright  \ottsym{\{}  \ottsym{\}}  \Vdash  \ottnt{e}  \ottsym{:}  \ottnt{A}                                          & (\ATAll)
    \end{array}
  \]
  \[
    \begin{array}{l@{\quad}l}
      \Gamma_{{\mathrm{1}}}  \ottsym{,}  \monoalpha  \ottsym{,}  \Gamma_{{\mathrm{2}}}  \Vdash   \lambda  \mathit{x}  .  \ottnt{e}   \ottsym{:}  \monoalpha                                                                                                 %
      \quad  \rightsquigarrow  \quad \Gamma_{{\mathrm{1}}}  \ottsym{,}  \monoalpha_{{\mathrm{1}}}  \ottsym{,}  \monoalpha_{{\mathrm{2}}}  \ottsym{,}   [   \monoalpha_{{\mathrm{1}}}   \rightarrow   \monoalpha_{{\mathrm{2}}}   /  \monoalpha  ]  \ottsym{(}  \Gamma_{{\mathrm{2}}}  \ottsym{,}  \mathit{x}  \ottsym{:}  \monoalpha_{{\mathrm{1}}}  \Vdash  \ottnt{e}  \ottsym{:}  \monoalpha_{{\mathrm{2}}}  \ottsym{)}              & (\ATAbsEx)   \\
      \Gamma_{{\mathrm{1}}}  \ottsym{,}  \triangleright  \ottsym{\{}  \vec{\polybeta}_{{\mathrm{1}}}  \ottsym{,}  \polyalpha  \ottsym{,}  \vec{\polybeta}_{{\mathrm{2}}}  \ottsym{\}}  \ottsym{,}  \Gamma_{{\mathrm{2}}}  \Vdash   \lambda  \mathit{x}  .  \ottnt{e}   \ottsym{:}   \boxed{ \polyalpha }                                             &              \\
      \quad  \rightsquigarrow  \quad \Gamma_{{\mathrm{1}}}  \ottsym{,}  \monoalpha  \ottsym{,}  \triangleright  \ottsym{\{}  \vec{\polybeta}_{{\mathrm{1}}}  \ottsym{,}  \polybeta  \ottsym{,}  \vec{\polybeta}_{{\mathrm{2}}}  \ottsym{\}}  \ottsym{,}   [  \monoalpha  \rightarrow  \polybeta  /  \polyalpha  ]  \ottsym{(}  \Gamma_{{\mathrm{2}}}  \ottsym{,}  \mathit{x}  \ottsym{:}  \monoalpha  \Vdash  \ottnt{e}  \ottsym{:}   \boxed{ \polybeta }   \ottsym{)}  & (\ATAbsBox)  \\
      \Gamma  \Vdash   \lambda  \mathit{x}  .  \ottnt{e}   \ottsym{:}  \ottnt{A}  \rightarrow  {B^\Box} %
      \quad  \rightsquigarrow  \quad \Gamma  \ottsym{,}  \mathit{x}  \ottsym{:}  \ottnt{A}  \Vdash  \ottnt{e}  \ottsym{:}  {B^\Box}                                                 & (\ATAbsFunA) \\
      \Gamma  \ottsym{,}  \triangleright  \ottsym{\{}  \vec{\polyalpha}  \ottsym{\}}  \ottsym{,}  \Delta  \Vdash   \lambda  \mathit{x}  .  \ottnt{e}   \ottsym{:}  {A^\Box}  \rightarrow  {B^\Box}                                                        &              \\
      \quad  \rightsquigarrow  \quad \Gamma  \ottsym{,}  \monoalpha  \ottsym{,}  \triangleright  \ottsym{\{}  \vec{\polyalpha}  \ottsym{\}}  \ottsym{,}  \Delta  \Vdash   \subR{ {A^\Box} }{ \monoalpha }   \ottsym{,}  \mathit{x}  \ottsym{:}  \monoalpha  \Vdash  \ottnt{e}  \ottsym{:}  {B^\Box}                  &
      (\ATAbsFunB)                                                                                               \\
      \quad \phantom{{} \rightsquigarrow {}} \quad \tif \mathsf{hasBox} \, \ottsym{(}  {A^\Box}  \ottsym{)}                                        &              \\[1ex]
      \Gamma_{{\mathrm{1}}}  \ottsym{,}  \monoalpha  \ottsym{,}  \Gamma_{{\mathrm{2}}}  \Vdash   \lambda  \mathit{x}  \ottsym{:}  \ottnt{A}  .  \ottnt{e}   \ottsym{:}  \monoalpha                                                         &              \\
      \quad  \rightsquigarrow  \quad \Gamma_{{\mathrm{1}}}  \ottsym{,}  \monoalpha_{{\mathrm{1}}}  \ottsym{,}  \monoalpha_{{\mathrm{2}}}  \ottsym{,}   [   \monoalpha_{{\mathrm{1}}}   \rightarrow   \monoalpha_{{\mathrm{2}}}   /  \monoalpha  ]  \ottsym{(}  \Gamma_{{\mathrm{2}}}  \Vdash   \subR{ \monoalpha_{{\mathrm{1}}} }{ \ottnt{A} }   \ottsym{,}  \mathit{x}  \ottsym{:}  \ottnt{A}  \Vdash  \ottnt{e}  \ottsym{:}  \monoalpha_{{\mathrm{2}}}  \ottsym{)}  & (\ATAAbsEx)  \\
      \Gamma_{{\mathrm{1}}}  \ottsym{,}  \triangleright  \ottsym{\{}  \vec{\polybeta}_{{\mathrm{1}}}  \ottsym{,}  \polyalpha  \ottsym{,}  \vec{\polybeta}_{{\mathrm{2}}}  \ottsym{\}}  \ottsym{,}  \Gamma_{{\mathrm{2}}}  \Vdash   \lambda  \mathit{x}  \ottsym{:}  \ottnt{A}  .  \ottnt{e}   \ottsym{:}   \boxed{ \polyalpha }                                       &              \\
      \quad  \rightsquigarrow  \quad \Gamma_{{\mathrm{1}}}  \ottsym{,}  \triangleright  \ottsym{\{}  \vec{\polybeta}_{{\mathrm{1}}}  \ottsym{,}  \polybeta  \ottsym{,}  \vec{\polybeta}_{{\mathrm{2}}}  \ottsym{\}}  \ottsym{,}   [  \ottnt{A}  \rightarrow  \polybeta  /  \polyalpha  ]  \ottsym{(}  \Gamma_{{\mathrm{2}}}  \ottsym{,}  \mathit{x}  \ottsym{:}  \ottnt{A}  \Vdash  \ottnt{e}  \ottsym{:}   \boxed{ \polybeta }   \ottsym{)}      & (\ATAAbsBox) \\
      \quad \phantom{{} \rightsquigarrow {}} \quad \tif   \mathrm{FTV}( \ottnt{A} )   \subseteq \mathrm{dom}( \Gamma_{{\mathrm{1}}} )                             &              \\
      \Gamma  \Vdash   \lambda  \mathit{x}  \ottsym{:}  \ottnt{A}  .  \ottnt{e}   \ottsym{:}  {A^\Box}  \rightarrow  {B^\Box}                                                                                                %
      \quad  \rightsquigarrow  \quad \Gamma  \Vdash   \subR{ {A^\Box} }{ \ottnt{A} }   \ottsym{,}  \mathit{x}  \ottsym{:}  \ottnt{A}  \Vdash  \ottnt{e}  \ottsym{:}  {B^\Box}                                       & (\ATAAbsFun)
    \end{array}
  \]\[
    \begin{array}{rcl@{\quad}l}
                 \Gamma  \ottsym{,}  \triangleright  \ottsym{\{}  \vec{\polyalpha}  \ottsym{\}}  \ottsym{,}  \Delta  \Vdash  \ottnt{e_{{\mathrm{1}}}}  \appright  \ottnt{e_{{\mathrm{2}}}}  \ottsym{:}  {A^\Box}                    &  \rightsquigarrow     &
                 \Gamma  \ottsym{,}  \triangleright  \ottsym{\{}  \vec{\polyalpha}  \ottsym{,}  \polyalpha  \ottsym{\}}  \ottsym{,}  \Delta  \Vdash  \ottnt{e_{{\mathrm{2}}}}  \ottsym{:}  \polyalpha  \Vdash  \ottnt{e_{{\mathrm{1}}}}  \ottsym{:}   \boxed{ \polyalpha }   \rightarrow  {A^\Box}  & (\ATAppR)                            \\
                                                                       &           & \tif \mathsf{not}\forall \, \ottsym{(}  {A^\Box}  \ottsym{)} & \\
                 \Gamma  \ottsym{,}  \triangleright  \ottsym{\{}  \vec{\polyalpha}  \ottsym{\}}  \ottsym{,}  \Delta  \Vdash  \ottnt{e_{{\mathrm{1}}}}  \appleft  \ottnt{e_{{\mathrm{2}}}}  \ottsym{:}  {A^\Box}                    &  \rightsquigarrow     &
                 \Gamma  \ottsym{,}  \triangleright  \ottsym{\{}  \vec{\polyalpha}  \ottsym{,}  \polyalpha  \ottsym{\}}  \ottsym{,}  \Delta  \Vdash  \ottnt{e_{{\mathrm{1}}}}  \ottsym{:}  \polyalpha  \rightarrow  {A^\Box}  \Vdash  \ottnt{e_{{\mathrm{2}}}}  \ottsym{:}   \boxed{ \polyalpha }  & (\ATAppL)                            \\
                                                                       &           & \tif \mathsf{not}\forall \, \ottsym{(}  {A^\Box}  \ottsym{)} &
    \end{array}
  \]
  \caption{Algorithmic typing rules}\label{fig:algo_typing}
\end{figure}
Many rules straightforwardly correspond to the declarative rules.
We mainly explain the difference between them. %

The side condition $\mathsf{not}\forall \, \ottsym{(}  {A^\Box}  \ottsym{)}$ in \ATUnit and \ATVar helps determine the rule to use uniquely.

The rule \ATAll additionally introduces the scope delimiter $\triangleright  \ottsym{\{}  \ottsym{\}}$
after the type variable $\ottmv{a}$ in the given worklist, as explained in \zcref{subsubsec:scope_delimiter}.

The rules \ATAbsEx, \ATAbsBox, \ATAbsFunA, and \ATAbsFunB are for lambda abstractions.
\ATAbsEx decomposes the given existential variable $\monoalpha$
into $\monoalpha_{{\mathrm{1}}}$ and $\monoalpha_{{\mathrm{2}}}$ for the argument and return type, respectively,
similarly to \ASLExFunA and \ASLExFunB.
\ATAbsBox corresponds to the declarative rule \TAbsBox.
Since \TAbsBox guesses the argument type $\tau$,
\ATAbsBox introduces the existential variable $\monoalpha$.
This $\monoalpha$ must be introduced just after the worklist $\Gamma_{{\mathrm{1}}}$
because the given polymorphic existential variable $\polyalpha$ must be solved to
a type well-formed under this worklist.
\ATAbsFunA is very similar to the declarative rule \TAbsFunA.
\ATAbsFunB has two notable points.
One is that it introduces the existential variable $\monoalpha$,
similarly to \ASLAllL or \ASRAllL.
The other is that the subtyping work item $ \subR{ {A^\Box} }{ \monoalpha } $ occurs just before the binding $\mathit{x}  \ottsym{:}  \monoalpha$.
This occurrence corresponds to the premise $ \Psi   \vdash  \subR{ {A^\Box} }{ \tau } $ in
the declarative rule \TAbsFunB.

The rules \ATAAbsEx, \ATAAbsBox, and \ATAAbsFun are for lambda abstractions with type annotations.
\ATAAbsEx
decomposes the given existential variable $\monoalpha$
into $\monoalpha_{{\mathrm{1}}}$ and $\monoalpha_{{\mathrm{2}}}$, similarly to \ATAbsEx.
It introduces the subtyping work item $ \subR{ \monoalpha }{ \ottnt{A} } $ just before the binding $\mathit{x}  \ottsym{:}  \ottnt{A}$,
similarly to \ATAbsFunB.
\ATAAbsBox has the side condition $  \mathrm{FTV}( \ottnt{A} )   \subseteq \mathrm{dom}( \Gamma_{{\mathrm{1}}} ) $
because the given polymorphic existential variable $\polyalpha$ must be solved to
a well-formed type under $\Gamma_{{\mathrm{1}}}$.
\ATAAbsFun
introduces the subtyping work item $ \subR{ {A^\Box} }{ \ottnt{A} } $ just before $\mathit{x}  \ottsym{:}  \ottnt{A}$,
similarly to \ATAbsFunB.

The rule \ATAppR introduces the polymorphic existential variable $\polyalpha$
in the rightmost scope delimiter.
Using the mini worklist $\Delta$ keeps the correspondence between
the polymorphic existential variable $\polyalpha$ and
the type $\ottnt{A}$ in the declarative typing rule \TAppR.
The rule \ATAppL is similar.

We show two examples to illustrate how our typing algorithm works.
The first example corresponds to the declarative judgment
$\mathit{f}  \ottsym{:}    \forall  \ottmv{a}  .  \ottmv{a}    \rightarrow   \ottmv{a}   \vdash  \mathit{f}  \appright  \ottsym{()}  \ottsym{:}   \boxed{  \mathsf{unit}  } $~.
\[
  \begin{array}{ll}
            & \triangleright  \ottsym{\{}  \polyalpha  \ottsym{\}}  \ottsym{,}  \mathit{f}  \ottsym{:}    \forall  \ottmv{a}  .  \ottmv{a}    \rightarrow   \ottmv{a}   \Vdash  \mathit{f}  \appright  \ottsym{()}  \ottsym{:}   \boxed{ \polyalpha }                                      \\
     \rightsquigarrow   & \triangleright  \ottsym{\{}  \polyalpha  \ottsym{,}  \polybeta  \ottsym{\}}  \ottsym{,}  \mathit{f}  \ottsym{:}    \forall  \ottmv{a}  .  \ottmv{a}    \rightarrow   \ottmv{a}   \Vdash  \ottsym{()}  \ottsym{:}  \polybeta  \Vdash  \mathit{f}  \ottsym{:}   \boxed{ \polybeta }   \rightarrow   \boxed{ \polyalpha }                    \\
     \rightsquigarrow   & \triangleright  \ottsym{\{}  \polyalpha  \ottsym{,}  \polybeta  \ottsym{\}}  \ottsym{,}  \mathit{f}  \ottsym{:}    \forall  \ottmv{a}  .  \ottmv{a}    \rightarrow   \ottmv{a}   \Vdash  \ottsym{()}  \ottsym{:}  \polybeta  \Vdash   \subL{   \forall  \ottmv{a}  .  \ottmv{a}    \rightarrow   \ottmv{a}  }{  \boxed{ \polybeta }   \rightarrow   \boxed{ \polyalpha }  }  \\
     \rightsquigarrow^*  & \monoalpha  \ottsym{,}  \mathit{f}  \ottsym{:}    \forall  \ottmv{a}  .  \ottmv{a}    \rightarrow   \ottmv{a}   \Vdash  \ottsym{()}  \ottsym{:}  \monoalpha                                                \\
     \rightsquigarrow   & \monoalpha  \ottsym{,}  \mathit{f}  \ottsym{:}    \forall  \ottmv{a}  .  \ottmv{a}    \rightarrow   \ottmv{a}   \Vdash   \subR{  \mathsf{unit}  }{ \monoalpha } 
  \end{array}
\]
The first step introduces the polymorphic existential variable $\polybeta$ using \ATAppR.
After the second reduction by \ATVar, this typing process proceeds similarly to
the aforementioned example of algorithmic subtyping.
The existential variable $\monoalpha$ is introduced via this subtyping process;
the unit expression is checked against it, and finally it is solved to $ \mathsf{unit} $.
The second example corresponds to the declarative judgment
$\mathit{f}  \ottsym{:}    \forall  \ottmv{a}  .  \ottmv{a}    \rightarrow   \ottmv{a}   \vdash  \mathit{f}  \appleft  \ottsym{()}  \ottsym{:}   \boxed{  \mathsf{unit}  } $~.
\[
  \begin{array}{ll}
           & \triangleright  \ottsym{\{}  \polyalpha  \ottsym{\}}  \ottsym{,}  \mathit{f}  \ottsym{:}    \forall  \ottmv{a}  .  \ottmv{a}    \rightarrow   \ottmv{a}   \Vdash  \mathit{f}  \appleft  \ottsym{()}  \ottsym{:}   \boxed{ \polyalpha }                         \\
     \rightsquigarrow  & \triangleright  \ottsym{\{}  \polyalpha  \ottsym{,}  \polybeta  \ottsym{\}}  \ottsym{,}  \mathit{f}  \ottsym{:}    \forall  \ottmv{a}  .  \ottmv{a}    \rightarrow   \ottmv{a}   \Vdash  \mathit{f}  \ottsym{:}  \polybeta  \rightarrow   \boxed{ \polyalpha }   \Vdash  \ottsym{()}  \ottsym{:}   \boxed{ \polybeta }       \\
     \rightsquigarrow  & \triangleright  \ottsym{\{}  \polyalpha  \ottsym{,}  \polybeta  \ottsym{\}}  \ottsym{,}  \mathit{f}  \ottsym{:}    \forall  \ottmv{a}  .  \ottmv{a}    \rightarrow   \ottmv{a}   \Vdash  \mathit{f}  \ottsym{:}  \polybeta  \rightarrow   \boxed{ \polyalpha }   \Vdash   \subL{  \mathsf{unit}  }{  \boxed{ \polybeta }  }  \\
     \rightsquigarrow  & \triangleright  \ottsym{\{}  \polyalpha  \ottsym{\}}  \ottsym{,}  \mathit{f}  \ottsym{:}    \forall  \ottmv{a}  .  \ottmv{a}    \rightarrow   \ottmv{a}   \Vdash  \mathit{f}  \ottsym{:}   \mathsf{unit}   \rightarrow   \boxed{ \polyalpha }                       \\
     \rightsquigarrow  & \triangleright  \ottsym{\{}  \polyalpha  \ottsym{\}}  \ottsym{,}  \mathit{f}  \ottsym{:}    \forall  \ottmv{a}  .  \ottmv{a}    \rightarrow   \ottmv{a}   \Vdash   \subL{   \forall  \ottmv{a}  .  \ottmv{a}    \rightarrow   \ottmv{a}  }{  \mathsf{unit}   \rightarrow   \boxed{ \polyalpha }  } 
  \end{array}
\]
The first step introduces the polymorphic existential variable $\polybeta$ using \ATAppL.
The two steps using \ATUnit and \ASLBox solve $\polybeta$ to $ \mathsf{unit} $, and
$\mathit{f}$ must have $ \mathsf{unit} $ as its argument type.
After using \ATVar, this typing process proceeds similarly to
the aforementioned example of algorithmic subtyping.

\subsection{Metatheory}

We show the metatheory of our typing algorithm. %
\zcref{subsubsec:algo_sound_app} shows that
our typing algorithm is sound with respect to the type system of \lang.
However, the algorithm is not complete with respect to the type system of \lang.
We show a counterexample to the completeness of our typing algorithm in \zcref{subsubsec:cyclic}.
However, our typing algorithm is expressive enough to subsume both the DK and the XO systems,
as shown in \zcref{subsubsec:algo_complete_app}.

\subsubsection{Preliminaries}\label{subsubsec:meta_pre}

We use \emph{declarative worklists} and \emph{worklist instantiation}
to relate our typing algorithm and the type system of \lang,
as Zhao et al.~\citet{zhao_mechanical_2019} do.
Declarative worklists and their reduction can be viewed as
the intermediate system between
our typing algorithm and the type system of \lang.
Every declarative typing rule of the type system of \lang
corresponds to the reduction of its declarative worklist counterpart.
An algorithmic work item can be instantiated into a declarative worklist
by appropriately instantiating the (polymorphic) existential variables in it.

A declarative worklist and its reduction are defined in \zcref{fig:decl_worklist}.
A declarative worklist is a worklist that does not contain variables or scope delimiters.
The subtyping/typing work item on top of the declarative worklist $\Omega$ is removed
only if it holds under the typing context $ |  \Omega  | $.
The translation $ |  \Omega  | $ just removes work items in the worklist $\Omega$, and is defined as follows.
\[
  \begin{array}{r@{\ }c@{\ }l@{\qquad}r@{\ }c@{\ }l@{\qquad}r@{\ }c@{\ }l@{\qquad}r@{\ }c@{\ }l}
                |   \emptyset   |  & = &  \emptyset      &
                |  \Omega  \ottsym{,}  \ottmv{a}  |                                                                         & = &  |  \Omega  |   \ottsym{,}  \ottmv{a}     &
                |  \Omega  \ottsym{,}  \mathit{x}  \ottsym{:}  \ottnt{A}  |                                                                     & = &  |  \Omega  |   \ottsym{,}  \mathit{x}  \ottsym{:}  \ottnt{A} &
                |  \Omega  \Vdash  \omega  |                                                                      & = &  |  \Omega  | 
  \end{array}
\]

\begin{figure}[t]
  \small
  \[
    \begin{array}{rrcl}
      \textbf{Declarative worklists} & \Omega &  \Coloneqq  &
       \emptyset   \mid  \Omega  \ottsym{,}  \ottmv{a}  \mid  \Omega  \ottsym{,}  \mathit{x}  \ottsym{:}  \ottnt{A}  \mid  \Omega  \Vdash  \omega
    \end{array}
  \]
  \noindent\fbox{$\Omega  \rightsquigarrow  \Omega'$}\hfill\mbox{} %
  \[
    \begin{array}{rcll@{\qquad}rcll}
                 \Omega  \ottsym{,}  \ottmv{a} &  \rightsquigarrow  & \Omega &                         &
                 \Omega  \ottsym{,}  \mathit{x}  \ottsym{:}  \ottnt{A}        &  \rightsquigarrow  & \Omega &                             \\
                 \Omega  \Vdash   \subL{ \ottnt{A} }{ {B^\Box} }    &  \rightsquigarrow  & \Omega & \tif   |  \Omega  |    \vdash  \subL{ \ottnt{A} }{ {B^\Box} }  &
                 \Omega  \Vdash   \subR{ {A^\Box} }{ \ottnt{B} }    &  \rightsquigarrow  & \Omega & \tif   |  \Omega  |    \vdash  \subR{ {A^\Box} }{ \ottnt{B} }      \\
                 \Omega  \Vdash  \ottnt{e}  \ottsym{:}  {A^\Box}     &  \rightsquigarrow  & \Omega & \tif  |  \Omega  |   \vdash  \ottnt{e}  \ottsym{:}  {A^\Box}
    \end{array}
  \]
  \caption{Declarative worklist and its reduction}\label{fig:decl_worklist}
\end{figure}

Worklist instantiation rules are defined in \zcref{fig:work_inst}.
The rule \InstRefl does nothing
because a declarative worklist $\Omega$ has no (polymorphic) existential variable.
The rule \InstEx instantiates the existential variable $\monoalpha$
with the monotype $\tau$ well-formed under the typing context $ |  \Omega  | $.
The rule \InstNil just removes the scope delimiter $\triangleright  \ottsym{\{}  \ottsym{\}}$.
The rule \InstPex instantiates the polymorphic existential variable $\polyalpha$
with the type $\ottnt{A}$ well-formed under the typing context $ |  \Omega  | $.

\begin{figure}
  \small
  \noindent\fbox{$\Gamma  \xrightarrow{ \mathrm{inst} }  \Omega$}\hfill\mbox{}
  \begin{mathpar}
    \inferrule{ }{
      \Omega  \xrightarrow{ \mathrm{inst} }  \Omega
    }\ \InstRefl

    \inferrule{
     |  \Omega_{{\mathrm{1}}}  |   \vdash  \tau \\ \Omega_{{\mathrm{1}}}  \ottsym{,}   [  \tau  /  \monoalpha  ]  \Gamma_{{\mathrm{2}}}   \xrightarrow{ \mathrm{inst} }  \Omega
    }{
      \Omega_{{\mathrm{1}}}  \ottsym{,}  \monoalpha  \ottsym{,}  \Gamma_{{\mathrm{2}}}  \xrightarrow{ \mathrm{inst} }  \Omega
    }\ \InstEx

    \inferrule{
      \Omega_{{\mathrm{1}}}  \ottsym{,}  \Gamma_{{\mathrm{2}}}  \xrightarrow{ \mathrm{inst} }  \Omega
    }{
      \Omega_{{\mathrm{1}}}  \ottsym{,}  \triangleright  \ottsym{\{}  \ottsym{\}}  \ottsym{,}  \Gamma_{{\mathrm{2}}}  \xrightarrow{ \mathrm{inst} }  \Omega
    }\ \InstNil

    \inferrule{
     |  \Omega_{{\mathrm{1}}}  |   \vdash  \ottnt{A} \\ \Omega_{{\mathrm{1}}}  \ottsym{,}  \triangleright  \ottsym{\{}  \vec{\polybeta}  \ottsym{\}}  \ottsym{,}   [  \ottnt{A}  /  \polyalpha  ]  \Gamma_{{\mathrm{2}}}   \xrightarrow{ \mathrm{inst} }  \Omega
    }{
      \Omega_{{\mathrm{1}}}  \ottsym{,}  \triangleright  \ottsym{\{}  \polyalpha  \ottsym{,}  \vec{\polybeta}  \ottsym{\}}  \ottsym{,}  \Gamma_{{\mathrm{2}}}  \xrightarrow{ \mathrm{inst} }  \Omega
    }\ \InstPex
  \end{mathpar}
  \caption{Worklist instantiation}\label{fig:work_inst}
\end{figure}

\subsubsection{Soundness}\label{subsubsec:algo_sound_app}

Our typing algorithm is sound with respect to the type system of \lang.
We denote the well-formedness of $\Gamma$ by $\vdash  \Gamma$.  We recall the soundness theorem and its corollary stated in \zcref{sec:algo_metatheory}.
\algsoundness*
\algsoundnesscor*

\subsubsection{A Counterexample to Completeness w.r.t.\ the Declarative System}\label{subsubsec:cyclic}

One may expect that our typing algorithm is complete with respect to
the declarative type system of \lang. %
\begin{quotation}
  If $\vdash  \Gamma$ and $\Gamma  \xrightarrow{ \mathrm{inst} }  \Omega$ and $\Omega  \rightsquigarrow^*   \emptyset $, then $\Gamma  \rightsquigarrow^*   \emptyset $.
\end{quotation}
However, this statement does not hold.
A counterexample is:
\[
  \Gamma = \triangleright  \ottsym{\{}  \polyalpha  \ottsym{,}  \polybeta  \ottsym{\}}  \Vdash   \subR{ \ottsym{(}   \boxed{ \polyalpha }   \rightarrow   \boxed{ \polybeta }   \ottsym{)}  \rightarrow   \mathsf{unit}  }{   \forall  \ottmv{a}  .  \ottsym{(}    \forall  \ottmv{b}  .  \ottmv{b}    \rightarrow   \ottmv{b}   \ottsym{)}    \rightarrow    \mathsf{unit}   }  ~.
\]
Since we have
\[
    \emptyset    \vdash  \subR{ \ottsym{(}   \boxed{  \mathsf{unit}  }   \rightarrow   \boxed{  \mathsf{unit}  }   \ottsym{)}  \rightarrow   \mathsf{unit}  }{   \forall  \ottmv{a}  .  \ottsym{(}    \forall  \ottmv{b}  .  \ottmv{b}    \rightarrow   \ottmv{b}   \ottsym{)}    \rightarrow    \mathsf{unit}   } ~,
\]
we must have $\Gamma  \rightsquigarrow^*   \emptyset $ for completeness.
However, the reduction from $\Gamma$ gets stuck as follows.
\[
  \begin{array}{rcl}
               \Gamma &  \rightsquigarrow  &
               \triangleright  \ottsym{\{}  \polyalpha  \ottsym{,}  \polybeta  \ottsym{\}}  \ottsym{,}  \ottmv{a}  \ottsym{,}  \triangleright  \ottsym{\{}  \ottsym{\}}  \Vdash   \subR{ \ottsym{(}   \boxed{ \polyalpha }   \rightarrow   \boxed{ \polybeta }   \ottsym{)}  \rightarrow   \mathsf{unit}  }{  \ottsym{(}    \forall  \ottmv{b}  .  \ottmv{b}    \rightarrow   \ottmv{b}   \ottsym{)}   \rightarrow    \mathsf{unit}   }  \\
                              &  \rightsquigarrow  &
               \triangleright  \ottsym{\{}  \polyalpha  \ottsym{,}  \polybeta  \ottsym{\}}  \ottsym{,}  \ottmv{a}  \ottsym{,}  \triangleright  \ottsym{\{}  \ottsym{\}}  \Vdash   \subR{  \mathsf{unit}  }{  \mathsf{unit}  }   \Vdash   \subL{   \forall  \ottmv{b}  .  \ottmv{b}    \rightarrow   \ottmv{b}  }{  \boxed{ \polyalpha }   \rightarrow   \boxed{ \polybeta }  }   \\
                              &  \rightsquigarrow  &
               \triangleright  \ottsym{\{}  \polyalpha  \ottsym{,}  \polybeta  \ottsym{\}}  \ottsym{,}  \ottmv{a}  \ottsym{,}  \monogamma  \ottsym{,}  \triangleright  \ottsym{\{}  \ottsym{\}}  \Vdash   \subR{  \mathsf{unit}  }{  \mathsf{unit}  }   \Vdash   \subL{  \monogamma   \rightarrow   \monogamma  }{  \boxed{ \polyalpha }   \rightarrow   \boxed{ \polybeta }  }        \\
                              &  \rightsquigarrow  &
               \triangleright  \ottsym{\{}  \polyalpha  \ottsym{,}  \polybeta  \ottsym{\}}  \ottsym{,}  \ottmv{a}  \ottsym{,}  \monogamma  \ottsym{,}  \triangleright  \ottsym{\{}  \ottsym{\}}  \Vdash   \subR{  \mathsf{unit}  }{  \mathsf{unit}  }   \Vdash   \subR{  \boxed{ \polybeta }  }{ \monogamma }   \Vdash   \subL{ \monogamma }{  \boxed{ \polyalpha }  } 
               \quad \not  \rightsquigarrow 
  \end{array}
\]
We cannot use the rule \ASLBox because
the existential variable $\monogamma$ occurs \emph{after}
the polymorphic existential variables $\polyalpha$ and $\polybeta$
in the worklist.
\mycomment{
  We explain why our typing algorithm confronts the above counterexample.
  The type variable $\ottmv{a}$ must occur later than
  the polymorphic existential variables $\polyalpha$ and $\polybeta$
  to reject the following worklist.
  \[
    \triangleright  \ottsym{\{}  \polyalpha  \ottsym{,}  \polybeta  \ottsym{\}}  \Vdash   \subR{ \ottsym{(}   \boxed{ \polyalpha }   \rightarrow   \boxed{ \polybeta }   \ottsym{)}  \rightarrow   \mathsf{unit}  }{  \ottsym{(}    \forall  \ottmv{a}  .  \ottmv{a}    \rightarrow   \ottmv{a}   \ottsym{)}   \rightarrow    \mathsf{unit}   } 
  \]
  The existential variable $\monogamma$ must occur later than the type variable $\ottmv{a}$
  to accept the following worklist.
  \[
     \emptyset   \Vdash   \subR{  \forall  \ottmv{a}  .  \ottsym{(}   \ottmv{a}   \rightarrow   \ottmv{a}   \ottsym{)}   \rightarrow   \mathsf{unit}  }{   \forall  \ottmv{a}  .  \ottsym{(}    \forall  \ottmv{b}  .  \ottmv{b}    \rightarrow   \ottmv{b}   \ottsym{)}    \rightarrow    \mathsf{unit}   } 
  \]
  Therefore, the existential variable $\monogamma$ must occur later than
  the polymorphic existential variables $\polyalpha$ and $\polybeta$.
  However, $\monogamma$ must occur earlier than $\polyalpha$ and $\polybeta$
  to accept the counterexample.
  Thus, we refer to the problem that the above counterexample exists
  as a \emph{cyclic scoping problem}.
}
This problem does not arise in existing approaches similar to boxy types:
Odersky et al.~\citet{odersky_colored_2001} do not allow for
the subtyping $\Psi  \vdash   \ottnt{A_{{\mathrm{1}}}}   \rightarrow   \ottnt{A_{{\mathrm{2}}}}   \mathbin{<:}   \forall  \ottmv{b}  .  \ottnt{B} $, and
Vytiniotis et al.~\citet{vytiniotis_boxy_2006} do not have
the subtyping corresponding to our judgment $ \Psi   \vdash  \subR{ {A^\Box} }{ \ottnt{B} } $.

\subsubsection{Completeness with respect to the DK and XO systems}\label{subsubsec:algo_complete_app}

Our typing algorithm is complete with respect to both the DK and XO systems,
while the problem presented in \zcref{subsubsec:cyclic} hinders the completeness
with respect to the type system of \lang.
To prove this completeness, we show that, under a restriction that excludes the aforementioned problem,
our typing algorithm is complete with respect to the type system of \lang.
Two judgments, $\vdash^\downarrow  \Gamma$ and $\ottsym{\{}  \ottsym{\}}  \vdash  \Gamma  \dashv  \ottsym{\{}  \vec{\polyalpha}  \ottsym{\}}$, represent this restriction.
The former judgment $\vdash^\downarrow  \Gamma$ excludes the problem discussed in \zcref{subsubsec:cyclic} and
rejects invalid inputs such as $ \emptyset   \Vdash   \subL{  \forall  \ottmv{a}  .  \ottnt{A}  }{ \ottnt{B} } $.
Because our typing algorithm introduces an existential variable
just before a scope delimiter,
we need at least one scope delimiter before any work item.
The latter judgment $\ottsym{\{}  \ottsym{\}}  \vdash  \Gamma  \dashv  \ottsym{\{}  \vec{\polyalpha}  \ottsym{\}}$ ensures that
any polymorphic existential variable occurs exactly once in a box and that
non-boxed occurrences are always before the boxed occurrence in $\Gamma$.
See \zcref{sec:full_defs} for the full definitions of these judgments.
\PartialComplete*
\mycomment{
  The statements $\vdash^\downarrow  \Gamma$ and $\ottsym{\{}  \ottsym{\}}  \vdash  \Gamma  \dashv  \ottsym{\{}  \vec{\polyalpha}  \ottsym{\}}$ restrict
  the input provided for our typing algorithm.
  The rejected input is classified into two groups.
  The former is just invalid for our typing algorithm.
  For example, one of them is $ \emptyset   \Vdash   \subL{  \forall  \ottmv{a}  .  \ottnt{A}  }{ \ottnt{B} } $.
  Because our typing algorithm introduces an existential variable
  just earlier than a scope delimiter,
  we need at least one scope delimiter earlier than any work.
  Another example is $\Gamma  \Vdash  \ottnt{e_{{\mathrm{1}}}}  \ottsym{:}   \boxed{ \polyalpha }   \Vdash  \ottnt{e_{{\mathrm{2}}}}  \ottsym{:}   \boxed{ \polyalpha } $.
  In this worklist, the types of $\ottnt{e_{{\mathrm{1}}}}$ and $\ottnt{e_{{\mathrm{2}}}}$ are synthesized,
  but the synthesized types will be oddly passed to the same polymorphic existential variable $\polyalpha$.
  The latter group includes the worklist that causes the cyclic scoping problem.
}

We show that
the restriction posed on the input of our typing algorithm in \zcref{thm:partial_complete}
is not so strict as to reject the well-typed program in the DK system or the XO system.
\algocompletenessDKXO*
This corollary indicates that
our typing algorithm successfully types the well-typed program under the DK or the XO system
by properly inserting the polymorphic existential variable.

\subsubsection{Decidability}\label{subsubsec:algo_decidability_app}

Finally, our typing algorithm is decidable.
\algodecidability*
The proof strategy follows that of
the decidability shown by Zhao et al.~\cite{zhao_mechanical_2019}.

  \section{Application Guide Insertion}\label{sec:marking}

\AI{Or, ``A Heuristics for Application Guide Insertion''?}
\TY{Fixed (cf. Section 1)}

This section presents a strategy to decide
whether to use the function- or argument-first bidirectional typing.
In \lang, the application guides $ \appright $ and $ \appleft $ lead a typing process
to apply typing rules \TAppR and \TAppL, respectively.
While writing these guides directly every time a function is applied
is cumbersome for programmers,
this design aligns with our intention that
a language designer define their own way to
insert application guides into their surface language.
We show a translation to help understand this intention.
Our prototype implementation includes this translation.
\AI{We should mention the existence of the implementation much earlier.}

Our translation aims to conservatively estimate
the degree of guessing required to synthesize the type of the given expression,
because we prefer to reduce guessing by using known types.
Guessing types in type synthesis arises mainly from
a lambda abstraction and a function application.
For example, type synthesis for $\ottsym{(}   \lambda  \mathit{x}  .  \mathit{x}   \ottsym{)}  \appright  \ottsym{()}$ includes type guessing.
In this case, changing the function-first guide $ \appright $ to
the argument-first guide $ \appleft $ removes type guessing.
Consider the function application $\mathit{f}  \appright  \ottsym{()}$ as another example,
supposing that $\mathit{f}$ has the type $  \forall  \ottmv{a}  .  \ottmv{a}    \rightarrow   \ottmv{a} $~.
Type synthesis for $\mathit{f}  \appright  \ottsym{()}$ guesses a monotype to instantiate $  \forall  \ottmv{a}  .  \ottmv{a}    \rightarrow   \ottmv{a} $ and
propagates the guessed type to the argument $\ottsym{()}$.
On the other hand, type synthesis for $\mathit{f}  \appleft  \ottsym{()}$ uses
the type $ \mathsf{unit} $ for such instantiation.
This example shows that properly selecting application guides reduces
the number of cases in which a typing process uses guessed types.
Therefore, a translation of a surface language to \lang needs to consider
the degree of guessing required to type the given expression.

We introduce \emph{guess values} and \emph{marks}, as shown in \zcref{fig:marking}, to
estimate how much guessing is required for typing expressions.
A guess value $\mu$ consists of $ \text{?} $,
which means that guessing a type may be required, and
$ \text{!} $, which means that type information is fully known.
We assign it to each expression to clarify our estimation result.
A mark $\ottnt{m}$ consists of $ \text{!} $ and a function-like form $\mu_{{\mathrm{1}}}  \rightarrow   \langle  \ottnt{m}  ,  \mu_{{\mathrm{2}}}  \rangle $.
The guess value $\mu_{{\mathrm{1}}}$ in $\mu_{{\mathrm{1}}}  \rightarrow   \langle  \ottnt{m}  ,  \mu_{{\mathrm{2}}}  \rangle $ indicates
whether the argument type is fully known.
The pair $ \langle  \ottnt{m}  ,  \mu_{{\mathrm{2}}}  \rangle $ is for the body of the lambda abstraction.
The guess value $\mu_{{\mathrm{2}}}$ indicates whether the body type is fully known or may not be.

\begin{figure}[t]
  \small
  \[
    \begin{array}{rrcl}
      \textbf{Guess Value} & \mu &  \Coloneqq  &  \text{?}   \mid   \text{!}                \\
      \textbf{Mark}        & \ottnt{m}  &  \Coloneqq  &  \text{!}   \mid  \mu_{{\mathrm{1}}}  \rightarrow   \langle  \ottnt{m}  ,  \mu_{{\mathrm{2}}}  \rangle 
    \end{array}
  \]
  \fbox{$\ottnt{e_{{\mathrm{1}}}}  \ottsym{:}   \langle  \ottnt{m}  ,  \mu  \rangle   \twoheadrightarrow  \ottnt{e_{{\mathrm{2}}}}$}\hfill\mbox{}
  \begin{mathpar}
    \inferrule{ }{
      \ottsym{()}  \ottsym{:}   \langle   \text{!}   ,   \text{!}   \rangle   \twoheadrightarrow  \ottsym{()}
    }\ \MUnit

    \inferrule{ }{
      \mathit{x}  \ottsym{:}   \langle   \text{!}   ,   \text{!}   \rangle   \twoheadrightarrow  \mathit{x}
    }\ \MVar

    \inferrule{
      \ottnt{e}  \ottsym{:}   \langle  \ottnt{m}  ,  \mu  \rangle   \twoheadrightarrow  \ottnt{e'}
    }{
       \lambda  \mathit{x}  .  \ottnt{e}   \ottsym{:}   \langle   \text{?}   \rightarrow   \langle  \ottnt{m}  ,  \mu  \rangle   ,   \text{?}   \rangle   \twoheadrightarrow   \lambda  \mathit{x}  .  \ottnt{e'} 
    }\ \MAbs

    \inferrule{
      \ottnt{e}  \ottsym{:}   \langle  \ottnt{m}  ,  \mu  \rangle   \twoheadrightarrow  \ottnt{e'}
    }{
       \lambda  \mathit{x}  \ottsym{:}  \ottnt{A}  .  \ottnt{e}   \ottsym{:}   \langle   \text{!}   \rightarrow   \langle  \ottnt{m}  ,  \mu  \rangle   ,  \mu  \rangle   \twoheadrightarrow   \lambda  \mathit{x}  \ottsym{:}  \ottnt{A}  .  \ottnt{e'} 
    }\ \MAAbs

    \inferrule{
    \ottnt{e_{{\mathrm{1}}}}  \ottsym{:}   \langle   \text{!}   ,  \mu_{{\mathrm{1}}}  \rangle   \twoheadrightarrow  \ottnt{e'_{{\mathrm{1}}}} \\ \ottnt{e_{{\mathrm{2}}}}  \ottsym{:}   \langle  \ottnt{m_{{\mathrm{2}}}}  ,   \text{!}   \rangle   \twoheadrightarrow  \ottnt{e'_{{\mathrm{2}}}}
    }{
    \ottnt{e_{{\mathrm{1}}}} \, \ottnt{e_{{\mathrm{2}}}}  \ottsym{:}   \langle   \text{!}   ,  \mu_{{\mathrm{1}}}  \rangle   \twoheadrightarrow  \ottnt{e'_{{\mathrm{1}}}}  \appleft  \ottnt{e'_{{\mathrm{2}}}}
    }\ \MAppA

    \inferrule{
    \ottnt{e_{{\mathrm{1}}}}  \ottsym{:}   \langle   \text{!}   ,  \mu_{{\mathrm{1}}}  \rangle   \twoheadrightarrow  \ottnt{e'_{{\mathrm{1}}}} \\ \ottnt{e_{{\mathrm{2}}}}  \ottsym{:}   \langle  \ottnt{m_{{\mathrm{2}}}}  ,   \text{?}   \rangle   \twoheadrightarrow  \ottnt{e'_{{\mathrm{2}}}}
    }{
    \ottnt{e_{{\mathrm{1}}}} \, \ottnt{e_{{\mathrm{2}}}}  \ottsym{:}   \langle   \text{!}   ,   \text{?}   \rangle   \twoheadrightarrow  \ottnt{e'_{{\mathrm{1}}}}  \appright  \ottnt{e'_{{\mathrm{2}}}}
    }\ \MAppB

    \inferrule{
    \ottnt{e_{{\mathrm{1}}}}  \ottsym{:}   \langle   \text{?}   \rightarrow   \langle  \ottnt{m_{{\mathrm{1}}}}  ,  \mu_{{\mathrm{1}}}  \rangle   ,  \mu'_{{\mathrm{1}}}  \rangle   \twoheadrightarrow  \ottnt{e'_{{\mathrm{1}}}} \\ \ottnt{e_{{\mathrm{2}}}}  \ottsym{:}   \langle  \ottnt{m_{{\mathrm{2}}}}  ,   \text{!}   \rangle   \twoheadrightarrow  \ottnt{e'_{{\mathrm{2}}}}
    }{
    \ottnt{e_{{\mathrm{1}}}} \, \ottnt{e_{{\mathrm{2}}}}  \ottsym{:}   \langle  \ottnt{m_{{\mathrm{1}}}}  ,  \mu_{{\mathrm{1}}}  \rangle   \twoheadrightarrow  \ottnt{e'_{{\mathrm{1}}}}  \appleft  \ottnt{e'_{{\mathrm{2}}}}
    }\ \MAppC

    \inferrule{
    \ottnt{e_{{\mathrm{1}}}}  \ottsym{:}   \langle   \text{?}   \rightarrow   \langle  \ottnt{m_{{\mathrm{1}}}}  ,  \mu_{{\mathrm{1}}}  \rangle   ,  \mu'_{{\mathrm{1}}}  \rangle   \twoheadrightarrow  \ottnt{e'_{{\mathrm{1}}}} \\ \ottnt{e_{{\mathrm{2}}}}  \ottsym{:}   \langle  \ottnt{m_{{\mathrm{2}}}}  ,   \text{?}   \rangle   \twoheadrightarrow  \ottnt{e'_{{\mathrm{2}}}}
    }{
    \ottnt{e_{{\mathrm{1}}}} \, \ottnt{e_{{\mathrm{2}}}}  \ottsym{:}   \langle  \ottnt{m_{{\mathrm{1}}}}  ,  \mu'_{{\mathrm{1}}}  \rangle   \twoheadrightarrow  \ottnt{e'_{{\mathrm{1}}}}  \appright  \ottnt{e'_{{\mathrm{2}}}}
    }\ \MAppD

    \inferrule{
    \ottnt{e_{{\mathrm{1}}}}  \ottsym{:}   \langle   \text{!}   \rightarrow   \langle  \ottnt{m_{{\mathrm{1}}}}  ,  \mu_{{\mathrm{1}}}  \rangle   ,  \mu'_{{\mathrm{1}}}  \rangle   \twoheadrightarrow  \ottnt{e'_{{\mathrm{1}}}} \\ \ottnt{e_{{\mathrm{2}}}}  \ottsym{:}   \langle  \ottnt{m_{{\mathrm{2}}}}  ,  \mu_{{\mathrm{2}}}  \rangle   \twoheadrightarrow  \ottnt{e'_{{\mathrm{2}}}}
    }{
    \ottnt{e_{{\mathrm{1}}}} \, \ottnt{e_{{\mathrm{2}}}}  \ottsym{:}   \langle  \ottnt{m_{{\mathrm{1}}}}  ,  \mu_{{\mathrm{1}}}  \rangle   \twoheadrightarrow  \ottnt{e'_{{\mathrm{1}}}}  \appright  \ottnt{e'_{{\mathrm{2}}}}
    }\ \MAppE
  \end{mathpar}
  \caption{Marking System}\label{fig:marking}
\end{figure}

We show our marking system in \zcref{fig:marking}.
Our marking judgment takes the form $\ottnt{e_{{\mathrm{1}}}}  \ottsym{:}   \langle  \ottnt{m}  ,  \mu  \rangle   \twoheadrightarrow  \ottnt{e_{{\mathrm{2}}}}$, which
means that the given expression $\ottnt{e_{{\mathrm{1}}}}$ has the mark $\ottnt{m}$ and the guess value $\mu$, and that
inserting application guides into $\ottnt{e_{{\mathrm{1}}}}$ results in $\ottnt{e_{{\mathrm{2}}}}$.
The rules \MUnit and \MVar assign $ \langle   \text{!}   ,   \text{!}   \rangle $ to
the unit value $\ottsym{()}$ and the variable $\mathit{x}$, respectively, since
we fully know the type information for them if they are well-typed.
The rule \MAbs uses $ \text{?} $ for the argument because
we must guess the argument type.
On the other hand, the rule \MAAbs uses $ \text{!} $ for the argument because
we know the argument type from the type annotation.
Both rules \MAbs and \MAAbs use $ \langle  \ottnt{m}  ,  \mu  \rangle $ for the body of the lambda abstraction,
similarly to an effect system.
The other rules are for function applications.
Our marking system prioritizes the argument-first style
if we fully know the type of the argument.
This decision comes from the fact that
we need to instantiate the polymorphic function types, such as $  \forall  \ottmv{a}  .  \ottmv{a}    \rightarrow   \ottmv{a} $,
using guessed types with the function-first style.
Therefore, the rules \MAppA and \MAppC insert the argument-first guide $ \appleft $.
In other cases, if we fully know the function type information without guessing,
the rules \MAppB and \MAppE insert the function-first guide $ \appright $.
Note that our marking system uses \MAppE even when we fully know the argument type,
because only expressions of the form $ \lambda  \mathit{x}  \ottsym{:}  \ottnt{A}  .  \ottnt{e} $ have $ \text{!}   \rightarrow   \langle  \ottnt{m}  ,  \mu  \rangle $, and because
we prioritize the type annotation over the synthesized type.
The insertion of the function-first guide in the rule \MAppD is arbitrary.
This choice depends heavily on the language designer.

We show how our marking system works using the following examples from \zcref{subsec:ffaf}.
\begin{itemize}
  \item $\ottsym{(}   \lambda  \mathit{f}  \ottsym{:}   \ottsym{(}    \forall  \ottmv{a}  .  \ottmv{a}    \rightarrow   \ottmv{a}   \ottsym{)}   \rightarrow   \ottsym{(}    \mathsf{int}   \times   \mathsf{bool}    \ottsym{)}   .  \mathit{f}   \ottsym{)} \, \ottsym{(}   \lambda  \mathit{g}  .  \ottsym{(}  \mathit{g} \, 42  \ottsym{,}  \mathit{g} \,  \mathsf{true}   \ottsym{)}   \ottsym{)}$
  \item $\ottsym{(}   \lambda  \mathit{f}  .  \mathit{f}   \ottsym{)} \, \ottsym{(}   \lambda  \mathit{g}  \ottsym{:}    \forall  \ottmv{a}  .  \ottmv{a}    \rightarrow   \ottmv{a}   .  \ottsym{(}  \mathit{g} \, 42  \ottsym{,}  \mathit{g} \,  \mathsf{true}   \ottsym{)}   \ottsym{)}$
\end{itemize}
Our marking system assigns the mark and guess value $ \langle   \text{!}   \rightarrow   \langle   \text{!}   ,   \text{!}   \rangle   ,   \text{!}   \rangle $ to the function
$ \lambda  \mathit{f}  \ottsym{:}   \ottsym{(}    \forall  \ottmv{a}  .  \ottmv{a}    \rightarrow   \ottmv{a}   \ottsym{)}   \rightarrow   \ottsym{(}    \mathsf{int}   \times   \mathsf{bool}    \ottsym{)}   .  \mathit{f} $.
Therefore, it chooses the function-first guide via the rule \MAppE.
On the other hand, our marking system assigns the mark and guess value $ \langle   \text{!}   ,   \text{!}   \rangle $
to the argument $ \lambda  \mathit{g}  \ottsym{:}    \forall  \ottmv{a}  .  \ottmv{a}    \rightarrow   \ottmv{a}   .  \ottsym{(}  \mathit{g} \, 42  \ottsym{,}  \mathit{g} \,  \mathsf{true}   \ottsym{)} $.
For simplicity, we assign the guess value $ \text{!} $ to the pair $\ottsym{(}  \mathit{g} \, 42  \ottsym{,}  \mathit{g} \,  \mathsf{true}   \ottsym{)}$
without introducing a mark for pairs, since both elements have $ \langle   \text{!}   ,   \text{!}   \rangle $.
Our marking system uses \MAppC to insert the argument-first guide.
Through our marking system, we get the following expressions
that are well-typed under the type system of \lang.
\begin{itemize}
  \item $\ottsym{(}   \lambda  \mathit{f}  \ottsym{:}   \ottsym{(}    \forall  \ottmv{a}  .  \ottmv{a}    \rightarrow   \ottmv{a}   \ottsym{)}   \rightarrow   \ottsym{(}    \mathsf{int}   \times   \mathsf{bool}    \ottsym{)}   .  \mathit{f}   \ottsym{)}  \appright  \ottsym{(}   \lambda  \mathit{g}  .  \ottsym{(}  \mathit{g} \, 42  \ottsym{,}  \mathit{g} \,  \mathsf{true}   \ottsym{)}   \ottsym{)}$
  \item $\ottsym{(}   \lambda  \mathit{f}  .  \mathit{f}   \ottsym{)}  \appleft  \ottsym{(}   \lambda  \mathit{g}  \ottsym{:}    \forall  \ottmv{a}  .  \ottmv{a}    \rightarrow   \ottmv{a}   .  \ottsym{(}  \mathit{g} \, 42  \ottsym{,}  \mathit{g} \,  \mathsf{true}   \ottsym{)}   \ottsym{)}$
\end{itemize}
Our marking system also properly inserts application guides to accept
the program $\ottsym{(}  \mathit{f_{{\mathrm{1}}}} \, \mathit{f_{{\mathrm{2}}}}  \ottsym{)} \, \mathit{f_{{\mathrm{3}}}}$ shown in \zcref{subsec:our_work}, which is ill-typed
under both the DK and XO systems, resulting in $\ottsym{(}  \mathit{f_{{\mathrm{1}}}}  \appright  \mathit{f_{{\mathrm{2}}}}  \ottsym{)}  \appleft  \mathit{f_{{\mathrm{3}}}}$.
We leave an empirical evaluation of how our marking system works in practice as future work.

  \section{Extensions}\label{sec:extensions}

This section discusses possible extensions for \lang.

\subsection{Let Polymorphism}\label{subsec:let-poly}

Let polymorphism is a practically important extension.
It helps achieve backward compatibility with the Hindley--Milner type system.
Furthermore, supporting let polymorphism enables \lang to subsume the original XO system.

A challenge in extending \lang with let polymorphism is the generalization of types.
Our formulation manages the scope of type variables.
This formulation does not fit with
the generalization of types $\mathsf{Gen} \, \ottsym{(}  \ottnt{A}  \ottsym{;}  \Psi  \ottsym{)}$
in the following traditional typing rule for let expressions.
\[
  \inferrule{
  \Psi  \vdash  \ottnt{e_{{\mathrm{1}}}}  \ottsym{:}  \ottnt{A} \\ \Psi  \ottsym{,}  \mathit{x}  \ottsym{:}  \mathsf{Gen} \, \ottsym{(}  \ottnt{A}  \ottsym{;}  \Psi  \ottsym{)}  \vdash  \ottnt{e_{{\mathrm{2}}}}  \ottsym{:}  \ottnt{B}
  }{
  \Psi  \vdash  \ottkw{let} \, \mathit{x}  \ottsym{=}  \ottnt{e_{{\mathrm{1}}}} \, \ottkw{in} \, \ottnt{e_{{\mathrm{2}}}}  \ottsym{:}  \ottnt{B}
  }
\]
The generalization $\mathsf{Gen} \, \ottsym{(}  \ottnt{A}  \ottsym{;}  \Psi  \ottsym{)}$ quantifies
every free type variable in $\ottnt{A}$ that does not occur in $\Psi$.
However, the type $\ottnt{A}$, which has a type variable that does not occur in $\Psi$,
is not well-formed under $\Psi$, because we manage type variables in a typing context.

To balance the strict management of type variables and the generalization of types,
we consider the following declarative typing rule for let expressions.
\[
  \inferrule{
  \Psi  \ottsym{,}  \Sigma  \vdash  \ottnt{e_{{\mathrm{1}}}}  \ottsym{:}   \boxed{ \ottnt{A} }  \\ \Psi  \ottsym{,}  \mathit{x}  \ottsym{:}  \mathsf{Gen} \, \ottsym{(}  \ottnt{A}  \ottsym{;}  \Sigma  \ottsym{)}  \vdash  \ottnt{e_{{\mathrm{2}}}}  \ottsym{:}  {A^\Box}
  }{
  \Psi  \vdash  \ottkw{let} \, \mathit{x}  \ottsym{=}  \ottnt{e_{{\mathrm{1}}}} \, \ottkw{in} \, \ottnt{e_{{\mathrm{2}}}}  \ottsym{:}  {A^\Box}
  }
\]
The metavariable $\Sigma$ denotes a context containing only type variables.
This rule allows us to use fresh type variables when synthesizing the type of $\ottnt{e_{{\mathrm{1}}}}$.
For example, we can have the following derivation using this rule.
\[
  \inferrule*{
  \ottmv{a}  \vdash   \lambda  \mathit{x}  .  \mathit{x}   \ottsym{:}   \boxed{  \ottmv{a}   \rightarrow   \ottmv{a}  }  \\ \mathit{f}  \ottsym{:}    \forall  \ottmv{a}  .  \ottmv{a}    \rightarrow   \ottmv{a}   \vdash  \ottnt{e}  \ottsym{:}  {A^\Box}
  }{
   \emptyset   \vdash  \ottkw{let} \, \mathit{f}  \ottsym{=}   \lambda  \mathit{x}  .  \mathit{x}  \, \ottkw{in} \, \ottnt{e}  \ottsym{:}  {A^\Box}
  }
\]
This idea of using another context $\Sigma$ only for let-bound expressions
is similar to incrementing
a \emph{rank}~\cite{remy1992extension} or \emph{level}~\cite{fan_practical_2025}.

\subsection{Explicit Type Application}\label{subsec:explicit-tapp}

Supporting explicit type application helps realize an impredicative type instantiation,
improving typeability.
Furthermore, Xie and Oliveira~\citet{xie_let_2018} show that
argument-first bidirectional typing with explicit type applications
encodes a type synonym, which is a new name for an existing type.

We have two major concerns about extending \lang with explicit type applications.
First, the XO system with explicit type applications extends an application context,
enabling it to stack not only types of arguments
but also explicitly applied type arguments.
Secondly, combining
polymorphic subtyping of Odersky et al.~\citet{odersky_putting_1996},
on which the DK, XO, and our type systems are based,
with explicit type applications
can break important properties \cite{zhao_elementary_2022}, as we discuss below.

Xie and Oliveira~\citet{xie_let_2018} develop an argument-first bidirectional type system
with explicit type application by using the known information about type arguments.
Their typing rule for type applications is as follows.
\[
  \inferrule{
    \Psi  \mid  \Xi  \ottsym{,}   @  \ottnt{A}   \vdash_{ \mathrm{XO} }  \ottnt{e}  \Rightarrow  \ottnt{B}
  }{
    \Psi  \mid  \Xi  \vdash_{ \mathrm{XO} }  \ottnt{e}  \mathbin{@}  \ottnt{A}  \Rightarrow  \ottnt{B}
  }
\]
This rule adds the applied type $\ottnt{A}$ into the application context $\Xi$.
Using the information about what types are supplied improves the typeability.
For example, the expression $\ottsym{(}    \Lambda  \ottmv{a}  .   \lambda  \mathit{x}  \ottsym{:}  \ottmv{a}  .  \mathit{x}     \ottsym{+}   1   \ottsym{)}  \mathbin{@}   \mathsf{int} $
is well-typed under the XO system with explicit type applications,
while it is ill-typed under the traditional type system for System F.
When typing the type abstraction $  \Lambda  \ottmv{a}  .   \lambda  \mathit{x}  \ottsym{:}  \ottmv{a}  .  \mathit{x}     \ottsym{+}   1 $,
the XO system knows that $ \mathsf{int} $ is supplied; consequently,
typing $ \mathit{x}   \ottsym{+}   1 $ succeeds.
This extended formalization inspires us to
support explicit type applications in \lang.
\mycomment{
  We extend the syntax of boxy types as follows.
  \[
    {A^\Box}, {B^\Box}  \Coloneqq  \cdots  \mid   @  \ottnt{A}   \rightarrow   {A^\Box} 
  \]
}
We add the new boxy type $ @  \ottnt{A}   \rightarrow   {A^\Box} $,
representing a polymorphic type that yields ${A^\Box}$ when supplied with $\ottnt{A}$.
The typing rules for type abstractions and type applications are as follows.

\begin{center}
  \begin{tabular}{cc}
    \begin{minipage}{0.45\linewidth}
      \begin{equation}
        \inferrule{
        \Psi  \vdash   [  \ottnt{A} / \ottmv{a}  ]  \ottnt{e}   \ottsym{:}  {A^\Box}
        }{
        \Psi  \vdash   \Lambda  \ottmv{a}  .  \ottnt{e}   \ottsym{:}   @  \ottnt{A}   \rightarrow   {A^\Box} 
        } \label{tabs}
      \end{equation}
    \end{minipage} &
    \begin{minipage}{0.45\linewidth}
      \begin{equation}
        \inferrule{
          \Psi  \vdash  \ottnt{e}  \ottsym{:}   @  \ottnt{A}   \rightarrow   {A^\Box} 
        }{
          \Psi  \vdash  \ottnt{e}  \mathbin{@}  \ottnt{A}  \ottsym{:}  {A^\Box}
        } \label{tapp}
      \end{equation}
    \end{minipage}
  \end{tabular}
\end{center}

\noindent
The former rule \eqref{tabs} substitutes the known type argument $\ottnt{A}$
into the type variable $\ottmv{a}$ in $\ottnt{e}$.
The latter rule \eqref{tapp} uses the known type argument $\ottnt{A}$
for typing the expression $\ottnt{e}$, similarly to how
the XO system adds a type argument to an application context.
These rules enable the type system of \lang to support
the aforementioned expression $\ottsym{(}    \Lambda  \ottmv{a}  .   \lambda  \mathit{x}  \ottsym{:}  \ottmv{a}  .  \mathit{x}     \ottsym{+}   1   \ottsym{)}  \mathbin{@}   \mathsf{int} $~.

However, straightforwardly combining
polymorphic subtyping of Odersky et al.~\citet{odersky_putting_1996} and
explicit type applications can break important properties,
as Zhao and Oliveira~\citet{zhao_elementary_2022} show.
An important property is the following lemma.
\begin{lemma}[Subsumption]
  If $\Psi  \vdash  \ottnt{e}  \Leftarrow  \ottnt{A}$ and $\Psi  \vdash  \ottnt{A}  \mathbin{<:}  \ottnt{B}$, then $\Psi  \vdash  \ottnt{e}  \Leftarrow  \ottnt{B}$.
\end{lemma}
This lemma ensures that
a programmer can annotate a program with a supertype of its type
without altering the typing result.
An explicit type application can break the subsumption lemma because
the result of a type application depends on the order of quantified type variables,
while polymorphic subtyping does not.
\mycomment{
  \[
     \emptyset   \vdash   \lambda  \mathit{x}  .  \ottsym{(}  \mathit{x}  \mathbin{@}   \mathsf{int}  \, 3 \, \ottsym{()}  \ottsym{)}   \Leftarrow   \ottsym{(}     \forall  \ottmv{a}  .   \forall  \ottmv{b}  .  \ottmv{a}     \rightarrow   \ottmv{b}    \rightarrow   \ottmv{a}   \ottsym{)}   \rightarrow    \mathsf{int}  
  \]
  and the subtyping judgment
  \begin{equation}
     \emptyset   \vdash   \ottsym{(}     \forall  \ottmv{a}  .   \forall  \ottmv{b}  .  \ottmv{a}     \rightarrow   \ottmv{b}    \rightarrow   \ottmv{a}   \ottsym{)}   \rightarrow    \mathsf{int}    \mathbin{<:}   \ottsym{(}     \forall  \ottmv{b}  .   \forall  \ottmv{a}  .  \ottmv{a}     \rightarrow   \ottmv{b}    \rightarrow   \ottmv{a}   \ottsym{)}   \rightarrow    \mathsf{int}   \label{bad_subty}
  \end{equation}
  hold, but the type checking judgment
  \[
     \emptyset   \vdash   \lambda  \mathit{x}  .  \ottsym{(}  \mathit{x}  \mathbin{@}   \mathsf{int}  \, 3 \, \ottsym{()}  \ottsym{)}   \Leftarrow   \ottsym{(}     \forall  \ottmv{b}  .   \forall  \ottmv{a}  .  \ottmv{a}     \rightarrow   \ottmv{b}    \rightarrow   \ottmv{a}   \ottsym{)}   \rightarrow    \mathsf{int}  
  \]
}
For example, the following type checking and subtyping judgments hold.
\begin{itemize}
  \item $ \emptyset   \vdash   \lambda  \mathit{x}  .  \ottsym{(}  \mathit{x}  \mathbin{@}   \mathsf{int}  \, 3 \, \ottsym{()}  \ottsym{)}   \Leftarrow   \ottsym{(}     \forall  \ottmv{a}  .   \forall  \ottmv{b}  .  \ottmv{a}     \rightarrow   \ottmv{b}    \rightarrow   \ottmv{a}   \ottsym{)}   \rightarrow    \mathsf{int}  $
  \item $ \emptyset   \vdash   \ottsym{(}     \forall  \ottmv{a}  .   \forall  \ottmv{b}  .  \ottmv{a}     \rightarrow   \ottmv{b}    \rightarrow   \ottmv{a}   \ottsym{)}   \rightarrow    \mathsf{int}    \mathbin{<:}   \ottsym{(}     \forall  \ottmv{b}  .   \forall  \ottmv{a}  .  \ottmv{a}     \rightarrow   \ottmv{b}    \rightarrow   \ottmv{a}   \ottsym{)}   \rightarrow    \mathsf{int}   \refstepcounter{equation}\hfill(\theequation) \label{bad_subty}$ %
\end{itemize}
However, the type checking judgment
\[
   \emptyset   \vdash   \lambda  \mathit{x}  .  \ottsym{(}  \mathit{x}  \mathbin{@}   \mathsf{int}  \, 3 \, \ottsym{()}  \ottsym{)}   \Leftarrow   \ottsym{(}     \forall  \ottmv{b}  .   \forall  \ottmv{a}  .  \ottmv{a}     \rightarrow   \ottmv{b}    \rightarrow   \ottmv{a}   \ottsym{)}   \rightarrow    \mathsf{int}  
\]
does not hold.
In the latter type checking judgment, since we know that
the variable $\mathit{x}$ has the type $   \forall  \ottmv{b}  .   \forall  \ottmv{a}  .  \ottmv{a}     \rightarrow   \ottmv{b}    \rightarrow   \ottmv{a} $,
the type application result $\mathit{x}  \mathbin{@}   \mathsf{int} $ has the type $   \forall  \ottmv{a}  .  \ottmv{a}    \rightarrow    \mathsf{int}     \rightarrow   \ottmv{a} $.
However, the expression $\mathit{x}  \mathbin{@}   \mathsf{int} $ takes the unit as its second argument,
resulting in a type error.
We need to address the violation of the subsumption lemma to support explicit type applications,
because our subtyping is based on that of Odersky et al.~\citet{odersky_putting_1996}.

We have two major directions %
to support explicit type application while preserving important properties.
The first direction is to update subtyping rules for polymorphic types,
following Zhao and Oliveira~\citet{zhao_elementary_2022}.
They forbid changing the order of quantified type variables
to add explicit type application to the DK system without violating the subsumption lemma.
The subtyping judgment \eqref{bad_subty} does not hold under this restriction.
The second direction is to forbid subtyping over an argument type.
The subtyping judgment \eqref{bad_subty} also does not hold under this restriction.
This approach is used in, for example,
the XO system with explicit type applications~\cite{xie_let_2018} and
Quick Look~\cite{serrano_quick_2020}.

  \section{Omitted Definitions}\label{sec:full_defs}

\subsection{Declarative System}

\deftitle{Well-formedness for the declarative system}{
  \fbox{$\vdash  \Psi$}\quad
  \fbox{$\Psi  \vdash  {A^\Box}$}
}

\begin{mathparpagebreakable}
  \small
  \inferrule{\mbox{}}{
    \vdash   \emptyset 
  }

  \inferrule{
  \vdash  \Psi \\  \ottmv{a}  \notin  \Psi 
  }{
    \vdash  \Psi  \ottsym{,}  \ottmv{a}
  }

  \inferrule{
  \Psi  \vdash  \ottnt{A} \\  \mathit{x}  \notin \mathrm{dom}( \Psi ) 
  }{
    \vdash  \Psi  \ottsym{,}  \mathit{x}  \ottsym{:}  \ottnt{A}
  }

  \inferrule{
    \Psi  \vdash  \ottnt{A}
  }{
    \Psi  \vdash   \boxed{ \ottnt{A} } 
  }

  \inferrule{
    \vdash  \Psi
  }{
    \Psi  \vdash   \mathsf{unit} 
  }

  \inferrule{
  \vdash  \Psi \\  \ottmv{a}  \in  \Psi 
  }{
    \Psi  \vdash  \ottmv{a}
  }

  \inferrule{
  \Psi  \vdash  {A^\Box} \\ \Psi  \vdash  {B^\Box}
  }{
    \Psi  \vdash  {A^\Box}  \rightarrow  {B^\Box}
  }

  \inferrule{
    \Psi  \ottsym{,}  \ottmv{a}  \vdash  \ottnt{A}
  }{
    \Psi  \vdash   \forall  \ottmv{a}  .  \ottnt{A} 
  }
\end{mathparpagebreakable}

\mycomment{
\begin{figure}[H]
  \small
  \noindent\fbox{$\vdash  \Psi$}\quad\fbox{$\Psi  \vdash  {A^\Box}$}\hfill\vspace{1ex}
  \begin{mathpar}
    \inferrule{\mbox{}}{
      \vdash   \emptyset 
    }

    \inferrule{
    \vdash  \Psi \\  \ottmv{a}  \notin  \Psi 
    }{
      \vdash  \Psi  \ottsym{,}  \ottmv{a}
    }

    \inferrule{
    \Psi  \vdash  \ottnt{A} \\  \mathit{x}  \notin \mathrm{dom}( \Psi ) 
    }{
      \vdash  \Psi  \ottsym{,}  \mathit{x}  \ottsym{:}  \ottnt{A}
    }

    \inferrule{
      \Psi  \vdash  \ottnt{A}
    }{
      \Psi  \vdash   \boxed{ \ottnt{A} } 
    }

    \inferrule{
      \vdash  \Psi
    }{
      \Psi  \vdash   \mathsf{unit} 
    }

    \inferrule{
    \vdash  \Psi \\  \ottmv{a}  \in  \Psi 
    }{
      \Psi  \vdash  \ottmv{a}
    }

    \inferrule{
    \Psi  \vdash  {A^\Box} \\ \Psi  \vdash  {B^\Box}
    }{
      \Psi  \vdash  {A^\Box}  \rightarrow  {B^\Box}
    }

    \inferrule{
      \Psi  \ottsym{,}  \ottmv{a}  \vdash  \ottnt{A}
    }{
      \Psi  \vdash   \forall  \ottmv{a}  .  \ottnt{A} 
    }
  \end{mathpar}
  \caption{Well-formedness for the declarative system}
\end{figure}
}

\subsection{Algorithmic System}

\mycomment{
  \begin{figure}[H]
    \small
    \[
      \begin{array}{rrcl}
        \textbf{Types}          & \ottnt{A}, \ottnt{B} &  \Coloneqq  &
         \mathsf{unit}   \mid  \ottmv{a}  \mid  \monoalpha \mycomment{ \mid  \polyalpha}  \mid   \ottnt{A}   \rightarrow   \ottnt{B}   \mid   \forall  \ottmv{a}  .  \ottnt{A}  \\
        \textbf{Boxy types}     & {A^\Box}, {B^\Box} &  \Coloneqq  &
         \boxed{ \polyalpha }   \mid   \mathsf{unit}   \mid  \ottmv{a}  \mid  \monoalpha  \mid  \polyalpha  \mid 
        {A^\Box}  \rightarrow  {B^\Box}  \mid   \forall  \ottmv{a}  .  \ottnt{A}                                \\
        \textbf{Work items}     & \omega        &  \Coloneqq  &
         \subL{ \ottnt{A} }{ {A^\Box} }   \mid   \subR{ {A^\Box} }{ \ottnt{A} }   \mid  \ottnt{e}  \ottsym{:}  {A^\Box}                                                      \\
        \textbf{Worklists}      & \Gamma        &  \Coloneqq  &
         \emptyset   \mid  \Gamma  \ottsym{,}  \mathit{x}  \ottsym{:}  \ottnt{A}  \mid  \Gamma  \Vdash  \omega  \mid 
        \Gamma  \ottsym{,}  \ottmv{a}  \mid  \Gamma  \ottsym{,}  \monoalpha  \mid  \Gamma  \ottsym{,}  \triangleright  \ottsym{\{}  \vec{\polyalpha}  \ottsym{\}}                                               \\
        \textbf{Mini worklists} & \Delta        &  \Coloneqq  &
         \emptyset   \mid  \Delta  \ottsym{,}  \mathit{x}  \ottsym{:}  \ottnt{A}  \mid  \Delta  \Vdash  \omega
      \end{array}
    \]
    \caption{Syntax for the algorithmic system}
  \end{figure}

  \begin{figure}[H]
    \small
    \[
      \begin{array}{rcl@{\quad}l@{\qquad}rcl@{\quad}l}
                   \Gamma  \ottsym{,}  \ottmv{a} &  \rightsquigarrow  & \Gamma & (\ATyVar) & \Gamma  \ottsym{,}  \monoalpha    &  \rightsquigarrow  & \Gamma & (\AEVar)              \\
                   \Gamma  \ottsym{,}  \triangleright  \ottsym{\{}  \vec{\polyalpha}  \ottsym{\}}                      &  \rightsquigarrow  & \Gamma & (\ADelim) & \Gamma  \ottsym{,}  \mathit{x}  \ottsym{:}  \ottnt{A} &  \rightsquigarrow  & \Gamma & (\AVar)
      \end{array}
    \]
    \caption{Binding removal rules}
  \end{figure}

  \begin{figure}[H]
    \renewcommand{\arraystretch}{1.2}
    \small
    \[
      \begin{array}{rcl@{\quad}l}
                   \Gamma_{{\mathrm{1}}}  \ottsym{,}  \triangleright  \ottsym{\{}  \vec{\polybeta}_{{\mathrm{1}}}  \ottsym{,}  \polyalpha  \ottsym{,}  \vec{\polybeta}_{{\mathrm{2}}}  \ottsym{\}}  \ottsym{,}  \Gamma_{{\mathrm{2}}}  \Vdash   \subL{ \ottnt{A} }{  \boxed{ \polyalpha }  }  &  \rightsquigarrow     &
                   \Gamma_{{\mathrm{1}}}  \ottsym{,}  \triangleright  \ottsym{\{}  \vec{\polybeta}_{{\mathrm{1}}}  \ottsym{,}  \vec{\polybeta}_{{\mathrm{2}}}  \ottsym{\}}  \ottsym{,}   [  \ottnt{A}  /  \polyalpha  ]  \Gamma_{{\mathrm{2}}}              & (\ASLBox)                \\
                                                                   &           &
                   \tif   \mathrm{FTV}( \ottnt{A} )   \subseteq \mathrm{dom}( \Gamma_{{\mathrm{1}}} )                 &                          \\
                   \Gamma_{{\mathrm{1}}}  \ottsym{,}  \triangleright  \ottsym{\{}  \vec{\polybeta}_{{\mathrm{1}}}  \ottsym{,}  \polyalpha  \ottsym{,}  \vec{\polybeta}_{{\mathrm{2}}}  \ottsym{\}}  \ottsym{,}  \Gamma_{{\mathrm{2}}}  \Vdash   \subR{  \boxed{ \polyalpha }  }{ \ottnt{A} }  &  \rightsquigarrow     &
                   \Gamma_{{\mathrm{1}}}  \ottsym{,}  \triangleright  \ottsym{\{}  \vec{\polybeta}_{{\mathrm{1}}}  \ottsym{,}  \vec{\polybeta}_{{\mathrm{2}}}  \ottsym{\}}  \ottsym{,}   [  \ottnt{A}  /  \polyalpha  ]  \Gamma_{{\mathrm{2}}}              & (\ASRBox)                \\
                                                                   &           &
                   \tif   \mathrm{FTV}( \ottnt{A} )   \subseteq \mathrm{dom}( \Gamma_{{\mathrm{1}}} )                  &                          \\
      \end{array}
    \]
    \[
      \begin{array}{rcl@{\quad}l@{\qquad}rcl@{\quad}l}
                   \Gamma  \Vdash   \subL{  \mathsf{unit}  }{  \mathsf{unit}  }  &  \rightsquigarrow      &
                   \Gamma                               & (\ASLUnit) &
                   \Gamma  \Vdash   \subR{  \mathsf{unit}  }{  \mathsf{unit}  }              &  \rightsquigarrow      &
                   \Gamma                               & (\ASRUnit)    \\
                   \Gamma  \Vdash   \subL{ \ottmv{a} }{ \ottmv{a} }                   &  \rightsquigarrow      &
                   \Gamma                               & (\ASLTVar) &
                   \Gamma  \Vdash   \subR{ \ottmv{a} }{ \ottmv{a} }                   &  \rightsquigarrow      &
                   \Gamma                               & (\ASRTVar)    \\
                   \Gamma  \Vdash   \subL{ \monoalpha }{ \monoalpha }                  &  \rightsquigarrow      &
                   \Gamma                               & (\ASLEVar) &
                   \Gamma  \Vdash   \subR{ \monoalpha }{ \monoalpha }                  &  \rightsquigarrow      &
                   \Gamma                               & (\ASLEVar)
      \end{array}
    \]
    \[
      \begin{array}{rcl@{\quad}l}
                   \Gamma  \Vdash   \subL{  \ottnt{A}   \rightarrow   \ottnt{B}  }{ {A^\Box}  \rightarrow  {B^\Box} }                             &  \rightsquigarrow      &
                   \Gamma  \Vdash   \subL{ \ottnt{B} }{ {B^\Box} }   \Vdash   \subR{ {A^\Box} }{ \ottnt{A} }                          & (\ASLFun)                  \\
                   \Gamma  \Vdash   \subR{ {A^\Box}  \rightarrow  {B^\Box} }{  \ottnt{A}   \rightarrow   \ottnt{B}  }                             &  \rightsquigarrow      &
                   \Gamma  \Vdash   \subR{ {B^\Box} }{ \ottnt{B} }   \Vdash   \subL{ \ottnt{A} }{ {A^\Box} }                          & (\ASRFun)                  \\
                   \Gamma  \Vdash   \subL{ \ottnt{A} }{  \forall  \ottmv{b}  .  \ottnt{B}  }                            &  \rightsquigarrow      &
                   \Gamma  \ottsym{,}  \ottmv{b}  \ottsym{,}  \triangleright  \ottsym{\{}  \ottsym{\}}  \Vdash   \subL{ \ottnt{A} }{ \ottnt{B} }                             & (\ASLAllR)                 \\
                   \Gamma  \Vdash   \subR{ {A^\Box} }{  \forall  \ottmv{b}  .  \ottnt{B}  }                           &  \rightsquigarrow      &
                   \Gamma  \ottsym{,}  \ottmv{b}  \ottsym{,}  \triangleright  \ottsym{\{}  \ottsym{\}}  \Vdash   \subR{ {A^\Box} }{ \ottnt{B} }  \quad \tif \mathsf{not}\square \, \ottsym{(}  {A^\Box}  \ottsym{)} & (\ASRAllR)                 \\
                   \Gamma  \ottsym{,}  \triangleright  \ottsym{\{}  \vec{\polyalpha}  \ottsym{\}}  \ottsym{,}  \Delta  \Vdash   \subL{  \forall  \ottmv{a}  .  \ottnt{A}  }{ {B^\Box} }               &  \rightsquigarrow      &
                   \Gamma  \ottsym{,}  \monoalpha  \ottsym{,}  \triangleright  \ottsym{\{}  \vec{\polyalpha}  \ottsym{\}}  \ottsym{,}  \Delta  \Vdash   \subL{  [  \monoalpha  /  \ottmv{a}  ]  \ottnt{A}  }{ {B^\Box} }                & (\ASLAllL)                 \\
                                                                          &            &
                   \tif \mathsf{not}\forall \, \ottsym{(}  {B^\Box}  \ottsym{)} \tand \mathsf{not}\square \, \ottsym{(}  {B^\Box}  \ottsym{)}            &                            \\
                   \Gamma  \ottsym{,}  \triangleright  \ottsym{\{}  \vec{\polyalpha}  \ottsym{\}}  \ottsym{,}  \Delta  \Vdash   \subR{  \forall  \ottmv{a}  .  \ottnt{A}  }{ \ottnt{B} }              &  \rightsquigarrow      &
                   \Gamma  \ottsym{,}  \monoalpha  \ottsym{,}  \triangleright  \ottsym{\{}  \vec{\polyalpha}  \ottsym{\}}  \ottsym{,}  \Delta  \Vdash   \subR{  [  \monoalpha  /  \ottmv{a}  ]  \ottnt{A}  }{ \ottnt{B} }                & (\ASRAllL)                 \\
                                                                          &            &
                   \tif \mathsf{not}\forall \, \ottsym{(}  \ottnt{B}  \ottsym{)}                                 &
      \end{array}
    \]
    \caption{Algorithmic subtyping rules}\label{fig:algo_subtyping}
  \end{figure}

  \begin{figure}[H]
    \renewcommand{\arraystretch}{1.2}
    \small
    \[
      \begin{array}{rcl@{\quad}l}
                   \Gamma_{{\mathrm{1}}}  \ottsym{,}  \monoalpha  \ottsym{,}  \Gamma_{{\mathrm{2}}}  \ottsym{,}  \monobeta  \ottsym{,}  \Gamma_{{\mathrm{3}}}  \Vdash   \subL{ \monoalpha }{ \monobeta }  &  \rightsquigarrow         &
                   \Gamma_{{\mathrm{1}}}  \ottsym{,}  \monoalpha  \ottsym{,}  \Gamma_{{\mathrm{2}}}  \ottsym{,}   [  \monoalpha  /  \monobeta  ]  \Gamma_{{\mathrm{3}}}            & (\ASLExExA)                  \\
                   \Gamma_{{\mathrm{1}}}  \ottsym{,}  \monoalpha  \ottsym{,}  \Gamma_{{\mathrm{2}}}  \ottsym{,}  \monobeta  \ottsym{,}  \Gamma_{{\mathrm{3}}}  \Vdash   \subL{ \monobeta }{ \monoalpha }  &  \rightsquigarrow         &
                   \Gamma_{{\mathrm{1}}}  \ottsym{,}  \monoalpha  \ottsym{,}  \Gamma_{{\mathrm{2}}}  \ottsym{,}   [  \monoalpha  /  \monobeta  ]  \Gamma_{{\mathrm{3}}}            & (\ASLExExB)                  \\
                   \Gamma_{{\mathrm{1}}}  \ottsym{,}  \ottmv{a}  \ottsym{,}  \Gamma_{{\mathrm{2}}}  \ottsym{,}  \monoalpha  \ottsym{,}  \Gamma_{{\mathrm{3}}}  \Vdash   \subL{ \ottmv{a} }{ \monoalpha }    &  \rightsquigarrow         &
                   \Gamma_{{\mathrm{1}}}  \ottsym{,}  \ottmv{a}  \ottsym{,}  \Gamma_{{\mathrm{2}}}  \ottsym{,}   [  \ottmv{a}  /  \monoalpha  ]  \Gamma_{{\mathrm{3}}}              & (\ASLExTVarA)                \\
                   \Gamma_{{\mathrm{1}}}  \ottsym{,}  \ottmv{a}  \ottsym{,}  \Gamma_{{\mathrm{2}}}  \ottsym{,}  \monoalpha  \ottsym{,}  \Gamma_{{\mathrm{3}}}  \Vdash   \subL{ \monoalpha }{ \ottmv{a} }    &  \rightsquigarrow         &
                   \Gamma_{{\mathrm{1}}}  \ottsym{,}  \ottmv{a}  \ottsym{,}  \Gamma_{{\mathrm{2}}}  \ottsym{,}   [  \ottmv{a}  /  \monoalpha  ]  \Gamma_{{\mathrm{3}}}              & (\ASLExTVarB)                \\
                   \Gamma_{{\mathrm{1}}}  \ottsym{,}  \monoalpha  \ottsym{,}  \Gamma_{{\mathrm{2}}}  \Vdash   \subL{  \mathsf{unit}  }{ \monoalpha }        &  \rightsquigarrow         &
                   \Gamma_{{\mathrm{1}}}  \ottsym{,}   [   \mathsf{unit}   /  \monoalpha  ]  \Gamma_{{\mathrm{2}}}                  & (\ASLExUnitA)                \\
                   \Gamma_{{\mathrm{1}}}  \ottsym{,}  \monoalpha  \ottsym{,}  \Gamma_{{\mathrm{2}}}  \Vdash   \subL{ \monoalpha }{  \mathsf{unit}  }        &  \rightsquigarrow         &
                   \Gamma_{{\mathrm{1}}}  \ottsym{,}   [   \mathsf{unit}   /  \monoalpha  ]  \Gamma_{{\mathrm{2}}}                  & (\ASLExUnitB)
      \end{array}
    \]
    \[
      \begin{array}{l@{\quad}l}
                   \Gamma_{{\mathrm{1}}}  \ottsym{,}  \monoalpha  \ottsym{,}  \Gamma_{{\mathrm{2}}}  \Vdash   \subL{  \ottnt{A}   \rightarrow   \ottnt{B}  }{ \monoalpha }                                       &                            \\
                   \quad  \rightsquigarrow  \quad
                                \Gamma_{{\mathrm{1}}}  \ottsym{,}  \monoalpha_{{\mathrm{1}}}  \ottsym{,}  \monoalpha_{{\mathrm{2}}}  \ottsym{,}   [   \monoalpha_{{\mathrm{1}}}   \rightarrow   \monoalpha_{{\mathrm{2}}}   /  \monoalpha  ]  \ottsym{(}  \Gamma_{{\mathrm{2}}}  \Vdash   \subL{ \ottnt{B} }{ \monoalpha_{{\mathrm{2}}} }   \Vdash   \subR{ \monoalpha_{{\mathrm{1}}} }{ \ottnt{A} }   \ottsym{)}  & (\ASLExFunA) \\
                   \quad \phantom{{} \rightsquigarrow {}} \quad \tif  \monoalpha  \notin   \mathrm{FTV}( \ottnt{A}  \rightarrow  \ottnt{B} )                                                     \\
                   \Gamma_{{\mathrm{1}}}  \ottsym{,}  \monoalpha  \ottsym{,}  \Gamma_{{\mathrm{2}}}  \Vdash   \subL{ \monoalpha }{ {A^\Box}  \rightarrow  {B^\Box} }                                       &                           \\
                   \quad  \rightsquigarrow  \quad
                                \Gamma_{{\mathrm{1}}}  \ottsym{,}  \monoalpha_{{\mathrm{1}}}  \ottsym{,}  \monoalpha_{{\mathrm{2}}}  \ottsym{,}   [   \monoalpha_{{\mathrm{1}}}   \rightarrow   \monoalpha_{{\mathrm{2}}}   /  \monoalpha  ]  \ottsym{(}  \Gamma_{{\mathrm{2}}}  \Vdash   \subL{ \monoalpha_{{\mathrm{2}}} }{ {B^\Box} }   \Vdash   \subR{ {A^\Box} }{ \monoalpha_{{\mathrm{1}}} }   \ottsym{)}   & (\ASLExFunB) \\
                   \quad \phantom{{} \rightsquigarrow {}} \quad \tif  \monoalpha  \notin   \mathrm{FTV}( {A^\Box}  \rightarrow  {B^\Box} )  
      \end{array}
    \]
    \caption{Algorithmic subtyping rules solving existential variables for $ \subtypingleft $}
  \end{figure}
}

\deftitle{Subtyping Rules Solving Existential Variables for $ \subtypingright $}{}

{
  \setlength{\LTpre}{10pt}
  \setlength{\LTpost}{0pt}
  \renewcommand{\arraystretch}{1.2}
  \small
  \begin{longtable}{RCL@{\quad}L}
                   \Gamma_{{\mathrm{1}}}  \ottsym{,}  \monoalpha  \ottsym{,}  \Gamma_{{\mathrm{2}}}  \ottsym{,}  \monobeta  \ottsym{,}  \Gamma_{{\mathrm{3}}}  \Vdash   \subR{ \monoalpha }{ \monobeta }  &  \rightsquigarrow         &
                   \Gamma_{{\mathrm{1}}}  \ottsym{,}  \monoalpha  \ottsym{,}  \Gamma_{{\mathrm{2}}}  \ottsym{,}   [  \monoalpha  /  \monobeta  ]  \Gamma_{{\mathrm{3}}}            & (\ASRExExA)                      \\
                   \Gamma_{{\mathrm{1}}}  \ottsym{,}  \monoalpha  \ottsym{,}  \Gamma_{{\mathrm{2}}}  \ottsym{,}  \monobeta  \ottsym{,}  \Gamma_{{\mathrm{3}}}  \Vdash   \subR{ \monobeta }{ \monoalpha }  &  \rightsquigarrow         &
                   \Gamma_{{\mathrm{1}}}  \ottsym{,}  \monoalpha  \ottsym{,}  \Gamma_{{\mathrm{2}}}  \ottsym{,}   [  \monoalpha  /  \monobeta  ]  \Gamma_{{\mathrm{3}}}            & (\ASRExExB)                      \\
                   \Gamma_{{\mathrm{1}}}  \ottsym{,}  \ottmv{a}  \ottsym{,}  \Gamma_{{\mathrm{2}}}  \ottsym{,}  \monoalpha  \ottsym{,}  \Gamma_{{\mathrm{3}}}  \Vdash   \subR{ \ottmv{a} }{ \monoalpha }    &  \rightsquigarrow         &
                   \Gamma_{{\mathrm{1}}}  \ottsym{,}  \ottmv{a}  \ottsym{,}  \Gamma_{{\mathrm{2}}}  \ottsym{,}   [  \ottmv{a}  /  \monoalpha  ]  \Gamma_{{\mathrm{3}}}              & (\ASRExTVarA)                    \\
                   \Gamma_{{\mathrm{1}}}  \ottsym{,}  \ottmv{a}  \ottsym{,}  \Gamma_{{\mathrm{2}}}  \ottsym{,}  \monoalpha  \ottsym{,}  \Gamma_{{\mathrm{3}}}  \Vdash   \subR{ \monoalpha }{ \ottmv{a} }    &  \rightsquigarrow         &
                   \Gamma_{{\mathrm{1}}}  \ottsym{,}  \ottmv{a}  \ottsym{,}  \Gamma_{{\mathrm{2}}}  \ottsym{,}   [  \ottmv{a}  /  \monoalpha  ]  \Gamma_{{\mathrm{3}}}              & (\ASRExTVarB)                    \\
                   \Gamma_{{\mathrm{1}}}  \ottsym{,}  \monoalpha  \ottsym{,}  \Gamma_{{\mathrm{2}}}  \Vdash   \subR{  \mathsf{unit}  }{ \monoalpha }        &  \rightsquigarrow         &
                   \Gamma_{{\mathrm{1}}}  \ottsym{,}   [   \mathsf{unit}   /  \monoalpha  ]  \Gamma_{{\mathrm{2}}}                  & (\ASRExUnitA)                    \\
                   \Gamma_{{\mathrm{1}}}  \ottsym{,}  \monoalpha  \ottsym{,}  \Gamma_{{\mathrm{2}}}  \Vdash   \subR{ \monoalpha }{  \mathsf{unit}  }        &  \rightsquigarrow         &
                   \Gamma_{{\mathrm{1}}}  \ottsym{,}   [   \mathsf{unit}   /  \monoalpha  ]  \Gamma_{{\mathrm{2}}}                  & (\ASRExUnitB)
  \end{longtable}
  \begin{longtable}{L@{\quad}L}
                   \Gamma_{{\mathrm{1}}}  \ottsym{,}  \monoalpha  \ottsym{,}  \Gamma_{{\mathrm{2}}}  \Vdash   \subR{ {A^\Box}  \rightarrow  {B^\Box} }{ \monoalpha }                                       &                            \\
                   \quad  \rightsquigarrow  \quad
                                \Gamma_{{\mathrm{1}}}  \ottsym{,}  \monoalpha_{{\mathrm{1}}}  \ottsym{,}  \monoalpha_{{\mathrm{2}}}  \ottsym{,}   [   \monoalpha_{{\mathrm{1}}}   \rightarrow   \monoalpha_{{\mathrm{2}}}   /  \monoalpha  ]  \ottsym{(}  \Gamma_{{\mathrm{2}}}  \Vdash   \subR{ {B^\Box} }{ \monoalpha_{{\mathrm{2}}} }   \Vdash   \subL{ \monoalpha_{{\mathrm{1}}} }{ {A^\Box} }   \ottsym{)}  & (\ASRExFunA) \\
                   \quad \phantom{{} \rightsquigarrow {}} \quad \tif  \monoalpha  \notin   \mathrm{FTV}( {A^\Box}  \rightarrow  {B^\Box} )           &                                             \\
                   \Gamma_{{\mathrm{1}}}  \ottsym{,}  \monoalpha  \ottsym{,}  \Gamma_{{\mathrm{2}}}  \Vdash   \subR{ \monoalpha }{  \ottnt{A}   \rightarrow   \ottnt{B}  }                                       &                                             \\
                   \quad  \rightsquigarrow  \quad
                                \Gamma_{{\mathrm{1}}}  \ottsym{,}  \monoalpha_{{\mathrm{1}}}  \ottsym{,}  \monoalpha_{{\mathrm{2}}}  \ottsym{,}   [   \monoalpha_{{\mathrm{1}}}   \rightarrow   \monoalpha_{{\mathrm{2}}}   /  \monoalpha  ]  \ottsym{(}  \Gamma_{{\mathrm{2}}}  \Vdash   \subR{ \monoalpha_{{\mathrm{2}}} }{ \ottnt{B} }   \Vdash   \subL{ \ottnt{A} }{ \monoalpha_{{\mathrm{1}}} }   \ottsym{)}   & (\ASRExFunB) \\
                   \quad \phantom{{} \rightsquigarrow {}} \quad \tif  \monoalpha  \notin   \mathrm{FTV}( \ottnt{A}  \rightarrow  \ottnt{B} )           &
  \end{longtable}
}

\mycomment{
  \begin{figure}[H]
    \renewcommand{\arraystretch}{1.2}
    \small
    \[
      \begin{array}{rcl@{\quad}l}
                   \Gamma  \Vdash  \ottsym{()}  \ottsym{:}  {A^\Box}                                                    &  \rightsquigarrow     &
                   \Gamma  \Vdash   \subL{  \mathsf{unit}  }{ {A^\Box} }  \quad \tif \mathsf{not}\forall \, \ottsym{(}  {A^\Box}  \ottsym{)}                   & (\ATUnit)                \\
                   \Gamma  \Vdash  \mathit{x}  \ottsym{:}  {A^\Box}                                                     &  \rightsquigarrow     &
                   \Gamma  \Vdash   \subL{ \ottnt{A} }{ {A^\Box} }  \quad \tif  \mathit{x}  \ottsym{:}  \ottnt{A}  \in  \Gamma  \tand \mathsf{not}\forall \, \ottsym{(}  {A^\Box}  \ottsym{)} & (\ATVar)                 \\
                   \Gamma  \Vdash  \ottnt{e}  \ottsym{:}   \forall  \ottmv{a}  .  \ottnt{A}                                           &  \rightsquigarrow     &
                   \Gamma  \ottsym{,}  \ottmv{a}  \ottsym{,}  \triangleright  \ottsym{\{}  \ottsym{\}}  \Vdash  \ottnt{e}  \ottsym{:}  \ottnt{A}                                          & (\ATAll)
      \end{array}
    \]
    \[
      \begin{array}{l@{\quad}l}
        \Gamma_{{\mathrm{1}}}  \ottsym{,}  \monoalpha  \ottsym{,}  \Gamma_{{\mathrm{2}}}  \Vdash   \lambda  \mathit{x}  .  \ottnt{e}   \ottsym{:}  \monoalpha                                                                                                 %
        \quad  \rightsquigarrow  \quad \Gamma_{{\mathrm{1}}}  \ottsym{,}  \monoalpha_{{\mathrm{1}}}  \ottsym{,}  \monoalpha_{{\mathrm{2}}}  \ottsym{,}   [   \monoalpha_{{\mathrm{1}}}   \rightarrow   \monoalpha_{{\mathrm{2}}}   /  \monoalpha  ]  \ottsym{(}  \Gamma_{{\mathrm{2}}}  \ottsym{,}  \mathit{x}  \ottsym{:}  \monoalpha_{{\mathrm{1}}}  \Vdash  \ottnt{e}  \ottsym{:}  \monoalpha_{{\mathrm{2}}}  \ottsym{)}              & (\ATAbsEx)   \\
        \Gamma_{{\mathrm{1}}}  \ottsym{,}  \triangleright  \ottsym{\{}  \vec{\polybeta}_{{\mathrm{1}}}  \ottsym{,}  \polyalpha  \ottsym{,}  \vec{\polybeta}_{{\mathrm{2}}}  \ottsym{\}}  \ottsym{,}  \Gamma_{{\mathrm{2}}}  \Vdash   \lambda  \mathit{x}  .  \ottnt{e}   \ottsym{:}   \boxed{ \polyalpha }                                             &              \\
        \quad  \rightsquigarrow  \quad \Gamma_{{\mathrm{1}}}  \ottsym{,}  \monoalpha  \ottsym{,}  \triangleright  \ottsym{\{}  \vec{\polybeta}_{{\mathrm{1}}}  \ottsym{,}  \polybeta  \ottsym{,}  \vec{\polybeta}_{{\mathrm{2}}}  \ottsym{\}}  \ottsym{,}   [  \monoalpha  \rightarrow  \polybeta  /  \polyalpha  ]  \ottsym{(}  \Gamma_{{\mathrm{2}}}  \ottsym{,}  \mathit{x}  \ottsym{:}  \monoalpha  \Vdash  \ottnt{e}  \ottsym{:}   \boxed{ \polybeta }   \ottsym{)}  & (\ATAbsBox)  \\
        \Gamma  \Vdash   \lambda  \mathit{x}  .  \ottnt{e}   \ottsym{:}  \ottnt{A}  \rightarrow  {B^\Box} %
        \quad  \rightsquigarrow  \quad \Gamma  \ottsym{,}  \mathit{x}  \ottsym{:}  \ottnt{A}  \Vdash  \ottnt{e}  \ottsym{:}  {B^\Box}                                                 & (\ATAbsFunA) \\
        \Gamma  \ottsym{,}  \triangleright  \ottsym{\{}  \vec{\polyalpha}  \ottsym{\}}  \ottsym{,}  \Delta  \Vdash   \lambda  \mathit{x}  .  \ottnt{e}   \ottsym{:}  {A^\Box}  \rightarrow  {B^\Box}                                                        &              \\
        \quad  \rightsquigarrow  \quad \Gamma  \ottsym{,}  \monoalpha  \ottsym{,}  \triangleright  \ottsym{\{}  \vec{\polyalpha}  \ottsym{\}}  \ottsym{,}  \Delta  \Vdash   \subR{ {A^\Box} }{ \monoalpha }   \ottsym{,}  \mathit{x}  \ottsym{:}  \monoalpha  \Vdash  \ottnt{e}  \ottsym{:}  {B^\Box}                  &
        (\ATAbsFunB)                                                                                               \\
        \quad \phantom{{} \rightsquigarrow {}} \quad \tif \mathsf{hasBox} \, \ottsym{(}  {A^\Box}  \ottsym{)}                                        &              \\[1ex]
        \Gamma_{{\mathrm{1}}}  \ottsym{,}  \monoalpha  \ottsym{,}  \Gamma_{{\mathrm{2}}}  \Vdash   \lambda  \mathit{x}  \ottsym{:}  \ottnt{A}  .  \ottnt{e}   \ottsym{:}  \monoalpha                                                         &              \\
        \quad  \rightsquigarrow  \quad \Gamma_{{\mathrm{1}}}  \ottsym{,}  \monoalpha_{{\mathrm{1}}}  \ottsym{,}  \monoalpha_{{\mathrm{2}}}  \ottsym{,}   [   \monoalpha_{{\mathrm{1}}}   \rightarrow   \monoalpha_{{\mathrm{2}}}   /  \monoalpha  ]  \ottsym{(}  \Gamma_{{\mathrm{2}}}  \Vdash   \subR{ \monoalpha_{{\mathrm{1}}} }{ \ottnt{A} }   \ottsym{,}  \mathit{x}  \ottsym{:}  \ottnt{A}  \Vdash  \ottnt{e}  \ottsym{:}  \monoalpha_{{\mathrm{2}}}  \ottsym{)}  & (\ATAAbsEx)  \\
        \Gamma_{{\mathrm{1}}}  \ottsym{,}  \triangleright  \ottsym{\{}  \vec{\polybeta}_{{\mathrm{1}}}  \ottsym{,}  \polyalpha  \ottsym{,}  \vec{\polybeta}_{{\mathrm{2}}}  \ottsym{\}}  \ottsym{,}  \Gamma_{{\mathrm{2}}}  \Vdash   \lambda  \mathit{x}  \ottsym{:}  \ottnt{A}  .  \ottnt{e}   \ottsym{:}   \boxed{ \polyalpha }                                       &              \\
        \quad  \rightsquigarrow  \quad \Gamma_{{\mathrm{1}}}  \ottsym{,}  \triangleright  \ottsym{\{}  \vec{\polybeta}_{{\mathrm{1}}}  \ottsym{,}  \polybeta  \ottsym{,}  \vec{\polybeta}_{{\mathrm{2}}}  \ottsym{\}}  \ottsym{,}   [  \ottnt{A}  \rightarrow  \polybeta  /  \polyalpha  ]  \ottsym{(}  \Gamma_{{\mathrm{2}}}  \ottsym{,}  \mathit{x}  \ottsym{:}  \ottnt{A}  \Vdash  \ottnt{e}  \ottsym{:}   \boxed{ \polybeta }   \ottsym{)}      & (\ATAAbsBox) \\
        \quad \phantom{{} \rightsquigarrow {}} \quad \tif   \mathrm{FTV}( \ottnt{A} )   \subseteq \mathrm{dom}( \Gamma_{{\mathrm{1}}} )                             &              \\
        \Gamma  \Vdash   \lambda  \mathit{x}  \ottsym{:}  \ottnt{A}  .  \ottnt{e}   \ottsym{:}  {A^\Box}  \rightarrow  {B^\Box}                                                                                                %
        \quad  \rightsquigarrow  \quad \Gamma  \Vdash   \subR{ {A^\Box} }{ \ottnt{A} }   \ottsym{,}  \mathit{x}  \ottsym{:}  \ottnt{A}  \Vdash  \ottnt{e}  \ottsym{:}  {B^\Box}                                       & (\ATAAbsFun)
      \end{array}
    \]\[
      \begin{array}{rcl@{\quad}l}
                   \Gamma  \ottsym{,}  \triangleright  \ottsym{\{}  \vec{\polyalpha}  \ottsym{\}}  \ottsym{,}  \Delta  \Vdash  \ottnt{e_{{\mathrm{1}}}}  \appright  \ottnt{e_{{\mathrm{2}}}}  \ottsym{:}  {A^\Box}                    &  \rightsquigarrow     &
                   \Gamma  \ottsym{,}  \triangleright  \ottsym{\{}  \vec{\polyalpha}  \ottsym{,}  \polyalpha  \ottsym{\}}  \ottsym{,}  \Delta  \Vdash  \ottnt{e_{{\mathrm{2}}}}  \ottsym{:}  \polyalpha  \Vdash  \ottnt{e_{{\mathrm{1}}}}  \ottsym{:}   \boxed{ \polyalpha }   \rightarrow  {A^\Box}  & (\ATAppR)                            \\
                                                                         &           & \tif \mathsf{not}\forall \, \ottsym{(}  {A^\Box}  \ottsym{)} & \\
                   \Gamma  \ottsym{,}  \triangleright  \ottsym{\{}  \vec{\polyalpha}  \ottsym{\}}  \ottsym{,}  \Delta  \Vdash  \ottnt{e_{{\mathrm{1}}}}  \appleft  \ottnt{e_{{\mathrm{2}}}}  \ottsym{:}  {A^\Box}                    &  \rightsquigarrow     &
                   \Gamma  \ottsym{,}  \triangleright  \ottsym{\{}  \vec{\polyalpha}  \ottsym{,}  \polyalpha  \ottsym{\}}  \ottsym{,}  \Delta  \Vdash  \ottnt{e_{{\mathrm{1}}}}  \ottsym{:}  \polyalpha  \rightarrow  {A^\Box}  \Vdash  \ottnt{e_{{\mathrm{2}}}}  \ottsym{:}   \boxed{ \polyalpha }  & (\ATAppL)                            \\
                                                                         &           & \tif \mathsf{not}\forall \, \ottsym{(}  {A^\Box}  \ottsym{)} &
      \end{array}
    \]
    \caption{Algorithmic typing rules}
  \end{figure}
}

\deftitle{Well-formedness}{
  \fbox{$\vdash  \Gamma$}\quad
  \fbox{$\Gamma  \vdash  {A^\Box}$}\quad
  \fbox{$\Gamma  \vdash  \ottnt{e}$}\quad
  \fbox{$\Gamma  \vdash  \omega$}
}

\begin{mathparpagebreakable}
  \inferrule{\mbox{}}{
    \vdash   \emptyset 
  }

  \inferrule{
  \vdash  \Gamma \\  \ottmv{a}  \notin  \Gamma 
  }{
    \vdash  \Gamma  \ottsym{,}  \ottmv{a}
  }

  \inferrule{
  \vdash  \Gamma \\  \monoalpha  \notin  \Gamma 
  }{
    \vdash  \Gamma  \ottsym{,}  \monoalpha
  }

  \inferrule{
  \vdash  \Gamma \\  \ottsym{\{}  \vec{\polyalpha}  \ottsym{\}}    \cap    \Gamma   \ottsym{=}   \emptyset 
  }{
    \vdash  \Gamma  \ottsym{,}  \triangleright  \ottsym{\{}  \vec{\polyalpha}  \ottsym{\}}
  }

  \inferrule{
  \vdash  \Gamma \\ \Gamma  \vdash  \ottnt{A}
  }{
    \vdash  \Gamma  \ottsym{,}  \mathit{x}  \ottsym{:}  \ottnt{A}
  }

  \inferrule{
    \Gamma  \vdash  \omega
  }{
    \vdash  \Gamma  \Vdash  \omega
  }

  \inferrule{
    \vdash  \Gamma
  }{
    \Gamma  \vdash   \mathsf{unit} 
  }

  \inferrule{
  \vdash  \Gamma \\  \ottmv{a}  \in  \Gamma 
  }{
    \Gamma  \vdash  \ottmv{a}
  }

  \inferrule{
  \vdash  \Gamma \\  \monoalpha  \in  \Gamma 
  }{
    \Gamma  \vdash  \monoalpha
  }

  \inferrule{
  \vdash  \Gamma \\  \triangleright  \ottsym{\{}  \vec{\polybeta}_{{\mathrm{1}}}  \ottsym{,}  \polyalpha  \ottsym{,}  \vec{\polybeta}_{{\mathrm{2}}}  \ottsym{\}}  \in  \Gamma 
  }{
    \Gamma  \vdash  \polyalpha
  }

  \inferrule{
    \Gamma  \vdash  \ottnt{A}
  }{
    \Gamma  \vdash   \boxed{ \ottnt{A} } 
  }

  \inferrule{
  \Gamma  \vdash  {A^\Box} \\ \Gamma  \vdash  {B^\Box}
  }{
    \Gamma  \vdash  {A^\Box}  \rightarrow  {B^\Box}
  }

  \inferrule{
    \Gamma  \ottsym{,}  \ottmv{a}  \vdash  \ottnt{A}
  }{
    \Gamma  \vdash   \forall  \ottmv{a}  .  \ottnt{A} 
  }

  \inferrule{
    \vdash  \Gamma
  }{
    \Gamma  \vdash  \ottsym{()}
  }

  \inferrule{
  \vdash  \Gamma \\  \mathit{x}  \ottsym{:}  \ottnt{A}  \in  \Gamma 
  }{
    \Gamma  \vdash  \mathit{x}
  }

  \inferrule{
    \Gamma  \ottsym{,}  \mathit{x}  \ottsym{:}  \ottnt{A}  \vdash  \ottnt{e}
  }{
    \Gamma  \vdash   \lambda  \mathit{x}  .  \ottnt{e} 
  }

  \inferrule{
    \Gamma  \ottsym{,}  \mathit{x}  \ottsym{:}  \ottnt{A}  \vdash  \ottnt{e}
  }{
    \Gamma  \vdash   \lambda  \mathit{x}  \ottsym{:}  \ottnt{A}  .  \ottnt{e} 
  }

  \inferrule{
  \Gamma  \vdash  \ottnt{e_{{\mathrm{1}}}} \\ \Gamma  \vdash  \ottnt{e_{{\mathrm{2}}}}
  }{
    \Gamma  \vdash  \ottnt{e_{{\mathrm{1}}}}  \appright  \ottnt{e_{{\mathrm{2}}}}
  }

  \inferrule{
  \Gamma  \vdash  \ottnt{e_{{\mathrm{1}}}} \\ \Gamma  \vdash  \ottnt{e_{{\mathrm{2}}}}
  }{
    \Gamma  \vdash  \ottnt{e_{{\mathrm{1}}}}  \appleft  \ottnt{e_{{\mathrm{2}}}}
  }

  \inferrule{
  \Gamma  \vdash  \ottnt{A} \\ \Gamma  \vdash  {A^\Box}
  }{
    \Gamma  \vdash   \subL{ \ottnt{A} }{ {A^\Box} } 
  }

  \inferrule{
  \Gamma  \vdash  {A^\Box} \\ \Gamma  \vdash  \ottnt{A}
  }{
    \Gamma  \vdash   \subR{ {A^\Box} }{ \ottnt{A} } 
  }

  \inferrule{
  \Gamma  \vdash  \ottnt{e} \\ \Gamma  \vdash  {A^\Box}
  }{
    \Gamma  \vdash  \ottnt{e}  \ottsym{:}  {A^\Box}
  }
\end{mathparpagebreakable}

\deftitle{Limited Well-formedness}{
  \fbox{$\vdash^\downarrow  \Gamma$}\quad
  \fbox{$\Gamma  \vdash^\downarrow  {A^\Box}$}\quad
  \fbox{$\Gamma  \vdash^\downarrow  \ottnt{e}$}\quad
  \fbox{$\Gamma  \vdash^\downarrow  \omega$}
}

\begin{mathparpagebreakable}
  \small
  \inferrule{\mbox{}}{
    \vdash^\downarrow   \emptyset 
  }

  \inferrule{
  \vdash^\downarrow  \Gamma \\  \ottmv{a}  \notin  \Gamma 
  }{
    \vdash^\downarrow  \Gamma  \ottsym{,}  \ottmv{a}
  }

  \inferrule{
  \vdash^\downarrow  \Gamma \\  \monoalpha  \notin  \Gamma 
  }{
    \vdash^\downarrow  \Gamma  \ottsym{,}  \monoalpha
  }

  \inferrule{
  \vdash^\downarrow  \Gamma \\  \ottsym{\{}  \vec{\polyalpha}  \ottsym{\}}    \cap    \Gamma   \ottsym{=}   \emptyset 
  }{
    \vdash^\downarrow  \Gamma  \ottsym{,}  \triangleright  \ottsym{\{}  \vec{\polyalpha}  \ottsym{\}}
  }

  \inferrule{
  \vdash^\downarrow  \Gamma \\ \Gamma  \vdash^\downarrow  \ottnt{S}
  }{
    \vdash^\downarrow  \Gamma  \ottsym{,}  \mathit{x}  \ottsym{:}  \ottnt{S}
  }

  \inferrule{
    \Gamma  \vdash^\downarrow  \omega
  }{
    \vdash^\downarrow  \Gamma  \Vdash  \omega
  }

  \inferrule{
    \vdash^\downarrow  \Gamma
  }{
    \Gamma  \vdash^\downarrow   \mathsf{unit} 
  }

  \inferrule{
  \vdash^\downarrow  \Gamma \\  \ottmv{a}  \in  \Gamma 
  }{
    \Gamma  \vdash^\downarrow  \ottmv{a}
  }

  \inferrule{
  \vdash^\downarrow  \Gamma \\  \monoalpha  \in  \Gamma 
  }{
    \Gamma  \vdash^\downarrow  \monoalpha
  }

  \inferrule{
    \vdash^\downarrow  \Gamma_{{\mathrm{1}}}  \ottsym{,}  \triangleright  \ottsym{\{}  \vec{\polybeta}_{{\mathrm{1}}}  \ottsym{,}  \polyalpha  \ottsym{,}  \vec{\polybeta}_{{\mathrm{2}}}  \ottsym{\}}  \ottsym{,}  \Delta_{{\mathrm{2}}}
  }{
    \Gamma_{{\mathrm{1}}}  \ottsym{,}  \triangleright  \ottsym{\{}  \vec{\polybeta}_{{\mathrm{1}}}  \ottsym{,}  \polyalpha  \ottsym{,}  \vec{\polybeta}_{{\mathrm{2}}}  \ottsym{\}}  \ottsym{,}  \Delta_{{\mathrm{2}}}  \vdash^\downarrow  \polyalpha
  }

  \inferrule{
    \vdash^\downarrow  \Gamma_{{\mathrm{1}}}  \ottsym{,}  \triangleright  \ottsym{\{}  \vec{\polybeta}_{{\mathrm{1}}}  \ottsym{,}  \polyalpha  \ottsym{,}  \vec{\polybeta}_{{\mathrm{2}}}  \ottsym{\}}  \ottsym{,}  \Delta_{{\mathrm{2}}}
  }{
    \Gamma_{{\mathrm{1}}}  \ottsym{,}  \triangleright  \ottsym{\{}  \vec{\polybeta}_{{\mathrm{1}}}  \ottsym{,}  \polyalpha  \ottsym{,}  \vec{\polybeta}_{{\mathrm{2}}}  \ottsym{\}}  \ottsym{,}  \Delta_{{\mathrm{2}}}  \vdash^\downarrow   \boxed{ \polyalpha } 
  }

  \inferrule{
  \Gamma  \vdash^\downarrow  \ottnt{A} \\ \Gamma  \vdash^\downarrow  {B^\Box}
  }{
    \Gamma  \vdash^\downarrow  \ottnt{A}  \rightarrow  {B^\Box}
  }

  \inferrule{
  \Gamma  \vdash^\downarrow   \boxed{ \polyalpha }  \\ \Gamma  \vdash^\downarrow  {B^\Box}
  }{
    \Gamma  \vdash^\downarrow   \boxed{ \polyalpha }   \rightarrow  {B^\Box}
  }

  \inferrule{
    \Gamma  \ottsym{,}  \ottmv{a}  \vdash^\downarrow  \ottnt{A}
  }{
    \Gamma  \vdash^\downarrow   \forall  \ottmv{a}  .  \ottnt{A} 
  }

  \inferrule{
    \vdash^\downarrow  \Gamma
  }{
    \Gamma  \vdash^\downarrow  \ottsym{()}
  }

  \inferrule{
  \vdash^\downarrow  \Gamma \\  \mathit{x}  \ottsym{:}  \ottnt{A}  \in  \Gamma 
  }{
    \Gamma  \vdash^\downarrow  \mathit{x}
  }

  \inferrule{
    \Gamma  \ottsym{,}  \mathit{x}  \ottsym{:}  \ottnt{A}  \vdash^\downarrow  \ottnt{e}
  }{
    \Gamma  \vdash^\downarrow   \lambda  \mathit{x}  .  \ottnt{e} 
  }

  \inferrule{
    \Gamma  \ottsym{,}  \mathit{x}  \ottsym{:}  \ottnt{A}  \vdash^\downarrow  \ottnt{e}
  }{
    \Gamma  \vdash^\downarrow   \lambda  \mathit{x}  \ottsym{:}  \ottnt{A}  .  \ottnt{e} 
  }

  \inferrule{
  \Gamma  \vdash^\downarrow  \ottnt{e_{{\mathrm{1}}}} \\ \Gamma  \vdash^\downarrow  \ottnt{e_{{\mathrm{2}}}}
  }{
    \Gamma  \vdash^\downarrow  \ottnt{e_{{\mathrm{1}}}}  \appright  \ottnt{e_{{\mathrm{2}}}}
  }

  \inferrule{
  \Gamma  \vdash^\downarrow  \ottnt{e_{{\mathrm{1}}}} \\ \Gamma  \vdash^\downarrow  \ottnt{e_{{\mathrm{2}}}}
  }{
    \Gamma  \vdash^\downarrow  \ottnt{e_{{\mathrm{1}}}}  \appleft  \ottnt{e_{{\mathrm{2}}}}
  }

  \inferrule{
  \Gamma  \ottsym{,}  \triangleright  \ottsym{\{}  \vec{\polyalpha}  \ottsym{\}}  \ottsym{,}  \Delta  \vdash^\downarrow  \ottnt{A} \\ \Gamma  \ottsym{,}  \triangleright  \ottsym{\{}  \vec{\polyalpha}  \ottsym{\}}  \ottsym{,}  \Delta  \vdash^\downarrow  {A^\Box}
  }{
    \Gamma  \ottsym{,}  \triangleright  \ottsym{\{}  \vec{\polyalpha}  \ottsym{\}}  \ottsym{,}  \Delta  \vdash^\downarrow   \subL{ \ottnt{A} }{ {A^\Box} } 
  }

  \inferrule{
  \Gamma  \ottsym{,}  \triangleright  \ottsym{\{}  \vec{\polyalpha}  \ottsym{\}}  \ottsym{,}  \Delta  \vdash^\downarrow  \ottnt{A} \\ \Gamma  \ottsym{,}  \triangleright  \ottsym{\{}  \vec{\polyalpha}  \ottsym{\}}  \ottsym{,}  \Delta  \vdash^\downarrow  \ottnt{B}
  }{
    \Gamma  \ottsym{,}  \triangleright  \ottsym{\{}  \vec{\polyalpha}  \ottsym{\}}  \ottsym{,}  \Delta  \vdash^\downarrow   \subR{ \ottnt{A} }{ \ottnt{B} } 
  }

  \inferrule{
  \Gamma  \ottsym{,}  \triangleright  \ottsym{\{}  \vec{\polyalpha}  \ottsym{\}}  \ottsym{,}  \Delta  \vdash^\downarrow   \boxed{ \polyalpha }  \\ \Gamma  \ottsym{,}  \triangleright  \ottsym{\{}  \vec{\polyalpha}  \ottsym{\}}  \ottsym{,}  \Delta  \vdash^\downarrow  \ottnt{B}
  }{
    \Gamma  \ottsym{,}  \triangleright  \ottsym{\{}  \vec{\polyalpha}  \ottsym{\}}  \ottsym{,}  \Delta  \vdash^\downarrow   \subR{  \boxed{ \polyalpha }  }{ \ottnt{B} } 
  }

  \inferrule{
  \Gamma  \ottsym{,}  \triangleright  \ottsym{\{}  \vec{\polyalpha}  \ottsym{\}}  \ottsym{,}  \Delta  \vdash^\downarrow  \ottnt{e} \\ \Gamma  \ottsym{,}  \triangleright  \ottsym{\{}  \vec{\polyalpha}  \ottsym{\}}  \ottsym{,}  \Delta  \vdash^\downarrow  {A^\Box}
  }{
    \Gamma  \ottsym{,}  \triangleright  \ottsym{\{}  \vec{\polyalpha}  \ottsym{\}}  \ottsym{,}  \Delta  \vdash^\downarrow  \ottnt{e}  \ottsym{:}  {A^\Box}
  }
\end{mathparpagebreakable}

\deftitle{Polymorphic existential variable counter}{
  \fbox{$\ottsym{\{}  \vec{\polyalpha}  \ottsym{\}}  \vdash  \Gamma  \dashv  \ottsym{\{}  \vec{\polybeta}  \ottsym{\}}$}\quad
  \fbox{$\Gamma  \mid  \ottsym{\{}  \vec{\polyalpha}  \ottsym{\}}  \vdash  {A^\Box}  \dashv  \ottsym{\{}  \vec{\polybeta}  \ottsym{\}}$}\quad
  \fbox{$\Gamma  \mid  \ottsym{\{}  \vec{\polyalpha}  \ottsym{\}}  \vdash  \omega  \dashv  \ottsym{\{}  \vec{\polybeta}  \ottsym{\}}$}
}

\begin{mathparpagebreakable}
  \small
  \inferrule{ }{
    \ottsym{\{}  \vec{\polyalpha}  \ottsym{\}}  \vdash   \emptyset   \dashv  \ottsym{\{}  \vec{\polyalpha}  \ottsym{\}}
  }

  \inferrule{
    \ottsym{\{}  \vec{\polyalpha}  \ottsym{\}}  \vdash  \Gamma  \dashv  \ottsym{\{}  \vec{\polybeta}  \ottsym{\}}
  }{
    \ottsym{\{}  \vec{\polyalpha}  \ottsym{\}}  \vdash  \Gamma  \ottsym{,}  \ottmv{a}  \dashv  \ottsym{\{}  \vec{\polybeta}  \ottsym{\}}
  }

  \inferrule{
    \ottsym{\{}  \vec{\polyalpha}  \ottsym{\}}  \vdash  \Gamma  \dashv  \ottsym{\{}  \vec{\polybeta}  \ottsym{\}}
  }{
    \ottsym{\{}  \vec{\polyalpha}  \ottsym{\}}  \vdash  \Gamma  \ottsym{,}  \monoalpha  \dashv  \ottsym{\{}  \vec{\polybeta}  \ottsym{\}}
  }

  \inferrule{
    \ottsym{\{}  \vec{\polyalpha}  \ottsym{\}}  \vdash  \Gamma  \dashv  \ottsym{\{}  \vec{\polybeta}  \ottsym{\}}
  }{
    \ottsym{\{}  \vec{\polyalpha}  \ottsym{\}}  \vdash  \Gamma  \ottsym{,}  \triangleright  \ottsym{\{}  \vec{\polygamma}  \ottsym{\}}  \dashv  \ottsym{\{}  \vec{\polybeta}  \ottsym{\}}
  }

  \inferrule{
    \ottsym{\{}  \vec{\polyalpha}  \ottsym{\}}  \vdash  \Gamma  \dashv  \ottsym{\{}  \vec{\polybeta}  \ottsym{\}}
  }{
    \ottsym{\{}  \vec{\polyalpha}  \ottsym{\}}  \vdash  \Gamma  \ottsym{,}  \mathit{x}  \ottsym{:}  \ottnt{A}  \dashv  \ottsym{\{}  \vec{\polybeta}  \ottsym{\}}
  }

  \inferrule{
  \Gamma  \mid  \ottsym{\{}  \vec{\polyalpha}  \ottsym{\}}  \vdash  \omega  \dashv  \ottsym{\{}  \vec{\polygamma}  \ottsym{\}} \\ \ottsym{\{}  \vec{\polygamma}  \ottsym{\}}  \vdash  \Gamma  \dashv  \ottsym{\{}  \vec{\polybeta}  \ottsym{\}}
  }{
    \ottsym{\{}  \vec{\polyalpha}  \ottsym{\}}  \vdash  \Gamma  \Vdash  \omega  \dashv  \ottsym{\{}  \vec{\polybeta}  \ottsym{\}}
  }

  \inferrule{ }{
    \Gamma  \mid  \ottsym{\{}  \vec{\polyalpha}  \ottsym{\}}  \vdash   \mathsf{unit}   \dashv  \ottsym{\{}  \vec{\polyalpha}  \ottsym{\}}
  }

  \inferrule{
     \ottmv{a}  \in  \Gamma 
  }{
    \Gamma  \mid  \ottsym{\{}  \vec{\polyalpha}  \ottsym{\}}  \vdash  \ottmv{a}  \dashv  \ottsym{\{}  \vec{\polyalpha}  \ottsym{\}}
  }

  \inferrule{
     \monoalpha  \in  \Gamma 
  }{
    \Gamma  \mid  \ottsym{\{}  \vec{\polyalpha}  \ottsym{\}}  \vdash  \monoalpha  \dashv  \ottsym{\{}  \vec{\polyalpha}  \ottsym{\}}
  }

  \inferrule{
     \triangleright  \ottsym{\{}  \vec{\polygamma}_{{\mathrm{1}}}  \ottsym{,}  \polyalpha  \ottsym{,}  \vec{\polygamma}_{{\mathrm{2}}}  \ottsym{\}}  \in  \Gamma 
  }{
    \Gamma  \mid  \ottsym{\{}  \vec{\polybeta}_{{\mathrm{1}}}  \ottsym{,}  \polyalpha  \ottsym{,}  \vec{\polybeta}_{{\mathrm{2}}}  \ottsym{\}}  \vdash  \polyalpha  \dashv  \ottsym{\{}  \vec{\polybeta}_{{\mathrm{1}}}  \ottsym{,}  \polyalpha  \ottsym{,}  \vec{\polybeta}_{{\mathrm{2}}}  \ottsym{\}}
  }

  \inferrule{
   \polyalpha  \notin  \ottsym{\{}  \vec{\polybeta}  \ottsym{\}}  \\  \triangleright  \ottsym{\{}  \vec{\polygamma}_{{\mathrm{1}}}  \ottsym{,}  \polyalpha  \ottsym{,}  \vec{\polygamma}_{{\mathrm{2}}}  \ottsym{\}}  \in  \Gamma 
  }{
    \Gamma  \mid  \ottsym{\{}  \vec{\polybeta}  \ottsym{\}}  \vdash   \boxed{ \polyalpha }   \dashv  \ottsym{\{}  \vec{\polybeta}  \ottsym{,}  \polyalpha  \ottsym{\}}
  }

  \inferrule{
  \Gamma  \mid  \ottsym{\{}  \vec{\polyalpha}  \ottsym{\}}  \vdash  {A^\Box}  \dashv  \ottsym{\{}  \vec{\polygamma}  \ottsym{\}} \\ \Gamma  \mid  \ottsym{\{}  \vec{\polygamma}  \ottsym{\}}  \vdash  {B^\Box}  \dashv  \ottsym{\{}  \vec{\polybeta}  \ottsym{\}}
  }{
    \Gamma  \mid  \ottsym{\{}  \vec{\polyalpha}  \ottsym{\}}  \vdash  {A^\Box}  \rightarrow  {B^\Box}  \dashv  \ottsym{\{}  \vec{\polybeta}  \ottsym{\}}
  }

  \inferrule{ }{
    \Gamma  \mid  \ottsym{\{}  \vec{\polyalpha}  \ottsym{\}}  \vdash   \forall  \ottmv{a}  .  \ottnt{A}   \dashv  \ottsym{\{}  \vec{\polyalpha}  \ottsym{\}}
  }

  \inferrule{
    \Gamma  \mid  \ottsym{\{}  \vec{\polyalpha}  \ottsym{\}}  \vdash  {B^\Box}  \dashv  \ottsym{\{}  \vec{\polybeta}  \ottsym{\}}
  }{
    \Gamma  \mid  \ottsym{\{}  \vec{\polyalpha}  \ottsym{\}}  \vdash   \subL{ \ottnt{A} }{ {B^\Box} }   \dashv  \ottsym{\{}  \vec{\polybeta}  \ottsym{\}}
  }

  \inferrule{
    \Gamma  \mid  \ottsym{\{}  \vec{\polyalpha}  \ottsym{\}}  \vdash  {A^\Box}  \dashv  \ottsym{\{}  \vec{\polybeta}  \ottsym{\}}
  }{
    \Gamma  \mid  \ottsym{\{}  \vec{\polyalpha}  \ottsym{\}}  \vdash   \subR{ {A^\Box} }{ \ottnt{B} }   \dashv  \ottsym{\{}  \vec{\polybeta}  \ottsym{\}}
  }

  \inferrule{
    \Gamma  \mid  \ottsym{\{}  \vec{\polyalpha}  \ottsym{\}}  \vdash  {A^\Box}  \dashv  \ottsym{\{}  \vec{\polybeta}  \ottsym{\}}
  }{
    \Gamma  \mid  \ottsym{\{}  \vec{\polyalpha}  \ottsym{\}}  \vdash  \ottnt{e}  \ottsym{:}  {A^\Box}  \dashv  \ottsym{\{}  \vec{\polybeta}  \ottsym{\}}
  }
\end{mathparpagebreakable}

\subsection{Declarative Worklist}

\deftitle{Syntax}{}

{
  \small
  \[
    \begin{array}{rrcl}
      \textbf{Declarative worklists} & \Omega &  \Coloneqq  &
       \emptyset   \mid  \Omega  \ottsym{,}  \ottmv{a}  \mid  \Omega  \ottsym{,}  \mathit{x}  \ottsym{:}  \ottnt{A}  \mid  \Omega  \Vdash  \omega
    \end{array}
  \]
}

\deftitle{Reduction Rules}{\fbox{$\Omega  \rightsquigarrow  \Omega'$}}

{
  \small
  \[
    \begin{array}{rcll@{\qquad}rcll}
                 \Omega  \ottsym{,}  \ottmv{a} &  \rightsquigarrow  & \Omega &                         &
                 \Omega  \ottsym{,}  \mathit{x}  \ottsym{:}  \ottnt{A}        &  \rightsquigarrow  & \Omega &                             \\
                 \Omega  \Vdash   \subL{ \ottnt{A} }{ {B^\Box} }    &  \rightsquigarrow  & \Omega & \tif   |  \Omega  |    \vdash  \subL{ \ottnt{A} }{ {B^\Box} }  &
                 \Omega  \Vdash   \subR{ {A^\Box} }{ \ottnt{B} }    &  \rightsquigarrow  & \Omega & \tif   |  \Omega  |    \vdash  \subR{ {A^\Box} }{ \ottnt{B} }      \\
                 \Omega  \Vdash  \ottnt{e}  \ottsym{:}  {A^\Box}     &  \rightsquigarrow  & \Omega & \tif  |  \Omega  |   \vdash  \ottnt{e}  \ottsym{:}  {A^\Box}
    \end{array}
  \]
}

\deftitle{Worklist Instantiation}{\fbox{$\Gamma  \xrightarrow{ \mathrm{inst} }  \Omega$}}

\begin{mathparpagebreakable}
  \small
  \inferrule{ }{
    \Omega  \xrightarrow{ \mathrm{inst} }  \Omega
  }\ \InstRefl

  \inferrule{
   |  \Omega_{{\mathrm{1}}}  |   \vdash  \tau \\ \Omega_{{\mathrm{1}}}  \ottsym{,}   [  \tau  /  \monoalpha  ]  \Gamma_{{\mathrm{2}}}   \xrightarrow{ \mathrm{inst} }  \Omega
  }{
    \Omega_{{\mathrm{1}}}  \ottsym{,}  \monoalpha  \ottsym{,}  \Gamma_{{\mathrm{2}}}  \xrightarrow{ \mathrm{inst} }  \Omega
  }\ \InstEx

  \inferrule{
    \Omega_{{\mathrm{1}}}  \ottsym{,}  \Gamma_{{\mathrm{2}}}  \xrightarrow{ \mathrm{inst} }  \Omega
  }{
    \Omega_{{\mathrm{1}}}  \ottsym{,}  \triangleright  \ottsym{\{}  \ottsym{\}}  \ottsym{,}  \Gamma_{{\mathrm{2}}}  \xrightarrow{ \mathrm{inst} }  \Omega
  }\ \InstNil

  \inferrule{
   |  \Omega_{{\mathrm{1}}}  |   \vdash  \ottnt{A} \\ \Omega_{{\mathrm{1}}}  \ottsym{,}  \triangleright  \ottsym{\{}  \vec{\polybeta}  \ottsym{\}}  \ottsym{,}   [  \ottnt{A}  /  \polyalpha  ]  \Gamma_{{\mathrm{2}}}   \xrightarrow{ \mathrm{inst} }  \Omega
  }{
    \Omega_{{\mathrm{1}}}  \ottsym{,}  \triangleright  \ottsym{\{}  \polyalpha  \ottsym{,}  \vec{\polybeta}  \ottsym{\}}  \ottsym{,}  \Gamma_{{\mathrm{2}}}  \xrightarrow{ \mathrm{inst} }  \Omega
  }\ \InstPex
\end{mathparpagebreakable}

\mycomment{
\begin{figure}[H]
  \small
  \[
    \begin{array}{rrcl}
      \textbf{Declarative worklists} & \Omega &  \Coloneqq  &
       \emptyset   \mid  \Omega  \ottsym{,}  \ottmv{a}  \mid  \Omega  \ottsym{,}  \mathit{x}  \ottsym{:}  \ottnt{A}  \mid  \Omega  \Vdash  \omega
    \end{array}
  \]
  \noindent\fbox{$\Omega  \rightsquigarrow  \Omega'$}\hfill\mbox{} %
  \[
    \begin{array}{rcll@{\qquad}rcll}
                 \Omega  \ottsym{,}  \ottmv{a} &  \rightsquigarrow  & \Omega &                         &
                 \Omega  \ottsym{,}  \mathit{x}  \ottsym{:}  \ottnt{A}        &  \rightsquigarrow  & \Omega &                             \\
                 \Omega  \Vdash   \subL{ \ottnt{A} }{ {B^\Box} }    &  \rightsquigarrow  & \Omega & \tif   |  \Omega  |    \vdash  \subL{ \ottnt{A} }{ {B^\Box} }  &
                 \Omega  \Vdash   \subR{ {A^\Box} }{ \ottnt{B} }    &  \rightsquigarrow  & \Omega & \tif   |  \Omega  |    \vdash  \subR{ {A^\Box} }{ \ottnt{B} }      \\
                 \Omega  \Vdash  \ottnt{e}  \ottsym{:}  {A^\Box}     &  \rightsquigarrow  & \Omega & \tif  |  \Omega  |   \vdash  \ottnt{e}  \ottsym{:}  {A^\Box}
    \end{array}
  \]
  \caption{Declarative worklist and its reduction}\label{fig:decl_worklist}
\end{figure}

\begin{figure}[H]
  \small
  \noindent\fbox{$\Gamma  \xrightarrow{ \mathrm{inst} }  \Omega$}\hfill\mbox{}
  \begin{mathpar}
    \inferrule{ }{
      \Omega  \xrightarrow{ \mathrm{inst} }  \Omega
    }\ \InstRefl

    \inferrule{
     |  \Omega_{{\mathrm{1}}}  |   \vdash  \tau \\ \Omega_{{\mathrm{1}}}  \ottsym{,}   [  \tau  /  \monoalpha  ]  \Gamma_{{\mathrm{2}}}   \xrightarrow{ \mathrm{inst} }  \Omega
    }{
      \Omega_{{\mathrm{1}}}  \ottsym{,}  \monoalpha  \ottsym{,}  \Gamma_{{\mathrm{2}}}  \xrightarrow{ \mathrm{inst} }  \Omega
    }\ \InstEx

    \inferrule{
      \Omega_{{\mathrm{1}}}  \ottsym{,}  \Gamma_{{\mathrm{2}}}  \xrightarrow{ \mathrm{inst} }  \Omega
    }{
      \Omega_{{\mathrm{1}}}  \ottsym{,}  \triangleright  \ottsym{\{}  \ottsym{\}}  \ottsym{,}  \Gamma_{{\mathrm{2}}}  \xrightarrow{ \mathrm{inst} }  \Omega
    }\ \InstNil

    \inferrule{
     |  \Omega_{{\mathrm{1}}}  |   \vdash  \ottnt{A} \\ \Omega_{{\mathrm{1}}}  \ottsym{,}  \triangleright  \ottsym{\{}  \vec{\polybeta}  \ottsym{\}}  \ottsym{,}   [  \ottnt{A}  /  \polyalpha  ]  \Gamma_{{\mathrm{2}}}   \xrightarrow{ \mathrm{inst} }  \Omega
    }{
      \Omega_{{\mathrm{1}}}  \ottsym{,}  \triangleright  \ottsym{\{}  \polyalpha  \ottsym{,}  \vec{\polybeta}  \ottsym{\}}  \ottsym{,}  \Gamma_{{\mathrm{2}}}  \xrightarrow{ \mathrm{inst} }  \Omega
    }\ \InstPex
  \end{mathpar}
  \caption{Worklist instantiation}\label{fig:work_inst}
\end{figure}

\subsection{Auxiliary Definitions}

\begin{figure}[H]
  \small
  \noindent
  \fbox{$\vdash  \Gamma$}\quad
  \fbox{$\Gamma  \vdash  {A^\Box}$}\quad
  \fbox{$\Gamma  \vdash  \ottnt{e}$}\quad
  \fbox{$\Gamma  \vdash  \omega$}\hfill\vspace{1ex}
  \begin{mathpar}
    \inferrule{\mbox{}}{
      \vdash   \emptyset 
    }

    \inferrule{
    \vdash  \Gamma \\  \ottmv{a}  \notin  \Gamma 
    }{
      \vdash  \Gamma  \ottsym{,}  \ottmv{a}
    }

    \inferrule{
    \vdash  \Gamma \\  \monoalpha  \notin  \Gamma 
    }{
      \vdash  \Gamma  \ottsym{,}  \monoalpha
    }

    \inferrule{
    \vdash  \Gamma \\  \ottsym{\{}  \vec{\polyalpha}  \ottsym{\}}    \cap    \Gamma   \ottsym{=}   \emptyset 
    }{
      \vdash  \Gamma  \ottsym{,}  \triangleright  \ottsym{\{}  \vec{\polyalpha}  \ottsym{\}}
    }

    \inferrule{
      \Gamma  \vdash  \omega
    }{
      \vdash  \Gamma  \Vdash  \omega
    }

    \inferrule{
      \vdash  \Gamma
    }{
      \Gamma  \vdash   \mathsf{unit} 
    }

    \inferrule{
    \vdash  \Gamma \\  \ottmv{a}  \in  \Gamma 
    }{
      \Gamma  \vdash  \ottmv{a}
    }

    \inferrule{
    \vdash  \Gamma \\  \monoalpha  \in  \Gamma 
    }{
      \Gamma  \vdash  \monoalpha
    }

    \inferrule{
    \vdash  \Gamma \\  \triangleright  \ottsym{\{}  \vec{\polybeta}_{{\mathrm{1}}}  \ottsym{,}  \polyalpha  \ottsym{,}  \vec{\polybeta}_{{\mathrm{2}}}  \ottsym{\}}  \in  \Gamma 
    }{
      \Gamma  \vdash  \polyalpha
    }

    \inferrule{
      \Gamma  \vdash  \ottnt{A}
    }{
      \Gamma  \vdash   \boxed{ \ottnt{A} } 
    }

    \inferrule{
    \Gamma  \vdash  {A^\Box} \\ \Gamma  \vdash  {B^\Box}
    }{
      \Gamma  \vdash  {A^\Box}  \rightarrow  {B^\Box}
    }

    \inferrule{
      \Gamma  \ottsym{,}  \ottmv{a}  \vdash  \ottnt{A}
    }{
      \Gamma  \vdash   \forall  \ottmv{a}  .  \ottnt{A} 
    }

    \inferrule{
      \vdash  \Gamma
    }{
      \Gamma  \vdash  \ottsym{()}
    }

    \inferrule{
    \vdash  \Gamma \\  \mathit{x}  \ottsym{:}  \ottnt{A}  \in  \Gamma 
    }{
      \Gamma  \vdash  \mathit{x}
    }

    \inferrule{
      \Gamma  \ottsym{,}  \mathit{x}  \ottsym{:}  \ottnt{A}  \vdash  \ottnt{e}
    }{
      \Gamma  \vdash   \lambda  \mathit{x}  .  \ottnt{e} 
    }

    \inferrule{
      \Gamma  \ottsym{,}  \mathit{x}  \ottsym{:}  \ottnt{A}  \vdash  \ottnt{e}
    }{
      \Gamma  \vdash   \lambda  \mathit{x}  \ottsym{:}  \ottnt{A}  .  \ottnt{e} 
    }

    \inferrule{
    \Gamma  \vdash  \ottnt{e_{{\mathrm{1}}}} \\ \Gamma  \vdash  \ottnt{e_{{\mathrm{2}}}}
    }{
      \Gamma  \vdash  \ottnt{e_{{\mathrm{1}}}}  \appright  \ottnt{e_{{\mathrm{2}}}}
    }

    \inferrule{
    \Gamma  \vdash  \ottnt{e_{{\mathrm{1}}}} \\ \Gamma  \vdash  \ottnt{e_{{\mathrm{2}}}}
    }{
      \Gamma  \vdash  \ottnt{e_{{\mathrm{1}}}}  \appleft  \ottnt{e_{{\mathrm{2}}}}
    }

    \inferrule{
    \Gamma  \vdash  \ottnt{A} \\ \Gamma  \vdash  {A^\Box}
    }{
      \Gamma  \vdash   \subL{ \ottnt{A} }{ {A^\Box} } 
    }

    \inferrule{
    \Gamma  \vdash  {A^\Box} \\ \Gamma  \vdash  \ottnt{A}
    }{
      \Gamma  \vdash   \subR{ {A^\Box} }{ \ottnt{A} } 
    }

    \inferrule{
    \Gamma  \vdash  \ottnt{e} \\ \Gamma  \vdash  {A^\Box}
    }{
      \Gamma  \vdash  \ottnt{e}  \ottsym{:}  {A^\Box}
    }
  \end{mathpar}
  \caption{Well-formedness for the algorithmic system}
\end{figure}

\begin{figure}[H]
  \small
  \noindent
  \fbox{$\vdash^\downarrow  \Gamma$}\quad
  \fbox{$\Gamma  \vdash^\downarrow  {A^\Box}$}\quad
  \fbox{$\Gamma  \vdash^\downarrow  \ottnt{e}$}\quad
  \fbox{$\Gamma  \vdash^\downarrow  \omega$}\hfill\vspace{1ex}
  \begin{mathpar}
    \inferrule{\mbox{}}{
      \vdash^\downarrow   \emptyset 
    }

    \inferrule{
    \vdash^\downarrow  \Gamma \\  \ottmv{a}  \notin  \Gamma 
    }{
      \vdash^\downarrow  \Gamma  \ottsym{,}  \ottmv{a}
    }

    \inferrule{
    \vdash^\downarrow  \Gamma \\  \monoalpha  \notin  \Gamma 
    }{
      \vdash^\downarrow  \Gamma  \ottsym{,}  \monoalpha
    }

    \inferrule{
    \vdash^\downarrow  \Gamma \\  \ottsym{\{}  \vec{\polyalpha}  \ottsym{\}}    \cap    \Gamma   \ottsym{=}   \emptyset 
    }{
      \vdash^\downarrow  \Gamma  \ottsym{,}  \triangleright  \ottsym{\{}  \vec{\polyalpha}  \ottsym{\}}
    }

    \inferrule{
    \vdash^\downarrow  \Gamma \\ \Gamma  \vdash^\downarrow  \ottnt{S}
    }{
      \vdash^\downarrow  \Gamma  \ottsym{,}  \mathit{x}  \ottsym{:}  \ottnt{S}
    }

    \inferrule{
      \Gamma  \vdash^\downarrow  \omega
    }{
      \vdash^\downarrow  \Gamma  \Vdash  \omega
    }

    \inferrule{
      \vdash^\downarrow  \Gamma
    }{
      \Gamma  \vdash^\downarrow   \mathsf{unit} 
    }

    \inferrule{
    \vdash^\downarrow  \Gamma \\  \ottmv{a}  \in  \Gamma 
    }{
      \Gamma  \vdash^\downarrow  \ottmv{a}
    }

    \inferrule{
    \vdash^\downarrow  \Gamma \\  \monoalpha  \in  \Gamma 
    }{
      \Gamma  \vdash^\downarrow  \monoalpha
    }

    \inferrule{
      \vdash^\downarrow  \Gamma_{{\mathrm{1}}}  \ottsym{,}  \triangleright  \ottsym{\{}  \vec{\polybeta}_{{\mathrm{1}}}  \ottsym{,}  \polyalpha  \ottsym{,}  \vec{\polybeta}_{{\mathrm{2}}}  \ottsym{\}}  \ottsym{,}  \Delta_{{\mathrm{2}}}
    }{
      \Gamma_{{\mathrm{1}}}  \ottsym{,}  \triangleright  \ottsym{\{}  \vec{\polybeta}_{{\mathrm{1}}}  \ottsym{,}  \polyalpha  \ottsym{,}  \vec{\polybeta}_{{\mathrm{2}}}  \ottsym{\}}  \ottsym{,}  \Delta_{{\mathrm{2}}}  \vdash^\downarrow  \polyalpha
    }

    \inferrule{
      \vdash^\downarrow  \Gamma_{{\mathrm{1}}}  \ottsym{,}  \triangleright  \ottsym{\{}  \vec{\polybeta}_{{\mathrm{1}}}  \ottsym{,}  \polyalpha  \ottsym{,}  \vec{\polybeta}_{{\mathrm{2}}}  \ottsym{\}}  \ottsym{,}  \Delta_{{\mathrm{2}}}
    }{
      \Gamma  \vdash^\downarrow   \boxed{ \polyalpha } 
    }

    \inferrule{
    \Gamma  \vdash^\downarrow  \ottnt{A} \\ \Gamma  \vdash^\downarrow  {B^\Box}
    }{
      \Gamma  \vdash^\downarrow  \ottnt{A}  \rightarrow  {B^\Box}
    }

    \inferrule{
    \Gamma  \vdash^\downarrow   \boxed{ \polyalpha }  \\ \Gamma  \vdash^\downarrow  {B^\Box}
    }{
      \Gamma  \vdash^\downarrow   \boxed{ \polyalpha }   \rightarrow  {B^\Box}
    }

    \inferrule{
      \Gamma  \ottsym{,}  \ottmv{a}  \vdash^\downarrow  \ottnt{A}
    }{
      \Gamma  \vdash^\downarrow   \forall  \ottmv{a}  .  \ottnt{A} 
    }

    \inferrule{
      \vdash^\downarrow  \Gamma
    }{
      \Gamma  \vdash^\downarrow  \ottsym{()}
    }

    \inferrule{
    \vdash^\downarrow  \Gamma \\  \mathit{x}  \ottsym{:}  \ottnt{A}  \in  \Gamma 
    }{
      \Gamma  \vdash^\downarrow  \mathit{x}
    }

    \inferrule{
      \Gamma  \ottsym{,}  \mathit{x}  \ottsym{:}  \ottnt{A}  \vdash^\downarrow  \ottnt{e}
    }{
      \Gamma  \vdash^\downarrow   \lambda  \mathit{x}  .  \ottnt{e} 
    }

    \inferrule{
      \Gamma  \ottsym{,}  \mathit{x}  \ottsym{:}  \ottnt{A}  \vdash^\downarrow  \ottnt{e}
    }{
      \Gamma  \vdash^\downarrow   \lambda  \mathit{x}  \ottsym{:}  \ottnt{A}  .  \ottnt{e} 
    }

    \inferrule{
    \Gamma  \vdash^\downarrow  \ottnt{e_{{\mathrm{1}}}} \\ \Gamma  \vdash^\downarrow  \ottnt{e_{{\mathrm{2}}}}
    }{
      \Gamma  \vdash^\downarrow  \ottnt{e_{{\mathrm{1}}}}  \appright  \ottnt{e_{{\mathrm{2}}}}
    }

    \inferrule{
    \Gamma  \vdash^\downarrow  \ottnt{e_{{\mathrm{1}}}} \\ \Gamma  \vdash^\downarrow  \ottnt{e_{{\mathrm{2}}}}
    }{
      \Gamma  \vdash^\downarrow  \ottnt{e_{{\mathrm{1}}}}  \appleft  \ottnt{e_{{\mathrm{2}}}}
    }

    \inferrule{
    \Gamma  \vdash^\downarrow  \ottnt{A} \\ \Gamma  \vdash^\downarrow  {A^\Box}
    }{
      \Gamma  \vdash^\downarrow   \subL{ \ottnt{A} }{ {A^\Box} } 
    }

    \inferrule{
    \Gamma  \ottsym{,}  \triangleright  \ottsym{\{}  \vec{\polyalpha}  \ottsym{\}}  \ottsym{,}  \Delta  \vdash^\downarrow  \ottnt{A} \\ \Gamma  \ottsym{,}  \triangleright  \ottsym{\{}  \vec{\polyalpha}  \ottsym{\}}  \ottsym{,}  \Delta  \vdash^\downarrow  \ottnt{B}
    }{
      \Gamma  \ottsym{,}  \triangleright  \ottsym{\{}  \vec{\polyalpha}  \ottsym{\}}  \ottsym{,}  \Delta  \vdash^\downarrow   \subR{ \ottnt{A} }{ \ottnt{B} } 
    }

    \inferrule{
    \Gamma  \ottsym{,}  \triangleright  \ottsym{\{}  \vec{\polyalpha}  \ottsym{\}}  \ottsym{,}  \Delta  \vdash^\downarrow   \boxed{ \polyalpha }  \\ \Gamma  \ottsym{,}  \triangleright  \ottsym{\{}  \vec{\polyalpha}  \ottsym{\}}  \ottsym{,}  \Delta  \vdash^\downarrow  \ottnt{B}
    }{
      \Gamma  \ottsym{,}  \triangleright  \ottsym{\{}  \vec{\polyalpha}  \ottsym{\}}  \ottsym{,}  \Delta  \vdash^\downarrow   \subR{  \boxed{ \polyalpha }  }{ \ottnt{B} } 
    }

    \inferrule{
    \Gamma  \ottsym{,}  \triangleright  \ottsym{\{}  \vec{\polyalpha}  \ottsym{\}}  \ottsym{,}  \Delta  \vdash^\downarrow  \ottnt{e} \\ \Gamma  \ottsym{,}  \triangleright  \ottsym{\{}  \vec{\polyalpha}  \ottsym{\}}  \ottsym{,}  \Delta  \vdash^\downarrow  {A^\Box}
    }{
      \Gamma  \ottsym{,}  \triangleright  \ottsym{\{}  \vec{\polyalpha}  \ottsym{\}}  \ottsym{,}  \Delta  \vdash^\downarrow  \ottnt{e}  \ottsym{:}  {A^\Box}
    }
  \end{mathpar}
  \caption{Limited well-formedness for the algorithmic system}
\end{figure}

\begin{figure}[H]
  \small
  \noindent
  \fbox{$\ottsym{\{}  \vec{\polyalpha}  \ottsym{\}}  \vdash  \Gamma  \dashv  \ottsym{\{}  \vec{\polybeta}  \ottsym{\}}$}\quad
  \fbox{$\Gamma  \mid  \ottsym{\{}  \vec{\polyalpha}  \ottsym{\}}  \vdash  {A^\Box}  \dashv  \ottsym{\{}  \vec{\polybeta}  \ottsym{\}}$}\quad
  \fbox{$\Gamma  \mid  \ottsym{\{}  \vec{\polyalpha}  \ottsym{\}}  \vdash  \omega  \dashv  \ottsym{\{}  \vec{\polybeta}  \ottsym{\}}$}\hfill\mbox{}
  \begin{mathpar}
    \inferrule{ }{
      \ottsym{\{}  \vec{\polyalpha}  \ottsym{\}}  \vdash   \emptyset   \dashv  \ottsym{\{}  \vec{\polyalpha}  \ottsym{\}}
    }

    \inferrule{
      \ottsym{\{}  \vec{\polyalpha}  \ottsym{\}}  \vdash  \Gamma  \dashv  \ottsym{\{}  \vec{\polybeta}  \ottsym{\}}
    }{
      \ottsym{\{}  \vec{\polyalpha}  \ottsym{\}}  \vdash  \Gamma  \ottsym{,}  \ottmv{a}  \dashv  \ottsym{\{}  \vec{\polybeta}  \ottsym{\}}
    }

    \inferrule{
      \ottsym{\{}  \vec{\polyalpha}  \ottsym{\}}  \vdash  \Gamma  \dashv  \ottsym{\{}  \vec{\polybeta}  \ottsym{\}}
    }{
      \ottsym{\{}  \vec{\polyalpha}  \ottsym{\}}  \vdash  \Gamma  \ottsym{,}  \monoalpha  \dashv  \ottsym{\{}  \vec{\polybeta}  \ottsym{\}}
    }

    \inferrule{
      \ottsym{\{}  \vec{\polyalpha}  \ottsym{\}}  \vdash  \Gamma  \dashv  \ottsym{\{}  \vec{\polybeta}  \ottsym{\}}
    }{
      \ottsym{\{}  \vec{\polyalpha}  \ottsym{\}}  \vdash  \Gamma  \ottsym{,}  \triangleright  \ottsym{\{}  \vec{\polygamma}  \ottsym{\}}  \dashv  \ottsym{\{}  \vec{\polybeta}  \ottsym{\}}
    }

    \inferrule{
      \ottsym{\{}  \vec{\polyalpha}  \ottsym{\}}  \vdash  \Gamma  \dashv  \ottsym{\{}  \vec{\polybeta}  \ottsym{\}}
    }{
      \ottsym{\{}  \vec{\polyalpha}  \ottsym{\}}  \vdash  \Gamma  \ottsym{,}  \mathit{x}  \ottsym{:}  \ottnt{A}  \dashv  \ottsym{\{}  \vec{\polybeta}  \ottsym{\}}
    }

    \inferrule{
    \Gamma  \mid  \ottsym{\{}  \vec{\polyalpha}  \ottsym{\}}  \vdash  \omega  \dashv  \ottsym{\{}  \vec{\polygamma}  \ottsym{\}} \\ \ottsym{\{}  \vec{\polygamma}  \ottsym{\}}  \vdash  \Gamma  \dashv  \ottsym{\{}  \vec{\polybeta}  \ottsym{\}}
    }{
      \ottsym{\{}  \vec{\polyalpha}  \ottsym{\}}  \vdash  \Gamma  \Vdash  \omega  \dashv  \ottsym{\{}  \vec{\polybeta}  \ottsym{\}}
    }

    \inferrule{ }{
      \Gamma  \mid  \ottsym{\{}  \vec{\polyalpha}  \ottsym{\}}  \vdash   \mathsf{unit}   \dashv  \ottsym{\{}  \vec{\polyalpha}  \ottsym{\}}
    }

    \inferrule{
       \ottmv{a}  \in  \Gamma 
    }{
      \Gamma  \mid  \ottsym{\{}  \vec{\polyalpha}  \ottsym{\}}  \vdash  \ottmv{a}  \dashv  \ottsym{\{}  \vec{\polyalpha}  \ottsym{\}}
    }

    \inferrule{
       \monoalpha  \in  \Gamma 
    }{
      \Gamma  \mid  \ottsym{\{}  \vec{\polyalpha}  \ottsym{\}}  \vdash  \monoalpha  \dashv  \ottsym{\{}  \vec{\polyalpha}  \ottsym{\}}
    }

    \inferrule{
       \triangleright  \ottsym{\{}  \vec{\polygamma}_{{\mathrm{1}}}  \ottsym{,}  \polyalpha  \ottsym{,}  \vec{\polygamma}_{{\mathrm{2}}}  \ottsym{\}}  \in  \Gamma 
    }{
      \Gamma  \mid  \ottsym{\{}  \vec{\polybeta}_{{\mathrm{1}}}  \ottsym{,}  \polyalpha  \ottsym{,}  \vec{\polybeta}_{{\mathrm{2}}}  \ottsym{\}}  \vdash  \polyalpha  \dashv  \ottsym{\{}  \vec{\polybeta}_{{\mathrm{1}}}  \ottsym{,}  \polyalpha  \ottsym{,}  \vec{\polybeta}_{{\mathrm{2}}}  \ottsym{\}}
    }

    \inferrule{
     \polyalpha  \notin  \ottsym{\{}  \vec{\polybeta}  \ottsym{\}}  \\  \triangleright  \ottsym{\{}  \vec{\polygamma}_{{\mathrm{1}}}  \ottsym{,}  \polyalpha  \ottsym{,}  \vec{\polygamma}_{{\mathrm{2}}}  \ottsym{\}}  \in  \Gamma 
    }{
      \Gamma  \mid  \ottsym{\{}  \vec{\polybeta}  \ottsym{\}}  \vdash   \boxed{ \polyalpha }   \dashv  \ottsym{\{}  \vec{\polybeta}  \ottsym{,}  \polyalpha  \ottsym{\}}
    }

    \inferrule{
    \Gamma  \mid  \ottsym{\{}  \vec{\polyalpha}  \ottsym{\}}  \vdash  {A^\Box}  \dashv  \ottsym{\{}  \vec{\polygamma}  \ottsym{\}} \\ \Gamma  \mid  \ottsym{\{}  \vec{\polygamma}  \ottsym{\}}  \vdash  {B^\Box}  \dashv  \ottsym{\{}  \vec{\polybeta}  \ottsym{\}}
    }{
      \Gamma  \mid  \ottsym{\{}  \vec{\polyalpha}  \ottsym{\}}  \vdash  {A^\Box}  \rightarrow  {B^\Box}  \dashv  \ottsym{\{}  \vec{\polybeta}  \ottsym{\}}
    }

    \inferrule{ }{
      \Gamma  \mid  \ottsym{\{}  \vec{\polyalpha}  \ottsym{\}}  \vdash   \forall  \ottmv{a}  .  \ottnt{A}   \dashv  \ottsym{\{}  \vec{\polyalpha}  \ottsym{\}}
    }

    \inferrule{
      \Gamma  \mid  \ottsym{\{}  \vec{\polyalpha}  \ottsym{\}}  \vdash  {B^\Box}  \dashv  \ottsym{\{}  \vec{\polybeta}  \ottsym{\}}
    }{
      \Gamma  \mid  \ottsym{\{}  \vec{\polyalpha}  \ottsym{\}}  \vdash   \subL{ \ottnt{A} }{ {B^\Box} }   \dashv  \ottsym{\{}  \vec{\polybeta}  \ottsym{\}}
    }

    \inferrule{
      \Gamma  \mid  \ottsym{\{}  \vec{\polyalpha}  \ottsym{\}}  \vdash  {A^\Box}  \dashv  \ottsym{\{}  \vec{\polybeta}  \ottsym{\}}
    }{
      \Gamma  \mid  \ottsym{\{}  \vec{\polyalpha}  \ottsym{\}}  \vdash   \subR{ {A^\Box} }{ \ottnt{B} }   \dashv  \ottsym{\{}  \vec{\polybeta}  \ottsym{\}}
    }

    \inferrule{
      \Gamma  \mid  \ottsym{\{}  \vec{\polyalpha}  \ottsym{\}}  \vdash  {A^\Box}  \dashv  \ottsym{\{}  \vec{\polybeta}  \ottsym{\}}
    }{
      \Gamma  \mid  \ottsym{\{}  \vec{\polyalpha}  \ottsym{\}}  \vdash  \ottnt{e}  \ottsym{:}  {A^\Box}  \dashv  \ottsym{\{}  \vec{\polybeta}  \ottsym{\}}
    }
  \end{mathpar}
  \caption{Polymorphic existential variable counter}
\end{figure}
}

\subsection{\texorpdfstring{\sysfi}{SystemFi}~\cite{cretin_power_2012}}

\deftitle{Syntax}{}

{
  \small
  \[
    \begin{array}{rrcl}
      \textbf{Expressions} & \ottnt{e} &  \Coloneqq  &
      \ottsym{()}  \mid  \mathit{x}  \mid   \lambda  \mathit{x}  \ottsym{:}  \ottnt{A}  .  \ottnt{e}   \mid  \ottnt{e_{{\mathrm{1}}}} \, \ottnt{e_{{\mathrm{2}}}}  \mid 
       \ottnt{C}  \langle  \ottnt{e}  \rangle   \mid   \Lambda  \ottmv{a}  .  \ottnt{e}                         \\
      \textbf{Coercions}   & \ottnt{C} &  \Coloneqq  &
       \Diamond^{ \ottnt{A} }   \mid   \ottnt{C_{{\mathrm{1}}}}  \rightarrow^{ \ottnt{A} }  \ottnt{C_{{\mathrm{2}}}}   \mid 
       \Lambda  \ottmv{a}  .  \ottnt{C}   \mid  \ottnt{C} \, \ottnt{A}  \mid   \ottnt{C_{{\mathrm{1}}}}  \langle  \ottnt{C_{{\mathrm{2}}}}  \rangle  \\
    \end{array}
  \]
}

\mycomment{
  \begin{figure}[H]
    \small
    \[
      \begin{array}{rrcl}
        \textbf{Expressions} & \ottnt{e} &  \Coloneqq  &
        \ottsym{()}  \mid  \mathit{x}  \mid   \lambda  \mathit{x}  \ottsym{:}  \ottnt{A}  .  \ottnt{e}   \mid  \ottnt{e_{{\mathrm{1}}}} \, \ottnt{e_{{\mathrm{2}}}}  \mid 
         \ottnt{C}  \langle  \ottnt{e}  \rangle   \mid   \Lambda  \ottmv{a}  .  \ottnt{e}                         \\
        \textbf{Coercions}   & \ottnt{C} &  \Coloneqq  &
         \Diamond^{ \ottnt{A} }   \mid   \ottnt{C_{{\mathrm{1}}}}  \rightarrow^{ \ottnt{A} }  \ottnt{C_{{\mathrm{2}}}}   \mid 
         \Lambda  \ottmv{a}  .  \ottnt{C}   \mid  \ottnt{C} \, \ottnt{A}  \mid   \ottnt{C_{{\mathrm{1}}}}  \langle  \ottnt{C_{{\mathrm{2}}}}  \rangle  \\
      \end{array}
    \]
    \caption{Syntax of \sysfi}
  \end{figure}
}

\deftitle{Typing Rules}{\fbox{$\Psi  \vdash_{ \mathrm{F}_{\iota} }  \ottnt{e}  \ottsym{:}  \ottnt{A}$}}

\begin{mathparpagebreakable}
  \small
  \inferrule{
    \Psi  \vdash   \mathsf{unit} 
  }{
    \Psi  \vdash_{ \mathrm{F}_{\iota} }  \ottsym{()}  \ottsym{:}   \mathsf{unit} 
  }

  \inferrule{
   \mathit{x}  \ottsym{:}  \ottnt{A}  \in  \Psi  \\ \vdash  \Psi
  }{
    \Psi  \vdash_{ \mathrm{F}_{\iota} }  \mathit{x}  \ottsym{:}  \ottnt{A}
  }

  \inferrule{
    \Psi  \ottsym{,}  \mathit{x}  \ottsym{:}  \ottnt{A}  \vdash_{ \mathrm{F}_{\iota} }  \ottnt{e}  \ottsym{:}  \ottnt{B}
  }{
    \Psi  \vdash_{ \mathrm{F}_{\iota} }   \lambda  \mathit{x}  \ottsym{:}  \ottnt{A}  .  \ottnt{e}   \ottsym{:}   \ottnt{A}   \rightarrow   \ottnt{B} 
  }

  \inferrule{
  \Psi  \vdash_{ \mathrm{F}_{\iota} }  \ottnt{e_{{\mathrm{1}}}}  \ottsym{:}   \ottnt{A}   \rightarrow   \ottnt{B}  \\ \Psi  \vdash_{ \mathrm{F}_{\iota} }  \ottnt{e_{{\mathrm{2}}}}  \ottsym{:}  \ottnt{A}
  }{
    \Psi  \vdash_{ \mathrm{F}_{\iota} }  \ottnt{e_{{\mathrm{1}}}} \, \ottnt{e_{{\mathrm{2}}}}  \ottsym{:}  \ottnt{B}
  }

  \inferrule{
    \Psi  \ottsym{,}  \ottmv{a}  \vdash_{ \mathrm{F}_{\iota} }  \ottnt{e}  \ottsym{:}  \ottnt{A}
  }{
    \Psi  \vdash_{ \mathrm{F}_{\iota} }   \Lambda  \ottmv{a}  .  \ottnt{e}   \ottsym{:}   \forall  \ottmv{a}  .  \ottnt{A} 
  }

  \inferrule{
  \Psi  \vdash_{ \mathrm{F}_{\iota} }  \ottnt{C}  \ottsym{:}  \ottnt{A}  \triangleright  \ottnt{B} \\ \Psi  \vdash_{ \mathrm{F}_{\iota} }  \ottnt{e}  \ottsym{:}  \ottnt{A}
  }{
    \Psi  \vdash_{ \mathrm{F}_{\iota} }   \ottnt{C}  \langle  \ottnt{e}  \rangle   \ottsym{:}  \ottnt{B}
  }
\end{mathparpagebreakable}

\mycomment{
\begin{figure}[H]
  \small
  \noindent\fbox{$\Psi  \vdash_{ \mathrm{F}_{\iota} }  \ottnt{e}  \ottsym{:}  \ottnt{A}$}\hfill\mbox{}
  \begin{mathpar}
    \inferrule{
      \Psi  \vdash   \mathsf{unit} 
    }{
      \Psi  \vdash_{ \mathrm{F}_{\iota} }  \ottsym{()}  \ottsym{:}   \mathsf{unit} 
    }

    \inferrule{
     \mathit{x}  \ottsym{:}  \ottnt{A}  \in  \Psi  \\ \vdash  \Psi
    }{
      \Psi  \vdash_{ \mathrm{F}_{\iota} }  \mathit{x}  \ottsym{:}  \ottnt{A}
    }

    \inferrule{
      \Psi  \ottsym{,}  \mathit{x}  \ottsym{:}  \ottnt{A}  \vdash_{ \mathrm{F}_{\iota} }  \ottnt{e}  \ottsym{:}  \ottnt{B}
    }{
      \Psi  \vdash_{ \mathrm{F}_{\iota} }   \lambda  \mathit{x}  \ottsym{:}  \ottnt{A}  .  \ottnt{e}   \ottsym{:}   \ottnt{A}   \rightarrow   \ottnt{B} 
    }

    \inferrule{
    \Psi  \vdash_{ \mathrm{F}_{\iota} }  \ottnt{e_{{\mathrm{1}}}}  \ottsym{:}   \ottnt{A}   \rightarrow   \ottnt{B}  \\ \Psi  \vdash_{ \mathrm{F}_{\iota} }  \ottnt{e_{{\mathrm{2}}}}  \ottsym{:}  \ottnt{A}
    }{
      \Psi  \vdash_{ \mathrm{F}_{\iota} }  \ottnt{e_{{\mathrm{1}}}} \, \ottnt{e_{{\mathrm{2}}}}  \ottsym{:}  \ottnt{B}
    }

    \inferrule{
      \Psi  \ottsym{,}  \ottmv{a}  \vdash_{ \mathrm{F}_{\iota} }  \ottnt{e}  \ottsym{:}  \ottnt{A}
    }{
      \Psi  \vdash_{ \mathrm{F}_{\iota} }   \Lambda  \ottmv{a}  .  \ottnt{e}   \ottsym{:}   \forall  \ottmv{a}  .  \ottnt{A} 
    }

    \inferrule{
    \Psi  \vdash_{ \mathrm{F}_{\iota} }  \ottnt{C}  \ottsym{:}  \ottnt{A}  \triangleright  \ottnt{B} \\ \Psi  \vdash_{ \mathrm{F}_{\iota} }  \ottnt{e}  \ottsym{:}  \ottnt{A}
    }{
      \Psi  \vdash_{ \mathrm{F}_{\iota} }   \ottnt{C}  \langle  \ottnt{e}  \rangle   \ottsym{:}  \ottnt{B}
    }
  \end{mathpar}
  \caption{Typing of \sysfi}
\end{figure}
}

\deftitle{Coercion Typing Rules}{\fbox{$\Psi  \vdash_{ \mathrm{F}_{\iota} }  \ottnt{C}  \ottsym{:}  \ottnt{A}  \triangleright  \ottnt{B}$}}

\begin{mathparpagebreakable}
  \small
  \inferrule{
    \Psi  \vdash  \ottnt{A}
  }{
    \Psi  \vdash_{ \mathrm{F}_{\iota} }   \Diamond^{ \ottnt{A} }   \ottsym{:}  \ottnt{A}  \triangleright  \ottnt{A}
  }

  \inferrule{
  \Psi  \vdash_{ \mathrm{F}_{\iota} }  \ottnt{C_{{\mathrm{1}}}}  \ottsym{:}  \ottnt{A_{{\mathrm{2}}}}  \triangleright  \ottnt{A_{{\mathrm{1}}}} \\ \Psi  \vdash_{ \mathrm{F}_{\iota} }  \ottnt{C_{{\mathrm{2}}}}  \ottsym{:}  \ottnt{B_{{\mathrm{1}}}}  \triangleright  \ottnt{B_{{\mathrm{2}}}}
  }{
    \Psi  \vdash_{ \mathrm{F}_{\iota} }   \ottnt{C_{{\mathrm{1}}}}  \rightarrow^{ \ottnt{A_{{\mathrm{2}}}} }  \ottnt{C_{{\mathrm{2}}}}   \ottsym{:}   \ottnt{A_{{\mathrm{1}}}}   \rightarrow   \ottnt{B_{{\mathrm{1}}}}   \triangleright   \ottnt{A_{{\mathrm{2}}}}   \rightarrow   \ottnt{B_{{\mathrm{2}}}} 
  }

  \inferrule{
  \Psi  \vdash  \ottnt{A} \\ \Psi  \ottsym{,}  \ottmv{a}  \vdash_{ \mathrm{F}_{\iota} }  \ottnt{C}  \ottsym{:}  \ottnt{A}  \triangleright  \ottnt{B}
  }{
    \Psi  \vdash_{ \mathrm{F}_{\iota} }   \Lambda  \ottmv{a}  .  \ottnt{C}   \ottsym{:}  \ottnt{A}  \triangleright   \forall  \ottmv{a}  .  \ottnt{B} 
  }

  \inferrule{
  \Psi  \vdash_{ \mathrm{F}_{\iota} }  \ottnt{C}  \ottsym{:}  \ottnt{A'}  \triangleright   \forall  \ottmv{a}  .  \ottnt{B}  \\ \Psi  \vdash  \ottnt{A}
  }{
    \Psi  \vdash_{ \mathrm{F}_{\iota} }  \ottnt{C} \, \ottnt{A}  \ottsym{:}  \ottnt{A'}  \triangleright   [  \ottnt{A}  /  \ottmv{a}  ]  \ottnt{B} 
  }

  \inferrule{
  \Psi  \vdash_{ \mathrm{F}_{\iota} }  \ottnt{C_{{\mathrm{1}}}}  \ottsym{:}  \ottnt{A_{{\mathrm{2}}}}  \triangleright  \ottnt{A_{{\mathrm{3}}}} \\ \Psi  \vdash_{ \mathrm{F}_{\iota} }  \ottnt{C_{{\mathrm{2}}}}  \ottsym{:}  \ottnt{A_{{\mathrm{1}}}}  \triangleright  \ottnt{A_{{\mathrm{2}}}}
  }{
    \Psi  \vdash_{ \mathrm{F}_{\iota} }   \ottnt{C_{{\mathrm{1}}}}  \langle  \ottnt{C_{{\mathrm{2}}}}  \rangle   \ottsym{:}  \ottnt{A_{{\mathrm{1}}}}  \triangleright  \ottnt{A_{{\mathrm{3}}}}
  }
\end{mathparpagebreakable}

\mycomment{
\begin{figure}[H]
  \small
  \noindent\fbox{$\Psi  \vdash_{ \mathrm{F}_{\iota} }  \ottnt{C}  \ottsym{:}  \ottnt{A}  \triangleright  \ottnt{B}$}\hfill\mbox{}
  \begin{mathpar}
    \inferrule{
      \Psi  \vdash  \ottnt{A}
    }{
      \Psi  \vdash_{ \mathrm{F}_{\iota} }   \Diamond^{ \ottnt{A} }   \ottsym{:}  \ottnt{A}  \triangleright  \ottnt{A}
    }

    \inferrule{
    \Psi  \vdash_{ \mathrm{F}_{\iota} }  \ottnt{C_{{\mathrm{1}}}}  \ottsym{:}  \ottnt{A_{{\mathrm{2}}}}  \triangleright  \ottnt{A_{{\mathrm{1}}}} \\ \Psi  \vdash_{ \mathrm{F}_{\iota} }  \ottnt{C_{{\mathrm{2}}}}  \ottsym{:}  \ottnt{B_{{\mathrm{1}}}}  \triangleright  \ottnt{B_{{\mathrm{2}}}}
    }{
      \Psi  \vdash_{ \mathrm{F}_{\iota} }   \ottnt{C_{{\mathrm{1}}}}  \rightarrow^{ \ottnt{A_{{\mathrm{2}}}} }  \ottnt{C_{{\mathrm{2}}}}   \ottsym{:}   \ottnt{A_{{\mathrm{1}}}}   \rightarrow   \ottnt{B_{{\mathrm{1}}}}   \triangleright   \ottnt{A_{{\mathrm{2}}}}   \rightarrow   \ottnt{B_{{\mathrm{2}}}} 
    }

    \inferrule{
    \Psi  \vdash  \ottnt{A} \\ \Psi  \ottsym{,}  \ottmv{a}  \vdash_{ \mathrm{F}_{\iota} }  \ottnt{C}  \ottsym{:}  \ottnt{A}  \triangleright  \ottnt{B}
    }{
      \Psi  \vdash_{ \mathrm{F}_{\iota} }   \Lambda  \ottmv{a}  .  \ottnt{C}   \ottsym{:}  \ottnt{A}  \triangleright   \forall  \ottmv{a}  .  \ottnt{B} 
    }

    \inferrule{
    \Psi  \vdash_{ \mathrm{F}_{\iota} }  \ottnt{C}  \ottsym{:}  \ottnt{A'}  \triangleright   \forall  \ottmv{a}  .  \ottnt{B}  \\ \Psi  \vdash  \ottnt{A}
    }{
      \Psi  \vdash_{ \mathrm{F}_{\iota} }  \ottnt{C} \, \ottnt{A}  \ottsym{:}  \ottnt{A'}  \triangleright   [  \ottnt{A}  /  \ottmv{a}  ]  \ottnt{B} 
    }

    \inferrule{
    \Psi  \vdash_{ \mathrm{F}_{\iota} }  \ottnt{C_{{\mathrm{1}}}}  \ottsym{:}  \ottnt{A_{{\mathrm{2}}}}  \triangleright  \ottnt{A_{{\mathrm{3}}}} \\ \Psi  \vdash_{ \mathrm{F}_{\iota} }  \ottnt{C_{{\mathrm{2}}}}  \ottsym{:}  \ottnt{A_{{\mathrm{1}}}}  \triangleright  \ottnt{A_{{\mathrm{2}}}}
    }{
      \Psi  \vdash_{ \mathrm{F}_{\iota} }   \ottnt{C_{{\mathrm{1}}}}  \langle  \ottnt{C_{{\mathrm{2}}}}  \rangle   \ottsym{:}  \ottnt{A_{{\mathrm{1}}}}  \triangleright  \ottnt{A_{{\mathrm{3}}}}
    }
  \end{mathpar}
  \caption{Coercion typing of \sysfi}
\end{figure}
}

\deftitle{Translation Rules in Soundness w.r.t.\ {\sysfi}}{
  \fbox{$ |  {A^\Box}  | _ \square $}\quad
  \fbox{$ |  \ottnt{e}  | $}\quad
  \fbox{$ |  \ottnt{e}  |_{ \mathrm{Coer} } $}
}

{
  \small
  \[
    \begin{array}{r@{\ }c@{\ }lr@{\ }c@{\ }lr@{\ }c@{\ }l}
                  |   \boxed{ \ottnt{A} }   | _ \square  & = & \ottnt{A}            &
                  |   \mathsf{unit}   | _ \square                                 & = &  \mathsf{unit}          &
                  |  \ottmv{a}  | _ \square                                    & = & \ottmv{a}              \\
                  |  {A^\Box}  \rightarrow  {B^\Box}  | _ \square                               & = &   |  {A^\Box}  | _ \square    \rightarrow    |  {B^\Box}  | _ \square     &
                  |   \forall  \ottmv{a}  .  \ottnt{A}   | _ \square                         & = &  \forall  \ottmv{a}  .  \ottnt{A}  &
                                                           &   &
    \end{array}
  \]
  \[
    \begin{array}{r@{\ }c@{\ }lr@{\ }c@{\ }lr@{\ }c@{\ }l}
                  |  \ottsym{()}  |  & = & \ottsym{()}        &
                  |  \mathit{x}  |                                    & = & \mathit{x}         &
                  |   \lambda  \mathit{x}  .  \ottnt{e}   |                                 & = &  \lambda  \mathit{x}  .   |  \ottnt{e}  |       \\
                  |   \lambda  \mathit{x}  \ottsym{:}  \ottnt{A}  .  \ottnt{e}   |                               & = &  \lambda  \mathit{x}  .   |  \ottnt{e}  |     &
                  |  \ottnt{e_{{\mathrm{1}}}}  \appright  \ottnt{e_{{\mathrm{2}}}}  |                             & = &  |  \ottnt{e_{{\mathrm{1}}}}  |  \,  |  \ottnt{e_{{\mathrm{2}}}}  |  &
                  |  \ottnt{e_{{\mathrm{1}}}}  \appleft  \ottnt{e_{{\mathrm{2}}}}  |                             & = &  |  \ottnt{e_{{\mathrm{1}}}}  |  \,  |  \ottnt{e_{{\mathrm{2}}}}  | 
    \end{array}
  \]
  \[
    \begin{array}{r@{\ }c@{\ }lr@{\ }c@{\ }lr@{\ }c@{\ }l}
                  |  \ottsym{()}  |_{ \mathrm{Coer} }                              & = & \ottsym{()}                 &
                  |  \mathit{x}  |_{ \mathrm{Coer} }                                                          & = & \mathit{x}                  &
                  |   \lambda  \mathit{x}  \ottsym{:}  \ottnt{A}  .  \ottnt{e}   |_{ \mathrm{Coer} }                                                  & = &  \lambda  \mathit{x}  .   |  \ottnt{e}  |_{ \mathrm{Coer} }                        \\
                  |  \ottnt{e_{{\mathrm{1}}}} \, \ottnt{e_{{\mathrm{2}}}}  |_{ \mathrm{Coer} }                                                      & = &  |  \ottnt{e_{{\mathrm{1}}}}  |_{ \mathrm{Coer} }  \,  |  \ottnt{e_{{\mathrm{2}}}}  |_{ \mathrm{Coer} }  &
                  |   \Lambda  \ottmv{a}  .  \ottnt{e}   |_{ \mathrm{Coer} }  & = &  |  \ottnt{e}  |_{ \mathrm{Coer} }           &
                  |   \ottnt{C}  \langle  \ottnt{e}  \rangle   |_{ \mathrm{Coer} }                                                      & = &  |  \ottnt{e}  |_{ \mathrm{Coer} } 
    \end{array}
  \]
}

\mycomment{
  \begin{figure}[H]
    \small
    \noindent\fbox{$ |  {A^\Box}  | _ \square $}\hfill\mbox{}
    \[
      \begin{array}{r@{\ }c@{\ }lr@{\ }c@{\ }lr@{\ }c@{\ }l}
                    |   \boxed{ \ottnt{A} }   | _ \square  & = & \ottnt{A}            &
                    |   \mathsf{unit}   | _ \square                                 & = &  \mathsf{unit}          &
                    |  \ottmv{a}  | _ \square                                    & = & \ottmv{a}              \\
                    |  {A^\Box}  \rightarrow  {B^\Box}  | _ \square                               & = &   |  {A^\Box}  | _ \square    \rightarrow    |  {B^\Box}  | _ \square     &
                    |   \forall  \ottmv{a}  .  \ottnt{A}   | _ \square                         & = &  \forall  \ottmv{a}  .  \ottnt{A}  &
                                                             &   &
      \end{array}
    \]
    \noindent\fbox{$ |  \ottnt{e}  | $}\hfill\mbox{}
    \[
      \begin{array}{r@{\ }c@{\ }lr@{\ }c@{\ }lr@{\ }c@{\ }l}
                    |  \ottsym{()}  |  & = & \ottsym{()}        &
                    |  \mathit{x}  |                                    & = & \mathit{x}         &
                    |   \lambda  \mathit{x}  .  \ottnt{e}   |                                 & = &  \lambda  \mathit{x}  .   |  \ottnt{e}  |       \\
                    |   \lambda  \mathit{x}  \ottsym{:}  \ottnt{A}  .  \ottnt{e}   |                               & = &  \lambda  \mathit{x}  .   |  \ottnt{e}  |     &
                    |  \ottnt{e_{{\mathrm{1}}}}  \appright  \ottnt{e_{{\mathrm{2}}}}  |                             & = &  |  \ottnt{e_{{\mathrm{1}}}}  |  \,  |  \ottnt{e_{{\mathrm{2}}}}  |  &
                    |  \ottnt{e_{{\mathrm{1}}}}  \appleft  \ottnt{e_{{\mathrm{2}}}}  |                             & = &  |  \ottnt{e_{{\mathrm{1}}}}  |  \,  |  \ottnt{e_{{\mathrm{2}}}}  | 
      \end{array}
    \]
    \noindent\fbox{$ |  \ottnt{e}  |_{ \mathrm{Coer} } $}\hfill\mbox{}
    \[
      \begin{array}{r@{\ }c@{\ }lr@{\ }c@{\ }lr@{\ }c@{\ }l}
                    |  \ottsym{()}  |_{ \mathrm{Coer} }                              & = & \ottsym{()}                 &
                    |  \mathit{x}  |_{ \mathrm{Coer} }                                                          & = & \mathit{x}                  &
                    |   \lambda  \mathit{x}  \ottsym{:}  \ottnt{A}  .  \ottnt{e}   |_{ \mathrm{Coer} }                                                  & = &  \lambda  \mathit{x}  .   |  \ottnt{e}  |_{ \mathrm{Coer} }                        \\
                    |  \ottnt{e_{{\mathrm{1}}}} \, \ottnt{e_{{\mathrm{2}}}}  |_{ \mathrm{Coer} }                                                      & = &  |  \ottnt{e_{{\mathrm{1}}}}  |_{ \mathrm{Coer} }  \,  |  \ottnt{e_{{\mathrm{2}}}}  |_{ \mathrm{Coer} }  &
                    |   \Lambda  \ottmv{a}  .  \ottnt{e}   |_{ \mathrm{Coer} }  & = &  |  \ottnt{e}  |_{ \mathrm{Coer} }           &
                    |   \ottnt{C}  \langle  \ottnt{e}  \rangle   |_{ \mathrm{Coer} }                                                      & = &  |  \ottnt{e}  |_{ \mathrm{Coer} } 
      \end{array}
    \]
    \caption{Translations in soundness w.r.t.\ {\sysfi}}
  \end{figure}
}

\subsection{The DK System~\cite{dunfield_complete_2013}}

\deftitle{Subtyping Rules}{\fbox{$\Psi  \vdash_{ \mathrm{DK} }  \ottnt{A}  \mathbin{<:}  \ottnt{B}$}}

\begin{mathparpagebreakable}
  \small
  \inferrule{
    \Psi  \vdash   \mathsf{unit} 
  }{
    \Psi  \vdash_{ \mathrm{DK} }   \mathsf{unit}   \mathbin{<:}   \mathsf{unit} 
  }

  \inferrule{
    \Psi  \vdash  \ottmv{a}
  }{
    \Psi  \vdash_{ \mathrm{DK} }  \ottmv{a}  \mathbin{<:}  \ottmv{a}
  }

  \inferrule{
  \Psi  \vdash_{ \mathrm{DK} }  \ottnt{A_{{\mathrm{2}}}}  \mathbin{<:}  \ottnt{A_{{\mathrm{1}}}} \\ \Psi  \vdash_{ \mathrm{DK} }  \ottnt{B_{{\mathrm{1}}}}  \mathbin{<:}  \ottnt{B_{{\mathrm{2}}}}
  }{
    \Psi  \vdash_{ \mathrm{DK} }   \ottnt{A_{{\mathrm{1}}}}   \rightarrow   \ottnt{B_{{\mathrm{1}}}}   \mathbin{<:}   \ottnt{A_{{\mathrm{2}}}}   \rightarrow   \ottnt{B_{{\mathrm{2}}}} 
  }

  \inferrule{
  \Psi  \vdash  \tau \\ \Psi  \vdash_{ \mathrm{DK} }   [  \tau  /  \ottmv{a}  ]  \ottnt{A}   \mathbin{<:}  \ottnt{B}
  }{
    \Psi  \vdash_{ \mathrm{DK} }   \forall  \ottmv{a}  .  \ottnt{A}   \mathbin{<:}  \ottnt{B}
  }

  \inferrule{
    \Psi  \ottsym{,}  \ottmv{a}  \vdash_{ \mathrm{DK} }  \ottnt{A}  \mathbin{<:}  \ottnt{B}
  }{
    \Psi  \vdash_{ \mathrm{DK} }  \ottnt{A}  \mathbin{<:}   \forall  \ottmv{a}  .  \ottnt{B} 
  }
\end{mathparpagebreakable}

\deftitle{Typing Rules}{
  \fbox{$\Psi  \vdash_{ \mathrm{DK} }  \ottnt{e}  \Rightarrow  \ottnt{A}$}\quad
  \fbox{$\Psi  \vdash_{ \mathrm{DK} }  \ottnt{e}  \Leftarrow  \ottnt{A}$}\quad
  \fbox{$ \Psi   \vdash_{ \mathrm{DK} }   \ottnt{A}  \bullet  \ottnt{e}  \Rightarrow\!\!\!\Rightarrow  \ottnt{B} $}
}

\begin{mathparpagebreakable}
  \small
  \inferrule{
   \mathit{x}  \ottsym{:}  \ottnt{A}  \in  \Psi  \\ \vdash  \Psi
  }{
    \Psi  \vdash_{ \mathrm{DK} }  \mathit{x}  \Rightarrow  \ottnt{A}
  }

  \inferrule{
    \Psi  \vdash_{ \mathrm{DK} }  \ottnt{e}  \Leftarrow  \ottnt{A}
  }{
    \Psi  \vdash_{ \mathrm{DK} }  \ottnt{e}  \ottsym{:}  \ottnt{A}  \Rightarrow  \ottnt{A}
  }

  \inferrule{
    \vdash  \Psi
  }{
    \Psi  \vdash_{ \mathrm{DK} }  \ottsym{()}  \Rightarrow   \mathsf{unit} 
  }

  \inferrule{
    \vdash  \Psi
  }{
    \Psi  \vdash_{ \mathrm{DK} }  \ottsym{()}  \Leftarrow   \mathsf{unit} 
  }

  \inferrule{
    \Psi  \ottsym{,}  \ottmv{a}  \vdash_{ \mathrm{DK} }  \ottnt{e}  \Leftarrow  \ottnt{A}
  }{
    \Psi  \vdash_{ \mathrm{DK} }  \ottnt{e}  \Leftarrow   \forall  \ottmv{a}  .  \ottnt{A} 
  }

  \inferrule{
    \Psi  \ottsym{,}  \mathit{x}  \ottsym{:}  \ottnt{A}  \vdash_{ \mathrm{DK} }  \ottnt{e}  \Leftarrow  \ottnt{B}
  }{
    \Psi  \vdash_{ \mathrm{DK} }   \lambda  \mathit{x}  .  \ottnt{e}   \Leftarrow   \ottnt{A}   \rightarrow   \ottnt{B} 
  }

  \inferrule{
  \Psi  \vdash  \tau  \rightarrow  \sigma \\ \Psi  \ottsym{,}  \mathit{x}  \ottsym{:}  \tau  \vdash_{ \mathrm{DK} }  \ottnt{e}  \Leftarrow  \sigma
  }{
    \Psi  \vdash_{ \mathrm{DK} }   \lambda  \mathit{x}  .  \ottnt{e}   \Rightarrow   \tau   \rightarrow   \sigma 
  }

  \inferrule{
  \Psi  \vdash_{ \mathrm{DK} }  \ottnt{e_{{\mathrm{1}}}}  \Rightarrow  \ottnt{A} \\  \Psi   \vdash_{ \mathrm{DK} }   \ottnt{A}  \bullet  \ottnt{e_{{\mathrm{2}}}}  \Rightarrow\!\!\!\Rightarrow  \ottnt{B} 
  }{
    \Psi  \vdash_{ \mathrm{DK} }  \ottnt{e_{{\mathrm{1}}}} \, \ottnt{e_{{\mathrm{2}}}}  \Rightarrow  \ottnt{B}
  }

  \inferrule{
  \Psi  \vdash_{ \mathrm{DK} }  \ottnt{e}  \Rightarrow  \ottnt{A} \\ \Psi  \vdash_{ \mathrm{DK} }  \ottnt{A}  \mathbin{<:}  \ottnt{B}
  }{
    \Psi  \vdash_{ \mathrm{DK} }  \ottnt{e}  \Leftarrow  \ottnt{B}
  }

  \inferrule{
  \Psi  \vdash  \tau \\  \Psi   \vdash_{ \mathrm{DK} }    [  \tau  /  \ottmv{a}  ]  \ottnt{A}   \bullet  \ottnt{e}  \Rightarrow\!\!\!\Rightarrow  \ottnt{B} 
  }{
     \Psi   \vdash_{ \mathrm{DK} }    \forall  \ottmv{a}  .  \ottnt{A}   \bullet  \ottnt{e}  \Rightarrow\!\!\!\Rightarrow  \ottnt{B} 
  }

  \inferrule{
  \Psi  \vdash_{ \mathrm{DK} }  \ottnt{e}  \Leftarrow  \ottnt{A} \\ \Psi  \vdash  \ottnt{B}
  }{
     \Psi   \vdash_{ \mathrm{DK} }    \ottnt{A}   \rightarrow   \ottnt{B}   \bullet  \ottnt{e}  \Rightarrow\!\!\!\Rightarrow  \ottnt{B} 
  }
\end{mathparpagebreakable}

\deftitle{Translation from the DK system to {\lang}}{\fbox{$ |  \ottnt{e}  |_{ \mathrm{DK} } $}}

{
  \small
  \[
    \begin{array}{r@{\ }c@{\ }lr@{\ }c@{\ }lr@{\ }c@{\ }l}
                  |  \ottsym{()}  |_{ \mathrm{DK} }  & = & \ottsym{()}                &
                  |  \mathit{x}  |_{ \mathrm{DK} }                                  & = & \mathit{x}                 &
                  |   \lambda  \mathit{x}  .  \ottnt{e}   |_{ \mathrm{DK} }                               & = &  \lambda  \mathit{x}  .   |  \ottnt{e}  |_{ \mathrm{DK} }              \\
                  |  \ottnt{e_{{\mathrm{1}}}} \, \ottnt{e_{{\mathrm{2}}}}  |_{ \mathrm{DK} }                              & = &  |  \ottnt{e_{{\mathrm{1}}}}  |_{ \mathrm{DK} }   \appright   |  \ottnt{e_{{\mathrm{2}}}}  |_{ \mathrm{DK} }   &
                  |  \ottnt{e}  \ottsym{:}  \ottnt{A}  |_{ \mathrm{DK} }                              & = & \ottsym{(}   \lambda  \mathit{x}  \ottsym{:}  \ottnt{A}  .  \mathit{x}   \ottsym{)}  \appright   |  \ottnt{e}  |_{ \mathrm{DK} }  &
                                                           &   &
    \end{array}
  \]
}

\mycomment{
\begin{figure}[H]
  \small
  \noindent\fbox{$\Psi  \vdash_{ \mathrm{DK} }  \ottnt{A}  \mathbin{<:}  \ottnt{B}$}\hfill\mbox{}
  \begin{mathpar}
    \inferrule{
      \Psi  \vdash   \mathsf{unit} 
    }{
      \Psi  \vdash_{ \mathrm{DK} }   \mathsf{unit}   \mathbin{<:}   \mathsf{unit} 
    }

    \inferrule{
      \Psi  \vdash  \ottmv{a}
    }{
      \Psi  \vdash_{ \mathrm{DK} }  \ottmv{a}  \mathbin{<:}  \ottmv{a}
    }

    \inferrule{
    \Psi  \vdash_{ \mathrm{DK} }  \ottnt{A_{{\mathrm{2}}}}  \mathbin{<:}  \ottnt{A_{{\mathrm{1}}}} \\ \Psi  \vdash_{ \mathrm{DK} }  \ottnt{B_{{\mathrm{1}}}}  \mathbin{<:}  \ottnt{B_{{\mathrm{2}}}}
    }{
      \Psi  \vdash_{ \mathrm{DK} }   \ottnt{A_{{\mathrm{1}}}}   \rightarrow   \ottnt{B_{{\mathrm{1}}}}   \mathbin{<:}   \ottnt{A_{{\mathrm{2}}}}   \rightarrow   \ottnt{B_{{\mathrm{2}}}} 
    }

    \inferrule{
    \Psi  \vdash  \tau \\ \Psi  \vdash_{ \mathrm{DK} }   [  \tau  /  \ottmv{a}  ]  \ottnt{A}   \mathbin{<:}  \ottnt{B}
    }{
      \Psi  \vdash_{ \mathrm{DK} }   \forall  \ottmv{a}  .  \ottnt{A}   \mathbin{<:}  \ottnt{B}
    }

    \inferrule{
      \Psi  \ottsym{,}  \ottmv{a}  \vdash_{ \mathrm{DK} }  \ottnt{A}  \mathbin{<:}  \ottnt{B}
    }{
      \Psi  \vdash_{ \mathrm{DK} }  \ottnt{A}  \mathbin{<:}   \forall  \ottmv{a}  .  \ottnt{B} 
    }
  \end{mathpar}
  \caption{Subtyping of the DK system}
\end{figure}

\begin{figure}[H]
  \small
  \noindent\fbox{$\Psi  \vdash_{ \mathrm{DK} }  \ottnt{e}  \Rightarrow  \ottnt{A}$}\quad
  \fbox{$\Psi  \vdash_{ \mathrm{DK} }  \ottnt{e}  \Leftarrow  \ottnt{A}$}\quad
  \fbox{$ \Psi   \vdash_{ \mathrm{DK} }   \ottnt{A}  \bullet  \ottnt{e}  \Rightarrow\!\!\!\Rightarrow  \ottnt{B} $}\quad\hfill\mbox{}
  \begin{mathpar}
    \inferrule{
     \mathit{x}  \ottsym{:}  \ottnt{A}  \in  \Psi  \\ \vdash  \Psi
    }{
      \Psi  \vdash_{ \mathrm{DK} }  \mathit{x}  \Rightarrow  \ottnt{A}
    }

    \inferrule{
      \Psi  \vdash_{ \mathrm{DK} }  \ottnt{e}  \Leftarrow  \ottnt{A}
    }{
      \Psi  \vdash_{ \mathrm{DK} }  \ottnt{e}  \ottsym{:}  \ottnt{A}  \Rightarrow  \ottnt{A}
    }

    \inferrule{
      \vdash  \Psi
    }{
      \Psi  \vdash_{ \mathrm{DK} }  \ottsym{()}  \Rightarrow   \mathsf{unit} 
    }

    \inferrule{
      \vdash  \Psi
    }{
      \Psi  \vdash_{ \mathrm{DK} }  \ottsym{()}  \Leftarrow   \mathsf{unit} 
    }

    \inferrule{
      \Psi  \ottsym{,}  \ottmv{a}  \vdash_{ \mathrm{DK} }  \ottnt{e}  \Leftarrow  \ottnt{A}
    }{
      \Psi  \vdash_{ \mathrm{DK} }  \ottnt{e}  \Leftarrow   \forall  \ottmv{a}  .  \ottnt{A} 
    }

    \inferrule{
      \Psi  \ottsym{,}  \mathit{x}  \ottsym{:}  \ottnt{A}  \vdash_{ \mathrm{DK} }  \ottnt{e}  \Leftarrow  \ottnt{B}
    }{
      \Psi  \vdash_{ \mathrm{DK} }   \lambda  \mathit{x}  .  \ottnt{e}   \Leftarrow   \ottnt{A}   \rightarrow   \ottnt{B} 
    }

    \inferrule{
    \Psi  \vdash  \tau  \rightarrow  \sigma \\ \Psi  \ottsym{,}  \mathit{x}  \ottsym{:}  \tau  \vdash_{ \mathrm{DK} }  \ottnt{e}  \Leftarrow  \sigma
    }{
      \Psi  \vdash_{ \mathrm{DK} }   \lambda  \mathit{x}  .  \ottnt{e}   \Rightarrow   \tau   \rightarrow   \sigma 
    }

    \inferrule{
    \Psi  \vdash_{ \mathrm{DK} }  \ottnt{e_{{\mathrm{1}}}}  \Rightarrow  \ottnt{A} \\  \Psi   \vdash_{ \mathrm{DK} }   \ottnt{A}  \bullet  \ottnt{e_{{\mathrm{2}}}}  \Rightarrow\!\!\!\Rightarrow  \ottnt{B} 
    }{
      \Psi  \vdash_{ \mathrm{DK} }  \ottnt{e_{{\mathrm{1}}}} \, \ottnt{e_{{\mathrm{2}}}}  \Rightarrow  \ottnt{B}
    }

    \inferrule{
    \Psi  \vdash_{ \mathrm{DK} }  \ottnt{e}  \Rightarrow  \ottnt{A} \\ \Psi  \vdash_{ \mathrm{DK} }  \ottnt{A}  \mathbin{<:}  \ottnt{B}
    }{
      \Psi  \vdash_{ \mathrm{DK} }  \ottnt{e}  \Leftarrow  \ottnt{B}
    }

    \inferrule{
    \Psi  \vdash  \tau \\  \Psi   \vdash_{ \mathrm{DK} }    [  \tau  /  \ottmv{a}  ]  \ottnt{A}   \bullet  \ottnt{e}  \Rightarrow\!\!\!\Rightarrow  \ottnt{B} 
    }{
       \Psi   \vdash_{ \mathrm{DK} }    \forall  \ottmv{a}  .  \ottnt{A}   \bullet  \ottnt{e}  \Rightarrow\!\!\!\Rightarrow  \ottnt{B} 
    }

    \inferrule{
    \Psi  \vdash_{ \mathrm{DK} }  \ottnt{e}  \Leftarrow  \ottnt{A} \\ \Psi  \vdash  \ottnt{B}
    }{
       \Psi   \vdash_{ \mathrm{DK} }    \ottnt{A}   \rightarrow   \ottnt{B}   \bullet  \ottnt{e}  \Rightarrow\!\!\!\Rightarrow  \ottnt{B} 
    }
  \end{mathpar}
  \caption{Typing of the DK system}
\end{figure}

\begin{figure}[H]
  \small
  \noindent\fbox{$ |  \ottnt{e}  |_{ \mathrm{DK} } $}\hfill\mbox{}
  \[
    \begin{array}{r@{\ }c@{\ }lr@{\ }c@{\ }lr@{\ }c@{\ }l}
                  |  \ottsym{()}  |_{ \mathrm{DK} }  & = & \ottsym{()}                &
                  |  \mathit{x}  |_{ \mathrm{DK} }                                  & = & \mathit{x}                 &
                  |   \lambda  \mathit{x}  .  \ottnt{e}   |_{ \mathrm{DK} }                               & = &  \lambda  \mathit{x}  .   |  \ottnt{e}  |_{ \mathrm{DK} }              \\
                  |  \ottnt{e_{{\mathrm{1}}}} \, \ottnt{e_{{\mathrm{2}}}}  |_{ \mathrm{DK} }                              & = &  |  \ottnt{e_{{\mathrm{1}}}}  |_{ \mathrm{DK} }   \appright   |  \ottnt{e_{{\mathrm{2}}}}  |_{ \mathrm{DK} }   &
                  |  \ottnt{e}  \ottsym{:}  \ottnt{A}  |_{ \mathrm{DK} }                              & = & \ottsym{(}   \lambda  \mathit{x}  \ottsym{:}  \ottnt{A}  .  \mathit{x}   \ottsym{)}  \appright   |  \ottnt{e}  |_{ \mathrm{DK} }  &
                                                           &   &
    \end{array}
  \]
  \caption{Translation from the DK system to {\lang}}
\end{figure}
}

\subsection{The XO System~\cite{xie_let_2018}}

\deftitle{Subtyping Rules}{
  \fbox{$\Psi  \vdash_{ \mathrm{XO} }  \ottnt{A}  \mathbin{<:}  \ottnt{B}$}\quad
  \fbox{$\Psi  \mid  \Xi  \vdash_{ \mathrm{XO} }  \ottnt{A}  \mathbin{<:}  \ottnt{B}$}
}

\begin{mathparpagebreakable}
  \small
  \inferrule{
    \Psi  \vdash   \mathsf{unit} 
  }{
    \Psi  \vdash_{ \mathrm{XO} }   \mathsf{unit}   \mathbin{<:}   \mathsf{unit} 
  }

  \inferrule{
    \Psi  \vdash  \ottmv{a}
  }{
    \Psi  \vdash_{ \mathrm{XO} }  \ottmv{a}  \mathbin{<:}  \ottmv{a}
  }

  \inferrule{
  \Psi  \vdash_{ \mathrm{XO} }  \ottnt{A_{{\mathrm{2}}}}  \mathbin{<:}  \ottnt{A_{{\mathrm{1}}}} \\ \Psi  \vdash_{ \mathrm{XO} }  \ottnt{B_{{\mathrm{1}}}}  \mathbin{<:}  \ottnt{B_{{\mathrm{2}}}}
  }{
    \Psi  \vdash_{ \mathrm{XO} }   \ottnt{A_{{\mathrm{1}}}}   \rightarrow   \ottnt{B_{{\mathrm{1}}}}   \mathbin{<:}   \ottnt{A_{{\mathrm{2}}}}   \rightarrow   \ottnt{B_{{\mathrm{2}}}} 
  }

  \inferrule{
  \Psi  \vdash  \tau \\ \Psi  \vdash_{ \mathrm{XO} }   [  \tau  /  \ottmv{a}  ]  \ottnt{A}   \mathbin{<:}  \ottnt{B}
  }{
    \Psi  \vdash_{ \mathrm{XO} }   \forall  \ottmv{a}  .  \ottnt{A}   \mathbin{<:}  \ottnt{B}
  }

  \inferrule{
    \Psi  \ottsym{,}  \ottmv{a}  \vdash_{ \mathrm{XO} }  \ottnt{A}  \mathbin{<:}  \ottnt{B}
  }{
    \Psi  \vdash_{ \mathrm{XO} }  \ottnt{A}  \mathbin{<:}   \forall  \ottmv{a}  .  \ottnt{B} 
  }

  \inferrule{
    \Psi  \vdash  \ottnt{A}
  }{
    \Psi  \mid   \emptyset   \vdash_{ \mathrm{XO} }  \ottnt{A}  \mathbin{<:}  \ottnt{A}
  }

  \inferrule{
  \Psi  \vdash  \tau \\ \Psi  \mid  \Xi  \ottsym{,}  \ottnt{B'}  \vdash_{ \mathrm{XO} }   [  \tau  /  \ottmv{a}  ]  \ottnt{A}   \mathbin{<:}  \ottnt{B}
  }{
    \Psi  \mid  \Xi  \ottsym{,}  \ottnt{B'}  \vdash_{ \mathrm{XO} }   \forall  \ottmv{a}  .  \ottnt{A}   \mathbin{<:}  \ottnt{B}
  }

  \inferrule{
  \Psi  \vdash_{ \mathrm{XO} }  \ottnt{A}  \mathbin{<:}  \ottnt{A_{{\mathrm{1}}}} \\ \Psi  \mid  \Xi  \vdash_{ \mathrm{XO} }  \ottnt{B_{{\mathrm{1}}}}  \mathbin{<:}  \ottnt{B_{{\mathrm{2}}}}
  }{
    \Psi  \mid  \Xi  \ottsym{,}  \ottnt{A}  \vdash_{ \mathrm{XO} }   \ottnt{A_{{\mathrm{1}}}}   \rightarrow   \ottnt{B_{{\mathrm{1}}}}   \mathbin{<:}   \ottnt{A}   \rightarrow   \ottnt{B_{{\mathrm{2}}}} 
  }
\end{mathparpagebreakable}

\deftitle{Typing}{\fbox{$\Psi  \mid  \Xi  \vdash_{ \mathrm{XO} }  \ottnt{e}  \Rightarrow  \ottnt{A}$}}

\begin{mathparpagebreakable}
  \small
  \inferrule{
    \vdash  \Psi
  }{
    \Psi  \mid   \emptyset   \vdash_{ \mathrm{XO} }  \ottsym{()}  \Rightarrow   \mathsf{unit} 
  }

  \inferrule{
   \mathit{x}  \ottsym{:}  \ottnt{A}  \in  \Psi  \\ \Psi  \mid  \Xi  \vdash_{ \mathrm{XO} }  \ottnt{A}  \mathbin{<:}  \ottnt{B}
  }{
    \Psi  \mid  \Xi  \vdash_{ \mathrm{XO} }  \mathit{x}  \Rightarrow  \ottnt{B}
  }

  \inferrule{
    \Psi  \ottsym{,}  \mathit{x}  \ottsym{:}  \tau  \mid   \emptyset   \vdash_{ \mathrm{XO} }  \ottnt{e}  \Rightarrow  \ottnt{B}
  }{
    \Psi  \mid   \emptyset   \vdash_{ \mathrm{XO} }   \lambda  \mathit{x}  .  \ottnt{e}   \Rightarrow   \tau   \rightarrow   \ottnt{B} 
  }

  \inferrule{
    \Psi  \ottsym{,}  \mathit{x}  \ottsym{:}  \ottnt{A}  \mid  \Xi  \vdash_{ \mathrm{XO} }  \ottnt{e}  \Rightarrow  \ottnt{B}
  }{
    \Psi  \mid  \Xi  \ottsym{,}  \ottnt{A}  \vdash_{ \mathrm{XO} }   \lambda  \mathit{x}  .  \ottnt{e}   \Rightarrow   \ottnt{A}   \rightarrow   \ottnt{B} 
  }

  \inferrule{
    \Psi  \ottsym{,}  \mathit{x}  \ottsym{:}  \ottnt{A}  \mid   \emptyset   \vdash_{ \mathrm{XO} }  \ottnt{e}  \Rightarrow  \ottnt{B}
  }{
    \Psi  \mid   \emptyset   \vdash_{ \mathrm{XO} }   \lambda  \mathit{x}  \ottsym{:}  \ottnt{A}  .  \ottnt{e}   \Rightarrow   \ottnt{A}   \rightarrow   \ottnt{B} 
  }

  \inferrule{
  \Psi  \vdash_{ \mathrm{XO} }  \ottnt{A'}  \mathbin{<:}  \ottnt{A} \\ \Psi  \ottsym{,}  \mathit{x}  \ottsym{:}  \ottnt{A}  \mid  \Xi  \vdash_{ \mathrm{XO} }  \ottnt{e}  \Rightarrow  \ottnt{B}
  }{
    \Psi  \mid  \Xi  \ottsym{,}  \ottnt{A'}  \vdash_{ \mathrm{XO} }   \lambda  \mathit{x}  \ottsym{:}  \ottnt{A}  .  \ottnt{e}   \Rightarrow   \ottnt{A'}   \rightarrow   \ottnt{B} 
  }

  \inferrule{
  \Psi  \mid   \emptyset   \vdash_{ \mathrm{XO} }  \ottnt{e_{{\mathrm{2}}}}  \Rightarrow  \ottnt{A} \\ \Psi  \mid  \Xi  \ottsym{,}  \ottnt{A}  \vdash_{ \mathrm{XO} }  \ottnt{e_{{\mathrm{1}}}}  \Rightarrow   \ottnt{A}   \rightarrow   \ottnt{B} 
  }{
    \Psi  \mid  \Xi  \vdash_{ \mathrm{XO} }  \ottnt{e_{{\mathrm{1}}}} \, \ottnt{e_{{\mathrm{2}}}}  \Rightarrow  \ottnt{B}
  }
\end{mathparpagebreakable}

\deftitle{Translation from the XO system to {\lang}}{\fbox{$ |  \ottnt{e}  |_{ \mathrm{XO} } $}}

{
  \small
  \[
    \begin{array}{r@{\ }c@{\ }lr@{\ }c@{\ }lr@{\ }c@{\ }l}
                  |  \ottsym{()}  |_{ \mathrm{XO} }  & = & \ottsym{()}               &
                  |  \mathit{x}  |_{ \mathrm{XO} }                                  & = & \mathit{x}                &
                                                           &   &                        \\
                  |   \lambda  \mathit{x}  .  \ottnt{e}   |_{ \mathrm{XO} }                               & = &  \lambda  \mathit{x}  .   |  \ottnt{e}  |_{ \mathrm{XO} }           &
                  |   \lambda  \mathit{x}  \ottsym{:}  \ottnt{A}  .  \ottnt{e}   |_{ \mathrm{XO} }                             & = &  \lambda  \mathit{x}  \ottsym{:}  \ottnt{A}  .   |  \ottnt{e}  |_{ \mathrm{XO} }         &
                  |  \ottnt{e_{{\mathrm{1}}}} \, \ottnt{e_{{\mathrm{2}}}}  |_{ \mathrm{XO} }                              & = &  |  \ottnt{e_{{\mathrm{1}}}}  |_{ \mathrm{XO} }   \appleft   |  \ottnt{e_{{\mathrm{2}}}}  |_{ \mathrm{XO} } 
    \end{array}
  \]
}

\mycomment{
\begin{figure}[H]
  \small
  \noindent\fbox{$\Psi  \vdash_{ \mathrm{XO} }  \ottnt{A}  \mathbin{<:}  \ottnt{B}$}\hfill\mbox{}
  \begin{mathpar}
    \inferrule{
      \Psi  \vdash   \mathsf{unit} 
    }{
      \Psi  \vdash_{ \mathrm{XO} }   \mathsf{unit}   \mathbin{<:}   \mathsf{unit} 
    }

    \inferrule{
      \Psi  \vdash  \ottmv{a}
    }{
      \Psi  \vdash_{ \mathrm{XO} }  \ottmv{a}  \mathbin{<:}  \ottmv{a}
    }

    \inferrule{
    \Psi  \vdash_{ \mathrm{XO} }  \ottnt{A_{{\mathrm{2}}}}  \mathbin{<:}  \ottnt{A_{{\mathrm{1}}}} \\ \Psi  \vdash_{ \mathrm{XO} }  \ottnt{B_{{\mathrm{1}}}}  \mathbin{<:}  \ottnt{B_{{\mathrm{2}}}}
    }{
      \Psi  \vdash_{ \mathrm{XO} }   \ottnt{A_{{\mathrm{1}}}}   \rightarrow   \ottnt{B_{{\mathrm{1}}}}   \mathbin{<:}   \ottnt{A_{{\mathrm{2}}}}   \rightarrow   \ottnt{B_{{\mathrm{2}}}} 
    }

    \inferrule{
    \Psi  \vdash  \tau \\ \Psi  \vdash_{ \mathrm{XO} }   [  \tau  /  \ottmv{a}  ]  \ottnt{A}   \mathbin{<:}  \ottnt{B}
    }{
      \Psi  \vdash_{ \mathrm{XO} }   \forall  \ottmv{a}  .  \ottnt{A}   \mathbin{<:}  \ottnt{B}
    }

    \inferrule{
      \Psi  \ottsym{,}  \ottmv{a}  \vdash_{ \mathrm{XO} }  \ottnt{A}  \mathbin{<:}  \ottnt{B}
    }{
      \Psi  \vdash_{ \mathrm{XO} }  \ottnt{A}  \mathbin{<:}   \forall  \ottmv{a}  .  \ottnt{B} 
    }
  \end{mathpar}
  \caption{Subtyping of the XO system}
\end{figure}

\begin{figure}[H]
  \small
  \noindent\fbox{$\Psi  \mid  \Xi  \vdash_{ \mathrm{XO} }  \ottnt{A}  \mathbin{<:}  \ottnt{B}$}\hfill\mbox{}
  \begin{mathpar}
    \inferrule{
      \Psi  \vdash  \ottnt{A}
    }{
      \Psi  \mid   \emptyset   \vdash_{ \mathrm{XO} }  \ottnt{A}  \mathbin{<:}  \ottnt{A}
    }

    \inferrule{
    \Psi  \vdash  \tau \\ \Psi  \mid  \Xi  \ottsym{,}  \ottnt{B'}  \vdash_{ \mathrm{XO} }   [  \tau  /  \ottmv{a}  ]  \ottnt{A}   \mathbin{<:}  \ottnt{B}
    }{
      \Psi  \mid  \Xi  \ottsym{,}  \ottnt{B'}  \vdash_{ \mathrm{XO} }   \forall  \ottmv{a}  .  \ottnt{A}   \mathbin{<:}  \ottnt{B}
    }

    \inferrule{
    \Psi  \vdash_{ \mathrm{XO} }  \ottnt{A}  \mathbin{<:}  \ottnt{A_{{\mathrm{1}}}} \\ \Psi  \mid  \Xi  \vdash_{ \mathrm{XO} }  \ottnt{B_{{\mathrm{1}}}}  \mathbin{<:}  \ottnt{B_{{\mathrm{2}}}}
    }{
      \Psi  \mid  \Xi  \ottsym{,}  \ottnt{A}  \vdash_{ \mathrm{XO} }   \ottnt{A_{{\mathrm{1}}}}   \rightarrow   \ottnt{B_{{\mathrm{1}}}}   \mathbin{<:}   \ottnt{A}   \rightarrow   \ottnt{B_{{\mathrm{2}}}} 
    }
  \end{mathpar}
  \caption{Subtyping with application contexts}
\end{figure}

\begin{figure}[H]
  \small
  \noindent\fbox{$\Psi  \mid  \Xi  \vdash_{ \mathrm{XO} }  \ottnt{e}  \Rightarrow  \ottnt{A}$}\hfill\mbox{}
  \begin{mathpar}
    \inferrule{
      \vdash  \Psi
    }{
      \Psi  \mid   \emptyset   \vdash_{ \mathrm{XO} }  \ottsym{()}  \Rightarrow   \mathsf{unit} 
    }

    \inferrule{
     \mathit{x}  \ottsym{:}  \ottnt{A}  \in  \Psi  \\ \Psi  \mid  \Xi  \vdash_{ \mathrm{XO} }  \ottnt{A}  \mathbin{<:}  \ottnt{B}
    }{
      \Psi  \mid  \Xi  \vdash_{ \mathrm{XO} }  \mathit{x}  \Rightarrow  \ottnt{B}
    }

    \inferrule{
      \Psi  \ottsym{,}  \mathit{x}  \ottsym{:}  \tau  \mid   \emptyset   \vdash_{ \mathrm{XO} }  \ottnt{e}  \Rightarrow  \ottnt{B}
    }{
      \Psi  \mid   \emptyset   \vdash_{ \mathrm{XO} }   \lambda  \mathit{x}  .  \ottnt{e}   \Rightarrow   \tau   \rightarrow   \ottnt{B} 
    }

    \inferrule{
      \Psi  \ottsym{,}  \mathit{x}  \ottsym{:}  \ottnt{A}  \mid  \Xi  \vdash_{ \mathrm{XO} }  \ottnt{e}  \Rightarrow  \ottnt{B}
    }{
      \Psi  \mid  \Xi  \ottsym{,}  \ottnt{A}  \vdash_{ \mathrm{XO} }   \lambda  \mathit{x}  .  \ottnt{e}   \Rightarrow   \ottnt{A}   \rightarrow   \ottnt{B} 
    }

    \inferrule{
      \Psi  \ottsym{,}  \mathit{x}  \ottsym{:}  \ottnt{A}  \mid   \emptyset   \vdash_{ \mathrm{XO} }  \ottnt{e}  \Rightarrow  \ottnt{B}
    }{
      \Psi  \mid   \emptyset   \vdash_{ \mathrm{XO} }   \lambda  \mathit{x}  \ottsym{:}  \ottnt{A}  .  \ottnt{e}   \Rightarrow   \ottnt{A}   \rightarrow   \ottnt{B} 
    }

    \inferrule{
    \Psi  \vdash_{ \mathrm{XO} }  \ottnt{A'}  \mathbin{<:}  \ottnt{A} \\ \Psi  \ottsym{,}  \mathit{x}  \ottsym{:}  \ottnt{A}  \mid  \Xi  \vdash_{ \mathrm{XO} }  \ottnt{e}  \Rightarrow  \ottnt{B}
    }{
      \Psi  \mid  \Xi  \ottsym{,}  \ottnt{A'}  \vdash_{ \mathrm{XO} }   \lambda  \mathit{x}  \ottsym{:}  \ottnt{A}  .  \ottnt{e}   \Rightarrow   \ottnt{A'}   \rightarrow   \ottnt{B} 
    }

    \inferrule{
    \Psi  \mid   \emptyset   \vdash_{ \mathrm{XO} }  \ottnt{e_{{\mathrm{2}}}}  \Rightarrow  \ottnt{A} \\ \Psi  \mid  \Xi  \ottsym{,}  \ottnt{A}  \vdash_{ \mathrm{XO} }  \ottnt{e_{{\mathrm{1}}}}  \Rightarrow   \ottnt{A}   \rightarrow   \ottnt{B} 
    }{
      \Psi  \mid  \Xi  \vdash_{ \mathrm{XO} }  \ottnt{e_{{\mathrm{1}}}} \, \ottnt{e_{{\mathrm{2}}}}  \Rightarrow  \ottnt{B}
    }
  \end{mathpar}
  \caption{Typing of the XO system}
\end{figure}

\begin{figure}[H]
  \small
  \noindent\fbox{$ |  \ottnt{e}  |_{ \mathrm{XO} } $}\hfill\mbox{}
  \[
    \begin{array}{r@{\ }c@{\ }lr@{\ }c@{\ }lr@{\ }c@{\ }l}
                  |  \ottsym{()}  |_{ \mathrm{XO} }  & = & \ottsym{()}               &
                  |  \mathit{x}  |_{ \mathrm{XO} }                                  & = & \mathit{x}                &
                                                           &   &                        \\
                  |   \lambda  \mathit{x}  .  \ottnt{e}   |_{ \mathrm{XO} }                               & = &  \lambda  \mathit{x}  .   |  \ottnt{e}  |_{ \mathrm{XO} }           &
                  |   \lambda  \mathit{x}  \ottsym{:}  \ottnt{A}  .  \ottnt{e}   |_{ \mathrm{XO} }                             & = &  \lambda  \mathit{x}  \ottsym{:}  \ottnt{A}  .   |  \ottnt{e}  |_{ \mathrm{XO} }         &
                  |  \ottnt{e_{{\mathrm{1}}}} \, \ottnt{e_{{\mathrm{2}}}}  |_{ \mathrm{XO} }                              & = &  |  \ottnt{e_{{\mathrm{1}}}}  |_{ \mathrm{XO} }   \appleft   |  \ottnt{e_{{\mathrm{2}}}}  |_{ \mathrm{XO} } 
    \end{array}
  \]
  \caption{Translation from the XO system to {\lang}}
\end{figure}
}
}{\relax}

\end{document}